%% file: PhD_thesis_ArXiv.tex
\documentclass[a4paper,11pt,times,numbered,print,index]{Classes/PhDThesisPSnPDF}

\usepackage{amsmath}
\usepackage{amssymb}
\usepackage{xcolor}
\usepackage{float}
\usepackage{enumerate}
\usepackage{slashed}

\usepackage{mathrsfs}
\usepackage{amsthm}
\usepackage{amstext}
\usepackage{graphicx}

\usepackage[rightcaption]{sidecap}
\usepackage{caption}
\usepackage{subcaption}

\makeatletter

\theoremstyle{plain}
\newtheorem{thm}{Theorem}[section]
\newtheorem{srule}[thm]{Selection rule}

\theoremstyle{definition} 
\newtheorem{defn}{Definition}[chapter] 

\theoremstyle{remark}

\newcommand{\bea}{\begin{eqnarray}}
\newcommand{\eea}{\end{eqnarray}}
\newcommand{\bit}{\begin{itemize}}
\newcommand{\eit}{\end{itemize}}

\def\nl{\nonumber \\}

\def\sgn{{\rm  sgn}}

\def\a{\alpha}

\def\b{\beta}

\def\l{\lambda}

\def\s{\sigma}

\def\p{\partial}

\def\le{\left(}
\def\ri{\right)}

\def\beq{\begin{equation}}
\def\eeq{\end{equation}}

\def\arr{{\rightarrow}}

\def\th {\tilde{h}}

\input{thesis-info}

\ifdefineAbstract
\fi

\ifdefineChapter
\fi

\begin{document}

\frontmatter

\maketitle


\begin{declaration}

This dissertation is a result of my own efforts. The work to which it refers is based on my PhD research projects:

\begin{enumerate}
 \item
 ``Trace anomaly for non-relativistic fermions,''
 with R.~Auzzi and G.~Nardelli,
  JHEP {\bf 1708} (2017) 042
  \href{https://arxiv.org/abs/1705.02229}{[arXiv:1705.02229 [hep-th]]}.
 \item
   ``Nonrelativistic trace and diffeomorphism anomalies in particle number background,''
   with R.~Auzzi and G.~Nardelli,
  Phys.\ Rev.\ D {\bf 97} (2018) no.8,  085010
  \href{https://arxiv.org/abs/1711.00910}{[arXiv:1711.00910 [hep-th]]}.
 \item
  ``Volume and complexity for warped AdS black holes,''
  with R.~Auzzi and G.~Nardelli,
  JHEP {\bf 1806} (2018) 063
  \href{https://arxiv.org/abs/1804.07521}{[arXiv:1804.07521 [hep-th]]}.
 \item
  ``Complexity and action for warped AdS black holes,''
  with R.~Auzzi, G.~Nardelli and N.~Zenoni,
  JHEP {\bf 1809} (2018) 013
  \href{https://arxiv.org/abs/1806.06216}{[arXiv:1806.06216 [hep-th]]}.
 \item
   ``Renormalization properties of a Galilean Wess-Zumino model,''
   with R.~Auzzi, G.~Nardelli and S.~Penati,
  JHEP {\bf 1906} (2019) 048
  \href{https://arxiv.org/abs/1904.08404}{[arXiv:1904.08404 [hep-th]]}.
  \item
   ``Subsystem complexity in warped AdS,''
   with R.~Auzzi, A.~Mitra, G.~Nardelli and N.~Zenoni,
  JHEP {\bf 1909} (2019) 114
 \href{https://arxiv.org/abs/1906.09345}{[arXiv:1906.09345 [hep-th]]}.
  \item
  ``On subregion action complexity in AdS$_3$ and in the BTZ black hole,''
  with R.~Auzzi, A.~Legramandi, G.~Nardelli, P.~Roy and N.~Zenoni,
  \href{https://arxiv.org/abs/1910.00526}{[arXiv:1910.00526 [hep-th]]}.
\end{enumerate}
I hereby declare that except where specific reference is made to the work of 
others, the contents of this dissertation are original and have not been 
submitted in whole or in part for consideration for any other degree or 
qualification in this, or any other university. 


\end{declaration}

\begin{acknowledgements}

There are a lot of people I need to thank a lot for the continuous support during the PhD and whose help was determinant to finish this work.
First of all, the biggest acknowledgments go to my supervisors\footnote{Someone was official, someone not, but this is irrelevant.} Roberto Auzzi, Giuseppe Nardelli and Silvia Penati, who explained me a lot of things and supported me during all these years.
The experience I lived in these years surely helped me to grow up professionally and, more importantly, as a person, also due to the various schools, workshops, trips where I met a lot of new people and I learnt new things.

In this sense, I thank for the professional and daily experiences all the professors and postdocs in Milan-Bicocca, \emph{i.e.} Alessandro Tomasiello, Alberto Zaffaroni, Sara Pasquetti, Noppadol Mekareeya, Carlo Oleari, Simone Alioli, Antonio Amariti,  Valentin Reys, Francesco Aprile, Paul Richmond, Yegor Zenkevich, Kate Eckerle, Vladimir Bashmakov \dots
Moreover, I thank a lot all the people working in Università Cattolica del Sacro Cuore di Brescia, where I studied during the bachelor and master degrees and who allowed me to work in the institute in a joint program with the Bicocca university: in particular I thank the professors Fausto Borgonovi, Alfredo Marzocchi, Marco Marzocchi, Silvia Pianta, Mauro Spera, Dario Mazzoleni, Stefania Pagliara, Marco Degiovanni, Claudio Giannetti \dots 

I also greatly thank Niels Obers and Shira Chapman, who accepted to evaluate the thesis and gave me a lot of valuable comments and insights, and the commitee coming for the oral exam, composed by Peter Horvathy and Alberto Lerda, both of them have been very nice with me.

To continue the acknowledgments, I find more convenient to switch now to italian.

I più grandi ringraziamenti vanno alle persone con cui ho passato più tempo, allietando incredibilmente le mie giornate e anche moltissime serate.
Parto dagli amici in Bicocca: grazie intanto ai miei coetanei con cui siamo entrati massivamente nello stesso anno del PhD: Andrea Legramandi, Giuseppe Bruno de Luca, Gabriele Lo Monaco, Carolina Gomez.
Grazie a chi si è aggiunto l'anno dopo, Ivan Garozzo e Marco Rocco, e poi due anni dopo, Matteo Sacchi, Lorenzo Coccia, Nicola Gorini e Simone (notevoli i memes prodotti per la temperatura nel nostro ufficio quest'estate).
Già presenti dagli anni prima, ci sono Morteza Seyed Hosseini, Luca Cassia, Silvia Ferrario Ravasio, il dottor Azzurli, Erika Colombo, Michele Turelli e Anton Nedelin (che era già postdoc, ma ha passato moltissimo tempo con noi).
Grazie a tutti voi\footnote{Ho sorvolato sui soprannomi, ma alcuni meritevoli sono EP, Cicci, cumpà, picciò \dots L'intepretazione della prima sigla è tuttora oggetto di ampi dibattiti.}! 
Rimarranno sempre tra i miei ricordi migliori i pranzi insieme, e soprattutto le serate cinema o giochi (Mario Kart o altri giochi retrò, anche se purtroppo abbiamo fatto solo una sessione \dots), le discussioni su Pokémon e su altri giochi\footnote{Menzione d'onore per i giochi del genere soulsborne e per le discussioni sullo splendido lavoro e sulle gaffe di Sabaku no Maiku.}, la costruzione della cucina e del \emph{trio} nel nostro ufficio, il pane tostato con la Nutella\footnote{Quanto mi manca! Ahah.} e il burro d'arachidi, la piantina di menta seguita da quella di zenzero, la visione a pranzo delle puntate di Death Note (per me era tipo la sesta volta!).
A questi si aggiungono i viaggi insieme alle varie scuole di dottorato, dove abbiamo alloggiato insieme e vissuto grandi e divertenti esperienze, tra cui potrei citare episodi come "\emph{Ma quell'orologio ha due lancette!}", lo streatching durante la pausa di una certa lezione, il test di dialetto bresciano che ho dovuto affrontare, la Nintendo Wii in appartamento a Trieste con il controller senza batterie, il viaggio su Flixbus giocando a Pokémon Showdown in competitivo per 5-6 ore, le partitelle a basket\footnote{A quando la partita a calcetto?} e tanto altro.
Tra gli studenti di PhD con cui ho condiviso momenti divertenti tra le varie scuole al GGI o all'ICTP ci sono Carlo Heissenberg, Sara Bonansea, Stefano Speziali, Luca Buoninfante, Alberto Merlano, Riccardo Conti, Luca Ciambelli, Lorenzo Menculini, Paolo Soresina, Paolo Milan \dots

Proseguendo cronologicamente a ritroso con i miei amici, troviamo i colleghi del DMF a Brescia: Tommaso Tosi, Giulia Zani, Silvia Bianchetti, Michele Zubani, Sara Zanini, Ilario Medaglia, Debora Coltrini, Eugenio Guarneri e Francesco Filippini\footnote{Scientificamente, menzione d'onore per gli ultimi due: abbiamo collaborato intensamente durante la tesi triennale e magistrale, rispettivamente.}.
Con loro posso ricordare tante serate di giochi da tavolo, da Maschera (uno dei miei preferiti, adoro troppo il bluff e gli azzardi sensati nei giochi da tavolo) a Seven Wonders, da Bang! a Citadels \dots
A questo si aggiungono i tanti compleanni festeggiati insieme, il weekend a Firenze durante il ponte del primo maggio qualche anno fa, i dibattiti sulle mie battute e il loro valore, gli episodi divertenti in Università (per chi può intendere, credo che negli ultimi mesi abbiamo dormito tutti sonni molto tranquilli), le varie camminate in montagna che ho saltato, le partitelle fisici contro matematici che avevamo fatto con clamorosi risultati, la tintura imperiale \dots
Tra le tante serate insieme, meritano sicuramente una menzione particolare Nick Fiini, Chiara Co', Gabriele Calzavara, Melania Fava, Pippo, e in dipartimento o altrove anche Mauro, Fede Armanti, Steve e poi Sara Rizzini, Mattia Angeli, Paolo Stornati, Paolo Franceschini, Andrea Tognazzi \dots
Un grazie per le giornate insieme ai miei compagni di ufficio a Brescia, Silvia Pagani, Sara Mastaglio, Antonio Miti, e anche a Chahan e Sonia Freddi.
In particolare, un grande ringraziamento va a Nicolò Zenoni, co-autore di vari paper e praticamente compagno di ufficio, con cui ci sono state varie discussioni scientifiche e riguardanti anime come Dragon Ball, Naruto, partite di tennis e altro. 

Continuando indietro nel tempo ci sono gli amici del liceo, tra cui merita una menzione particolare Lavneet Banger detto Bunch, con i suoi straordinari e divertenti racconti di episodi realmente accaduti nella sua permanenza in Canada, che mi hanno notevolmente rallegrato.
E qui la lista arriva ai membri della Fellowship, che seppure abbia iniziato a diffondersi per il mondo, resta saldamente un punto di riferimento: Fede, Mix, Ken, Rocco (e le loro sorelle ahah), Paolo, Riccardo (da Ric, aggiungerei per chi può intendere), Marci, Sam.
Anche qui gli episodi esaltanti sono davvero molti (a malapena numerabili): citerei sicuramente Hulk, partite a Tekken, le grondaie, i \emph{tutti tutti}, i nobili tornei degli scacchi, le partite a Risiko, Monopoli, cooperative al Signore degli Anelli\footnote{Qui vanno necessariamente citati un "\emph{Giammai mi sacrificherò per tutti}" e il ruolo di causalità e casualità nel gioco in questione.}, alla Serenissima, a Diplomacy (molto tempo fa ormai), le coperte durante il soggiorno a Monaco di Baviera, frasi come "\emph{è tratto da un fumetto!}", le chiamate della Farnesina in Crimea, il Pedro all'eredità, le discussioni sulle partite del Milan (ahimé, sono brutti tempi), le discussioni su anime di grande qualità e i consigli su giochi come Undertale (assolutamente straordinario, Toby Fox è un genio)  \dots
Mi viene anche nostalgia ripensando alle discussioni sulle classifiche degli anime o sulle mono tech da giocare nei mazzi di Yu-gi-oh!, cose che ormai risalgono a una decina di anni fa.
Un grande ringraziamento anche a Giulia Ferrari, che ha letto e commentato il poster che ho preparato sulla partecipazione delle donne nel mondo scientifico! A ciò si aggiungono diversi consigli in tema di anime.

Infine, un ringraziamento particolare a mia mamma, che mi ha sempre supportato, sopportato e sostenuto con affetto in tutte le mie scelte (oltre a far da mangiare e tutte le altre cose tipiche delle mamme! Mi piace molto vivere da solo perché posso organizzarmi come voglio, ma trovo noioso far sempre da mangiare ahah), e a mio papà, tra le innumerevoli partite a carte e i consigli di vita.
Grazie anche a tutti gli zii, cugini e nonni!

Concludo questi lunghi ringraziamenti (sono stato prolisso anche in questa parte della tesi, yeah!) ancora una volta con un "\emph{Grazie, ragazzi!}" come direbbe Seb Vettel\footnote{A proposito di sport, come per il Milan, anche in F1 sono tempi magri \dots}, senza tutti voi, questa tesi non sarebbe stata possibile.

\end{acknowledgements}

\begin{abstract}

This thesis focuses on the investigation of two research areas: non-relativistic field theories and holographic complexity.

In the first part we review the general classification of the trace anomaly for 2+1 dimensional field theories coupled to a Newton-Cartan background and we also review the heat kernel method, which is used to study one-loop effective actions and then allows to compute anomalies for a given theory.
We apply this technique to extract the exact coefficients of the curvature terms of the trace anomaly for both a non-relativistic free scalar and a fermion, finding a relation with the conformal anomaly of the 3+1 dimensional relativistic counterpart which suggests the existence of a non-relativistic version of the \emph{a}-theorem on which we comment.  
We continue the analysis of non-relativistic free scalar and fermion with the heat kernel method by turning on a source for the particle mass: on this background, we find that there is no gravitational anomaly, but the trace anomaly is not gauge invariant.

We then consider a specific model realizing a $\mathcal{N}=2$ supersymmetric extension of the Bargmann group in 2+1 dimensions with non-vanishing superpotential, obtained by null reduction of a relativistic Wess-Zumino model.
We check that the superpotential is protected against quantum corrections as in the relativistic parent theory, thus finding a non-relativistic version of the non-renormalization theorem.
Moreover, we find strong evidence that the theory is one-loop exact, due to the causal structure of the non-relativistic propagator together with mass conservation.

In the second part of the thesis we review the holographic conjectures proposed by Susskind to describe the time-evolution of the Einstein-Rosen bridge in gravitational theories: the complexity=volume and complexity=action.
These quantities may be used as a tool to investigate dualities, and we investigate both the volume and the action for black holes living in warped $\mathrm{AdS}_3 $ spacetime, which is a non-trivial modification of usual $\mathrm{AdS}_3$ with non-relativistic boundary isometries.
In particular, we analytically compute the time dependence of complexity finding an asymptotic growth rate proportional to the product of Hawking temperature and Bekenstein-Hawking entropy.
In this context, there exist extensions of the holographic proposals when the dual state from the field theory side is mixed, \emph{i.e.} we consider only a subregion on the boundary.
We study the structure of UV divergences, the sub/super-additivity behaviour of complexity and its temperature dependence for warped black holes in 2+1 dimensions when the subregion is taken to be one of the two disconnected boundaries.
Finally, we analytically compute the subregion action complexity for a general segment on the boundary in the BTZ black hole background, finding that it is equal to the sum of a linearly divergent term proportional to the size of the subregion and of a term
proportional to the entanglement entropy.
While this result suggests a strong relation of complexity with entanglement entropy, we find after investigating the case of two disjoint segments in the BTZ background that there are additional finite contributions: 
as a consequence, mutual holographic complexity carries a different content compared to mutual information.
This means that entropy is not enough!

\end{abstract}


\tableofcontents




\printnomenclature

\mainmatter

\chapter{Introduction}

Symmetries play a crucial role in modern physics: when a law of physics does not change upon some transformation, this is said to
exhibit an invariance.
The use of symmetries allows to obtain conserved quantities which simplify the description of the system, or imposes restrictions on the way a model needs to be formulated.
An important example of a symmetry and of the consequences of its existence is Poincaré invariance, which was understood to be a fundamental symmetry of Nature after the formulation of the theory of relativity.
Translation and rotation invariances imply the existence of a symmetric and conserved energy-momentum tensor, and requiring that the symmetry holds gives restrictions on the theoretical model describing a physical system, \emph{e.g.} it constrains the action and the form of correlation functions.

A broader way in which the concept of symmetry can be applied is in the context of the Renormalization Group: it refers to an invariance of the observables under changes of the scales at which physical quantities are defined.
In this case, the independence of the theory from such an arbitrary scale implies the existence of a differential equation of the kind 
\beq
\frac{d}{d \mu} \mathcal{O} = 0 \, ,
\eeq
where $\mathcal{O}$ is a physical observable and $\mu$ is a mass scale.
The solutions of such a relation consist in trajectories in the space of Quantum Field Theories.
The application of these ideas allows to understand the asymptotic behaviour of gauge theories and to find relevant physical quantities like the critical exponents of second order phase transitions.

In this context, the search for emergent symmetries has recently acquired great relevance: a new symmetry
may arise in the infrared, even if absent from the microscopic Hamiltonian, due to the presence of an interacting infrared fixed point in the renormalization group flow.

The material contained in the Introduction to this thesis is organized as follows.
In section \ref{int-Non-relativistic trace anomalies} we will discuss the relevance of the conformal symmetry and of the corresponding quantum anomaly in relation to universal properties of the Renormalization Group flow, \emph{i.e.} the irreversibility of a trajectory in the space of Quantum Field Theories.
Moreover, we will justify why non-relativistic symmetry can play a meaningful role in the discussion, and which insights can provide in the context of a better understanding of the laws of physics. 
In section \ref{int-Non-relativistic supersymmetry} the special role played by Supersymmetry will be discussed, in particular in relation to the powerful exact results that this invariance provides, such as non-renormalization theorems.
We will be interested in studying the implementation of Supersymmetry as a graded extension of the Galilean algebra in order to analyze if similar results apply to non-relativistic models.
In section \ref{int-Holographic complexity} we will tackle a different problem related to quantum information and geometry, a nice realization of the $\mathrm{AdS/CFT}$ correspondence.
In particular, we will discuss the relation between the evolution in time of the Einstein-Rosen bridge in connection with computational complexity, and we will study this quantity for black holes in spacetimes which do not contain the Lorentz group as an isometry at the boundary, thus providing insights on the investigation of non-relativistic realizations of holography.

\section{Non-relativistic trace anomalies}
\label{int-Non-relativistic trace anomalies}

Weyl invariance gives important restrictions on relativistic field theories, implying that the classical energy-momentum tensor is traceless.
However Weyl symmetry is in general lost after quantization, and the trace of the energy-momentum tensor is non-vanishing when the system is coupled to curved backgrounds (\textit{trace anomaly})
\beq
\langle T^{\mu}_{\,\,\, \mu}    \rangle \equiv \mathcal{A} \ne 0 \, .
\eeq
It is possible to write the most general expression of the trace anomaly in \emph{d} spacetime dimensions consistent with diffeomorphism invariance and satisfying the Wess-Zumino consistency conditions
\beq
\Delta_{\sigma_{1}\sigma_{2}}^{\mathrm{WZ}} W = ( \delta_{\sigma_{1}} \delta_{\sigma_{2}} - \delta_{\sigma_{2}}  \delta_{\sigma_{1}} ) W = 0 \, ,
\label{Wess-Zumino consistency conditions}
\eeq
where $W$ is the generating functional of connected diagrams.
In particular, it is well known that in 2 dimensions
\beq
\mathcal{A}_{\mathrm{d=2}} = c R \,  ,
\eeq
where \emph{c} is the \emph{central charge} of the corresponding Conformal Field theory and it is related to the Lorentz structure of the matter fields.
In 4 dimensions, the anomaly is
\beq
\mathcal{A}_{\mathrm{d=4}} = a E_{4} - c W_{\mu \nu \rho \sigma}^{2} + \mathcal{A}_{\mathrm{ct}} \, , 
\eeq
where $ E_{4} $ and $ W_{\mu \nu \rho \sigma}^{2} $ are the Euler density and the square of the Weyl tensor, respectively. The term $ \mathcal{A}_{\mathrm{ct}}  $ refers the scheme-dependent part\footnote{More precisely, the classification of the terms entering the trace anomaly is a cohomological problem and the scheme-dependent part refers to expressions which are not only closed, but also exact under Weyl variations.}.

Trace anomalies allow to characterize in the relativistic case the irreversibility properties of the Renormalization Group.
In the case of relativistic (1+1)-dimensional theories, this is established by Zamolodchikov's \emph{c}-theorem \cite{Zamolodchikov:1986gt}: there exists a function defined in the space of Quantum Field Theories which is monotonically decreasing along a Renormalization Group trajectory and coincides at fixed points with the central charge \emph{c} of the corresponding Conformal Field Theory.

Some of these results can be extended to four dimensional theories. In particular it was conjectured in \cite{Cardy:1988cwa} that 
such monotonically decreasing function exists and coincides with the conformal anomaly coefficient $a$ at the fixed point (\emph{a}-theorem).
A perturbative proof of this conjecture based on local Renormalization Group flow equations was given by Osborn \cite{Osborn:1989td}, while a non-perturbative proof based on 't Hooft anomaly matching was given by Komargodski and Schwimmer in 2011 \cite{Komargodski:2011vj}.

The terms composing the anomaly can be divided into type A and type B terms depending from their Weyl variation\footnote{Type B anomalies are invariant under Weyl transformations, while type A are not.} \cite{Deser:1993yx}. It turns out that the coefficients of type A anomalies\footnote{In the relativistic case there exists a general procedure to show that type A anomalies must give scale-free contributions to the effective action in dimensional regularization, which in turn implies that they are related to topological invariants \cite{Deser:1993yx}.} are candidates for an \emph{a}-theorem.
This can be understood in the framework of local Renormalization Group followed in \cite{Osborn:1989td}: since type B anomalies are Weyl invariant scalars, they are trivial solutions of the Wess-Zumino consistency conditions (\ref{Wess-Zumino consistency conditions}), and then they do not give any non-trivial contraint on their coefficients.
On the other hand, type A anomalies have non-vanishing Weyl variation and then the local Renormalization Group equations following from application of (\ref{Wess-Zumino consistency conditions}) give meaningful constraints.

One can asks if the previous results are related  to the relativistic content of the theory, and if an analogue of the Weyl group exists for Galilean-invariant field theories.
At first sight, the two cases look very different.
First of all, the Klein-Gordon equation
\begin{equation}
(-\hbar^2 c^2 \square + m^{2} c^{4} ) \Phi =  0  
\end{equation}
is evidently invariant under a dilatation parametrized by a constant factor $\sigma$ as
\begin{equation}
x^{\mu}  \rightarrow   e^{2 \sigma} x^{\mu}  \, .
\label{3.0 Other conformal transformation}
\end{equation}
This behaviour is a consequence of the fact that the coordinates of Minkowski spacetime can be described in terms of a common four-vector containing both time and space.

On the other hand, the Schr\"odinger equation for a free particle is
\begin{equation}
i \hbar \frac{\partial}{\partial t} \psi (\vec{x} , t) =- \frac{\bigtriangleup}{2m} \psi (\vec{x} , t)   \,  ,
\end{equation}
which is not invariant under scale transformations (\ref{3.0 Other conformal transformation}).
Indeed, in non-relativistic theories the scaling of time and spatial coordinates must be different in order to keep the kinetic term invariant.

We can interpret the discrepancy between the two cases due to the appearance in the Klein-Gordon equation of both the speed of light and the mass of the particle, which allows to interpret the latter as an inverse length.
On the other hand, this is not true for the Schr\"odinger equation and ultimately allows the mass to not be interpreted as an inverse length. 
In this way we can rescale space and time while retaining quantities (such as the mass) which have inequivalent dimensions and no scaling properties.

This allows to consider transformations of kind 
\begin{equation}
x^{i} \rightarrow e^{\sigma} x^{i}  \,   ,   \,\,\,\,\,\,  t \rightarrow e^{z \sigma} t  \, ,
\label{Dynamical exponent}
\end{equation}
where the dynamical exponent \emph{z} parametrizes the anisotropy between space and time.
The Lifshitz case is characterized by invariance under the previous transformations for a general value of \emph{z}.
The relativistic conformal case can be recovered for $ z=1 , $ whereas by requiring symmetry under Galilean boosts we need to choose $ z=2 , $ which leads to the invariance of the Schr\"odinger equation.

In the case of Lifshitz theories, a detailed study of trace anomalies for various dimensions
and values of $z$ was  carried on in \cite{Adam:2009gq,Baggio:2011ha,Griffin:2011xs,Arav:2014goa}. 
The result does not give any reasonable candidate for a decreasing $a$-function; several anomalies are indeed
possible at the scale-invariant fixed points, but their Weyl variation vanishes identically (type B anomalies). 
An analysis like the one developed  in \cite{Osborn:1989td,Jack:1990eb,Osborn:1991gm} for relativistic  
theories would suggest that no monotonically-decreasing anomaly coefficient is present in the Lifshitz case.

A more promising arena for searching decreasing $a-$functions along a Renormalization Group flow is the Schr\"odinger case.
The algebra contains the generators $H$ for time translations and $P_i$ for spatial translations, $L_{ij}$ for spatial rotations, $K_i$ for Galilean boosts, $D$ for dilatations and $C$ for special conformal transformations, satisying the commutation relations
\bea
\nonumber
& [P_j, K_k] = i \delta_{jk} M \, , \qquad
[H, K_j] = i P_j \, ,   & \\
\nonumber
& \left[  L_{ij} , P_{k}  \right] = i \left( \delta_{ik} P_{j} -\delta_{jk} P_{i} \right) \, , \qquad
\left[  L_{ij} , K_{k}  \right] = i \left( \delta_{ik} K_{j} - \delta_{jk} K_{i} \right) \, , \\
&\left[  L_{ij} , L_{kl}  \right] = i \left( \delta_{ik} L_{jl} - \delta_{jk} L_{il} + \delta_{il} L_{kj} - \delta_{jl} L_{ki} \right) \, , \\
\nonumber
& \left[  P_{i} , D  \right] = i P_{i} \, , \qquad \left[  P_{i} , C  \right] = i K_{i} \, , \qquad
 \left[  K_{i} , D  \right] = -i K_{i} \, , \\
 \nonumber
& \left[  H , D  \right] = 2i H \, , \qquad
 \left[  H , C  \right] = i D \, , \qquad
 \left[  C , D  \right] = -2 i C \, .
 \label{bosonic part Schr\"odinger algebra in d+1 dimensions}
\eea
Physical representations of this algebra require the mass to be conserved, which is implemented via the introduction of a $U(1)$ central extension (called Bargmann algebra) with the generator $M.$

Galilean invariance is usually thought as a low-energy approximation of theories with Poincar\'e
 invariance, and as such it can be found by performing the $c \rightarrow \infty $ limit 
  in the corresponding relativistic setting\footnote{When performing this procedure, divergent expressions in the speed of light appear and we need to introduce some subtraction terms via a chemical potential and
  by appropriately rescaling the fields \cite{Jensen:2014wha}.}. 
On the other hand, it is possible to obtain the Galilean group by discrete light cone quantization, 
which consists in a dimensional reduction along a null direction of a relativistic theory living in one higher dimension \cite{Duval:1984cj}. 
A simple example where the procedure can be shown is the null reduction of the Klein-Gordon equation for a massless scalar field in $d+2$ dimensional Minkowski spacetime \cite{Son:2008ye}
\begin{equation}
\square \Phi = - \partial_{0}^{2} \Phi + \sum_{i=1}^{d+1} \partial_{i}^{2} \Phi = 0 \, ,
\end{equation}
which is conformally-invariant.

We define the light-cone coordinates
\begin{equation}
x^{\pm} = \frac{x^{d+1} \pm x^{0}}{\sqrt{2}} \, ,
\end{equation}
so that the Klein-Gordon equation becomes
\begin{equation}
\left(  2 \frac{\partial}{\partial x^{-}} \frac{\partial}{\partial x^{+}} + \sum_{i=1}^{d} \partial_{i}^{2} \right) \Phi = 0 \, . 
\end{equation}
Making the identification $ \partial / \partial x^{-}  = im $ we obtain
\begin{equation}
\left(   2im \frac{\partial}{\partial x^{+}} + \sum_{i=1}^{d} \partial_{i}^{2} \right) \Phi = 0 \, ,
\end{equation}
also written as
\begin{equation}
i \frac{\partial}{\partial x^{+}} \Phi = - \frac{1}{2m} \sum_{i=1}^{d} \partial_{i}^{2}  \Phi \, .
\end{equation}
This is the Schr\"odinger equation with the interpretation of the coordinate$\  x^{+} $\ with time.

This relation is clearly invariant under transformations of the Schr\"odinger group in $d+1$ dimensions, and it was derived from the Klein-Gordon equation, invariant under the conformal group in the enlarged $d+2$ dimensional spacetime.
This means that the Schr\"odinger group in \emph{d} spatial dimensions is a subgroup of the conformal group in $d+2$ spacetime dimensions, \emph{i.e.} $ O(d+2 , 2) . $

In nonrelativistic theories the mass spectrum is usually discrete, because there is not a direct relation with energy and the gap corresponds to the mass of the lightest particle of the system. 
A way to obtain the discreteness of the mass spectrum consists in requiring periodicity along a light-cone coordinate, so that the field $  \Phi $ can be decomposed as
\begin{equation}
\Phi (x^M) =  e^{im x^{-}} \phi (x^{\mu})  \, ,
\label{3.4 relativistic and non-relativistic fields}
\end{equation}
where $  \phi $ does not depend on the $ x^{-} $ coordinate, in fact $x^M=(x^-,x^{\mu})=(x^-,x^+,x^i) .$
This decomposition of the field allows to interpret $ \partial / \partial x^{-}  = im ,$ where $m$ is the eigenvalue of the $U(1)$ mass generator $M.$

The previous case is an explicit realization of Discrete Light Cone Quantization which shows how the Galilean-invariant case is related to the Lorentz-invariant case in one higher dimension.
In this way we understand that there is a relation between the relativistic trace anomaly in even dimensional spacetimes and the non-relativistic anomaly in odd dimensional ones\footnote{It is well known that the relativistic trace anomaly is non-vanishing only in even spacetime dimensions. This result can be found \emph{e.g.} by dimensional analysis.}.
On the other hand, the fact that many tensorial quantities in the non-relativistic framework can be found by null reduction does not imply that the relativistic results can be immediately imported from the parent theory.
In fact, the quantization of a theory does not commute in general with the non-relativistic limit, and this gives rise to meaningful results that we will investigate in this thesis.

In order to study the trace anomaly, the field theory must be coupled to a curved background whose metric acts as a source for the definition of the energy-momentum tensor.
In the relativistic case the natural candidate is a pseudo-Riemannian manifold, while the analog concept in the non-relativistic setting is the Newton-Cartan geometry, a coordinate-independent way to describe Newtonian gravity.
The main properties of this geometry will be described in chapter \ref{chapt-Non-relativistic actions}, where we will also address the problem of defining Weyl invariance in this context.

Due to the relation between Lorentz and Galilean-invariant quantities given by null reduction, the minimal non-trivial case where a Newton-Cartan trace anomaly can be investigated is $2+1$ dimensions\footnote{It can be shown that in $0+1$ dimensions there is not enough structure to obtain non-vanishing curvature invariants, while in even spacetime dimensions arguments similar to the odd-dimensional relativistic case forbid the existence of curvature invariants with the correct scaling dimension.}.
The analysis of the Newton-Cartan conformal anomaly was initiated in \cite{Jensen:2014hqa}, where an infinite number
of possible terms entering the anomaly was found. In this situation, it is difficult to figure out
the existence of an $a$-theorem, due to the infinitely many coefficients that are in principle present, and the infinite number of Wess-Zumino consistency conditions to solve.
With these premises, the natural conclusion would be that non-relativistic theories cannot admit
an $a$-theorem: either there are not type A anomalies (Lifshitz theories) or there are too many (Schr\"odinger theories).

It turns out that there is a selection rule which splits the possible
scalars entering the trace anomaly into distinct sectors, each with a finite numbers of terms \cite{Arav:2016xjc}.
The structure of the anomaly critically depends whether causal backgrounds are or not allowed. If backgrounds satisying the causality condition are considered, the possible scalars collapse to only one sector and there is just a finite number of terms in the anomaly \cite{Auzzi:2015fgg}.
However only one term with vanishing Weyl variation (type B term) survives, spoiling the possibility of an \emph{a}-theorem in this case.

On the other hand, the coupling to Newton-Cartan gravity may be seen as a formal trick to introduce sources
for the energy-momentum tensor. 
We can then decide to study non-causal backgrounds and consider each sector composing the anomaly separately.
It is also possible to study the local Renormalization Group flow equations using Wess-Zumino consistency conditions: the idea is to consider arbitrary local rescaling of the lengths via a Weyl transformation and to introduce a space-time dependence for the couplings, that act as sources for local operators \cite{Osborn:1989td}.
The result is that there is a sector which is the analogue of the 3+1 dimensional relativistic case, and the coefficient of the corresponding type A anomaly is the natural candidate for a non relativistic \emph{a}-theorem \cite{Auzzi:2016lrq}.

The analysis of the Newton-Cartan trace anomaly by means of the classification of terms satisying the dimensional requirements plus the Wess-Zumino consistency conditions gives a general expression, but does not ensure that the coefficients multiplying the curvature terms are non-vanishing.
When considering specific models, functional techniques such as the Fujikawa method can be used to determine the exact expression of the trace anomaly.
In chapter \ref{chapt-The heat kernal technique} we will face the problem with the heat kernel procedure.

From a condensed-matter perspective, there are many motivations for studying field theories with non-relativistic symmetries, in particular using descriptions in terms of a Schr\"odinger  conformal field theory \cite{Hagen:1972pd,Jackiw:1990mb,Mehen:1999nd}.
Such an example is given by fermions at unitarity in 3+1 dimensions, which interact in a fine-tuned way such that their scattering length is infinite\footnote{This kind of system can be realized experimentally, but it is very difficult to treat theoretically due to the absence of a perturbative parameter which allows to perform a series expansion. On the other hand, the power of effective field theory and the symmetry arguments coming from the Schr\"odinger invariance allow to investigate properties of the model which were not accesible before.} \cite{Son:2005rv, Nishida:2007pj}.
Another interesting class of non-relativistic conformal field theories involves anyons in 2+1 dimensions.
They play an important role in the fractional quantum Hall effect, where a theoretical treatment requires a diffeomorphism invariance for the model, then naturally leading to a coupling with torsional Newton-Cartan geometry \cite{Son:2013rqa, Geracie:2014nka}.
The technique of effective actions to analyze non-relativistic systems has become very useful in many other context: in nuclear physics \emph{e.g}. \cite{Kaplan:1998tg}, for cold atoms \cite{Nishida:2010tm}, and even for quantum mechanical problems like the Efimov effect \cite{Bedaque:1998kg, Bedaque:1998km, Braaten:2004rn}.

\section{Non-relativistic supersymmetry}
\label{int-Non-relativistic supersymmetry}

Supersymmetry is a special invariance which rotates bosonic into fermionic degrees of freedom and that has been studied for several decades, mostly from high energy physicist's perspective.
Introduced in the context of extensions to the Standard Model as a symmetry able to explain the hierarchy problem of the Higgs mass, supersymmetry gives a strong analytic control on several quantum physical quantities, which in some cases can be exactly
computed.
Indeed, when the effective action or the superpotential have a holomorphic dependence on the quantum fields and coupling constants, it is possible to get restrictions on the flow of these quantities under renormalization, leading to the non-renormalization theorem \cite{Grisaru:1979wc, Seiberg:1993vc}.

To get a feeling of the power of holomorphicity, we briefly review the original argument by Seiberg.
We consider the high-energy physics of a system at scale $ \mu_0 $ to be described by a SUSY-invariant theory with bare action $ S_{\mu_0}$.
We assume that the bare action contains a superpotential  $ W_{\mu_0}(g_i, \Phi_a)$ that depends on a set of chiral superfields $ \Phi_a $ and coupling constants $ g_i$.
The key observation is that each coupling in the Lagrangian can be interpreted as the vacuum expectation value of the scalar component of a heavy chiral superfield.

This makes manifest that the superpotential of the bare action is holomorphic not only in the chiral superfields, but also in the coupling constants.
This can also be proven in terms of a supersymmetric Ward identity. 
Therefore, the low-energy Wilsonian effective action with superpotential $ W_{\mu< \mu_0} $ must be holomorphic in the coupling constants. 

Moreover the coupling constants, viewed as the expectation value of heavy chiral superfields, spontaneously break global symmetries of the free action. Assuming that these symmetries do not acquire anomalies at quantum level, the Wilsonian effective action must be invariant under these symmetries. 
Holomorphicity and global symmetries can then be used to infer the exact expression of the effective superpotential. 

We now apply this general statement to the specific case in which the UV physics is described by the WZ model for a single massive chiral superfield $ \Phi$. Thus 
\beq
S_{\mathrm{int}}= \int d^4 x d^2 \theta \, W_{\mu_0} + \mathrm{c.c.} =
\int d^4 x d^2 \theta \, \le \frac{m}{2} \Phi^2 + \frac{\lambda}{3!} \Phi^3 \ri + \mathrm{h.c.} 
\label{Superpotential at high energy}
\eeq
where $ m, \lambda $ are promoted to background chiral superfields. 

This action is invariant under the global group $ U(1)_G \times  U(1)_R$ if we assign the following set of charges
\begin{table}[H]
\begin{center}
\begin{tabular}{|c|c|c|} \hline
Superfields  & $ U(1)_G $ & $ U(1)_R $  \\ \hline
$ \Phi $ & $  1 $  & $ 1 $ \\ \hline
$ m $ & $ -2 $  & $ 0 $ \\ \hline
$ \lambda $ & $ -3 $ & $ -1 $  \\ \hline
\end{tabular}
\end{center}
\end{table}
The $U(1)_R$ factor is the ordinary R-symmetry of $ \mathcal{N}=1 $ SUSY theories in four dimensions, under which the $ (\theta, \bar{\theta} )$ coordinates conventionally carry charge $ (-1,1) $.  Spinorial coordinates are instead neutral under $U(1)_G $.

Due to the previous discussion, and assuming global symmetries to be not anomalous at low energies, the form of the superpotential $ W_{\mu< \mu_0} $  is constrained by holomorphicity and invariance under $  U(1)_G \times  U(1)_R $ group to be of the form
\beq
W_{\mu} = m \Phi^2 \, f \le \frac{\lambda \Phi}{m} \ri 
\eeq
This is in fact the most general expression which has charge 0 under $ U(1)_G$ and charge 2 under $ U(1)_R$. 

Now, taking the Laurent expansion of $f$ we can write
\beq
W_{\mu} = m \Phi^2 \, \sum_n  a_n \le \frac{\lambda \Phi}{m} \ri^n = \sum_n a_n m^{1-n} \lambda^n \Phi^{n+2} 
\eeq
However, the holomorphic dependence of the superpotential on the couplings $m, \lambda$ requires
\beq
n \geq 0 \, , \qquad n \leq 1 \quad \Rightarrow \quad
n= 0,1 
\eeq
This fixes the superpotential at any scale to be  
\beq
W_{\mu} =  \frac{m}{2} \Phi^2 + \frac{\lambda}{3!} \Phi^3 
\eeq
The bare superpotential is quantum exact, and has not received any loop correction.
This shows how the homolorphicity of the effective action in the superfields and in the coupling constants produced a non-perturbative result with simple arguments and without any loop computation.

There are several interesting settings where supersymmetry also appears as an emergent symmetry in condensed matter systems.
For example, superconformal invariance in two dimensions arises in the tricritical
Ising model \cite{Friedan:1984rv}. Supersymmetry also appears in the description of quantum phase transitions
at the boundary of topological superconductors \cite{Grover:2013rc}, in optical lattices \cite{Yu:2010zv},
and in many other settings \cite{Huijse:2014ata,Jian:2014pca,Rahmani:2015qpa,Yu:2019opk,Lee:2006if}.
It is then a natural question to investigate
non-relativistic incarnations of supersymmetry, since this kind of invariance
might be emergent in the infrared of some real world systems.

In addition, even if supersymmetry plays an indirect role in holography, most of the explicit examples where the AdS/CFT correspondence is verified by quantitative checks are supersymmetric.
So, in order to find the precise holographic dual of a given gravity background
which geometrically realizes the Schr\"odinger symmetry \cite{Son:2008ye},
it may be useful to focus on an explicitly supersymmetric theoretical setting.

Supersymmetric extensions of the Galilean algebra were first introduced in 3+1 dimensions \cite{Puzalowski:1978rv}, where two super-Galilean algebras were constructed, $ \mathcal{S}_1 \mathcal{G} $ which includes a single two-component spinorial supercharge and 
$ \mathcal{S}_2 \mathcal{G} $, which contains two supercharges. They can be obtained as the non-relativistic limit of ${\cal N}=1$ and ${\cal N}=2$ Super-Poincar\'e algebras, respectively. Alternatively, $\mathcal{S}_2 \mathcal{G}$ can be obtained performing a null reduction of the super-Poincar\`e algebra in 4+1 dimensions. It turns out that $ \mathcal{S}_1 \mathcal{G} \subset \mathcal{S}_2 \mathcal{G}$.

We give an explicit example of these supersymmetric extensions of the Galilean group in the $2+1$ dimensional case.
The bosonic part of the algebra is simply given by eq. (\ref{bosonic part Schr\"odinger algebra in d+1 dimensions}) with the identification $L_{12}=J,$ since the angular momentum is a pseudo-scalar on the plane.
The fermionic part is
\bea
& [Q,J] = \frac12 Q \, , \qquad
\lbrace Q, Q^{\dagger}  \rbrace = \sqrt{2} M \, , \nonumber \\
& [\tilde{Q},J] = - \frac12 \tilde{Q} \, , \quad
[\tilde{Q}, K_1 - i K_2] = -i Q \, , \quad
\lbrace \tilde{Q}, \tilde{Q}^{\dagger}  \rbrace = \sqrt{2} H \, ,    \\
& \lbrace Q, \tilde{Q}^{\dagger} \rbrace = -  (P_1 -i P_2) \, , \quad
\lbrace \tilde{Q}, Q^{\dagger} \rbrace = -  (P_1 + i P_2) \, ,
\nonumber
\eea
where $Q, \tilde{Q}$ are two complex supercharges.
This is the non-relativistic $ \mathcal{N}=2 $ SUSY algebra in 2+1 dimensions, which first appeared in the non-relativistic SUSY extension of Chern-Simons matter systems, where an enhanced superconformal symmetry arises \cite{Leblanc:1992wu}. 
Removing $\tilde{Q}$ from \eqref{commu superGalileo2} we obtain the $ \mathcal{S}_1 \mathcal{G}$ algebra.
 
In 3+1 dimensions theories with  $ \mathcal{S}_1 \mathcal{G} $  and $\mathcal{S}_2 \mathcal{G}$ invariance have been considered in \cite{Puzalowski:1978rv,Clark:1983ne,deAzcarraga:1991fa,Meyer:2017zfg}, 
while in 2+1 dimensions Chern-Simons theories with   $ \mathcal{S}_2 \mathcal{G} $ symmetry
were studied in  \cite{Leblanc:1992wu,Meyer:2017zfg,Bergman:1995zr}. Moreover, supersymmetric generalizations of the 
Schr\"odinger  algebra have been investigated \cite{Leblanc:1992wu, Beckers:1986ty, Gauntlett:1990nk, Duval:1993hs}, as well as Lifshitz supersymmetry \cite{Chapman:2015wha}. 
Recent developments about the power of holomorphicity applied to the renormalization of supersymmetric Lifshitz theories are treated in \cite{Arav:2019tqm}.

In chapter \ref{chapt-Non-relativistic Supersymmetry} we will build  an example of a theory with $ \mathcal{S}_2 \mathcal{G} $
Supersymmetry in $2+1$ dimensions, which we obtain by null reduction from a $3+1$ dimensional ${\cal N}=1$ Wess-Zumino model, and we will investigate its renormalization properties.


\section{Holographic complexity}
\label{int-Holographic complexity}

The AdS/CFT correspondence gives a non-perturbative formulation of quantum gravity in asymptotically AdS spacetimes in terms of the Quantum Field theory living on the boundary.
The geometry of the gravitational theory in the bulk hiddenly encodes quantum information properties: for example the Bekenstein-Hawking entropy is proportional to the area of the event horizon of a black hole
\beq
S_{\rm BH} = \frac{A}{4 G} \, ,
\eeq
and the area of a minimal surface in AdS is dual to the entanglement entropy of the boundary subregion \cite{Ryu:2006bv}.

\begin{figure}[h]
\centering
{\includegraphics[scale=0.30]{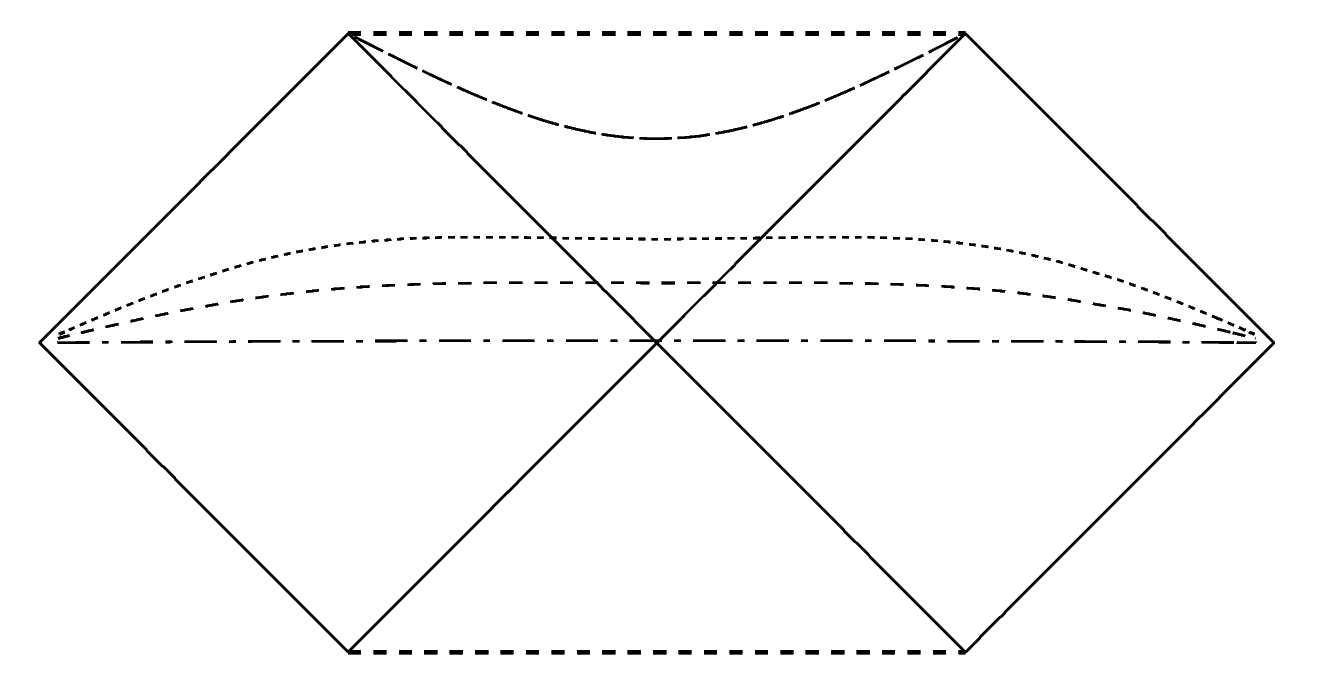}}  
\caption{Set of extremal slices for an eternal black hole.
The top slice is obtained when time goes to infinity in this effective decription.}
\label{fig-ERB gravitational collapse}
\end{figure}

The entropy is related to the counting of degrees of freedom in the dual quantum description of a black hole
and the microscopic interpretation was given in the context of string theory \cite{Strominger:1996sh}, where the number of microstates is identified with
\beq
n_{\rm microstates} = e^{S_{\rm BH}} \, .
\eeq
However, entropy does not seem the right quantity in order to describe the evolution of the Einstein-Rosen bridge in the interior of a black hole because it grows with time far after the black hole reaches thermal equilibrium \cite{Susskind:2014moa}.
We can indeed follow the time evolution of the Einstein-Rosen bridge in the context of general relativity by considering a foliation of spacetime with global spacelike slices satisfying some regularity properties, \emph{i.e.}
\begin{itemize}
\item Geodesically complete causal curves must intersect these slices once.
\item Slices must stay away from curvature singularities.
\item The entire region outside the horizon must be foliated by these slices.
\end{itemize}
Given the set of spacelike slices anchored on a spatial sphere with infinite radius, it can be proven that there exists one with maximum volume.
After choosing this one, we let the time $t$ to vary and this gives a foliation of spacetime with maximal slices.
An example of such a procedure is shown in fig. \ref{fig-ERB gravitational collapse} for a kind of eternal black hole that will be considered in chapter \ref{chapt-Complexity for warped AdS black holes}.

We observe that as time increases, the maximal slices go even further in the interior of the black hole until when $t=\infty$ we find the top final slice, which is completely inside the horizon.

In the AdS/CFT correspondence, a two-sided eternal black hole is dual to a thermofield double
state, in which the two Conformal Field Theories living on the left and right boundaries are
entangled \cite{Maldacena:2001kr}. Taking the two boundary times going in the same direction, this entangled
state is time-dependent, and the geometry of the Einstein-Rosen bridge connecting the two sides grows
linearly with time. 
This suggests that the investigation of the properties of the Einstein-Rosen bridge can give insights on the internal part of a black hole, which is expected to be related to quantum gravity aspects.
Moreover, we notice that the Einstein-Rosen bridge grows for a much longer timescale compared to the thermalization time, and then entropy appears not a valid quantity to describe this process.

In order to find a boundary dual to such behaviour, a new
quantum information tool has joined the discussion: computational complexity. For a
quantum-mechanical system, it is defined as the minimal number of basic unitary operations
which are needed to prepare a given state starting from a simple reference state.

There is a simple example which shows how the order of magnitude of entropy and complexity differ \cite{Susskind:2014moa}.
Consider a system composed by $K$ classical bits, which are identified by associating the binary values 0 or 1 for each of them. 
We identify:
\begin{itemize}
\item A simple state\footnote{We could as well identify the simple state as $(1,1, \dots ,1) .$ Since there is not much difference between the two choices, we assume to identify states under a global $\mathbb{Z}_2$ transformation acting simultanesouly on all the classical bits.} as $ (0,0, \dots, 0). $
\item A generic state as a random collection of 0 and 1.
\item A simple operation as the flip of a single bit $ (0 \leftrightarrow 1). $
\end{itemize}
In this case, the maximum entropy is the logarithm of the number of microstates (which are $2^K$) 
\beq
S^{\rm cl}_{\rm max} = K \log 2 \, ,
\eeq
while the maximal complexity corresponds to the least number of flips to perform in order to go from the reference state $(0,0, \dots , 0)$ to the most complex one, which is
\beq
(\underbrace{0, \dots, 0}_{K/2},\underbrace{1, \dots, 1}_{K/2}) \, .
\eeq
This shows that 
\beq
\mathcal{C}^{\rm cl}_{\rm max} = K/2 \, .
\eeq
We observe that the classical entropy and complexity are both linear in the number $K$ of classical bits.

Things drastically change at the quantum level.
First of all, we need to take an Hilbert state instead of a generic set of states, and operations are required to be unitary.
Furthermore, we identify
\begin{itemize}
\item A simple state\footnote{In this case we identify states under a global $SU(2)$ transformation.} as $  | 00 \dots 0 \rangle.  $
\item A generic state as a generic superposition of qubits with complex coefficients $ | \psi \rangle = \sum_{i=1}^{2^K} \alpha_i |i \rangle.  $
\item A simple operation as the action on 2 qubits, which is the simplest procedure which creates a non-vanishing entanglement in the system.
\end{itemize}
While the maximum entropy is the same (the number of microstates does not change between the classical and quantum cases)
\beq
S^{\rm qu}_{\rm max} = K \log 2 \, ,
\eeq
now the most complex state is obtained by changing the coefficients of the generic superpositions, which are in number $2^K .$
This implies that the number of operations to perform is
\beq
\mathcal{C}^{\rm qu}_{\rm max} \sim e^{K} \, .
\eeq
We observe that in this case we have an exponential behaviour for complexity instead of the power-law dependence for the entropy. 
Correspindingly, the time to get maximal entropy and maximal complexity are very different at quantum level, justifying heuristically the proposal that complexity can describe the time evolution of the Einstein-Rosen bridge.

\begin{figure}[h]
\centering
{\includegraphics[scale=0.45]{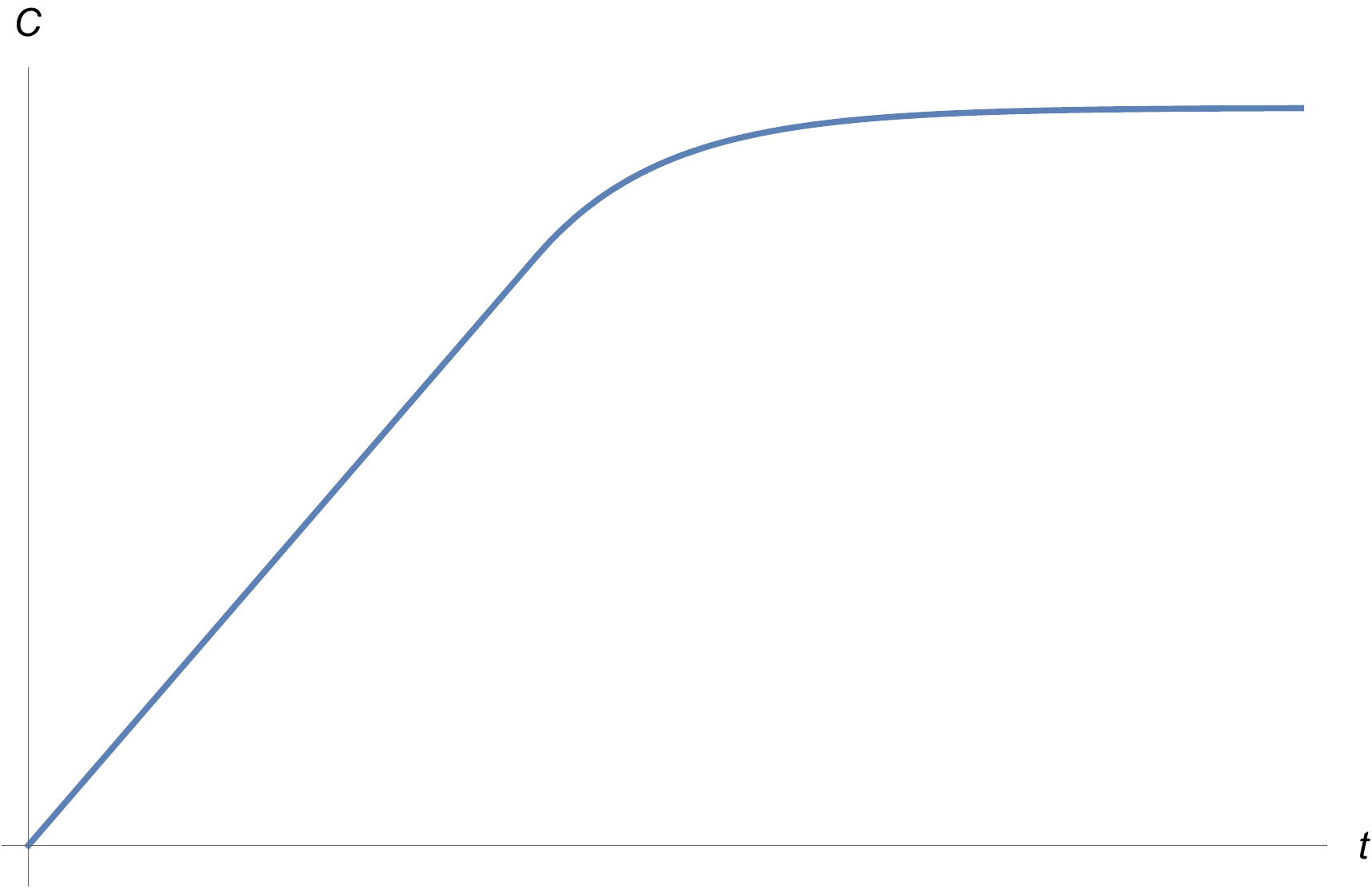}}  
\caption{Expected time evolution of complexity in a typical chaotic system.
The period of time where the effective description for the black holes is expected to be valid corresponds to the linear behaviour of complexity with respect to time in the graph.}
\label{fig-evolution_complexity_Susskind}
\end{figure}

In addition, from tensor network expectations the computational complexity is thought to behave as in fig. \ref{fig-evolution_complexity_Susskind}: there should be a short period of time when complexity grows linearly, and it reaches a constant value of saturation after an exponential time in the order of the size of the system, when quantum effects arise.

After a time of order $e^{e^K}$ (not shown in the previous graph) Poincaré recurrencies are expected to arise, leading to a decreasing of complexity to the original value, and a periodic behaviour should manifest. 

A proper definition of complexity in quantum field theory has several subtleties: 
the choice of the reference state, the allowed set of elementary quantum gates and
the amount of tolerance which is introduced in order to specify the accuracy with
which the state should be produced. 
Two different gravity duals of the quantum complexity of a state have been proposed so
far: the complexity=volume \cite{Stanford:2014jda} and the complexity=action \cite{Brown:2015lvg} conjectures\footnote{Various other similar versions exist, but they are all based on the Volume and Action proposals.}. 

In the Volume conjecture, complexity is proportional to the volume of a maximal codimension-one sub-manifold hanging from the boundary
\beq
C_V \sim   \frac{\textrm{Max}(V)}{G l} \, .
\eeq
While this proposal is a natural generalization of the entanglement entropy and has a physical interpretation as the volume of the Einstein-Rosen bridge, it requires the introduction of an \emph{ad hoc} length scale $l,$ which can be the $\mathrm{AdS}$ or the Schwarzschild radius or other relevant quantities dependent from the holographic dictionary.

In the Action conjecture, complexity is proportional to the gravitational action $I$ evaluated in the Wheeler-De Witt  patch, \emph{i.e.} the bulk domain of dependence of a Cauchy surface anchored at the boundary state
\beq
C_A = \frac{I}{\pi \hbar} \, .
\eeq
In this case the action has several contributions beyond the traditional bulk Einstein-Hilbert and boundary Gibbons-Hawking-York terms: they come from the null surfaces and from the joints at the intersection of boundary segments, which are necessary to compute the full time dependence of the Wheeler-De Witt action in AdS spacetime \cite{Lehner:2016vdi}.

This conjecture appears more universal than the volume one because of the absence of the length scale in the definition of complexity, and the late time behaviour is the same of the volume case.
On the other hand, the behaviour for intermediate times of the two proposals is different, which is a reason why it is interesting to investigate both of them.

There are some minimal requirements that we ask complexity to satisfy.
Based on dimensional grounds and on the observation that complexity is an extensive quantity, we require that the linear behaviour in time should have the rate
\beq
\frac{d \mathcal{C}}{dt} \sim T S \, ,
\eeq
being $T$ the temperature of the black hole and $S$ the entropy\footnote{This regime corresponds to late times in the semiclassical effective description where the black hole is studied; instead phenomena like the saturation of complexity are expected to arise after the Page time, when the effects of the Hawking radiation become important.}.

Moreover, extremal black holes are ground states and therefore static, which means that
\beq
\le \frac{d \mathcal{C}}{dt} \ri_{\rm extr} = 0 \, .
\eeq
This is also expected from the fact that extremal black holes usually have vanishing temperature.

Quantum complexity can access some informations that the entopy by itself cannot.
First of all, from the gravity side, complexity should be able to access to regimes of the black hole evolution which are much longer than the thermalization time.
We may also hope that an investigation of complexity for evaporating black holes can shed light on the information paradox: while the usual way to follow the process is by means of the Page curve for the entropy, the relation between the internal of the black hole and the Hawking radiation can be better understood in the context of complexity (in the spirit of ER=EPR interpretation \cite{Maldacena:2013xja}).
In this sense, since complexity is an object that investigate the interior of the black hole, it goes in principle beyond the territory of entanglement entropy computations, whose holographic dual is given by the Ryu-Takayanagi curve which usually stays ouside the horizon.
Another interesting topic that we can hope to better understand with complexity concerns the technique of bulk reconstruction.
In particular, proposals like \cite{Papadodimas:2012aq, Papadodimas:2013jku, Papadodimas:2015jra} aim to investigate the inside of a black hole and even part of the other asymptotic region by starting from one of the boundaries of the spacetime.
We think that complexity can give some hints to tackle this kind of problems.

It is interesting to consider extensions of holography to spacetimes that are not asymptotically AdS.
A non-trivial deformation of $ \mathrm{AdS}_3 $ which only preserves the isometries $ SL(2,\mathbb{R}) \times U(1) $ is given by Warped $ \mathrm{AdS}_3 . $ 
This spacetime is conjectured to be dual to a class of non-relativistic theories in 1+1 dimensions, called 
Warped Conformal Field Theories.
They can be interpreted to be Lifshitz-invariant with dynamical exponent $z=\infty$ and in curved backgrounds they naturally couple to Newton-Cartan geometry.
The entanglement entropy was studied in this context and an analog of the Cardy formula was found \cite{Detournay:2012pc}.

The conjectured duality is still far from being understood, in particular the field theory side is still in its infancy: it is then important to pursue the study of the subject in order to gain valuable insights when the duality involves non-AdS
asymptotic. 
Furthermore explicit realizations of Warped Conformal Field Theories seem to be pathologic or at the brink of non-locality: they admit an infinite number of exactly marginal non-local deformations which must be tuned away \cite{Jensen:2017tnb}.
In this framework, it is useful to analyze quantities which do not require the introduction of an explicit action, like anomalies, entanglement entropy or complexity.
We will test the holographic proposals for complexity in chapter \ref{chapt-Complexity for warped AdS black holes} by computing the Volume and the Action for black holes in Warped $ \mathrm{AdS}_3 .$

When the state on the boundary is mixed, \emph{i.e.} we anchor the extremal slice to a subregion, holographic proposals for complexity similar to the case of entanglement entropy exist \cite{Carmi:2016wjl}.
We will apply both these proposals for black holes in Warped $ \mathrm{AdS}_3 $ in chapter \ref{chapt-Subregion complexity for warped AdS black holes} in a specific case where the subregion is taken to be one of the asymptotic boundaries.
In chapter \ref{chapt-Subregion action complexity of the BTZ black hole} we will further investigate the properties of subregion complexity=action for more general mixed states in the context of the BTZ black hole. 

Conclusions and discussions on the results obtained in this thesis are collected in chapter \ref{chapt-Conclusions and outlook}.
We put technical details of computations and conventions to the Appendices.


\part{Non-relativistic quantum field theory}


\chapter{Non-relativistic actions}
\label{chapt-Non-relativistic actions}

In this chapter we will describe all the ingredients necessary for the investigation of the non-relativistic trace anomaly: a local version of the Galilean group (the Newton-Cartan geometry), how the Weyl transformations act on such a background, and the action for Galilean-invariant bosons and fermions.
This material is intended as the set up necessary to undergo the investigation of the trace anomaly in explicit cases with the heat kernel technique that will be developed in chapter \ref{chapt-The heat kernal technique}.

The discussion will be mostly referred to a general $d+1$ dimensional spacetime with non-relativistic symmetry, but in the derivation of the action for a fermion coupled to Newton-Cartan (NC) geometry we will focus on the case of interest, \emph{i.e.} $2+1$ dimensions.
While there are various methods to approach the problem of defining a local version of the Galilean group, we will focus on the Discrete Light-Cone Quantization (DLCQ) technique, which consists in the dimensional reduction of a $d+2$ dimensional spacetime along a null direction.
The reason is that such procedure automatically implements all the non-relativistic symmetries\footnote{This procedure is particularly convenient when dealing with the $U(1)$ gauge invariance and the local version of Galilean boosts (Milne boosts), which are very difficult to implement simultaneously.}, thus overcoming various problems that must be treated carefully with an intrinsic  formulation without referring to a relativistic parent theory.

The NC geometry was first introduced as a tool to write newtonian gravity in a diffeomorphism-invariant fashion; for a review see  \cite{gravitation}.
Recently, works by Son and collaborators \cite{Geracie:2014nka,Son:2005rv,Hoyos:2011ez,Son:2013rqa}
showed that it can be used as a powerful tool to study condensed matter systems
with galilean invariance; the main idea is to use it
 as source  for energy-momentum tensor for quantum field theory
 description of several condensed matter systems.
Strongly-coupled system with Galilean invariance can be studied 
holographically \cite{Son:2008ye,Balasubramanian:2008dm};
 also in this approach the NC geometry is a natural
formalism \cite{Christensen:2013lma,
Hartong:2014oma,Hartong:2014pma}.
A theoretical approach to fermions invariant under the Galilean group was firstly faced in \cite{LevyLeblond:1967zz}; the coupling to a NC background by using the $c \rightarrow \infty$ limit was done in \cite{Geracie:2014nka,Fuini:2015yva}, while other studies on fermions with the null reduction were performed in \cite{Duval:1995fa}.


\section{Newton-Cartan geometry}
\label{sect-Newton-Cartan geometry}

We consider as a starting point a $d+2$ dimensional Lorentzian manifold whose coordinates are denoted with late capital latin indices $ x^M . $
In order to deal with spinors, we introduce early capital latin indices $x^A$ to denote the tangent space, where the metric is locally flat.
We introduce light-cone coordinates
\beq
x^{\pm} = \frac{x^{d+1} \pm x^0}{\sqrt{2}}
\label{light-cone coordinates}
\eeq
which allow to decompose the spacetime and tangent space structures as
\beq
\begin{aligned}
&  x^M=(x^-,x^{\mu})=(x^-,x^+,x^i)    \qquad (i=1,\dots,d)  \\
&  x^A=(x^-,x^{\alpha})=(x^-,x^+,x^a)  \qquad  (a=1,\dots,d) \, .
\end{aligned}
\label{decomposition of coordinates in the parent theory}
\eeq
Latin lower-case letters refer to the spatial indices, while greek letters refer to the spacetime content of the $d+1$ dimensional non-relativistic theory.
As for the relativistic parent, early and late letters refer to flat and curved indices, respectively.

Ambiguities can arise since the light-cone indices appear in the curved manifold and in the tangent space; in these cases we distinguish them by adding a subscript 
\beq
\underset{(A)}{\pm} \, , \qquad \underset{(M)}{\pm}  \, .
\eeq
The null reduction is realized by compactifying $x^-$ on a small circle of radius $ R$.
For convenience,  we rescale $x^- \to x^-/R$ in such a way that the rescaled 
coordinate is adimensional. In order to keep the metric tensor adimensional,
we also rescale $x^+ \to R \, x^+$.
In the DLCQ dictionary, the light-cone direction $x^+$ after the compactification is interpreted as the time of the $d+1$ dimensional non-relativistic theory.

On a curved manifold, this operation is performed by taking the most general $d+2$ dimensional metric with null Killing vector
\beq
n^M = (1, \bold{0}) \, , \qquad
n_M = (0, n_{\mu}) \, ,
\eeq
which turns out to be of the form
\beq
\label{null-red metric}
G_{MN}= \begin{pmatrix}
0  & n_{\nu}  \\
n_{\mu}  & n_{\mu} A_{\nu} + n_{\nu} A_{\mu} + h_{\mu\nu}
\end{pmatrix} \, ,  \qquad
G^{MN}= \begin{pmatrix}
A^2 - 2 v \cdot A  & v^{\nu}-h^{\nu\sigma} A_{\sigma}  \\
v^{\mu}-h^{\mu\sigma} A_{\sigma}  & h^{\mu\nu}
\end{pmatrix} \, . 
\eeq
In order to parametrize all the degrees of freedom of a metric with the required isometry, we introduced the vector fields $ A_{\mu}, v^{\mu} $ and the semipositive-definite symmetric tensors $h_{\mu\nu}, h^{\mu\nu} . $  
Their interpretation as NC data will be clear soon.

We denote the determinant of the metric as
\beq
\sqrt{g}= \sqrt{- \det G_{AB}}=
\sqrt{{\rm det}(h_{\mu \nu}+ n_\mu n_\nu)} \, .
\eeq
Being $G_{MN}$ a non-degenerate metric on a pseudo-Riemannian manifold, we can define the usual Levi-Civita connection and we call $\mathcal{D}_M$ the covariant derivative associated to it.

The null reduction prescription also requires that any local field is decomposed as 
\beq
\Phi(x^M)=\varphi(x^{\mu}) e^{imx^{-}} \, .
\label{decomposition fields null reduction}
\eeq

The quantities appearing in the decomposition of the metric (\ref{null-red metric}) define the basic ingredients of a $d+1$ dimensional NC geometry: $ n= n_{\mu} dx^{\mu} $ is a nowhere-vanishing one-form which locally gives the time direction, $ h^{\mu\nu} $ is a semipositive definite symmetric tensor of rank $d$ which satisfies the condition 
\beq
 n_{\mu} h^{\mu\nu} = 0  
 \label{1st condition NC geometry}
 \eeq 
and is interpreted as an inverse metric on spatial slices.

In analogy with the pseudo-Riemannian case, we would like to define a torsionless connection whose induced covariant derivative $ \hat{D}_{\mu}$ preserves the constancy of the metric.
In the case of NC geometry, a similar condition would be to require
\beq
\hat{D}_{\mu} n_{\nu} = 0 \, , \qquad
\hat{D}_{\mu} h^{\nu \rho} = 0 \, .
\eeq 
A connection can be introduced by defining a velocity vector $v^{\mu}$ subject to the constraint 
\beq
n_{\mu} v^{\mu} = 1 \, ,
\label{2nd condition NC geometry}
\eeq
and a covariant symmetric tensor $ h_{\mu\nu} $ which satisfies
\beq
h^{\mu \rho}h_{\rho \nu}=\delta^\mu_\nu -v^\mu n_\nu\equiv P^\mu_\nu \, , \qquad
h_{\mu \nu} v^{\nu} =0 \, ,
\eeq
where $P^\mu_\nu$ is the projector onto spatial directions.
 
However, it turns out that the constancy of $(n_{\mu},h^{\mu\nu})$ can be fulfilled only by introducing a non-vanishing torsion in the $d+1$ dimensional connection, and furthermore the covariant derivative is determined only up to a two-form $F.$
More precisely, the Christoffel symbol can be taken to be
\beq
\hat{\Gamma}^\mu_{\nu \rho}=
v^\mu \p_\rho n_\nu +\frac12 h^{\mu \s} (\p_\nu h_{\rho \s}
+\p_\rho h_{\nu \s}-\p_\s h_{\nu \rho})
+ h^{\mu \s} n_{(\nu} F_{\rho) \s} \, ,
\label{nonrel connection with torsion}
\eeq
which has a purely temporal torsion.
Moreover, it is not restrictive to take the two-form $F$ to be close, which allows to locally define a gauge connection such that $F=dA.$
This quantity enters the $d+2$ dimensional metric on the Lorentzian manifold and is naturally associated to the $U(1)$ mass or particle number, which is a conserved quantity in non-relativistic theories.
The gauge transformations of the field $A_{\mu}$ are naturally interpreted from the $d+2$ dimensional point of view as additive reparametrizations along the null direction $x^- .$

The ambiguity in the definition of the connection is related to the fact that the velocity vector, the covariant spatial metric and the gauge connection are not uniquely defined.
The following set of transformations (called Milne boosts) leaves the metric in the form\footnote{Modified Milne transformation may also be considered, 
but then the null reduction trick can not be used (see e.g. \cite{Jensen:2014aia}).} (\ref{null-red metric})
\bea
v'^\mu & = & v^\mu+h^{\mu \nu} \psi_\nu \, \nl
h'_{\mu \nu} & = & h_{\mu \nu} -(n_\mu P_\nu^\rho+ n_\nu P_\mu^\rho) \psi_\rho
+n_\mu n_\nu h^{\rho \sigma} \psi_\rho \psi_\sigma \, , \nl
A'_\mu & = & A_\mu+P^\rho_\mu \psi_\rho -\frac{1}{2} n_\mu h^{\a \b} \psi_\a \psi_\b \, , 
\eea
where $ \psi= \psi_{\mu} dx^{\mu} $ is a one-form parametrizing the transformation, while $n_\mu$ and $h^{\mu \nu}$ are invariant.
There is not a convenient intrinsic $d+1$ dimensional way to build Milne boost-invariant quantities; the invariants that we can build by direct computation are
\bea
&&v_A^\mu = v^\mu - h^{\mu \xi} A_\xi \, , \qquad
(h_A)_{\mu \nu}=h_{\mu \nu} +A_\mu n_\nu + A_\nu n_\mu \, , \qquad
\phi_A = A^2 - 2 v \cdot A \, ,
\nl
&& (Q_A)_{\mu \nu \sigma} = (\p_\mu (h_A)_{\nu \s} +\p_\nu (h_A)_{\mu \s} - \p_\s (h_A)_{\mu \nu}) \, ,
\label{definition of Milne boost invariants}
\eea
where $A^2=h^{\mu \nu}A_\mu A_\nu$ and $A \cdot v = v^\mu A_\mu$.
The subscript $A$ is a notation to identify the invariance of the object under Milne boosts.

It is not possible to find a $d+1$ dimensional connection which is invariant both under $U(1)$ gauge transformations and Milne boosts, but only under one of them. 
From the point of view of the relativistic parent, this is the statement that the Christoffel symbol is not invariant under reparametrizations along the null direction $x^-,$ which represent the gauge variation under a local $U(1)$ transformation of the system.
Moreover, it is important to observe that the Christoffel symbol in eq. (\ref{nonrel connection with torsion}) is not the Levi-Civita connection corresponding to the metric (\ref{null-red metric}), which instead is torsionless and given by
\bea
& & \Gamma^{-}_{--}=\Gamma^\mu_{--}=0 \, , \qquad
\Gamma^-_{\mu -}=\frac{1}{2} v_A^\s \tilde{F}_{\mu \s}
 \, ,
\qquad
\Gamma^\mu_{\nu -}= \frac{1}{2} h^{\mu \s} 
\tilde{F}_{\nu \s}
\, ,
\nl
&&
\Gamma^-_{\mu \nu}= \frac{1}{2} \le \phi_A 
(\p_\mu n_\nu + \p_\nu n_\mu)
+v_A^\s 
(Q_A)_{\mu \nu \s}
\ri \, ,
\nl
&&
\Gamma^\mu_{\nu \rho}= \frac{1}{2} \le v_A^\mu  (\p_\nu n_\rho + \p_\rho n_\nu)
+h^{\mu \s} (Q_A)_{\nu \rho \s}
  \ri \, .
\eea
In particular, the relation with the $d+1$ dimensional Cristopphel symbol is given by
\beq
\Gamma^\mu_{\nu \rho}=\hat{\Gamma}^\mu_{(\nu \rho)}+
\frac12 h^{\mu \sigma} (Q_A)_{\nu \rho \s} \, ,
\eeq
while $\hat{\Gamma}^\mu_{[\nu \rho]}$ is not directly related to
$\Gamma^\mu_{\nu \rho}$.

The frame fields defining the locally flat metric in $d+1$ dimensions can be derived as well from the $d+2$ dimensional relativistic framework.
The tangent space in light-cone coordinates is equipped with the metric
\beq
G_{AB}= G^{AB} = \begin{pmatrix}
0  & 1 & 0   & \dots & 0 \\
1  & 0 & 0  & \dots & 0 \\
0  & 0 & 1   & \dots & 0 \\
\dots & \dots & \dots & \dots  & \dots \\
0 & 0 & 0 & 0 & 1
\end{pmatrix} \, ,
\eeq
and this induces the usual definition of the $d+2$ dimensional vielbein with the relations
\bea
& G_{MN}= e^{A}_{\,\,\, M} G_{AB} e^{B}_{\,\,\, N}  \, ,  
\qquad  G_{AB}= e^{M}_{\,\,\, A} G_{MN} e^{N}_{\,\,\, B}  \, ,  
\nl
&  e^{A}_{\,\,\, M} e^{M}_{\,\,\, B} = \delta^{A}_{\,\,\, B}  \, , \qquad
  e^{M}_{\,\,\, A} e^{A}_{\,\,\, N} = \delta^{M}_{\,\,\, N}  \, .
\eea
The corresponding $d+1$ dimensional vielbein defined by dimensional reduction  is not unique, but we take the following convenient choice 
\beq
e^A_{\,\,\, M} = \begin{pmatrix}
e^-_{\,\,\, M}  \\
e^+_{\,\,\, M}  \\
e^a_{\,\,\, M}
\end{pmatrix} = 
 \begin{pmatrix}
e^-_{\,\,\, -} \ \ & e^-_{\,\,\, \mu}  \\
e^+_{\,\,\, -} \ \ & e^+_{\,\,\, \mu} \\
e^a_{\,\,\, -} \ \ & e^a_{\,\,\, \mu}
\end{pmatrix}
=
 \begin{pmatrix}
1 \ \ & A_{\mu} \\
0 \ \ & n_{\mu} \\
\mathbf{0} \ \ & e^a_{\mu}
\end{pmatrix}
\, .
\eeq
Using the consistency relations
\beq
e^{M}_{\,\,\, A} e^{B}_{\,\,\, M} = \delta_{A}^{\,\,\,B}  \, , \qquad  
e^{A}_{\,\,\, M} e^{N}_{\,\,\, A} = \delta_{M}^{\,\,\,N}  \, ,
\label{4 consistenza tra vielbein}
\eeq
we can derive a simple expression for the inverse vielbein
\beq
e^M_{\,\,\, A} = \begin{pmatrix} e^M_{\,\,\, -}\ \ & e^M_{\,\,\, +} \ \ & e^M_{\,\,\, a}   \end{pmatrix} =
 \begin{pmatrix} e^-_{\,\,\, -}\ \ & e^-_{\,\,\, +} \ \ & e^-_{\,\,\, a}  \\
 e^{\mu}_{\,\,\, -}\ \ & e^{\mu}_{\,\,\, +} \ \ & e^{\mu}_{\,\,\, a}  \end{pmatrix}
= \begin{pmatrix}1 \ \ & -v^{\sigma} A_{\sigma} \ \ &  - h^{\nu \sigma} A_{\sigma} e^a_{\nu}  \\
\bold{0} \ \ &  v^{\mu} \ \ &   h^{\mu\nu} e^{a}_{\nu} \end{pmatrix} \, .
\eeq
The previous construction of the Newton-Cartan geometry in $d+1$ dimensions from a relativistic parent allows to obtain a structure which is automatically invariant under
\begin{itemize}
\item Diffeomorphisms in the $d+1$ dimensional spacetime
\item $U(1)$ gauge transformations
\item Milne boosts
\end{itemize}
In fact, diffeomorphisms along the $d+1$ dimensions of the non-relativistic theory are obviously inherited from the diffeomorphisms of the higher dimensional theory, while the gauge transformations come from coordinate reparametrizations along the $x^-$ direction.
Furthermore, Milne boosts invariance is built-in from the choice of the metric (\ref{null-red metric}).

For these reasons, it is convenient to build tensors and scalars starting using the null reduction: the non-relativistic symmetries are automatically implemented and the classification of terms entering the trace anomaly is easier and under control.


\section{Null reduction of the Klein-Gordon action}
\label{sect-Null reduction of the Klein-Gordon action}

We apply the null reduction prescription to the case of a relativistic free scalar in a curved background.
The action with minimal coupling to gravity is given by
\beq
S= \int d^{d+2} x \, \sqrt{-\mathrm{det} \, G_{MN}} \, \le -G^{MN} \p_M \Phi^{\dagger} \p_N \Phi - \xi R \Phi^{\dagger} \Phi \ri \, .
\eeq
If we take the metric (\ref{null-red metric}) and the decomposition of fields (\ref{decomposition fields null reduction}), we obtain the $d+1$ dimensional non-relativistic action
\beq
S= \int d^{d+1} x \, \sqrt{g} \, \left\lbrace i m v^{\mu} \le \varphi^{\dagger} D_t \varphi - D_{\mu} \varphi^{\dagger} \varphi  \ri - h^{\mu\nu} D_{\mu} \varphi^{\dagger} D_{\nu} \varphi - \xi R \varphi^{\dagger} \varphi \right\rbrace \, ,
\label{action non-relativistic scalar real time}
\eeq
where the derivative is covariant only with respect to the gauge connection 
\beq
D_{\mu} \varphi = \p_{\mu} \varphi - i m A_{\mu} \varphi \, .
\label{derivative covariant with respect to gauge connection, free scalar}
\eeq
We can get more a better understanding of the system by considering the case of flat space 
\beq
n_{\mu} = (1, \mathbf{0}) \, , \quad  h_{\mu\nu} = {\rm diag}(0,\mathbf{1}) \, ,
 \quad v^{\mu} = (1,\mathbf{0}) 
\label{flatland}
\eeq
and $A_{\mu}=0, $ which brings the action to the form
\beq
S= 
\int d^{d+1} x \, \le 2 i m \varphi^{\dagger} \p_t \varphi - |\p_i \varphi|^2 \ri = \int d^{d+1} x \, \varphi^{\dagger} \le 2 i m \p_t + \p_i^2 \ri \varphi \, .
\label{free Schr\"odinger action}
\eeq
As expected, the Euler-Lagrange equations of motion are immediately identified with the Schr\"odinger equation for the free scalar field
\beq
i \p_t \varphi = - \frac{1}{2m} \p_i^2 \varphi  \, .
\eeq
If we add a non-vanishing gauge field $A_{\mu}\ne 0$ to the system, the result is simply the action (\ref{free Schr\"odinger action}) with the minimal coupling replacement $\p_{\mu} \rightarrow D_{\mu} .$

Finally, we consider the case where the gauge field is set to 0 and the background is curved.
For future analysis it is convenient to write the action as a differential operator of a quadratic form using integration by parts to get
\beq
S= \int d^{d+1} x \, \sqrt{g} \, \varphi^{\dagger} \left\lbrace i m v^{\mu} \p_{\mu} \varphi + \frac{i m \p_{\mu} (\sqrt{g} v^{\mu} \varphi)}{\sqrt{g}} + \frac{\p_{\mu} (\sqrt{g} h^{\mu\nu} \p_{\nu} \varphi)}{\sqrt{g}} - \xi R \varphi \right\rbrace \, .
\label{non-relativistic free scalar with A=0}
\eeq


\section{Null reduction of the Dirac action}
\label{sect-Null reduction of the Dirac action}

The way fermions are treated in Quantum Mechanics (QM) is very different from Quantum Field Theory (QFT): while in the latter case they satisfy the first-order Dirac equation (contrarily to the second-order Klein-Gordon equation), in the former case they satisfy the same Schr\"odinger equation as bosons.
Moreover, properties such as spin are attached \emph{by hands}, contrarily to the machinery of the Clifford algebra in the QFT treatment.
This very different behaviour seems a consequence of the non-relativistic nature of QM, which describes theories at low energies and speeds,
but we can see that it is instead a consequence of the framework of first quantization.

Following \cite{LevyLeblond:1967zz}, it is possible to find a first-order differential equation for fermions inspired by the Dirac's method used for relativistic QFT.
This procedure allows to derive from first principles the same result which is found from the $c \rightarrow \infty$ limit of the Dirac equation, where the Weyl spinors are recognized to split into an auxiliary and a dynamical doublet.
While they are mixed in the Dirac equation, when integrating out the auxiliary Weyl fermion we obtain a single Schr\"odinger equation for the dynamical one.

In the spirit of this procedure, and following the null reduction prescription, we are led to consider the $d+2$ dimensional Dirac action as the starting point.
The  Dirac operator is expressed as
\beq
\slashed D= \gamma^{M} D_{M} = \gamma^{A} e^{M}_{\,\,\, A} D_{M}  \, ,
\eeq
where the covariant derivative contains
\beq
D_{M} \Psi =\left( \p_{M} + \frac{1}{4} \omega_{MAB} \gamma^{AB} \right) \Psi = \left( \p_{M} + \frac{1}{8} \omega_{MAB} [\gamma^{A}, \gamma^{B}] \right) \Psi \, .
\eeq
Conventions about the Dirac matrices in light-cone coordinates and the spin connection $ \omega_{MAB} $ are summarized in Appendix \ref{app-conv}.

The Dirac action in curved spacetime is not uniquely defined, but there are various prescriptions which differ when the connection is torsionful.
By taking the torsionless Levi-Civita connection in $d+2$ dimension, the Lagrangian can be made hermitian by means of partial integration, and it is not ambiguous to consider the action
\beq
S= \int d^4 x \sqrt{g} \, i \, \bar{\Psi} \slashed D \Psi \, .
\label{didirac}
\eeq
From now on, we will consider specifically the case of a null reduction from $3+1$ dimensions to get a $2+1$ dimensional non-relativistic theory.
In order to perform the DLCQ technique, we take the metric (\ref{null-red metric}) as the background and we specify the components of the Dirac spinor in $3+1$ dimensions
\beq
\Psi=  \begin{pmatrix}
\Psi_L \\ \Psi_R 
\end{pmatrix} \, .
\eeq
Since we consider a massless Dirac action in $3+1$ dimensions, the Weyl spinors decouple and we can restrict our analysis to the action for the left-handed part
\beq
S_{L}=\int d^4 x \sqrt{g} \,  \mathcal{L}_W = \int d^4 x \sqrt{g} \,  i \, \Psi_L^{\dagger} \bar{\sigma}^A D_A \Psi_L  \, .
\label{Weyl action 3+1 dimensions}
\eeq
This technicism allows to obtain the correct number of degrees of freedom to describe the non-relativistic fermion: indeed, the Dirac spinor in $2+1$ dimensions only contains 2 complex components as opposed to the 4 components of the higher-dimensional parent.
In this way the dictionary of null reduction requires a decomposition of the relativistic field $ \Psi_L $ into a non-relativistic field times a phase along the compact direction
\beq
\Psi_L (x^M)=\begin{pmatrix}
\xi (x^{\mu}) \\ \chi (x^{\mu})
\end{pmatrix}  e^{im x^-}\, ,
\eeq
where $\xi, \chi$ are complex numbers. 

We decompose the covariant derivative into the light-cone and the spatial directions
\beq
\begin{aligned}
& D_{\underset{(A)}{-}} = e^M_{\,\,\, \underset{(A)}{-}} D_M = 
\begin{pmatrix} 1 & \bold{0} \end{pmatrix} \begin{pmatrix} D_{\underset{(M)}{-}} \\ D_{\mu} \end{pmatrix} = D_{\underset{(M)}{-}} \, , &
\\
& D_{\underset{(A)}{+}} = e^M_{\,\,\, \underset{(A)}{+}} D_M = 
\begin{pmatrix} -v^{\sigma} A_{\sigma} & v^{\mu} \end{pmatrix} \begin{pmatrix} D_{\underset{(M)}{-}} \\ D_{\mu} \end{pmatrix} = - v^{\sigma} A_{\sigma} D_{\underset{(M)}{-}} + v^{\mu} D_{\mu}  \, , &
\\
& D_{a} = e^M_{\,\,\, a} D_M = 
\begin{pmatrix} -e^{\sigma}_{\,\,\, a} A_{\sigma} & e^{\mu}_{\,\,\, a} \end{pmatrix} \begin{pmatrix} D_{\underset{(M)}{-}} \\ D_{\mu} \end{pmatrix} =  -e^{\sigma}_{\,\,\, a} A_{\sigma}  D_{\underset{(M)}{-}} + e^{\mu}_{\,\,\, a} D_{\mu}  \, . &
\end{aligned}
\eeq
In this way we decompose the sum in eq.~(\ref{Weyl action 3+1 dimensions}) as
\beq
\begin{aligned}
\mathcal{L}_W  & =
i e^{-imx^-} \begin{pmatrix}  \xi^{\dagger} & \chi^{\dagger}  \end{pmatrix} \bar{\sigma}^-  D_{\underset{(A)}{-}} \left[  \begin{pmatrix}  \xi \\ \chi  \end{pmatrix}  e^{imx^-}  \right] + i e^{-imx^-} \begin{pmatrix}  \xi^{\dagger} & \chi^{\dagger}  \end{pmatrix}  \bar{\sigma}^+ D_{\underset{(A)}{+}} \left[  \begin{pmatrix}  \xi \\ \chi  \end{pmatrix}  e^{imx^-}  \right] + \\
& + i e^{-imx^-} \begin{pmatrix}  \xi^{\dagger} & \chi^{\dagger}  \end{pmatrix} \sigma^a  e^M_{\,\,\, a}  D_{M} \left[  \begin{pmatrix}  \xi \\ \chi  \end{pmatrix}  e^{imx^-}  \right] \, ,
\end{aligned}
\eeq
and the explicit expressions of the covariant derivatives give
\beq
 \begin{aligned}
\mathcal{L}_W & = - \sqrt{2} m \xi^{\dagger} \xi - \sqrt{2} i \chi^{\dagger} \hat{D}_{t} \chi + i \chi^{\dagger} (\hat{D}_{1} +i \hat{D}_2 ) \xi + i \xi^{\dagger}(\hat{D}_{1} - i \hat{D}_2 ) \chi + \\
& +  \frac{i}{4} \begin{pmatrix}  \xi^{\dagger} & \chi^{\dagger}  \end{pmatrix} 
(\bar{\sigma}^+  v^{\mu} +\sigma^a  e^{\mu}_{\,\,\, a} )
\omega_{\mu AB} \sigma^{AB}  \begin{pmatrix}  \xi \\ \chi  \end{pmatrix}  \\
& +  \frac{i}{4}  \begin{pmatrix}  \xi^{\dagger} & \chi^{\dagger}  \end{pmatrix} \left( \bar{\sigma}^-  - v^{\sigma} A_{\sigma} \bar{\sigma}^+ - \sigma^a e^{\sigma}_{\,\,\, a} A_{\sigma}  \right) \omega_{ {\underset{(M)}{-}} AB} \sigma^{AB}   \begin{pmatrix}  \xi \\ \chi  \end{pmatrix}  \, .
\end{aligned} 
\label{Lagrangiana-ferm}
\eeq
In this formula we introduced derivatives which are covariant with respect to the local U(1) symmetry
\beq
\hat{D}_t = v^{\mu} \left( \p_{\mu} -i m A_{\mu}  \right) \, , \qquad
\hat{D}_a = e^{\mu}_{\,\,\, a} \left( \p_{\mu} -i m A_{\mu}  \right) \, . 
\eeq
In order to write explicitly the last two lines of eq. (\ref{Lagrangiana-ferm}), we need the use the precise expression of the components of the spin connection.
In fact the sum implicitly contains a summation over spinorial objects, whose matricial content depends from the particular Pauli matrix we are summing over.
Using the results in Appendix \ref{app-conv} we can re-write the Lagrangian in the compact form
\beq
\mathcal{L}_W=
 \begin{pmatrix}  \xi^{\dagger} & \chi^{\dagger}  \end{pmatrix} 
  \begin{pmatrix} A  & B \\ C & D \end{pmatrix} 
 \begin{pmatrix}  \xi \\ \chi  \end{pmatrix} \, , 
 \label{Lagrangiana-ferm2}
\eeq
where
\bea
A &=& -\sqrt{2}\le m+\frac14 \tilde{F}_{\mu \nu} e^\mu_1 e^\nu_2\ri\, ,
\nl
B &=& (e_1^\mu-i e^\mu_2) (i \tilde{D}_\mu +\frac{i}{4} \tilde{F}_{\mu \nu} v^\nu) \, ,
 \qquad
C =(e_1^\mu+i e^\mu_2) (i \tilde{D}_\mu +i \frac{ 3}{4} \tilde{F}_{\mu \nu} v^\nu) \,  ,
\nl
D &=& \sqrt{2} \left[  v^\mu (-i \tilde{D}_\mu 
-\frac{i}{4} h^{\rho \sigma} \p_\mu h_{\rho \sigma})
-\frac{i}{2} (v^\mu v^\nu \p_\mu n_\nu +\p_\mu v^\mu)
-\frac14 F_{\mu \nu}  e^\mu_1 e^\nu_2  
\right] \, .
\eea
Here $\tilde{D}_\mu$  denotes a covariant derivative
which includes only the gauge and the curved space spin
connection  $\tilde{\omega}_{\mu  a b}$ built 
with the spatial tetrad $e^a_\mu$; this derivative acts
on the matter fields $\xi$ and $\chi$ as follows
\beq
\tilde{D}_\mu \xi=
\left[ \p_\mu + \frac{i}{2} \tilde{\omega}_{\mu 1 2} 
- i m A_\mu \right] \xi  \, ,  \qquad
\tilde{D}_\mu \chi=
\left[ \p_\mu -\frac{i}{2} \tilde{\omega}_{\mu 1 2} 
- i m A_\mu \right] \chi  \, ,  
\eeq
where
\beq
\tilde{\omega}_{\mu  a b}  = \frac{1}{2}
\le e^{\nu}_{\,\,\,a} \left(  \p_{\mu} e^{b}_{\,\,\, \nu} -  \p_{\nu} e^{b}_{\,\,\, \mu}  \right)  -   e^{\nu}_{\,\,\, b} \left(  \p_{\mu} e^{a}_{\,\,\, \nu} -  \p_{\nu} e^{a}_{\,\,\, \mu}   \right)  
 -e^{\nu}_{\,\,\,a} e^{\rho}_{\,\,\, b}    e^c_{\,\,\, \mu} \left( \p_{\nu} e^{c}_{\,\,\, \rho}-
   \p_{\rho} e^{c}_{\,\,\, \nu}  \right) \ri \, . 
\eeq
It is important to observe that the field $\xi$ is auxiliary and can be integrated out by means of the Euler-Lagrange equations of motion 
\beq
\xi=\frac{i(e^\mu_1-i e^\mu_2)\le \tilde{D}_\mu+\frac14 v^\nu \tilde{F}_{\mu \nu}\ri \chi}
{\sqrt{2} \le m+\frac{\tilde{F}_{\mu \nu} e_1^\mu e_2^\nu}{4}\ri}
\label{chichi} \, .
\eeq
Replacing it into the action in eq.~(\ref{Lagrangiana-ferm2}),
we could obtain a cumbersome
Lagrangian written only in terms of $\chi$.
We can have a better understanding of the system by considering some limiting cases. 
 
 
\subsection{Flat spacetime}
\label{sect-Flat spacetime}

The simplest limit to consider is the case where the background is flat, described by eq. (\ref{flatland}) plus $ A_{\mu}=0. $
In this case the covariant derivative reduces to a simple partial derivative and the action for the left-handed Weyl spinor becomes
\beq
\begin{aligned}
\mathcal{L}_W  =  - \sqrt{2} m \xi^{\dagger} \xi 
- \sqrt{2}i \chi^{\dagger} \p_{t} \chi
 +  i \chi^{\dagger} (\p_1 + i \p_2) \xi
+  i \xi^{\dagger} (\p_1 - i \p_2) \chi   \, .
\end{aligned} 
\eeq
The equations of motion are
\beq
\xi = \frac{i}{m} \frac{1}{\sqrt{2}} (\p_1 - i \p_2)  \chi \, , \qquad
\p_t \chi =    \frac{1}{\sqrt{2}} (\p_1 + i \p_2) \xi \, .
 \label{eq-free}
\eeq
In particular, the auxiliary field $\xi$ can be easily integrated out, giving the Schr\"odinger equation for the dynamical component $\chi,$ with action
\beq
S= \int d^3 x \, \le i \chi^{\dagger} \p_t \chi - \frac{1}{2m} |\p_i \chi|^2 \ri \, .
\eeq
The set of equations of motion and the Lagrangian  obtained via null reduction from the $3+1$ dimensional right-handed Weyl spinor are analog to this result, giving another Schr\"odinger equation decoupled from the left-handed component.


\subsection{Gyromagnetic ratio}
\label{sect-Gyromagnetic ratio}

The next limit that we consider is flat spacetime (\ref{flatland}), plus a non-trivial gauge field $A_{\mu} \ne 0$ which accounts for a generic particle number background\footnote{More correctly, this $U(1)$ symmetry in the presence of different 
species of fields $\psi_i$ corresponds to the mass, because in the minimal coupling it enters the action as
$-\sum_i m_i A_0 |\psi_i |^2$, where $m_i$ is the mass of the field $\psi_i$. In the presence of a single species, mass
and particle number are proportional to each other. For simplicity, we refer to this $U(1)$ symmetry as particle number.}.
In this case the covariant derivative contains a gauge connection which arises from the presence of the non-trivial background gauge field in the metric (\ref{null-red metric}). Specializing the general formulas in Appendix \ref{app-conv} to this case, we find that the non-vanishing components of the spin connection are
\beq
\omega_{++i}=-F_{0 i}=-E_i \, , \qquad
\omega_{i+j}=-\frac{1}{2} F_{ij} = -\frac{B}{2} \, , \qquad
\omega_{0ij}=-\frac12 F_{ij} = -\frac{B}{2} \, .
\eeq
The action for the dynamical field $\chi,$ obtained by integrating out the auxiliary component, is given by
\beq
S = \int d^3 x \, \left[ \frac{i}{2} \chi^{\dagger} \overset{\leftrightarrow}{\p_t} \chi - \frac{1}{2m} \delta^{ij} (D_i \chi)^{\dagger} (D_j \chi)  -  \frac{1}{4}B \chi^{\dagger} \chi  \right] \, .  
\eeq
This expression allows to extract the gyromagnetic ratio of a non-relativistic fermion.
First of all, the analogous computation for the decoupled right-handed component gives as the only difference an opposite sign for the $B \chi^{\dagger} \chi$ coupling.
Then the generic form of the gyromagnetic coupling in $2+1$ dimensions is
\beq
\mp g \frac{q}{4 m} B \varphi^\dagger \varphi
\eeq
where $q$ is the charge and 
the $\mp$ sign refers to 
left or right-handed spinor, respectively.
Since in our conventions the charge associated to the particle number 
symmetry is $q=m$, we find a gyromagnetic ratio $g=1$.
This is consistent with the form of the Milne boost
transformations which come from null reduction,
which are valid for $g=2s$ \cite{Geracie:2014nka}:
this is the simplest way in which Galilean covariance can be realized.


\section{Non-relativistic Weyl invariance}
\label{sect-Non-relativistic Weyl invariance}

The actions for the free non-relativistic bosons and fermions that we derived via null reduction are not only invariant under the Galilean group, but also under dilatations and special conformal transformations, which enlarge the symmetries to the Schr\"odinger group.
While this is true in the limiting case of flat space, we expect that the system is also invariant under the local version of this group, obtained with the minimal coupling of the matter fields to a Newton-Cartan background.

In this generic situation, we need to define Weyl transformations for the objects in the curved geometry. In the Lifshitz case (\ref{Dynamical exponent}) we consider the variations
\bea
& n_\mu \arr e^{z \s} n_\mu \, , \qquad
v^\mu \arr e^{-z \s} v^\mu \, , \qquad
A_\mu \arr e^{(2-z)\s} A_\mu \, , & \\
& \sqrt{g} \,  \arr \, e^{(d+z) \s}\sqrt{g} \, , \qquad
h_{\mu \nu} \arr e^{2 \s} h_{\mu \nu} \, , \qquad
h^{\mu \nu} \arr e^{-2 \s} h^{\mu \nu} \, , &
\label{weylbis}
\eea
where $ \sigma=\sigma (x^{\mu}) $ is a spacetime-dependent quantity.

In particular, the Schr\"odinger case is achieved when $z=2$ with corresponding Weyl variations of the NC data given by
\beq
n_\mu \arr e^{2 \s} n_\mu \, , \qquad
v^\mu \arr e^{-2 \s} v^\mu \, , \qquad
h_{\mu \nu} \arr e^{2 \s} h_{\mu \nu} \qquad
h^{\mu \nu} \arr e^{-2 \s} h^{\mu \nu} \, .
\label{weyl1}
\eeq
The action (\ref{action non-relativistic scalar real time}) for the free non-relativistic boson can be seen to be invariant under this set of transformation rules.

A Weyl transformation on the Newton-Cartan background is equivalent to a  Weyl transformation in the extra-dimensional metric
in eq.~(\ref{null-red metric}) which is independent from the $x^-$ coordinate:
\beq
n^A D_A \s =0 \, .
\eeq
The transformations in the set (\ref{weyl1}) can also be derived from the null reduction method by requiring
\beq
G_{MN} \arr e^{2 \s} G_{MN} \, , 
\qquad
G^{MN} \arr e^{-2 \s} G^{MN} \, ,
\qquad
n^A \arr n^A \, , \qquad 
n_A \arr  e^{2 \s} n_A \, .
\label{weyl2}
\eeq
In fact, this transformation of the DLCQ metric is exactly the same which is required in the context of relativistic conformal transformations.

We can also find a corresponding set of variations for the frame fields 
\bea
e^-_M &\arr& e^-_M \, , \qquad
e^+_M \arr e^+_M \, e^{2 \sigma} \, , \qquad
e^a_M \arr e^a_M \, e^{ \sigma}
\nl
e^M_- &\arr& e^M_- \, , \qquad
e^M_+ \arr e^M_+ \, e^{-2 \sigma} \, , \qquad
e^M_a \arr e^M_a \, e^{ -\sigma} \, ,
\eea
and consequently for the spin connection
\bea
\label{weyl-spinc}
\omega_{-ab} & \arr &  \omega_{-ab} \, , \qquad 
 \omega_{-+ a}  \arr e^{-\s} ( \omega_{-+ a} + e^\nu_a \p_\nu \s ) \, , \qquad
 \omega_{\mu-a} \arr e^\s (  \omega_{\mu-a}+ n_\mu e^\nu_a \p_\nu \s) \, ,
\nl
 \omega_{\mu - + } & \arr & \omega_{\mu - + } -\p_\mu \s +n_\mu v^\nu \p_\nu \s \, ,
\qquad
\omega_{\mu+a}\arr e^{-\s} \left( \omega_{\mu+a} +\le - v^\nu  e^a_\mu 
+e^\nu_a A_\mu \ri  \p_\nu \s \right)  \, ,
\nl
\omega_{\mu a b} & \arr &
\omega_{\mu a b}+  \le 
 e^a_\mu e^\nu_b 
-e^b_\mu e^\nu_a 
\ri \p_\nu \s \,  .
\eea
It is evident that the various components of the frame fields and of the spin connection change differently under Weyl transformations.
A similar situation happens for the components of the left-handed Weyl spinor, contrarily to the relativistic case.
The transformation of the $(\xi,\chi)$ components is as follows:
 \beq
\xi \arr e^{-2 \s} \xi \, , \qquad \chi \arr e^{-\s} \chi \, .
\label{weyl-weight-fermions}
\eeq
 This can be derived from dimensional analysis in the flat case,
see eq.~(\ref{eq-free}): in units of length, $[\varphi] = -1$  and $[\chi] = -2 $.
In the case of a Dirac fermion
\beq
\Psi = \begin{pmatrix}  \xi_L \\ \chi_L \\ \chi_R \\ \xi_R    \end{pmatrix} \, ,
\qquad  {\rm length \,\,\, dimensions \,\,\, are } \qquad
[\Psi ] = \begin{pmatrix} -2 \\ -1 \\ -1 \\ -2   \end{pmatrix} \, .
\eeq
The different length dimension of the components arises due to the particular behviour of the tetrads, and because one of them is auxiliary and the other is dynamical. 

We note that this Weyl weight choice
 is crucial in order to assign to the term $\bar{\Psi} \Psi $
a well-defined Weyl weight. 
A conformal coupling term such as
$R \bar{\Psi} \Psi$ would have mass dimension $5$,
spoiling conformal invariance.

It is possible to check that the action in eq.~(\ref{Lagrangiana-ferm}) for the non-relativistic free fermion is Weyl invariant, provided that eqs~(\ref{weyl-weight-fermions}) and (\ref{chichi}) are used.
One can also verify that this is consistent with eq.~(\ref{chichi}): if we insert $\chi \arr e^{-\s} \chi$, we indeed find that
$\xi \arr e^{-2 \s} \xi$.


\section{General classification of the non-relativistic trace anomaly}
\label{sect-General classification of the non-relativistic trace anomaly}

In this section we consider the problem of classifying the terms entering the trace anomaly for a Schr\"odinger-invariant field theory in 2+1 dimensions coupled to a NC geometry.
We will briefly review the procedure to determine such classification and we will summarize the results found in the literature \cite{Jensen:2014hqa, Arav:2016xjc, Auzzi:2015fgg, Auzzi:2016lrq}.

The general form of the trace anomaly on a curved background can be found by solving a cohomological problem.
The following steps need to be applied:
\begin{enumerate}
\item We parametrized the most generic Weyl variation as
\beq
\delta_{\sigma} W = \int d^{d+1} x \, \sqrt{g} \, \sigma (x) \mathcal{A} (x) \, ,
\eeq
where $\mathcal{A}$ is a scalar built from NC data which is invariant under the non-relativistic symmetries: diffeomorphisms, gauge transformations and Milne boosts.
\item We express the most general expression for $\mathcal{A}$ as a linear combination of a basis of independent terms.
\item We impose the Wess-Zumino consistency conditions
\begin{equation}
 \Delta_{\sigma_{1}\sigma_{2}}^{\mathrm{WZ}} W =  \delta_{\sigma_{1}(x)} 
\int d^{4}y \, \sqrt{-G} \, \mathcal{A} (y) \sigma_{2} (y)   
-  \delta_{\sigma_{2}(x)} 
\int d^{4}y \, \sqrt{-G} \, \mathcal{A} (y) \sigma_{1} (y)  =0  \, .
\end{equation}
\item We eliminate from the basis the terms that are exact in the cohomology (\emph{i.e.} they can be written as the Weyl variation of other terms in the basis).
\end{enumerate}

The power of the null reduction method is that we can apply this procedure using the tensors in the relativistic parent theory, and the scalars built in this way automatically satisfy the invariances required in the point 1 of the previous procedure.
While we can take various results from the 3+1 dimensional relativistic case\footnote{In fact the scalars built from the metric and the corresponding Levi-Civita connection are formally the same of the usual relativistic case. We only need to remark that curvature invariants secretely contain the NC data, since they appear in the metric used for the null reduction.}, here the important difference stays in the additional vector $n_M .$ 

It turns out that the space of expressions with uniform scaling dimension and invariant under the
symmetries of the non-relativistic theory can be divided into distinct sectors invariant under Weyl transformations.
These sectors are distinguished by the number of appearances of $n :$ all the terms with a fixed number of factors of $n$ transform into each other under Weyl transformations.
The cohomological problem can be studied separately for each sector.

The classification of the trace anomaly changes drastically if a causal structure on the NC geometry is required.
This technically amounts to imposing the Frobenius condition on the one-form identifying the local time direction
\beq
n \wedge dn = 0 \, ,
\eeq
which defines an integrable structure.

If Frobenius condition is applied, the possible scalars entering the anomaly collapse to only one sector, and then they
compose a finite set. Unfortunately, the Euler density $E_4$ can be written as a linear combination of other DLCQ scalars, and type A anomalies disappear, precluding the existence of an $a-$theorem.
The only independent term can be chosen to be the null reduction of the squared Weyl tensor, plus scheme-dependent scalars that can be eliminated with an appopriate choice of counterterms
\beq
\mathcal{A} = b W^2_{MNPQ} + \mathcal{A}_{\rm ct} \, .
\eeq
If the Frobenius condition is not required, there are still infinite sectors and we can study them separately.
There is a minimal sector without appearances of $n$ which is the null reduction of the 3+1 dimensional relativistic case
\beq
\mathcal{A}^{0} = a E_{4} - c W_{MNPQ}^{2} + \mathcal{A}_{\mathrm{ct}} \, , 
\eeq
The coefficient of the Euler density is then a good candidate for a non-relativistic version of the $a-$theorem.

Instead it is possible to prove that the next sector with a single appearance of $n$ has vanishing trace anomaly 
\beq
\mathcal{A}^{1} = 0 \, . 
\eeq
The situation in the successive sectors is still not clear: by  dimensional analysis, for each $n_M$ we can add one extra DLCQ covariant derivatives\footnote{Since the DLCQ Riemann tensor is the commutator of two covariant derivatives, two $n_M$ are needed in order to buy a curvature} $D_M .$
Examples of such terms which can enter the anomaly are
 \beq
 n^M D_M R_{NP} R^{NP} \, , \qquad
 R_{MNPQ} R^{MNPS} R^Q_{\, \, \, \, TSU} n^T n^U \, .
\label{anoma-esempio}
 \eeq
However, the cohomological problem in these sectors is not studied and then we do not know if type A anomalies appear.


\section{Trace anomaly near a flat background}
\label{sect-Trace anomaly near a flat background}

The general procedure for the classification of the trace anomaly allows to find a basis of curvature invariants, but does not identify the coefficients with which they appear in specific systems, in particular we do not know if some terms do not enter at all the trace anomaly.
In principle, it is also possible that all the coefficients of the linear combination vanish and that the trace anomaly is exactly zero!
In order to avoid or to investigate this possibility, we will study in chapter \ref{chapt-The heat kernal technique} the trace anomaly in specific cases with the heat kernel technique, a method which gives precisely the coefficients of the terms entering the trace anomaly.
Tipically the heat kernel procedure is performed going nearby flat space; here we show how to treat variations of the background fields of the NC geometry.

Due to the conditions (\ref{1st condition NC geometry}) and (\ref{2nd condition NC geometry}), arbitrary variations of the geometric data are not allowed but we can parametrize them with an arbitrary $\delta n_{\mu},$ and the transverse perturbations $\delta u^{\mu}$ and $\delta \tilde{h}^{\mu\nu}$ such that
\beq
\delta u^{\mu} \, n_{\mu} = 0 \, , \qquad
\delta \tilde{h}^{\mu\nu} n_{\nu} = 0 \, .
\eeq
This means that the variation of the metric fields are
 \beq
 \delta n_\mu \, , \qquad
 \delta v^\mu = - v^\mu v^\a \delta n_\a + \delta u^\mu \, , \qquad
 \delta h^{\mu \nu} =-v^{\mu} \delta n^{\nu}-\delta n^\mu v^\nu 
 -\delta \th^{\mu \nu} \, .
 \label{varii}
 \eeq
If we specialize to a variation around flat space, which is the case of interest for the heat kernel expansion, these variations take the form
 \bea
 n_\mu  &=&  (1+\delta n_0 , \delta n_i) \, , \qquad
 v^\mu =(1-\delta n_0, \delta u_i) \, , \qquad \delta \th^{0 i}=0 \, ,
 \nl
  h_{\mu \nu} &=&
  \left(
\begin{array}{cc}
 0& -\delta u_i \\ -\delta u_i  &  \delta_{ij} + \delta \th_{ij} \\
  \end{array}\right) \, , \qquad
h^{\mu \nu}=
  \left(
\begin{array}{cc}
 0& -\delta n_i \\ -\delta n_i  &  \delta_{ij} - \delta \th_{ij} \\
  \end{array}\right) \, ,
\eea
which can be written in terms of the parent metric as
 \bea
G_{MN} & = &
  \left(
\begin{array}{ccc}
0 & 1 + \delta n_0 &  \delta n_i \\
1+\delta n_0 & 2 \delta A_0 & \delta A_i -\delta u_i \\ 
\delta n_i  & \delta A_i -\delta u_i &\delta_{ij} + \delta \th_{ij} \\
  \end{array}\right) \, ,
  \nl
  G^{MN} & = &
  \left(
\begin{array}{ccc}
-2 A_0 & 1- \delta n_0 & -\delta A_i+ \delta u_i \\
1-\delta n_0 &  0& -\delta n_i \\ 
-\delta A_i+ \delta u_i  &  -\delta n_i&\delta_{ij} + \delta \th_{ij} \\
  \end{array}\right) \, .
  \label{perturba}
\eea
These sources are used to define conserved currents, in particular the ones entering the energy-momentum tensor multiplet through the variation of the vacuum functional
\beq
\delta W = \int d^d x\sqrt{-g} \le
\frac{1}{2} T_{i j} \delta \th_{i j} + j^\mu \delta A_\mu
-\epsilon^\mu \delta n_\mu - p_i \delta u_i 
\ri \, ,
\eeq
where $p_i$ is the momentum density,
$T_{i j}$ is the spatial stress tensor, $j^\mu=(j^0,j^i)$ contains
the number density and current and $\epsilon^\mu=(\epsilon^0,\epsilon^i)$
 the energy density and current. The $U(1)$ number current is proportional to  the momentum density\footnote{This is a direct consequence of
eq.~(\ref{perturba}), because only the combination
 $\delta A_i - \delta u_i$ enters the DLCQ metric}.

This decomposition allows to find the Ward identities associated to the various symmetries of the non-relativistic theory.
Particle number conservation implies the conservation of the $U(1)$ current
\beq
\langle \p_\mu  j^\mu \rangle =0  \ .
\eeq
Associated to diffeomorphism invariance there are the conservation of the spatial stress tensor and of the energy current 
\beq
\label{diffcurrent}
\langle \p_t p^j  + \p_i T^{ij }\rangle =0 \ , \quad \langle \p_\mu  \epsilon^\mu\rangle=0 \, .
\eeq 
Finally, local Weyl transformations entail the Ward identity associated to the conservation of the scale current,  
which is found to be\footnote{Strictly speaking, the scale current has an additional term proportional to the scaling dimension $\Delta$ of the matter field. However, such term is a total derivative and can always be reabsorbed by a current redefinition. }
\beq
\label{defscalecurrent}
J_S^0 = p_i x^i - 2t \epsilon^0 \, , \quad
J_S^i =  x^j T^i_{\,\,\, j} -2t \epsilon^i  \, ,  \quad \langle \p_\mu  J_S^\mu \rangle =0\ .
\eeq
By expanding explicitly the scale Ward identity we have
\beq
\langle \p_\mu J_S^\mu \rangle=
\langle T^i_{\,\,\, i} - 2 \epsilon^0\rangle - 2t \langle \p_\mu  \epsilon^\mu\rangle  + 
x^j \langle  \p_t  p_j + \p_i  T^{i }_{\ j}\rangle  = 0 \, .
\label{Classical scale nonrel}
\eeq
Equation (\ref{Classical scale nonrel}) is interesting, because it explicitly shows the relations intertwining 
 between tracelessness of the energy-momentum tensor,
conservation of the energy momentum tensor and scale invariance. 
A quantum violation of the scale symmetry manifests 
as a non conservation of the scale current $J_S^\mu$ which, in turn,  is equivalent to  a violation of the tracelessness condition
 $\langle T^i_{\,\,\, i} - 2 \epsilon^0\rangle =0$ only if the energy-momentum tensor does not have a
 diffeomorphism anomaly, {\it i.e.} only if the conditions (\ref{diffcurrent}) are satisfied. 
 On the other hand, if the energy momentum tensor  is not conserved at the quantum level, not only the trace anomaly, but also the diffeomorphism anomaly  contribute to the scale anomaly. 

If the diffeomorphisms ar chosen to be preserved\footnote{Since the diffeomorphisms are a gauge transformation, it is essential to preserve them in order to avoid the loss of unitarity in the theory.}, we can derive the Ward identity giving the tracelessness condition for the energy-momentum tensor by performing a Weyl variation nearby flat space
\beq
\delta W= \s G_{MN} \frac{\delta W}{\delta G_{MN}} = 
\s \le \delta^{ij} \frac{\delta W}{\delta (\delta \th_{ij}) } +
2 \frac{\delta W}{\delta (\delta n_0)} \ri =
\s( T^i_i -2 \epsilon^0)  \, ,
\eeq
which vanishes in the classical case or in flat space.
In general the non-vanishing of this expression is the trace anomaly, which in $2+1$ dimensions for $z=2$ can be parametrized in terms of DLCQ quantities as
\beq
\Delta W=\int \sqrt{g} d^3 x\  \s 
\left( -a E_4 + c W^2 + b R^2 + d D_A D^A R +e R^{AB}_{\,\,\,\,\,\,\, CD} 
R_{ABEF} \frac{\epsilon^{CDEF}}{\sqrt{g}}
\right) + \dots
\label{anoma}
\eeq  
Apart from the fact that we are performing a null reduction of the higher-dimensional tensors, the terms in parenthesis are exactly the same of the $3+1$ dimensional relativistic case, and $a,c,e$ correspond to anomaly coefficients, while $b=0$ from the Wess-Zumino consistency conditions \cite{Bonora:1983ff} and $d$ can be removed by local counterterms.
The dots in eq.~(\ref{anoma}) correspond to an infinite number of possible
terms with a higher number of derivatives, which however belong to 
other Weyl sectors.


\chapter{The heat kernel technique}
\label{chapt-The heat kernal technique}

\begin{center}
\emph{Most of the content of this chapter appeared previously in \cite{Auzzi:2017jry,Auzzi:2017wwc}.}
\end{center}

The heat kernel is a mathematical tool which has powerful applications in QFT to study the Casimir effect, effective actions, quantum anomalies and many other quantities.
In this chapter we will start with a review of how the heat kernel can be used to renormalize the one-loop effective action and to extract quantum anomalies.
Then we will treat how this procedure applies in the non-relativistic case, and we will use the technique to investigate the trace anomaly for a NC background in specific examples.
The main references are \cite{Vassilevich:2003xt, Mukhanov:2007zz, Solodukhin:2009sk}.


\section{General procedure}
\label{sect-General procedure}

In this section we set up the general procedure to find the vacuum generating functional in curved space $ W[J=0, g_{\mu\nu}] $ corresponding to the kinetic operator of a given Lagrangian action.
This will be the starting point to derive a relation between the energy-momentum tensor and a set of coefficients coming from a series expansion of the so-called heat kernel operator.
Ultimately, this leads to a precise way to compute the trace anomaly.
In this section we review this procedure in the relativistic case by following the approach of  \cite{Vassilevich:2003xt}.


\subsection{The heat kernel and the zeta function operators}
\label{sect-The heat kernel and the zeta function operators}

\begin{defn}
Given a \emph{n}-dimensional  Riemannian manifold $ \mathcal{M} $  where is defined a  self-adjoint and elliptic\footnote{An elliptic operator has at most a finite number of zero and negative modes.} differential operator $\mathcal{D}, $  the heat kernel operator is defined as
\beq
K(\tau, \mathcal{D}) =  \exp \le -\tau \mathcal{D} \ri \, ,
\eeq
and the zeta function of the operator $\mathcal{D}$ as
\beq
\zeta(s, \mathcal{D}) = \mathrm{Tr} ( \mathcal{D}^{-s} ) \, .
\eeq
\end{defn}
The zeta function and the heat kernel operators are related by a Mellin transformation via the Euler Gamma function
\beq
\zeta(s, \mathcal{D}) =
\Gamma(s)^{-1} \int_0^{\infty} d \tau \, \tau^{s-1} \mathrm{Tr} \left[ K(\tau, \mathcal{D}) \right]  \, .
\eeq
This relation can be inverted giving
\beq
\mathrm{Tr}\left[  K(\tau,\mathcal{D})\right]  = \frac{1}{2 \pi i} \int ds \, \tau^{-s} \Gamma(s) \zeta(s,\mathcal{D}) \, ,
\label{inverse Mellin heat kernel and zeta function}
\eeq
where the integration contour encircles all the poles of the integrand.

If $\mathcal{M}$ is a manifold without boundaries, there exists an asymptothic expansion of the trace of the heat kernel operator which takes the form
\beq
\mathrm{Tr}\left[  K(\tau,\mathcal{D})\right] = \sum_{k=0}^{\infty} \tau^{\frac{k-n}{2}} a_k (\mathcal{D}) \, ,
\label{definition Seeley-De Witt coefficients}
\eeq
where the $a_k$ are called Seeley-De Witt coefficients.
This set of coefficients will play a fundamental role in determining the trace anomaly, because they can be locally computed in most physical cases in terms of the volume and of boundary integrals of local invariants.

Using this series expansion inside the integral transformation (\ref{inverse Mellin heat kernel and zeta function}), we find a representation of the Seleey-De Witt coefficients in terms of the zeta function 
\beq
a_k (\mathcal{D}) = \mathrm{Res}_{s=\frac{n-k}{2}} \left[ \Gamma(s) \zeta(s,\mathcal{D}) \right] \, ,
\eeq 
with the particular case
\beq
 a_n (\mathcal{D}) = \zeta(0, \mathcal{D}) \, . 
 \label{Seeley-De Witt coefficient a_n}
\eeq


\subsection{Renormalization of the effective action}
\label{sect-Renormalization of the effective action}

Let's turn to the physical part of the problem.
We consider a QFT for a generic field $\phi$ with generating functional
\beq
Z[J] = e^{-W[J]} = \int \mathcal{D} \phi \, \exp \left( - S[\phi,J] \right) \, .
\label{general path integral of the theory}
\eeq
The heat kernel method is most suited for the investigation of one-loop properties of a system, and as such we take an approximation where we expand the action up to second order in the quantum fluctuations of the field
\beq
S [\phi, J]= S_{\mathrm{cl}} + \langle \phi, J \rangle + \langle \phi , \mathcal{D} \phi  \rangle + \mathcal{O}(\phi^3) \, ,
\label{gaussian approximation action}
\eeq
where $S_{\mathrm{cl}}$ is the action on the classical background and the bracket $\langle \dots \rangle$ denotes an inner product in the space of quantum fields. For example, the inner product involving a real scalar field $\varphi$ is
\beq
\langle \varphi J  \rangle = \int_{\mathcal{M}} d^n x \, \sqrt{g} \, \varphi(x) J(x) \, .
\eeq
The linear term in the expansion (\ref{gaussian approximation action}) contains both a contribution from external sources and the first variation of the action, the latter piece vanishing on the classical equations of motion.
The differential operator $\mathcal{D}$ modulates the quadratic quantum fluctuations of the action.

Within the gaussian approximation (\ref{gaussian approximation action}), the path integral (\ref{general path integral of the theory}) can be solved exactly. 
In the specific case of a real scalar field the result is
\beq
Z_{\mathrm{scalar}}[J] = e^{-S_{\mathrm{cl}}} \left[ \det(\mathcal{D})\right] ^{- \frac12} \exp \le \frac{1}{4} J \mathcal{D}^{-1} J \ri \, ,
\eeq
while for other theories containing complex scalars, Dirac spinors or other fields the result usually involves different powers of the functional determinant of the differential operator $\mathcal{D}.$
From now on, we will use the numerical factors referred to the case of a real scalar field, while the other aspects of the derivation will be completely general.

The part of the gaussian expansion (\ref{gaussian approximation action}) involving the differential operator $\mathcal{D}$ contributes to the so-called one-loop effective action
\beq
W_{\mathrm{eff}} = \frac12 \log \le \det \mathcal{D} \ri \, .
\eeq
There exists an integral representation of the logarithm which relates the one-loop effective action to the heat kernel. Given a generic positive eigenvalue $\lambda$ of the differential operator\footnote{Since the operator is required to be elliptic, there exists at most a finite number of non-positive modes.} $\mathcal{D},$ we have
\beq
\log \lambda = - \int_0^{\infty} \frac{d \tau}{\tau} e^{- \tau \lambda} \, . 
\eeq
This relation holds up to an infinite constant which, being independent from $\lambda,$ we will ignore.
Using the identity $ \log \le \det \mathcal{D} \ri = \mathrm{Tr} \le \log \mathcal{D} \ri , $ we can extend the previous equation to the trace of the heat kernel operator, finding
\beq
W_{\mathrm{eff}} = - \frac12 \int_{0}^{\infty} \frac{d\tau}{\tau} \, \mathrm{Tr} \left[K(\tau, \mathcal{D}) \right]  \, .
\eeq
As it happens for a particular eigenvalue $\lambda,$ this identity is formally true up to an infinite constant.
In order to regularize the effective action, we consider a shift of the power in the denominator 
\beq
W_{\mathrm{reg}} (s) = - \frac12 \mu^{2s} \int_0^{\infty} \frac{d \tau}{\tau^{1-s}} \, \mathrm{Tr} \left[ K(\tau, \mathcal{D}) \right] \, , 
\eeq
where $\mu$ is a mass scale introduced to account for the different dimensions of the integrand with respect to the case $s=0.$
Using the inverse Mellin transformation (\ref{inverse Mellin heat kernel and zeta function}), the regularized effective action can be expressed as
\beq
W_{\mathrm{reg}} (s)= - \frac12 \mu^{2s} \Gamma(s) \zeta(s,\mathcal{D}) \, , 
\eeq
from which the name \emph{zeta function regularization}.
The Laurent expansion of the Euler Gamma function around 0
\beq
\Gamma(s) = \frac{1}{s} - \gamma_E + \mathcal{O}(s) 
\eeq
allows to determine 
\beq
W_{\mathrm{reg}}(s)= - \frac12 \le \frac{1}{s} - \gamma_E + \log \mu^2 \ri \zeta(0,\mathcal{D}) - \frac12 \zeta' (0,\mathcal{D}) + \mathcal{O}(s) \, ,
\eeq
where $ f' \equiv \frac{d}{ds} f . $
This shows that the divergence of the one-loop effective action arises due to a simple pole located at $s=0, $ and precisely this contribution gives a divergence in the Seeley-De Witt coefficient $a_n(\mathcal{D})$ via eq. (\ref{Seeley-De Witt coefficient a_n}).
This result requires a renormalization procedure which eliminates the divergent part, giving the renormalized effective action as the remaining part at $s=0,$ \emph{i.e.}
\beq
W^{\mathrm{ren}} = - \frac12 \zeta' (0,\mathcal{D}) - \frac12 \log(\tilde{\mu}^2)\zeta(0,\mathcal{D}) \, ,
\label{renormalized effective action via zeta function regularization}
\eeq
where   $ \tilde{\mu}^2 = e^{-\gamma_E} \mu^2 .  $


\subsection{Relation with the trace anomaly}
\label{sect-Relation with the trace anomaly}

The energy-momentum tensor of a QFT coupled to a curved background is defined as
\beq
T_{\mu\nu} =  \frac{2}{\sqrt{g}} \frac{\delta W}{\delta g^{\mu\nu}} \, .
\eeq
Let's consider a conformal transformation
\beq
g_{\mu\nu} \rightarrow e^{2 \sigma (x)} g_{\mu\nu} \, ,
\eeq
its infinitesimal version allows to express the variation of the effective action under this change of the metric as
\beq
\delta W =  \frac12 \int_{\mathcal{M}} d^n x \, \sqrt{g} \, T^{\mu\nu} \delta g_{\mu\nu} =
-  \int_{\mathcal{M}} d^n x \, \sqrt{g} \, \sigma (x) T^{\mu}_{\,\,\, \mu} \, .
\label{variation of the effective action under a conformal transformation}
\eeq
This result clearly shows that the invariance of a theory under conformal transformations can be expressed as the vanishing of the trace of the energy-momentum tensor.
The opposite case, when $T^{\mu}_{\,\,\, \mu} \ne 0,$ signals the quantum breaking of the conformal symmetry and goes under the name of \emph{trace anomaly}.
 
Our final aim is to relate the trace anomaly to the Seeley-de Witt coefficients.
We start with a general variation of the zeta function operator
\beq
\delta \zeta(s,\mathcal{D}) = -s \mathrm{Tr}\left[  (\delta \mathcal{D}) \mathcal{D}^{-s-1} \right]   \, .
\eeq
If the classical action is conformally invariant, the differential operator is conformally covariant and then transforms under a Weyl variation as\footnote{This assertion is not so simple to derive as it may appear, due to the classical conformal invariance of the problem.
In fact, the relation is strictly speaking true only after a similarity transformation $ \mathcal{D} \rightarrow e^{\alpha \sigma(x)} \mathcal{D} e^{- \alpha \sigma (x)}  $ which leaves the functional determinant invariant. A careful treatment of this aspect will be done in the specific computation of the diffeomorphism anomaly, to which the same method applies.}
\beq
\mathcal{D} \rightarrow e^{-2 \sigma(x)} \mathcal{D} \, .
\eeq
In this way, the variation of the zeta function under a conformal transformation is
\beq
\delta \zeta(s,\mathcal{D}) =
2 s \sigma(x) \mathrm{Tr} \mathcal{D}^{-s} = 2 s \sigma (x)  \zeta(s,\mathcal{D})  \, ,
\eeq
and this induces a change of the one-loop renormalized effective action (\ref{renormalized effective action via zeta function regularization}) given by
\beq
\delta W_{\mathrm{ren}} = -  \sigma (x) \zeta(0,\mathcal{D}) = - \sigma (x) a_n (\mathcal{D}) = - \int_{\mathcal{M}} d^n x \, \sqrt{g} \, \sigma  a_n (x,\mathcal{D}) \, .
\eeq
Comparing with eq. (\ref{variation of the effective action under a conformal transformation}) we obtain
\beq
T^{\mu}_{\,\,\, \mu} (x)=  a_n (x,\mathcal{D}) \, . 
\eeq
Stated in this way, the problem to determine the trace anomaly is reduced to the computation of the Seeley-De Witt coefficient $ a_n (x, \mathcal{D}). $

\subsection{Computation of the Seeley-De Witt coefficients}

We restrict to a class of second order operators of the Laplace type, \emph{i.e.} that can be represented as 
\beq
\mathcal{D} = - \le g^{\mu\nu} \p_{\mu} \p_{\nu} + a^{\mu} \p_{\mu} + b \ri 
\label{differential operator of the Laplace type}
\eeq
with an appropriate choice of the matrix valued functions $a^{\mu}, b .$
The last expression can be further decomposed as a perturbation of a reference operator $\mathcal{D}_0$ as
\beq
\mathcal{D} = \mathcal{D}_0 + \delta \mathcal{D} \, ,
\label{decomposition heat kernel general}
\eeq
where a convenient choice is $\mathcal{D}_0 = -\square,$ when expanding the metric around flat space as
\beq
g_{\mu\nu} = \delta_{\mu\nu} + h_{\mu\nu} \, .
\eeq
In order to study the heat kernel of the differential operator $\mathcal{D}$ defined on the curved manifold $\mathcal{M},$ it is convenient to work with an inner product which does not involve the determinant of the metric in the measure.
This requirement leads to a normalization of the eigenstates given by
\beq
\langle  x t | x' t' \rangle_g = \frac{\delta(x-x') \delta(t-t')}{\sqrt{g}} \, ,
\eeq
and to the definition of another differential operator
\beq
O = g^{1/4} (x) \le \mathcal{D}_0 + \delta \mathcal{D} \ri g^{-1/4} (x) = - \le \square + V \ri  \, ,
\eeq
where $g$ is the determinant of the metric.
The last decomposition of the new hermitian operator $O$ in terms of a perturbation of flat space is always possible going around a locally inertial frame. 
An important implication of this choice for the inner product and the Hilbert space where the operator $O$ lives is that the expansion giving the Seeley-De Witt coefficients takes a factor in the determinant of the metric
\beq
\mathrm{Tr}\left[  K(\tau,O)\right] = \sum_{k=0}^{\infty} \tau^{\frac{k-n}{2}} a_k (O) =
\sum_{k=0}^{\infty} \sqrt{g} \, \tau^{\frac{k-n}{2}} a_k (\mathcal{D}) =
\frac{\sqrt{g}}{\tau^{n/2}} \left[ 1+ a_2 (\mathcal{D}) \tau + a_4(\mathcal{D}) \tau^2 + \mathcal{O}(\tau^3) \right]
\, .
\label{new definition Seeley-De Witt coefficients}
\eeq

We find the implications of this perturbative expansion on the heat kernel operator.
Given the requirements of $\mathcal{D}$ to be self-adjoint and elliptic, the heat kernel is analytic and can then be Taylor-expanded around $\tau=0,$ giving
\begin{equation}
K(\tau) = \sum_{i=0}^{\infty} K_i (\tau)  \, ,
\end{equation}
where $ K_n (\tau) $ is an operator of \emph{n}-th order in the perturbation $V.$
By construction, the heat kernel of such a differential operator solves the exact differential equation
\begin{equation}
\frac{dK}{d\tau} = (\square + V) K \, .
\end{equation}
The assumption that the deviation from flat metric and the potential are small translates into the fact that we can solve the differential equation order by order and we can truncate the series.
The leading term is
\begin{equation}
\frac{dK_0}{d\tau} = \bigtriangleup K_0  \, ,
\end{equation}
while the first-order term is
\begin{equation}
\frac{dK_1}{d\tau} = \bigtriangleup K_1 + V K_0 \, .
\label{2.9 Equazione calore al I ordine}
\end{equation}
Remembering that the initial condition is $ K(\tau) = \mathbf{1} , $ we infer the initial condition for the 0-th and 1-st order of the heat kernel expansion, which are $ K_0 (\tau) = \mathbf{1} $ and $ K_1 (\tau) =0 . $ 
Thus the formal solution of the equation for $ K_0 $ is
\begin{equation}
K_0 (\tau ) = \exp(\tau \square)  \, .
\end{equation}
The equation for the first-order term can be solved using the method of variation of constants by searching solutions of the form
\begin{equation}
K_1 (\tau) = K_0 (\tau) C(\tau)  \, .
\end{equation}
Putting this trial function in eq. (\ref{2.9 Equazione calore al I ordine}) gives
\begin{equation}
K_0 (\tau) \frac{dC(\tau)}{d \tau} = V K_0 (\tau)  \, ,
\end{equation}
whose solution is
\begin{equation}
C(\tau) = \int_0^{\tau} d\tau' \, K_0^{-1} (\tau') \, V \, K_0 (\tau')  \, . 
\end{equation}
In order to avoid using the inverse of the heat kernel at zero-th order to appear in the solution, we observe that the solution for $ K_0 (\tau) $ satisfies the relation
\begin{equation}
K_0 (\tau) K_0 (\tau') = K_0 (\tau + \tau') \, .
\end{equation}
We can in particular choose the proper times $ \tau+ \tau' = 0 , $ so that $ K_0 (0) = \mathbf{1} $ and we derive
\begin{equation}
K_0^{-1} (\tau) = K_0 (- \tau)  \, .
\end{equation}
This allows to write the final result for $ K_1 (\tau) $ as
\begin{equation}
K_1 (\tau) = \int_0^{\tau} d \tau' \, K_0 (\tau - \tau') \, V \, K_0 (\tau')  \, . 
\end{equation}
The procedure can be applied recursively order by order in the perturbative expansion.
The result for the $i$-th order of the series is
\beq
K_i (\tau) = \int_0^{\tau} d \tau_i  \int_0^{\tau_i} d \tau_{i-1}  \dots 
\int_0^{\tau_2} d \tau_1
 \, K_0 (\tau - \tau_i) \, V \, K_0 (\tau_i - \tau_{i-1}) V \dots
K_0 (\tau_2 - \tau_1)  V K_0 (\tau_1)    \, .
\label{Dyson expansion heat kernel}
\eeq
Using this perturbative expansion, we can determine the Seeley-De Witt coefficients by means of eq. (\ref{definition Seeley-De Witt coefficients}) order by order in $V.$
In particular, the $a_n$ coefficient gives the trace anomaly. 

Here we put a remark: using this series expansion we find the expression of the quantum anomaly up to a chosen order in the parameter of the expansion.
However, in the case of the trace anomaly, we know that the exact expression must be a scalar under diffeomorphisms.
Having this hint, we will interpret the perturbative results in terms of the curvature invariants, and then we will infer that the expression obtained up to the chosen order of the expansion is valid at all orders.

\subsection{Heat kernel in flat space}

The zero-th order term of the heat kernel expansion refers to the flat space solution, which can be solved analytically.
In configuration space we have
\begin{equation}
\langle x | K_0 (\tau) | y \rangle = 
\langle x | e^{\tau \bigtriangleup} | y \rangle = 
e^{\tau \bigtriangleup_x} \delta (x-y) \, .
\end{equation}
If we replace the Delta function with its integral representation
\begin{equation}
\delta(x-y) = \int \frac{d^n k}{(2 \pi)^n} \, e^{ik(x-y)} \, ,
\end{equation}
we find
\begin{equation}
e^{\tau \square_x} \delta (x-y)= e^{\tau \square_x} \int \frac{d^n k}{(2 \pi)^n} \, e^{ik(x-y)} =
\int \frac{d^n k}{(2 \pi)^n} \, e^{-\tau k^2 + ik(x-y)}  \, .
\end{equation}
Since the final expression is a Gaussian integral, we can directly compute it finding
\begin{equation}
G_0 (x,y, \tau) = \langle x | K_0 (\tau) | y \rangle = 
\frac{1}{(4 \pi \tau)^{n/2}} \exp \left[ - \frac{(x-y)^2}{4 \tau} \right]  \, .
\end{equation} 
This is simply the Green function of the heat kernel equation in \emph{n} dimensions. 
In particular it contributes to the trace via the expression
\begin{equation}
G_0 (x,x,\tau) = \langle x | K_0 (\tau) | x \rangle = 
\frac{1}{(4 \pi \tau)^{n/2}}  \, .
\label{Trace of flat space rel heat kernel}
\end{equation}
Putting this exact solution into the Dyson expansion (\ref{Dyson expansion heat kernel}), we can find the perturbative corrections order by order in $V.$

\section{Non-relativistic heat kernel}

The derivation of the heat kernel technique in section \ref{sect-General procedure} assumes the existence of a differential operator of the Laplace type which we can expand as in eq. (\ref{differential operator of the Laplace type}).

What happens in the non-relativistic case? Is it always possible to express a Galilean-invariant action as an hermitian and elliptic operator, and expand with respect to a well-defined flat operator?
The first problem to face is that Euclidean space is not well-defined in the non-relativistic case, and then we need to give a prescription for a consistent way to perform the analogous of a Wick rotation.
We consider as an example the action for a Galilean-invariant free scalar in $d+1$ dimensions coupled to a NC gravity with $A_{\mu}=0,$ see eq. (\ref{non-relativistic free scalar with A=0}). 
The analog of the Wick rotation in this case would require to send
\beq
t \rightarrow - i t_E \, , \qquad
m \rightarrow i m_E \, ,
\eeq
according to the prescription of \cite{Solodukhin:2009sk}.
In the following, we will omit the subscript $E$ referring to Euclidean space.
The equivalent prescription to apply in curved space is to send
\beq
v^{\mu} \rightarrow i v^{\mu} \, , \qquad
m \rightarrow i m \, , \qquad
n_{\mu} \rightarrow - i n_{\mu} \, , \qquad
\sqrt{g} \rightarrow i \sqrt{g} \, .
\label{Wick rotation non-relativistic case}
\eeq
In the previous section we were able to consider a local inertial frame to expand the differential operator as in eq. (\ref{decomposition heat kernel general}). 
In the non-relativistic case, the Schr\"odinger operator in Euclidean space around which we expand the solution is not an elliptic operator, being defined as
\beq
- 2 i m \p_t + \p_i^2 \, .
\eeq 
The only possible positive-definite operator that we can build out of it arises from the interpretation
\beq
 \bigtriangleup_E  \equiv - 2 m \sqrt{- \p_t^2} + \p_i^2 \, , \qquad
 \mathcal{D}_0 = -\bigtriangleup_E
 \label{Euclidean Schr\"odinger operator}
\eeq
where we only take the positive branch cut of the square root.
Using the rules (\ref{Wick rotation non-relativistic case}), we obtain an Euclidean version for the non-relativistic free scalar given by
\beq
S_E = \int d^{d+1} x \, \sqrt{g} \, \varphi^{\dagger} \left\lbrace  m v^{\mu} \sqrt{-\p^2_{\mu}} \varphi + \frac{ m \sqrt{-\p^2_{\mu}} (\sqrt{g} v^{\mu} \varphi)}{\sqrt{g}} - \frac{\p_{\mu} (\sqrt{g} h^{\mu\nu} \p_{\nu} \varphi)}{\sqrt{g}} + \xi R \varphi \right\rbrace \, .
\label{action free scalar Euclidean space}
\eeq
Expanding the differential operator around flat space, it can be written as
\beq
\mathcal{D} = -  \bigtriangleup_E + \delta \mathcal{D}  \quad
\Rightarrow  \quad
O = g^{1/4} (x) \, \mathcal{D} \, g^{-1/4} (x) = - \le \bigtriangleup_E + V \ri \, .
\label{splitting of the non-relativistic heat kernel operator}
\eeq
Given this decomposition, the perturbative expansion works exactly in the same way of the relativistic case.

The Euclidean prescription that we chose in eq. (\ref{Euclidean Schr\"odinger operator}) deserves more comments.
First of all, the heat kernel procedure is well-defined only when an hermitian and elliptic operator is used \cite{Vassilevich:2003xt}.
Our prescription (which was introduced in \cite{Solodukhin:2009sk}) is the only possible way to define an operator which such requirements starting from the real time Schr\"odinger one, which contains only a single partial derivative of time.
While it is true that this procedure changes the spectrum of the theory, it seems to us a prescription similar to the way in which a Dirac fermion is treated in the relativistic case.
In fact, in that situation the Dirac operator by itself is not elliptic, and the standard way to proceed is to evaluate the heat kernel for the squared operator $\slashed D^2,$ which is instead of the Laplace type and has a positive spectrum.
At the end, the Binet theorem is used and the square root of the result is extracted.
Our procedure for the non-relativistic case appears in spirit the same.
We remark that the results given in the following subsections are consistent with \cite{Fernandes:2017nvx}, where the Fujikawa technique is used, without the necessity to perform a Euclidean continuation.
A different approach with the heat kernel computation is considered in \cite{Pal:2017ntk}, where the final result is proportional to the Dirac delta function of the mass of the non-relativistic particle.
We disagree with their result, in particular we point out that their Euclidean prescription for the Schr\"odinger operator defines a parabolic operator instead of an elliptic one, and then the corresponding heat kernel operator is not well-defined.

\subsection{Flat space}

We compute the zero-th order contribution to the heat kernel coming from the flat space Schr\"odinger operator.
We can write the Euclidean version in eq. (\ref{Euclidean Schr\"odinger operator}) as a sum of hermitian and elliptic operators via the integral expansion \cite{Solodukhin:2009sk}
\beq
\label{expansion Solodukhin}
e^{-2 m   \sqrt{-\p_t^2}  }
=\int_0^\infty d \s \frac{m}{\sqrt{\pi}} 
\frac{1}{\s^{3/2}} e^{-\frac{ m^2}{\s}} e^{-\s (-\p_t^2)} \, ,
\eeq
in such a way that we find
\beq
G_{-\bigtriangleup_t} (t,t',\tau) = \langle t | e^{-2 m \tau \sqrt{-\p_t^2}}  |  t' \rangle = \int_0^{\infty} d \s \frac{m}{2 \pi} 
\frac{\tau}{\s^{3/2}}  \exp \left[ - \frac{4 \tau^2 m^2 + (t-t')^2}{4 \s} \right] = \frac{m \tau}{2 \pi} \frac{1}{m^2 \tau^2 + \frac{(t-t')^2}{4}} \, .
\eeq
The spatial part of the flat heat kernel factorizes and is the same of the relativistic case.
Putting the results together, the flat non-relativistic heat kernel in $d+1$ dimensions becomes
\beq
G_{-\bigtriangleup_E} (x,t,x',t',\tau) = \langle x,t | e^{\tau \bigtriangleup_E} |x',t' \rangle = \frac{m \tau}{2 \pi} \frac{1}{m^2 \tau^2 + \frac{(t-t')^2}{4}} \frac{1}{(4 \pi \tau)^{d/2}} \exp \left[ - \frac{(x-x')^2}{4 \tau} \right] \, .
\eeq
In particular, the trace of the heat kernel operator is
\beq
 \mathrm{Tr} \, K(- \bigtriangleup_E, \tau) =  \langle x,t | e^{\tau \bigtriangleup_E} |x,t \rangle =
\frac{2}{m (4 \pi \tau)^{d/2}} \, .
\eeq
It is interesting to compare this expression with the corresponding one (\ref{Trace of flat space rel heat kernel}) in the relativistic case.
A comparison shows that the Schr\"odinger operator in $d+1$ dimensions feels the same spectral dimension
\beq
d_{O} = - 2 \frac{\p \log \mathrm{Tr} \, K_{\mathcal{O}} (\tau)}{\p \log \tau}
\label{spectral dimension operator in heat kernel}
\eeq
as the Laplace operator in $d+2$ dimensions.
This is a pleasant result in the light of the null reduction procedure, which relates the Schr\"odinger group with the conformal group in one higher dimension.

When performing the heat kernel expansion in a $d+1$ dimensional non-relativistic background, we will find in the next Section a behaviour of kind
\beq
\begin{aligned}
\mathrm{Tr}\left[  K(\tau, O)\right] &=
\frac{1}{\tau^{d/2 +1}} \left[a_0 (O) + a_2 (O) \tau + a_4 (O) \tau^2  + \mathcal{O}(\tau^3) \right] = \\
& = \frac{\sqrt{g}}{\tau^{d/2 +1}} \left[a_0 (\mathcal{D}) + a_2 (\mathcal{D}) \tau + a_4 (\mathcal{D}) \tau^2  + \mathcal{O}(\tau^3) \right]
 \, .
 \end{aligned}
\label{definition Seeley-De Witt coefficients non-relativistic case}
\eeq
Due to the spectral dimension (\ref{spectral dimension operator in heat kernel}), we will need to take the $a_4 (O)$ coefficient to find the trace anomaly in $2+1$ dimensions.


\subsection{Heat kernel expansion}

It turns out that the operator $ O = g^{1/4} \mathcal{D} g^{-1/4} $ entering the heat kernel expansion for a non-relativistic free scalar and fermion can be put into the form 
\beq
\begin{aligned}
\langle x, t | O | x',t' \rangle  = &
\langle x, t | \left[ \bigtriangleup_E \, \mathbf{1} + P(x,t)  \delta(x-x') \delta(t-t')  + S(x,t) \sqrt{- \p_t^2} \delta(x-x') \delta(t-t') +  \right. \\ & \left. + Q_i (x,t) \, \p_i  \delta(x-x') \delta(t-t')     \right]  | x', t' \rangle  \, .
\end{aligned} 
\label{Operatore heat kernel in termini di P,S,Q_i}
\eeq
At the first order the Dyson series is
\beq
K_{1} (\tau) = \mathrm{tr} \int_0^\tau d\tau' \, \langle x,t | e^{(\tau-\tau') \bigtriangleup_E} V(x,t) e^{\tau' \bigtriangleup_E} | x', t' \rangle \, ,
\eeq
where here the trace is only evaluated on the internal indices of the differential operator (such as spinorial ones).
According to eq.~(\ref{Operatore heat kernel in termini di P,S,Q_i}), we can decompose the expression as
\beq
\begin{aligned}
K_{1} (\tau)  &= K_{1P} (\tau) + K_{1S} (\tau) + K_{1Q_i} (\tau)   = \mathrm{tr} \int_0^\tau d\tau' \, \langle xt | e^{(\tau-\tau') \bigtriangleup_E} P(x,t) e^{\tau' \bigtriangleup_E} | x' t' \rangle + \\
& + \mathrm{tr} \int_0^\tau d\tau' \, \langle xt | e^{(\tau-\tau') \bigtriangleup_E} S(x,t) \sqrt{-\p_t^2} e^{\tau' \bigtriangleup_E} | x' t' \rangle  + \mathrm{tr} \int_0^\tau d\tau' \, \langle xt | e^{(\tau-\tau') \bigtriangleup_E} Q_i(x,t) \p_i e^{\tau' \bigtriangleup} | x' t' \rangle \, .
\label{1st order splitting heat kernel}
\end{aligned}
\eeq
These integrals are explicitly computed in Appendix \ref{sect-First order expansion of the heat kernel operator} for time-independent operators and the result is
\beq
\label{K_1P 1st order heat kernel}
\mathrm{Tr} \, K_{1P} (\tau) =
\frac{2}{m (4 \pi \tau)^{d/2+1}} \le \tau P(x) + \frac{1}{6} \tau^2 \p_x^2 P(x) + \mathcal{O} (\tau^3) \ri \, ,
\eeq
\beq
\label{K_1S 1st order heat kernel}
\mathrm{Tr} \, K_{1S} (\tau) =
\frac{2}{m(4 \pi \tau)^{d/2+1}} \, \mathrm{tr} \, \left( \frac{S}{2m} + \frac{\tau}{12m} \p_i^2 S + \frac{\tau^2}{120m} \p_i^4 S + \mathcal{O} (\tau^3) \right) \, ,
\eeq
\beq
\label{K_1Q_i 1st order heat kernel}
\mathrm{Tr} \, K_{1Q_i} (\tau) =
\frac{2}{m(4 \pi \tau)^{d/2+1}} \, \mathrm{tr} \, \left( - \frac{\tau}{2} \p_i Q_i - \frac{\tau^2}{12} \p_i \p^2_k Q_i + \mathcal{O} (\tau^3) \right) \, .
\eeq
At the second order the heat kernel expansion is
\beq
K_{2} (s) = \mathrm{tr}
\int_0^\tau d\tau' \int_0^{\tau'} d\tau''  \, \langle x,t | e^{(\tau-\tau') \bigtriangleup_E} V(x,t) e^{(\tau'-\tau'') \bigtriangleup_E}
V(x,t) e^{\tau'' \bigtriangleup_E} | x', t' \rangle \, .
\eeq
$K_{2}$ splits into the sum of several contributions:
\beq
K_{2} (\tau)  = 
 \sum_{X} K_{2X} (\tau) =
 \label{2nd order splitting heat kernel}
 \eeq
 \[=
K_{2 P P} (\tau) + K_{2 S S} (\tau)  +  K_{2 P S} (\tau) + K_{2 S P} (\tau) + K_{2Q_i a_j} (\tau) + 
K_{2Q_i P} (\tau) + K_{2P Q_i} (\tau) + K_{2Q_i S} (\tau) + K_{2 S Q_i} (\tau)  \, .
\]
Their expressions are computed in Appendix \ref{sect-Second order expansion of the heat kernel operator} for time-independent operators and they are given by
\beq
\label{K_2PP 2nd order heat kernel}
\mathrm{Tr} \, K_{2PP} = \frac{2}{m(4 \pi \tau)^{d/2+1}} \mathrm{tr} \left( \frac{\tau^2}{2} P(x)^2 + \mathcal{O}(\tau^3) \right)  \, ,
\eeq
\beq
\begin{aligned}
\mathrm{Tr} \, K_{2SS}  = & \frac{2}{m(4 \pi \tau)^{d/2+1}} \mathrm{tr} \left( \frac{S^2}{4m^2} + \frac{\tau}{12m^2} S \p^2 S +
\frac{\tau}{24m^2} \p_k S \p_k S + \frac{\tau^2}{120m^2} S \p^4 S + \right. \\
& \left. +\frac{\tau^2}{144m^2} \p^2 S \p^2 S +
\frac{\tau^2}{60m^2} \p_i\p^2 S \p_i S + \frac{\tau^2}{180m^2} \p_{ij} S \p_{ij} S  + \mathcal{O}(\tau^3) \right)  \, ,
\end{aligned}
\eeq
\beq
\mathrm{Tr} \, K_{2PS} = \tilde{K}_{2SP}   =  \frac{1}{m(4 \pi \tau)^{d/2+1}} \mathrm{tr} \left( \frac{\tau}{2m} SP + \frac{\tau^2}{12m} S \p^2 P + \frac{\tau^2}{12m}\p^2 S  P + \frac{\tau^2}{12m} \p_i S \p_i P + \mathcal{O}(\tau^3) \right)  \, ,
\eeq
\beq
\begin{aligned}
\label{a-p1}
 \mathrm{Tr} \, K_{2 Q_j Q_i } & = \frac{2 }{m (4 \pi \tau)^{d/2+1}}  
\textrm{tr}  \left[ -  \frac{\tau}{4}  Q_i Q_i    
- \frac{\tau^2}{24} (\p_{j} Q_i) (\p_i a_j) \right. \\
& \left. +  \frac{\tau^2}{8} (\p_{i} Q_i) (\p_{j} a_j)   -  \frac{\tau^2}{12} Q_i (\p^2 Q_i) 
 -  \frac{\tau^2}{24} (\p_i a_j)^2 
+  \mathcal{O}(\tau^3) \right]  \, ,
\end{aligned}
\eeq
\beq
\label{a-p2}
\mathrm{Tr} \, K_{2 Q_i P}  =  \frac{2}{m(4 \pi \tau)^{d/2+1}} \mathrm{tr} \left( - \frac{\tau^2}{3} P (\p_i Q_i)  - \frac{\tau^2}{6} (\p_i P) Q_i  + \mathcal{O}(\tau^3) \right)  \, ,
\eeq
\beq
\label{a-p3}
\mathrm{Tr} \, K_{2  P Q_i}  =  \frac{2}{m(4 \pi \tau)^{d/2+1}} \mathrm{tr} \left(   \frac{\tau^2}{6} Q_i (\p_i P) - \frac{\tau^2}{6} (\p_i Q_i) P + \mathcal{O}(\tau^3) \right)  \, ,
\eeq
\beq
\label{a-s1}
\begin{aligned}
& \mathrm{Tr} \, K_{2 Q_i S}  =  \frac{2}{m(4 \pi \tau)^{d/2+1}} \mathrm{tr} \left[ - \frac{\tau}{24 m^2} \le   S \p_k^2 S + \frac{1}{2} (\p_k S)^2 \ri + \right. \\ 
& \left.  -\frac{\tau^2}{80 m^2}  \le \frac{1}{2} S \p_k^2 \p_j^2  S + \frac{7}{12} (\p_k^2 S)^2 + \frac{13}{12} \p_k S (\p_k \p^2_j S) + \frac{1}{3} (\p_k \p_j S)^2 \ri + \mathcal{O}(\tau^3) \right]  \, ,
\end{aligned}
\eeq
\beq
\label{a-s2}
\begin{aligned}
& \mathrm{Tr} \, K_{2 S Q_i}  =  \frac{2}{m(4 \pi \tau)^{d/2+1}} \mathrm{tr} \left[  \frac{\tau}{48 m^2} \le  (\p_k S)^2  - S \p_k^2 S \ri + \right. \\ 
& \left.  + \frac{\tau^2}{80m^2 }  \le - \frac{1}{3} S \p_k^2 \p_j^2  S - \frac{1}{4} (\p_k^2 S)^2 + \frac{1}{4} \p_k S (\p_k \p^2_j S) + \frac{1}{3} (\p_k \p_j S)^2 \ri + \mathcal{O}(\tau^3) \right]  \, .
\end{aligned}
\eeq


\subsection{A specific perturbation of flat space}

We consider a specific perurbation of $2+1$ dimensional flat space which will be the reference background for the computation of the heat kernel for non-relativistic free scalars and fermions.
It is described by the conditions
\beq
n_{\mu} = \left( \frac{1}{1- \eta(x^{i})} , 0, 0 \right) \, , \qquad 
v^{\mu} = \left( 1- \eta(x^{i}) , 0, 0 \right) \, , \qquad 
h_{ij}= \delta_{ij}  \, , \qquad   A_{\mu} = 0 \, ,
\label{metric perturbation heat kernel computations}
\eeq
which gives a constraint on the spatial part of the frame fields
\beq
e^a_i=e^i_a=\delta^i_a \, .
\eeq
Notice that we consider a time-independent background, as it is sufficient to find non-trivial curvature invariants appearing in the trace anomaly.

The determinant of the metric is
\beq
\sqrt{g} = \frac{1}{1- \eta (x^i)} \, .
\eeq
The non-vanishing components of the spin connection and Cristoffel symbols are:
\bea
\omega_{\underset{(M)}{-} + a}  = \frac{1}{2} \frac{\p_a \eta}{1-\eta}   \, , \qquad 
\omega_{\mu \underset{(A)}{-}\underset{(A)}{+} } = - \frac{1}{2} \delta_{\mu i} \frac{\p_i \eta}{1- \eta}  \, , \quad
\omega_{\mu \underset{(A)}{-} a } = \frac{1}{2} \delta_{\mu +} \frac{\p_a \eta}{(1- \eta)^2}  \, ,
\nl
\Gamma^{-}_{\,\,\, \mu -} = \frac{1}{2} \delta_{\mu i} \frac{\p_i \eta}{1- \eta}   \, , \quad
\Gamma^{\rho}_{\,\,\, \mu-} =  - \frac{1}{2}  \delta_{\mu +} \delta^{\rho i} \frac{\p_i \eta}{(1- \eta)^2}  \, , \quad
\Gamma^{\rho}_{\,\,\, \mu \nu} = \frac{1}{2} \delta^{\rho +} \delta_{\mu +} \delta_{\nu i} \frac{\p_i \eta}{1- \eta}  \, .
\eea
Using eq. (\ref{metric perturbation heat kernel computations}) , we compute the Ricci scalar
\beq
R= -2 \p_i^2 \eta - 2 \eta \p_i^2 \eta - \frac{7}{2} \p_i \eta \p_i \eta + \mathcal{O} (\eta^3)  \, .
\label{Ricci scalar perturbation of flat space}
\eeq
The dimension 4 curvature invariants are given up to second order in the perturbation parameter $\eta$ by
\bea
\label{leadcurv}
& R^2 = 4  (\p_i^2  \eta)^2 + \mathcal{O}(\eta^3) \, , \qquad
 W_{ABCD}^2 = \frac{1}{3} (\p_i^2  \eta)^2 + \mathcal{O}(\eta^3)
\, , \qquad 
E_4 = 2 ( \p_i^2  \eta)^2-2 \p_{ij}  \eta \p_{ij}  \eta + \mathcal{O}(\eta^3) \, , &
\nl
& D_A D^A R =
-2 \p_i^2 \p_j^2 \eta
-2 (\p_i^2 \eta)^2 
-2 \eta \p_i^2 \p_j^2 \eta
-13\p_k \eta \p_k \p_i^2 \eta  
-7 \p_{ij}  \eta \p_{ij}  \eta + \mathcal{O}(\eta^3) \, . &
\eea
In the next section, we will compute the heat kernel expansion and the Seeley-De Witt coefficients in terms of the curvature invariants.


\section{Trace anomaly in specific examples}
\label{sect-Trace anomaly in specific examples}


\subsection{Trace anomaly for a non-relativistic free scalar}
\label{sect-Trace anomaly for a non-relativistic free scalar}

The differential operator $\mathcal{D}_{\mathrm{bos}}$ to consider for a non-relativistic free scalar coupled to NC gravity is given in eq. (\ref{action free scalar Euclidean space}).
From this expression we find that the functions appearing in eq. (\ref{Operatore heat kernel in termini di P,S,Q_i}) for the specific background (\ref{metric perturbation heat kernel computations}) are given by
\beq
S(x) = 2 m \eta \, , \qquad
P(x)= - \le \frac{\p_i^2 \eta}{2} + \frac12 \eta \p_i^2 \eta + \frac{3}{4} (\p_i \eta)^2  \ri  + \xi \le 2 \p_i^2 \eta + 2 \eta \p_i^2 \eta  + \frac{7}{2} (\p_i \eta)^2 \ri \, ,
\eeq
and $Q_i (x)=0 .$
Putting these expressions inside the first and second order heat kernel expansion given in eqs. (\ref{K_1P 1st order heat kernel})-(\ref{K_1Q_i 1st order heat kernel}) and in eqs. (\ref{K_2PP 2nd order heat kernel})-(\ref{a-s2}), we obtain the $a_4(\mathcal{D}_{\mathrm{bos}})$ coefficient in $2+1$ dimensions: 
\beq
\begin{aligned}
a_4 (O_{\mathrm{bos}})  = \sqrt{g} a_4 (\mathcal{D}_{\mathrm{bos}}) = \frac{1}{8 m \pi^2} & \left[ 
\frac{7- 180 \xi +720 \xi^2}{360} (\p^2 \eta)^2
+  \frac{210 \xi-41}{180} (\p_{ij} \eta  )^2  - \frac{1-5\xi}{15} \p^4 \eta    \right.\\
& \left.    
 - \frac{2(1-5\xi)}{15} \eta \p^4 \eta  
-  \frac{13}{30} (1- 5 \xi) (\p_i \eta) (\p_i \p^2 \eta)  + \mathcal{O}(\eta^3)  \right] \, .
\end{aligned}
\eeq
Using the leading order expansions (\ref{Ricci scalar perturbation of flat space}) and (\ref{leadcurv}), we are finally able to re-write the Seeley-De Witt coefficient in terms of the curvature invariants finding the trace anomaly as
\beq
\mathcal{A}_{\rm bos} =  a_4 (\mathcal{D}_{\mathrm{bos}}) = \frac{1}{8m\pi^2} \left[ -\frac{1}{360} E_4 + \frac{3}{360} W^2 + \frac12 \le \xi - \frac{1}{6} \ri^2 R^2  + \frac{1-5 \xi}{30} \, D^2 R   \right]  \, . 
\eeq


\subsection{Trace anomaly for a non-relativistic free fermion}
\label{sect-Trace anomaly for a non-relativistic free fermion}

In the fermionic case there are two main differences with respect to the free scalar: the path integral contains a Berezin integration  and the Dirac operator $\slashed D$ is not elliptic after Wick rotation, even in the relativistic framework.
The first change simply influences the sign of the functional determinant. In fact, in the bosonic case the effective action is
\beq
e^{i W} = \int {\cal D}\varphi^\dagger\,  {\cal D}\varphi\  e^{i \int d^d x \, \varphi^{\dagger} \mathcal{D}_{\mathrm{bos}} \varphi  } \, ,
\eeq
whose solution is 
\beq
i W_{\mathrm{bos}} = - \log \det (\mathcal{D}_{\mathrm{bos}}) \, .
\eeq
In the fermionic case, the vacuum functional is instead
\beq
e^{i W} = \int {\cal D}\bar{\psi} \,  {\cal D}\psi\  e^{i \int d^d x \, \bar{\psi} \slashed D \psi  } \, ,
\eeq
and the anticommuting nature of the spinorial objects gives an additional minus sign
\beq
i W_{\mathrm{ferm}} =  \log \det (\slashed D) \, .
\eeq
While this procedure would in principle work, the fact that the Dirac operator is not elliptic forbids the possibility to apply the heat kernel method to this differential operator.
The problem is solved by studying the square of the Dirac operator $\slashed D^2,$ which is instead both hermitian and elliptic after Wick rotation.
In fact, this operator is also of the Laplace type because it can be written as (see e.g. \cite{Christensen:1978md},  \cite{Freedman}):
\beq
 \left( i \slashed D \right)^2  =  -\square + \frac{1}{4} R  \, , \qquad
 \Box=D_A D^A \, . 
 \label{dirac2}
\eeq
At the level of the functional determinant, to recover the original differential operator we need to extract the square root, which simply gives a factor of $1/2$ in the functional determinant
\beq
i W_{\mathrm{ferm}} = \frac12 \log \det (\slashed D^2) \, .
\eeq
The same trick applies to the non-relativistic case, where we perform the null reduction of the differential operator (\ref{dirac2}).
The result in the background (\ref{metric perturbation heat kernel computations}), after applying the rules for the Euclidean space formulation
\beq
t \rightarrow -i t \, , \qquad
\p_t \rightarrow i \p_t \, , \qquad
m \rightarrow i m \, ,
\eeq
is the differential operator
\bea
& & g^{1/4} \left( \square - \frac{1}{4} R \right)_{\! \! \!E} g^{-1/4} \Psi    =  \nl   
& =&  \left[ -2 i m  \, \bold{1} +2 i m \eta   \, \bold{1} + \frac{i}{2} (\p_a \eta) \gamma^{+a} \right] \p_t \Psi
  + \left[ -\frac{1}{2} (\p_a \eta) \gamma^{-+}  -\frac{1}{2} \eta (\p_a \eta) \gamma^{-+} \right] \p_a \Psi +  \nl
&+ &   \left[ \frac{1}{8} (\p_a \eta)^2 \, \bold{1}- \frac{1}{4} \p^2 \eta \gamma^{-+} - \frac{1}{4} \eta (\p^2 \eta) \gamma^{-+} 
- \frac{1}{4} (\p_a \eta)^2 \gamma^{-+} 
 - \frac{1}{2} m (\p_a \eta) \gamma^{-a} - \frac{1}{2} m \eta (\p_a \eta) \gamma^{-a} \right] \Psi 
+ \nl
& + &  \left[ \frac{1}{16} (\p_a \eta)^2  \bold{1}
+ \frac{1}{16} (\p_a \eta)(\p_b \eta) \lbrace \gamma^{-a} , \gamma^{+b} \rbrace \right] \Psi + \p^2 \Psi  \, . 
\eea
This expression can be put in the form (\ref{Operatore heat kernel in termini di P,S,Q_i}) with the identifications
\beq
\begin{aligned}
P(x) & =   \frac{3}{16} (\p_i \eta)^2 \, \bold{1} - \frac{1}{4} (\p^2 \eta) \gamma^{-+} - \frac{1}{4} \eta (\p^2 \eta) \gamma^{-+} 
- \frac{1}{4} (\p_i \eta)^2 \gamma^{-+}  \\
& - \frac{1}{2} m (\p_i \eta) \gamma^{-i} - \frac{1}{2} m \eta (\p_i \eta) \gamma^{-i} 
+ \frac{1}{16} (\p_i \eta)(\p_j \eta) \lbrace \gamma^{-i} , \gamma^{-j} \rbrace \, ,  \\
S(x) & = 2 m \eta \, \bold{1} + \frac{1}{2} (\p_i \eta) \gamma^{+i} \, ,  \\
Q_i (x) & = - \frac{1}{2} (\p_i \eta) \gamma^{-+} - \frac{1}{2} \eta (\p_i \eta) \gamma^{-+} \, .
\end{aligned} 
\label{Parti heat kernel}
\eeq
In particular, more explicitly the matrix content is given by
\beq
P(x)= \begin{pmatrix}
P_{11} (x) & 0 & 0 & 0 \\ P_{21} (x) & P_{22} (x) & 0 & 0 \\ 0 & 0 & P_{22} (x) & P_{32}(x) \\ 0 & 0 & 0 & P_{11} (x) 
\end{pmatrix} \, ,
\eeq
where
\beq
\begin{aligned}
 P_{11} (x) & = \frac{5}{16} (\p_i \eta)^2 + \frac{1}{4} (\p^2 \eta) + \frac{1}{4} \eta (\p^2 \eta) \, , \\
  P_{22} (x)  & = -\frac{3}{16} (\p_i \eta)^2 - \frac{1}{4} (\p^2 \eta) - \frac{1}{4} \eta (\p^2 \eta) \, ,  \\
P_{21} (x) & =  \frac{\sqrt{2}}{2} m \left[ (\p_1 + i \p_2) \eta + \eta  (\p_1 + i \p_2) \eta \right] \, ,  \\
P_{32} (x) & = \frac{\sqrt{2}}{2} m \left[ (-\p_1 + i \p_2) \eta + \eta  (-\p_1 + i \p_2) \eta \right]  \, .
\end{aligned}
\eeq
Moreover
\beq
S(x)= \begin{pmatrix}
S_{11} (x) & S_{12} (x) & 0 & 0 \\ 0 & S_{11} (x) & 0 & 0 \\ 0 & 0 & S_{11} (x) & 0 \\ 0 & 0 & S_{43} (x) & S_{11} (x) 
\end{pmatrix} \, ,
\eeq
where
\beq
S_{11} (x) = 2 m \eta \, , \qquad
S_{12} (x) = \frac{\sqrt{2}}{2} (\p_1 - i \p_2 ) \eta \, , \qquad
S_{43} (x) = -  \frac{\sqrt{2}}{2} (\p_1 + i \p_2 ) \eta \, .
\eeq
Finally
\beq
Q_i (x)= Q_{11} (x) \begin{pmatrix}
1 & 0 & 0 & 0 \\ 0 & -1 & 0 & 0 \\ 0 & 0 & -1 & 0 \\ 0 & 0 & 0 & 1
\end{pmatrix} \, ,
\qquad
Q_{11} (x)= \frac{1}{2} (\p_i \eta) +  \frac{1}{2} \eta (\p_i \eta) \, .
\eeq
The perturbative expansion works as in the bosonic case, apart from an additional factor in the flat space heat kernel coming from the trace over the spinorial indices of the operators.
This gives
\beq
\mathrm{Tr} \, K_{\bigtriangleup} (s) = \mathrm{Tr} \langle x t | e^{\tau \bigtriangleup \, \bold{1}} | x t \rangle =
\frac{2}{m (4 \pi \tau)^{d/2+1}} \, \mathrm{Tr} (\bold{1}) = \frac{8}{m (4 \pi \tau)^{d/2+1}}   \, .
\eeq
The formulae (\ref{K_1P 1st order heat kernel})-(\ref{K_1Q_i 1st order heat kernel}) and (\ref{K_2PP 2nd order heat kernel})-(\ref{a-s2}) still apply and we obtain after several computations that in $2+1$ dimensions
\beq
 \sqrt{g} a_4(\slashed D^2_E) = 
 \frac{2}{m(4 \pi)^{2}} \left[ \frac{1}{15} \p^4 \eta +  \frac{2}{15} \eta (\p^4 \eta)       
+  \frac{13}{30} (\p_i \eta) (\p_i \p^2 \eta) + \frac{1}{9} (\p^2 \eta)^2 +  \frac{31}{180} (\p_{ij} \eta  )^2 + \mathcal{O}(\eta^3)  \right] \, , 
\eeq
which is recognized to be 
\beq
  a_4 ( \slashed{D}^2_E) = \frac{1}{8m\pi^2} \left( \frac{11}{360} E_4 - \frac{1}{20} W^2  - \frac{1}{30} D^2 R   \right)   \, . 
\eeq
The trace anomaly is finally given by
\beq
 \mathcal{A}_{\rm ferm}  = - \frac{1}{2} a_4 (\slashed D_E^2)  \, . 
\eeq


\section{Trace anomaly with particle number background}
\label{sect-Trace anomaly with particle number background}

The background (\ref{metric perturbation heat kernel computations}) is rich enough to allow the computation of non-vanishing curvature invariants, but on the other hand it hides the presence of many terms which can enter the trace anomaly and depend from the gauge field.
In particular, in \cite{Fernandes:2017nvx} it was shown by means of the Fujikawa technique in a NC background with a non-vanishing
particle number that the trace anomaly was not $U(1)$ gauge invariant. 

In order to investigate these issues, we consider a NC background whose data are
\beq
n_{\mu} = (1, \mathbf{0}) \, , \qquad
v^{\mu} = (1, \mathbf{0}) \, , \qquad
h_{ij} = \delta_{ij} \, , \qquad
A_{\mu} = (A_0 (t,x^i), A_i (t,x^i)) \, ,
\label{NC background heat kernel with particle number}
\eeq
which in terms of the higher-dimensional metric give
\beq
G_{MN} = \begin{pmatrix}
0 & 1 & 0 & 0 \\
1 & 2A_0 & A_1 & A_2 \\
0 & A_1 & 1 & 0 \\
0 & A_2 & 0 & 1 \\
\end{pmatrix} \, , \qquad
G^{MN} = \begin{pmatrix}
-2 A_0+ A_i A_i & 1 & -A_1 & -A_2 \\
1 & 0 & 0 & 0 \\
-A_1 & 0 & 1 & 0 \\
-A_2 & 0 & 0 & 1 \\
\end{pmatrix} \, ,
\eeq
with vielbein
\beq
e^A_{\,\,\, M} = \begin{pmatrix}
1 & A_0 & A_1 & A_2 \\
0 & 1 & 0 & 0 \\
0 & 0 & 1 & 0 \\
0 & 0 & 0 & 1 \\
\end{pmatrix} \, , \qquad
e^M_{\,\,\, A} = \begin{pmatrix}
1 & -A_0 & -A_1 & -A_2 \\
0 & 1 & 0 & 0 \\
0 & 0 & 1 & 0 \\
0 & 0 & 0 & 1 \\
\end{pmatrix}\, .
\eeq
In this case the metric is flat apart from the contribution from the gauge connection.
This greatly simplifies the spin connection, whose only non-vanishing components are
\beq
\omega_{++i} = - F_{0i} \, ,  \qquad
\omega_{+ij} = - \frac12 F_{ij} \, , \qquad \omega_{i+ j} = - \frac12 F_{ij} \, ,
\eeq
while the non-vanishing components of the Christoffel symbol are
\beq
\Gamma^{-}_{\,\,\, \mu \nu} =\frac12 (v_A)^{\sigma} (Q_A)_{\mu\nu\sigma} \, , \qquad  \Gamma^{\rho}_{\,\,\, \mu\nu} =\frac12 h^{\rho \sigma} (Q_A)_{\mu\nu\sigma} \, ,
\eeq
In this notation, $F=dA$ is the usual field strength while the other quantities are Milne-boost invariants defined in eq. (\ref{definition of Milne boost invariants}).

The perturbative expansion of the heat kernel applies to the differential operators computed on this background as well, but now the gauge field is considered to be time-dependent.
This requires a generalization of the insertion operators which is performed in Appendices \ref{app-Time-dependent insertion contributions to the heat kernel (first order)} and \ref{app-Time-dependent insertion contributions to the heat kernel (second order)}, but surprisingly the result is that the formal expression is unchanged and given again by eqs. (\ref{K_1P 1st order heat kernel})-(\ref{K_1Q_i 1st order heat kernel}) and (\ref{K_2PP 2nd order heat kernel})-(\ref{a-s2}). 

\subsection{Boson}

The action (\ref{action non-relativistic scalar real time}) for a Galilean-invariant scalar field reduces on the background (\ref{NC background heat kernel with particle number}) to
\beq
\label{actionscalar with gauge connection}
S = \int d^{3} x \, \left[ 2 im \varphi^{\dagger} \p_t \varphi
 +    \varphi^{\dagger} \p_i^2 \varphi - 2 im A_i \varphi^{\dagger} \p_i \varphi 
 + \left( 2 m^2 A_0  - m^2 A_i A_i  - i m \p_i A_i \right) 
 \varphi^{\dagger} \varphi  \right] \, .
\eeq
In order to get the Euclidean space version of this functional, we need to perform a rotation of the gauge field according to\footnote{The unconventional redefinition of 
the gauge field  in the imaginary time formalism is required by consistency with  $[D_\mu, D_\nu] = - i m F_{\mu \nu}$ and the prescription $m \rightarrow i m$. The imaginary mass is required in order to get a positive definite euclidean action.}
\beq
A_0 \rightarrow A_0 \, \qquad \, A_i \rightarrow -i A_i \, ,
\eeq
which gives the imaginary time action
\beq
S_E   = -  \int d^{3} x \,  \varphi^{\dagger} \left[ \bigtriangleup - 2 i m A_i  \p_i   - 2  m^2 A_0 - m^2 A_i A_i  - i m (\p_i A_i)   \right] \varphi \, .
\eeq
The differential operator in parenthesis is naturally splitted as
\beq
\label{decomposition differential operator scalar with gauge field}
P(t,x^i) = -2  m^2 A_0 - m^2 A_i A_i  - i  m (\p_i A_i)  \, , \quad
S(t,x^i) = 0 \, , \quad
Q_i (t,x^i) = - 2 im A_i \, .
\eeq
In particular, all insertions containing at least one factor of $S(t,x^i)$ trivially vanish.
Summing all the contributions to the $a_4$ Seeley-De Witt coefficient, we find
\beq
\label{a4scalar with particle number}
{\cal A}_{\mathrm{bos}} =a_4 (\mathcal{D}_{\mathrm{bos}})  = - \frac{m}{8 \pi^2} \left( \frac13 \p^2 A_0 + \frac16  B^2 - {2 m^2} A_0^2 + \mathcal{O} (A_{\mu}^3)\right)  \, ,
\eeq
where $B= F_{12}$.

\subsection{Fermion}

Similar steps can be followed for the null reduction of the Dirac spinor, whose differential operator in the imaginary time space in the background (\ref{NC background heat kernel with particle number}) is
\beq
\slashed D^2_E \Psi = \bigtriangleup \Psi - 2  m^2 A_0 \Psi -  m^2 A_k A_k \Psi 
- i  m (\p_i A_i) \Psi - 2 i m A_i (\p_i \Psi) +
\eeq
\[
- m F_{i0} \gamma^{+i} \Psi  - \frac{1}{4} i  m F_{ij} \gamma^{ij} \Psi + \frac12  m A_i F_{ij} \gamma^{+j} \Psi + \frac12 i  F_{ij} \gamma^{+j}(\p_i \Psi) + \frac{1}{4} i  (\p_i F_{ij}) \gamma^{+j} \Psi \, .
\] 
In this way we identify
\beq
P(t,x^i) = \left[ -2  m^2 A_0 - m^2 A_k A_k - i m (\p_i A_i) \right] \mathbf{1} - m F_{i0} \gamma^{+i} -
 \frac{1}{4} i m F_{ij} \gamma^{ij}  + \frac12  m A_i F_{ij} \gamma^{+j} + \frac{1}{4} i (\p_i F_{ij}) \gamma^{+j} \, ,
\eeq
\beq
S(t,x^i) = 0 \, ,  \qquad
Q_i (t,x^i) = \le - 2 i m A_i \ri \mathbf{1} + \frac12 i F_{ij} \gamma^{+j} \, ,
\eeq
where the explicit Dirac matrices are
\beq
\gamma^{+1} = \begin{pmatrix}
0 & \sqrt{2} & 0 & 0 \\
0 & 0 & 0 & 0 \\
0 & 0 & 0 & 0 \\
0 & 0 & - \sqrt{2} & 0 \\
\end{pmatrix} \, , \ 
\gamma^{+2} = \begin{pmatrix}
0 & -\sqrt{2} i & 0 & 0 \\
0 & 0 & 0 & 0 \\
0 & 0 & 0 & 0 \\
0 & 0 & - \sqrt{2} i & 0 \\
\end{pmatrix} \, , \ 
\gamma^{12} = \begin{pmatrix}
-i & 0 & 0 & 0 \\
0 & i & 0 & 0 \\
0 & 0 & -i & 0 \\
0 & 0 & 0 & i \\
\end{pmatrix} \, .
\eeq
From these expressions we extract again the trace anomaly finding
\beq
a_4 (\slashed D^2_E) 
= - \frac{m}{48 \pi^2} B^2 - \frac{m}{6 \pi^2} \p^2 A_0 + \frac{m^3}{\pi^2} A_0^2  + \mathcal{O}(A_{\mu}^3) \, .
\eeq
The trace of the stress-energy tensor is finally given by
\beq
\mathcal{A}_{\mathrm{ferm}}  = - \frac12 a_4 (\slashed D^2_E)   = \frac{m}{12 \pi^2} \p^2 A_0 - \frac{m^3}{2 \pi^2} A_0^2 + \frac{m}{96 \pi^2} B^2   + \mathcal{O}(A_{\mu}^3)  \, . 
\label{Dirac trace anomaly with gauge field}
\eeq


\section{Diffeomorphism anomaly}
The heat kernel technique allows not only to compute the trace anomaly corresponding to a given differential operator, but also any other quantum anomaly using the same $\zeta$-function regularization introduced in section \ref{sect-Renormalization of the effective action}.
The key point is to investigate the variation
\beq
\delta \zeta (s, \mathcal{D}) = - s \, \mathrm{Tr} \, \le (\delta \mathcal{D}) \mathcal{D}^{-s-1} \ri 
\label{variation zeta function, repeated}
\eeq
under the classical symmetry whose quantum anomaly we want to compute.

For simplicity, we are interested to study the diffeomorphism anomaly for a non-relativistic free scalar, whose associated Euclidean differential operator is
\beq
\mathcal{D}\varphi  = i m v^{\mu} D_{\mu} \varphi + \frac{i m }{\sqrt{g}} D_{\mu} \le \sqrt{g} v^{\mu} \varphi \ri - \frac{1}{\sqrt{g}} D_{\mu} \le \sqrt{g} h^{\mu\nu} D_{\nu} \varphi\ri \, .
\eeq
We specialize again to the flat background with generic gauge field, in which case the differential operator becomes
\beq
\mathcal{D}_0 = 2 i m \p_0 - \p_i^2 +2 m^2 A_0 + m^2 A_i A_i + 2 i m A_i \p_i + i m (\p_i A_i) \, .
\eeq
Its variation under diffeomorphisms is
\bea
\delta \mathcal{D}_0   &=& - 2 i m (\p_0 \varepsilon^{\mu}) \p_{\mu} + 2 (\p_i \varepsilon^{\mu}) \p_i \p_{\mu} + (\p_i^2 \varepsilon^{\mu}) \p_{\mu} + 2 m^2 \varepsilon^{\mu} (\p_{\mu} A_0)
\nonumber  \\
& & + 2 i m \varepsilon^{\mu} (\p_{\mu} A_i) \p_i - 2 i m A_i (\p_i \varepsilon^{\mu}) \p_{\mu} + i m \varepsilon^{\mu} (\p_i \p_{\mu}A_i) + 2 m^2 A_i \varepsilon^{\mu} (\p_{\mu} A_i) \, .
\label{variazione sotto diffeo operatore}
\eea
Using these expressions inside eq. (\ref{variation zeta function, repeated}), we would be able to find after some steps the corresponding diffeomorphism anomaly from the heat kernel expansion.
Before proceeding with the calculation, it is convenient to perform a trick. We notice that eq. (\ref{variation zeta function, repeated}) is a trace, and as such it is invariant under the similarity transformation
\beq
\label{dtilde}
\zeta (s, \tilde{\mathcal{D}}) = \zeta (s, \mathcal{D}) \, \qquad \text{ if} \quad\tilde{\mathcal{D}} = e^{\hat{O}} \mathcal{D} e^{-\hat{O}} \, ,
\eeq 
because the operators $\mathcal{D}$ and $\tilde{\mathcal{D}}$ have the same functional determinant.

In particular, when dealing with diffeomorphisms, it is convenient to consider 
\beq
\hat{O} = \alpha \xi^{\mu} \p_{\mu} \, ,
\eeq
with $ \alpha $  a real coefficient and $ \xi^{\mu} $ transforming under diffeomorphisms as
$\delta \xi^{\mu} = \varepsilon^{\mu}. $

In this way we obtain 
\begin{align} \nonumber
\tilde{\mathcal{D}_0} = e^{\alpha \xi^{\mu} \p_{\mu}} \mathcal{D}_0 e^{- \alpha \xi^{\nu} \p_{\nu}} & = \mathcal{D}_0 - 2 i m \alpha (\p_0 \xi^{\mu}) \p_{\mu} + 2 \alpha (\p_i \xi^{\mu}) \p_i \p_{\mu} + \alpha (\p_i^2 \xi^{\mu}) \p_{\mu} \\
& + 2 m^2 \alpha \xi^{\mu} (\p_{\mu} A_0) + 2 m^2 \alpha A_i \xi^{\mu} (\p_{\mu} A_i) + 2 i m \alpha \xi^{\mu} (\p_{\mu} A_i) \p_i \\
\nonumber
& + i m \alpha \xi^{\mu} (\p_{\mu} \p_i A_i) - 2 i m \alpha A_i (\p_i \xi^{\mu}) \p_{\mu} + \mathcal{O} (\xi^2) \, .
\end{align}
Using eq. (\ref{variazione sotto diffeo operatore}) and setting   $ \alpha=-1 $, we obtain
$
\delta \tilde{\mathcal{D}_0} =0,
$
which in turn implies $ \delta W^{\mathrm{ren}}=0, $ and therefore there is no gravitational anomaly
\beq
\langle \p_{\mu} T^{\mu}_{\,\,\, \nu} \rangle =0 \, .
\eeq


\section{Comments and discussion}

It is interesting to compare the results for the non-relativistic complex scalar and the fermion among themselves and with the $3+1$ dimensional relativistic parents.
The general structure of the trace anomaly is
\begin{equation}
\mathcal{A} =
\frac{1}{360( 4 \pi^2)} \le -a E_{4} +c W^{2}_{MNPQ} + b R^2 + d D^2 R  \ri + \dots \, ,
\end{equation}
where the coefficients for the cases of interest are reported in the table below. 
The dots refer to the possible terms in the other sectors of the non-relativistic trace anomaly and to the possible $U(1)$ gauge and Milne-boost violating terms, which we comment later.

\begin{table}[h] 
\begin{center}    
\begin{tabular}  {|l|c|c|c|c|} \hline \textbf{Spin} & \textbf{a} & \textbf{c} & \textbf{b} & \textbf{d}  \\
\hline 
0 (relativistic real scalar) & $  1 $ & $ 3 $ & $ 180 \le \xi - \frac{1}{6} \ri^2 $  & $ 12 \le 5 \xi - 1 \ri $  \\[0.5ex] 
0 (non-rel. complex scalar) & $  \frac{2}{m} $ & $  \frac{6}{m} $ & $ \frac{360}{m} \le \xi - \frac{1}{6} \ri^2 $  & $ \frac{24}{m} \le 5 \xi - 1 \ri $  \\[0.5ex] \hline
$ \frac12 $ (relativistic Dirac) & $ 11 $  & $ 18 $ & 0  & 12   \\[0.5ex] 
$ \frac12 $ (non-rel. fermion) & $ \frac{11}{m} $  & $ \frac{18}{m} $ & 0  & $ \frac{12}{m} $  \\[0.5ex] \hline
\end{tabular}   
\caption{\footnotesize Coefficients of the trace anomaly for free fields with various spins.} 
\end{center}  
\label{Table with coefficients heat kernel}  
\end{table}

First of all, we stress that the curvature invariants entering the trace anomaly are computed in $3+1$ dimensions in the relativistic case, while they are null reduced to $2+1$ dimensions in the non-relativistic one.
In particular, in the latter case they can be expressed only in terms of NC geometric data, without referring in any way to higher-dimensional quantities.
In the bosonic case, the parameter $\xi$ is the coupling with the Ricci scalar, which both in the conformal and in the Schr\"odinger cases turns out to be $\xi=1/6.$
If $\xi$ assumes this value, we observe that the  $b$ coefficient vanishes both for the scalar and the fermionic cases, because the $R^2$ term is forbidden by Wess-Zumino consistency conditions.
The coefficient $d$ is instead non-vanishing but refers to the curvature invariant $D^2 R,$ which is scheme-dependent and then less interesting because it does not contain universal informations of the RG flow.
Finally, the $a,c$ coefficients are scheme-independent anomalies and the first one is a candidate for a non-relativistic version of the $a$-theorem.

The comparison with the relativistic parent shows that the coefficients are the same, apart from an overall normalization $1/m$ depending only from the $U(1)$ mass of the non-relativistic particle\footnote{The scalar field also differs for a factor of 2 because in the relativistic case we put the result for a real field, while the Galilean case necessarily requires a complex field, due to the $U(1)$ mass conservation.}.
A natural guess for an $a$-theorem in the case where the only degrees of freedom involved in the physical system are scalars and spin $1/2$ fermions would be
\beq
a_{\rm UV} \propto \sum_{\rm scalars}^{\rm UV} \frac{1}{m} +
\frac{11}{2}  \sum_{\rm fermions}^{\rm UV} \frac{1}{m} 
\geq
\sum_{\rm scalars}^{\rm IR} \frac{1}{m} +
\frac{11}{2}  \sum_{\rm fermions}^{\rm IR} \frac{1}{m}
\propto a_{\rm IR} \, .
\label{aaath}
\eeq
 In Galilean-invariant theories the mass of a bound state is equal to the sum of the masses of the elementary constituents: no bound-state
contribution is present as in the relativistic case.
The $1/m$ dependence may give a quantitative formulation to the physical intuition that bound states should form in the IR: as energy is added, bound states tend to be broken.

Now we comment on the additional terms found in eqs. (\ref{a4scalar with particle number}) and (\ref{Dirac trace anomaly with gauge field}) when a particle number background is turned on.
First of all, we observe that the terms entering the trace anomaly are the same for the bosonic and the fermionic case, which confirms that the structure of the expression is universal.
In both cases, the result is not invariant either under Milne boosts nor under gauge transformations.
Since there is a strong relation between these symmetries\footnote{In the Bargmann algebra, the commutator of the momentum and a boost is the particle number generator.}, it is not surprising that the violation of one of them entails the violation of the other.

As was done in the case with background (\ref{metric perturbation heat kernel computations}), we should try to express the resulting anomaly as a combination of terms which are invariant under the non-relativistic symmetries, but since two of them are violated, we should at least try to obtain a diffeomorphism-invariant combination.
A possibility is that the terms containing only the temporal component of the gauge field $A_0$ arise from the combination $v^{\mu}A_{\mu},$ which is a scalar and has the correct scaling dimension.
In this way, the term
\beq
\p^2 A_0 = \p^2 (v^{\mu} A_{\mu})
\eeq
can be reabsorbed by a local counterterm proportional to $R v^{\mu} A_{\mu}$ in the vacuum functional.
This is not possible for the term $A_0^2 = (v^{\mu} A_{\mu})^2 ,$ which is instead a genuine type B anomaly.

Both in the free scalar and free fermion examples, the field $A_0$ plays the role of an external chemical potential
for the particle number $J_0$; in the multiple species case, $J_0$ plays the role of mass density.
Moreover, studying geodesics in a NC background, one sees that $A_0$ can also be identified 
 as the Newtonian gravitational potential. On physical ground one would expect mass conservation
in an external gravitational field. On the other hand, the breaking of gauge invariance
in eqs. (\ref{a4scalar with particle number}) and (\ref{Dirac trace anomaly with gauge field})  may hint a violation of the conservation of the $U(1)$ current, which would not be consistent with the physical intuition.
For these reasons, we think that this topic deserves further future investigations to understand the precise mechanism responsible for these results.


\chapter{Non-relativistic Supersymmetry}
\label{chapt-Non-relativistic Supersymmetry}

\begin{center}
\emph{The work in this chapter has previously appeared in \cite{Auzzi:2019kdd}.}
\end{center}

Anomalies are a powerful tool to investigate dualities and non-trivial properties of a physical system, such as the irreversibility properties of the RG flow controlled by the trace anomaly which we studied in chapters \ref{chapt-Non-relativistic actions}, \ref{chapt-The heat kernal technique}.
In this context, another powerful tool which allows to investigate dualities and to obtain non-perturbative results is Supersymmetry (SUSY), an invariance which rotates bosons into fermions and viceversa. 
In the relativistic case, SUSY gives strong constraints on the quantum corrections of physical theories, giving rise to the non-renormalization theorem and many other exact results \cite{Grisaru:1979wc, Seiberg:1993vc}.

In the previous chapters we introduced the group of symmetries for non-relativistic systems and we observed that many interesting results about the irreversibility of the RG flow could be investigated for these systems.
Similarly, it is interesting to understand if some exact results coming from the SUSY invariance can be inherited from the relativistic case, and if the intertwining with the Schr\"odinger group can give even further restrictions on the quantum corrections of such systems.

The study of such systems can also be interesting from the point of view of holography: in fact, most of the examples where $\mathrm{AdS/CFT}$ correspondence is verified are supersymmetric.
SUSY may allow to gain control on non-relativistic holography and suggest the way to build explicit top-down examples.

Furthermore, we know that non-relativistic systems are well understood from the point of view of experimental realizations: we may hope to find some hints to understand if SUSY is a fundamental invariance that is broken at low energies, if it arises at low energies as an enhanced symmetry or if it plays any other important role in high energy physics.

SUSY extensions of the Galilean algebra were first introduced in 3+1 dimensions \cite{Puzalowski:1978rv}, where two super-Galilean algebras were constructed, $ \mathcal{S}_1 \mathcal{G} $ which includes a single two-component spinorial supercharge and 
$ \mathcal{S}_2 \mathcal{G} $, which contains two supercharges. 
There are various ways to find such gradations of the Galilean algebra.
One method consists in performing the In\"{o}n\"{u}-Wigner contraction of the ${\cal N}=1$ and ${\cal N}=2$ Super-Poincar\'e algebras in the $c \to \infty$ limit\footnote{When performing this procedure, divergent expressions in the speed of light appear 
 and we need to introduce some subtraction terms via a chemical potential and by appropriately rescaling the fields \cite{Jensen:2014wha}.}. 
On the other hand, $\mathcal{S}_2 \mathcal{G}$ can be obtained performing a null reduction of the super-Poincar\`e algebra in 4+1 dimensions. It turns out that $ \mathcal{S}_1 \mathcal{G} \subset \mathcal{S}_2 \mathcal{G}$.

In this chapter we will apply the DLCQ prescription to build a theory with  
 $ \mathcal{S}_2 \mathcal{G} $ invariance in $2+1$ dimensions, obtained from the $3+1$ dimensional ${\cal N}=1$ relativistic Wess-Zumino model.
We will show that a non-vanishing superpotential is allowed only when considering at least two species of chiral fields and in this context we will investigate the renormalization properties of this Galilean Wess-Zumino model in order to compare with the relativistic non-renormalization theorem.
We anticipate the remarkable UV properties of this model:
\begin{itemize}
 \item  like the relativistic parent, the model is renormalizable and the superpotential 
 term does not acquire quantum corrections,
 \item unlike the relativistic parent, there is strong evidence that the whole renormalization of the two-point function is just at one loop (we check this claim explicitly up to four loops and discuss in general higher orders). This remarkable property is due to the $U(1)$  symmetry associated to the non-relativistic particle number conservation, which constrains the number of diagrams at a given perturbative order. Moreover, the causal structure strongly reduces the number of non-vanishing diagrams.
 \end{itemize}
As a consequence of these two properties, we will build a set of selection rules which will simplify the analysis of the allowed quantum corrections.

\section{Non-relativistic supersymmetry algebra}
\label{sect-Non-relativistic supersymmetry algebra}

We start with the graded generalization of the Galilean algebra in $2+1$ dimensions containing two complex supercharges, which leads to the invariance under the $ \mathcal{S}_2 \mathcal{G} $ group. 
The algebra is\footnote{We remark that in the model we are going to consider, the presence of an R-symmetry is not necessary in the algebra. In fact, in the upcoming non-perturbative proof on the non-renormalization theorem (see sections \ref{sect-The non-renormalization theorem}, \ref{sect-Non-relativistic non-renormalization theorem}), we will need to assume that the theory in the UV after promoting the couplings to background superfields will possess a $U(1)_R$ symmetry, but nothing is required on the original model.
Since in all the evaluations for the Galilean WZ model the grassmannian part of the superfields and of the supercharges is unaffected, this requirement is analog with the same argument applied in the relativistic case.}
\begin{eqnarray}
& [P_j, K_k] = i \delta_{jk} M \, , \qquad
[H, K_j] = i P_j \, ,  \nonumber \\ 
& [P_j, J] = -i \epsilon_{jk} P_k \, , \qquad
[K_j, J] = -i \epsilon_{jk} K_k  \, ,  \qquad \qquad j,k=1,2  \label{Bargmann} \\
& \nonumber \\
& [Q,J] = \frac12 Q \, , \qquad
\lbrace Q, Q^{\dagger}  \rbrace = \sqrt{2} M \, , \nonumber \\
& [\tilde{Q},J] = - \frac12 \tilde{Q} \, , \quad
[\tilde{Q}, K_1 - i K_2] = -i Q \, , \quad
\lbrace \tilde{Q}, \tilde{Q}^{\dagger}  \rbrace = \sqrt{2} H \, ,   \nonumber \\
& \lbrace Q, \tilde{Q}^{\dagger} \rbrace = -  (P_1 -i P_2) \, , \quad
\lbrace \tilde{Q}, Q^{\dagger} \rbrace = -  (P_1 + i P_2) \, .
\label{commu superGalileo2}
\end{eqnarray}
Here $P_j$ are the spatial components of the momentum, $ K_j $ are the generators of Galilean boosts, $ J $ is the planar angular momentum and $Q, \tilde{Q}$ are the two complex supercharges. The central charge $M$ corresponds to the mass or particle number conservation\footnote{When we consider the coupling with a curved NC background, this is the responsibile for the appeareance of the $U(1)$ gauge field in the metric data.}.

The  $\mathcal{S}_2 \mathcal{G}$ algebra can be obtained in different ways. We can start with the non-relativistic Galilean algebra, add two supercharges and impose consistency conditions (this was done in $3+1$ dimensions \cite{Puzalowski:1978rv}). Alternatively, we can perform the In\"{o}n\"{u}-Wigner contraction of the $2+1$ dimensional super-Poincar\`e algebra in the $c \to \infty$ limit \cite{Bergshoeff:2015uaa}. Finally, it can be obtained by null reduction of ${\cal N}=1$ super-Poincar\'e in $3+1$ dimensions.
We will follow the last approach, since it is the most convenient for constructing the non-relativistic ${\cal N}=2$ superspace.


\subsection{Null reduction from the $\mathcal{N}=1$ SUSY algebra in $3+1$ dimensions}

In order to perform the null reduction of the $\mathcal{N}=1$ SUSY algebra in $3+1$ dimensions, we introduce light-cone coordinates in the parent theory in such a way that the spacetime is parametrized by
\beq
x^M = (x^-, x^+ , x^1, x^2)  \equiv (x^- , x^{\mu})  \qquad \quad x^{\pm} = \frac{x^3 \pm x^0}{\sqrt{2}} \, .
\label{light-cone coordinates, 2nd appearance}
\eeq
In flat Minkowski space, the DLCQ method is realized  by compactifying $x^-$ on a small circle of radius $ R$.
For convenience,  we rescale $x^- \to x^-/R$ in such a way that the rescaled coordinate is adimensional, and consequently we also rescale $x^+ \to R \, x^+$ in order for the metric tensor to be dimensionless.

It is well known that the bosonic part of the super-Poincar\'e algebra reduces to the Bargmann algebra \eqref{Bargmann} 
by identifying some components of the linear and angular momenta with the central charge and the boost operator \cite{Duval:1984cj,Son:2008ye}.
The fermionic part of the algebra requires to consider the relativistic anticommutator $\{\mathcal{Q}_\alpha , \bar{\mathcal{Q}}_{\dot{\beta}} \} = i \sigma^M_{\alpha \dot{\beta}} \partial_M$ and write the r.h.s. in terms of light-cone derivatives\footnote{For conventions on four-dimensional spinors and light-cone coordinates, see appendix \ref{app-conv}.} $\partial_{\pm} = \frac{1}{\sqrt{2}}( \partial_3 \pm \partial_0)$ 
\beq
\lbrace \mathcal{Q} , \bar{\mathcal{Q}} \rbrace = i \begin{pmatrix}
\sqrt{2} \p_+  &   \p_1 - i \p_2 \\
 \p_1 + i \p_2 & - \sqrt{2} \p_{-} 
\end{pmatrix}  \, .
\label{SUSY algebra relativistica}
\eeq 
We also require that any local function of the spacetime coordinates is decomposed into a non-relativistic field plus a phase according to eq. (\ref{decomposition fields null reduction}), \emph{m} being the dimensionless parameter associated to the $U(1)$ mass conservation.
In this way we have a dictionary for the spacetime derivatives 
\beq
\p_+ \rightarrow \p_t \, , \qquad
\p_- \rightarrow i m \, .
\eeq 
In addition, we interpret the two-components complex Weyl spinors in 3+1 dimensions as complex Grassmann scalars in 2+1 dimensions, according to the identification $\mathcal{Q}_{\alpha} \to Q_\alpha$, 
 $\bar{\mathcal{Q}}_{\dot{\beta}} \to Q_\beta^\dagger .$
Combining all these rules we obtain the algebra
\beq
\begin{aligned}
& \lbrace Q_{1} , Q_1^{\dagger} \rbrace =  \sqrt{2} i \p_{t} =   \sqrt{2}  H \, , \qquad \qquad \qquad \quad
 \lbrace Q_{1} , Q_{2}^{\dagger} \rbrace =   i (\p_{1}-i\p_2) = - (P_1 - i P_2)  \, ,  \\
& \lbrace Q_{2} , Q_{1}^{\dagger} \rbrace =   i (\p_{1}+i\p_2) = - (P_1 + i P_2)  \, , \qquad 
\lbrace Q_{2} , Q_{2}^{\dagger} \rbrace = -  \sqrt{2} i \p_{-} =    \sqrt{2} m  \, ,
\end{aligned} 
\label{SUSY algebra dalla DLCQ}
\eeq
which is identical to \eqref{commu superGalileo2} upon the identifications\footnote{We note that identification \eqref{identification} is required to obtain the correct anticommutators $\{ Q,  Q^\dagger \}$ and  $\{ \tilde{Q},\tilde{Q}^\dagger \}$, but it interchanges  $ (P_1 + i P_2) $ and $ (P_1 - i P_2) $ in the mixed anticommutators. This is simply due to the fact that we chose $x^-$ as the compact light-cone coordinate. Had we chosen $x^+$ we would have obtained exactly the algebra in \eqref{commu superGalileo2}. Since having $ (P_1 + i P_2) $ and $ (P_1 - i P_2) $ interchanged does not affect our construction, we take $ x^- $ as the compact direction being this a more conventional choice in the literature.}
\beq
\label{identification}
Q_1 =  \tilde{Q} \, , \qquad
Q_1^{\dagger} =  \tilde{Q}^{\dagger} \, , \qquad
Q_2 = Q \, , \qquad
Q_2^{\dagger} = Q^{\dagger} \, ,
\eeq
and $m$ with the eigeinvalue of the U(1) generator $M$.

It is instructive to look at the similar case of a Kaluza-Klein reduction 
from $3+1$ dimensional $\mathcal{N}=1$ SUSY algebra to the $2+1$ dimensional $\mathcal{N}=2$ SUSY algebra.
In this case we start from the same anticommutation rules $\{ \mathcal{Q}_\alpha, \bar{\mathcal{Q}}_{\dot{\beta}} \} = -\sigma^M_{\alpha \dot{\beta}} P_M$, but we compactify along the $ x^3 $ direction with momentum $p_3 \equiv Z.$
The result is \cite{Aharony:1997bx}
\beq
\lbrace \mathcal{Q}_{\alpha} , \mathcal{Q}_{\beta}^\dagger \rbrace =  -\sigma^{\mu}_{\,\, \alpha \beta} P_{\mu} +  i \epsilon_{\alpha \beta} Z 
\eeq 
where $ \mu=0,1,2 $ and \emph{Z} plays the role of a central term.
This expression is very similar to the one found in the non-relativistic case, eq. \eqref{SUSY algebra dalla DLCQ} with $m$ playing the role of a central charge. However, while in the relativistic reduction a central term appears in the fermionic part of the algebra when we reduce the number of dimensions, in the non-relativistic case the central charge is produced already in the bosonic sector (without requiring any SUSY extension) and accounts for the physical fact that in non-relativistic theories the particle number is a conserved quantity.


\subsection{Non-relativistic superspace}
\label{sect-Non relativistic superfields}

In the context of relativistic QFT, the superfield formulation is a nice tool which allows to build actions and quantities automatically invariant under the SUSY transformations.
In this context, fields belonging to the same multiplet are organized in a unique superfield, and the spacetime coordinates are supplemented by additional Grassmann coordinates which compose the superspace.
We want to apply this technique to the $\mathcal{N}=2$ non-relativistic superspace by applying the null reduction to the $\mathcal{N}=1$ formulation in $3+1$ dimensions.
The non-relativistic superspace was first introduced in four dimensions \cite{Clark:1983ne,deAzcarraga:1991fa}, whereas previous constructions in three dimensions based on different techniques can be found in \cite{Bergman:1995zr,Nakayama:2009ku}.

We start with the set of superspace coordinates  $(x^M, \theta^\alpha, \bar{\theta}^{\dot{\alpha}})$ and we take the explicit realization of the super-Poincar\'e algebra in terms of the following supercharges\footnote{Superspace conventions are discussed in Appendix \ref{app-conv}.}
\beq
{\cal Q}_{\alpha} = i \frac{\p}{\p \theta^{\alpha}} - \frac12 \bar{\theta}^{\dot{\beta}}  \p_{\alpha \dot{\beta}}  \, , \qquad
\bar{\cal Q}_{\dot{\alpha}} = - i \frac{\p}{\p \bar{\theta}^{\dot{\alpha}}} + \frac12 \theta^{\beta} \p_{\beta \dot{\alpha}} 
\label{supercariche rel 4d}
\eeq
and SUSY covariant derivatives 
\beq
{\cal D}_{\alpha} = \frac{\p}{\p \theta^{\alpha}} - \frac{i}{2} \bar{\theta}^{\dot{\beta}} \p_{\alpha \dot{\beta}}   \, , \qquad
\bar{\cal D}_{\dot{\alpha}} =  \frac{\p}{\p \bar{\theta}^{\dot{\alpha}}} - \frac{i}{2} \theta^{\beta}  \p_{\beta \dot{\alpha}} \, ,
\label{derivate covarianti rel 4d}
\eeq
which act on local superfields $\Psi(x^M, \theta^\alpha, \bar{\theta}^{\dot{\alpha}})$.

The ${\cal N}=2$ non-relativistic superspace in 2+1 dimensions can be obtained with a suitable generalization of the null reduction prescription that we used to treat separately bosons and fermions in chapter \ref{chapt-Non-relativistic actions}, and to obtain the algebra in the previous section.
We consider light-cone coordinates and rewrite $\p_{\alpha \dot{\beta}}  = \sigma^M_{\alpha \dot{\beta}} \p_M$ in (\ref{supercariche rel 4d}, \ref{derivate covarianti rel 4d}) in terms of $\p_{\pm}, \p_1,\p_2$. 
Since supersymmetry requires each field component of a multiplet to be an eigenfunction of the $\p_-$ operator with the same eigenvalue $m$, the reduction can be done directly at the level of superfields, by requiring a decomposition of the fields as
\beq
\Psi(x^M, \theta^\alpha, \bar{\theta}^{\dot{\alpha}}) = e^{im x^{-}} \tilde{\Psi}(x^+ \equiv t, x^{i}, \theta^{\alpha}, (\theta^\alpha)^\dagger) \, .
\label{decomposition rel and non-rel superfield}
\eeq
Acting on these superfields with supercharges and covariant derivatives (\ref{supercariche rel 4d}, \ref{derivate covarianti rel 4d}) rewritten in terms of light-cone derivatives, and performing the identification $\p_+ \equiv \p_t$ and $\p_- \equiv iM$ (with eigenvalue  $m$), we obtain\footnote{From now on we rename $(\theta^\a)^\dagger \equiv \bar{\theta}^\a$ and similarly for the other grassmannian quantities.}
\beq  \label{nonrelQD}
\begin{cases}
Q_1 = i \frac{\p}{\p \theta^1} - \frac12  \bar{\theta}^2 (\p_1 - i \p_2) - \frac{1}{\sqrt{2}} \bar{\theta}^1  \p_t \\
\bar{Q}_1 = -i \frac{\p}{\p \bar{\theta}^1} + \frac12 \theta^2 (\p_1 + i \p_2) + \frac{1}{\sqrt{2}} \theta^1 \p_t \\
Q_2 = i \frac{\p}{\p \theta^2} - \frac12  \bar{\theta}^1 (\p_1 + i \p_2) - \frac{i}{\sqrt{2}} \bar{\theta}^2 M \\
\bar{Q}_2 = -i  \frac{\p}{\p \bar{\theta}^2} + \frac12 \theta^1 (\p_1 - i \p_2) - \frac{i}{\sqrt{2}} \theta^2 M  \\
\end{cases}
\begin{cases}
D_1 =  \frac{\p}{\p \theta^1} - \frac{i}{2} \bar{\theta}^2 (\p_1 - i \p_2) - \frac{i}{\sqrt{2}} \bar{\theta}^1 \p_t \\
\bar{D}_1 =   \frac{\p}{\p \bar{\theta}^{1}} - \frac{i}{2} \theta^2 (\p_1 + i \p_2) - \frac{i}{\sqrt{2}} \theta^1 \p_t \\
D_2 =  \frac{\p}{\p \theta^2} - \frac{i}{2} \bar{\theta}^1 (\p_1 + i \p_2) - \frac{1}{\sqrt{2}} \bar{\theta}^2 M \\
\bar{D}_2 =   \frac{\p}{\p \bar{\theta}^2} - \frac{i}{2} \theta^1 (\p_1 - i \p_2) - \frac{1}{\sqrt{2}} \theta^2 M  \, . \\
\end{cases}
\eeq
These operators realize a representation of the non-relativistic algebra \eqref{commu superGalileo2} and can be interpreted as the supercharges and the covariant derivatives of a three-dimensional ${\cal N}=2$ superspace described by coordinates $(t, x^1, x^2,  \theta^1, \theta^2, \bar{\theta}^1,  \bar{\theta}^2)$. Correspondingly, the functions $\tilde{\Psi}$ in \eqref{decomposition rel and non-rel superfield} are three-dimensional ${\cal N}=2$ superfields realizing a representation of the non-relativistic SUSY algebra\footnote{While the structure of the non-relativistic supercharges may suggest the appearance of an interesting complex structure, we did not find any useful consequence of their particular form.
In fact, the dependence from $z=x^1+i x^2$ and $\bar{z}=x^1-ix^2$ does not seem separable, and we did not find any remarkable holomorphic dependence among the relevant objects of the theory.
The change of variables $(x^1,x^2) \rightarrow (z,\bar{z})$ seems only a useful computational way to gather the spatial dependence, but does not play a meaningful role in the supergraph formulation.}. 

We point out that the non-relativistic superspace and the corresponding supercharges could be alternatively constructed by quotienting the SUSY extension of the Bargmann algebra by the subgroup of spatial rotations and Galilean boosts, in analogy with the construction of the relativistic superspace as the quotient super-Poincar\'e$/SO(1,3)$.
However, we used the null reduction method for convenience because it refers to the quotient algebra already found in $3+1$ dimensions, and the analogy with this case is also helpful in view of the investigation of the renormalization pattern of the model and of the holomorphic properties of the superpotential.
 
As in the relativistic case, imposing suitable constraints we can reduce the number of superfield components and realize irreducible representations of the superalgebra. In particular, we are interested in the construction of (anti)chiral superfields. These can be obtained either by null reduction of the four-dimensional (anti)chiral superfields,  $\bar{\cal D}_{\dot{\alpha}} \Sigma = 0$ (${\cal D}_{\alpha} \bar{\Sigma} = 0$), or directly in three-dimensional superspace by imposing 
\beq
\bar{D}_\a \Sigma = 0 , \qquad \qquad D_\a \bar{\Sigma} = 0 
\eeq
where the covariant derivatives are given in \eqref{nonrelQD}. 

Defining coordinates 
\beq
x_{L,R}^{\mu} = x^{\mu} \mp i \theta^{\alpha} (\bar{\sigma}^\mu)_{\a \beta} \bar{\theta}^{\beta}  \qquad \qquad \mu = +, 1,2
\eeq
which satisfy $\bar{D}_\a x_L^\mu =0, D_\a x_R^\mu =0$, the (anti)chiral superfields have the following expansion
\begin{align} 
\label{supercampi x_L x_R}
& \Sigma(x_L^\mu, \theta^{\alpha}) = \varphi(x_L^\mu) +  \theta^{\alpha} \tilde \psi_{\alpha} (x_L^\mu) - \theta^2 F(x_L^\mu) & \\
& \bar{\Sigma}(x_R^\mu,  \bar{\theta}^{\beta}) = \bar{\varphi}(x_R^\mu) +  \bar{\theta}_{\gamma} \Bar{\tilde \psi}^{\gamma} (x_R^\mu) - \bar{\theta}^2 \bar{F}(x_R^\mu) &
\end{align}
We are now ready to build automatically SUSY-invariant actions using the Berezin integration on spinorial coordinates. In the relativistic setting, we can define for a generic superfield $\Psi$ the term 
\beq \label{relberezin}
\int d^4x d^4\theta \, \Psi = \int d^4x \, {\cal D}^2 \bar{\cal D}^2 \Psi \Big|_{\theta = \bar{\theta}=0}
\eeq
with covariant derivatives given in \eqref{derivate covarianti rel 4d}.
Performing the null reduction and extracting the $x^-$ dependence of the superfield by setting $\Psi = e^{imx^-} \tilde{\Psi}$, we obtain the prescription for the Berezin integrals in the non-relativistic superspace
\beq \label{Berezin integration null reduction}
\begin{aligned}
& \int d^4x d^4\theta \, \Psi = \int d^4x \, {\cal D}^2 \bar{\cal D}^2 \Psi \Big|_{\theta = \bar{\theta}=0} \;  \longrightarrow  \\
&  \int d^3x D^2 \bar{D}^2 \tilde{\Psi} \Big|_{\theta = \bar{\theta}=0} \; \times \frac{1}{2\pi} \int_0^{2\pi}  dx^- \,  e^{imx^-}   \equiv \int d^3x d^4\theta \, \tilde{\Psi}  \; \times \frac{1}{2\pi} \int_0^{2\pi}  dx^- \,  e^{imx^-}
\end{aligned}
\eeq
 where in the r.h.s. $d^3x \equiv dt \,dx^1 dx^2$ and the spinorial derivatives are the ones in eq. \eqref{nonrelQD}. 
It is immediate to observe that whenever $m \neq 0$ we obtain a trivial reduction due to the $x^-$ integral. Non-vanishing expressions arise only if the super-integrand $\Psi$ is uncharged respect to the mass generator.
In the construction of SUSY invariant actions this is equivalent to require the action to be invariant under one extra global U(1) symmetry \cite{deAzcarraga:1991fa}.  
This immediately shows that a superpotential term is admitted in non-relativistic theories only when at least two species of chiral superfields are considered in the matter content. 


\section{Review of the relativistic Wess-Zumino model}
\label{sect-Review of the relativistic Wess-Zumino model}

In this Section we will briefly review the renormalization properties of the $3+1$ dimensional relativistic Wess-Zumino (WZ) model.
In particular, we will set up the conventions for the study of the system both in superspace and component formulation, and we will recall the supergraph technique, which is very useful to simplify the computation of the quantum corrections.
The same techniques will be adopted in the non-relativistic context in the next section.

The classical action \cite{Wess:1974tw} is given by
\beq
S= \int d^4 x \, d^4 \theta  \,  \bar{\Sigma} \Sigma + \int d^4 x \, d^2 \theta \, \le \frac{m}{2} \Sigma^2  + \frac{\lambda}{3!} \Sigma^3  \ri  + \mathrm{h.c.}
\label{classical rel WZ model}
\eeq
and describes the dynamics of the field components of a chiral superfield $\Sigma = (\phi, \psi, F)$.
In anticipation of the non-relativistic case, we focus on the massless model $m=0,$ which is simpler to study but is general enough to find interesting results.
With this choice, the model is classically scale invariant ($\lambda$ is a dimensionless coupling).

When using the explicit decomposition of the chiral superfield into its components via eq. \eqref{components}, the action becomes
\beq
S =  
\int d^4x \Big[ - \p^M \bar{\phi} \, \p_M \phi + i \bar{\psi}  \bar{\sigma}^M \p_M \psi + \bar{F}  F   
+ \left( 3 \lambda F \phi^2 - 3 \lambda \psi^{\alpha} \psi_{\alpha} \phi + \mathrm{h.c.} \right) \Big] \, .
\label{component action}
\eeq
This is 
manifestly invariant under ${\cal N}=1$ SUSY transformations
\beq
\delta_{\varepsilon} \Sigma = \left[i \varepsilon^{\alpha} {\cal Q}_{\alpha} + i \bar{\varepsilon}_{\dot{\alpha}} \bar{\cal Q}^{\dot{\alpha}}  , \Sigma \right] \, ,
\eeq
which in component form read
\beq
\begin{cases}
\delta_{\varepsilon} \phi =  -\varepsilon^{\alpha} \psi_{\alpha} \\
\delta_{\varepsilon} \psi_{\alpha} =  i \bar{\varepsilon}^{\dot{\alpha}} (\p_{\alpha {\dot{\alpha}}} \phi) + \varepsilon_{\alpha} F \\
\delta_{\varepsilon} F =  - i \bar{\varepsilon}^{\dot{\alpha}} \p_{\alpha {\dot{\alpha}}} \psi^{\alpha} \, .
\label{Rel SUSY variations in components}
\end{cases}
\eeq


\subsection{Renormalization in superspace}
\label{sect-Renormalization in superspace}

The quantization of the theory is performed by considering the generating functional
\beq
Z [J, \bar{J}] = \int \mathcal{D} \Sigma \, \mathcal{D}  \bar{\Sigma}  \, \exp \left\lbrace i \le S + \int d^2 \theta \, J \Sigma + \int d^2 \bar{\theta} \, \bar{J}  \bar{\Sigma} \ri  \right\rbrace  \, ,
\eeq
where $J, \bar{J}$ are chiral and anti-chiral superfields acting as sources, respectively.
Since they are constrained by the requirement of being annihilated by half of the supercharges, when performing the functional differentiation we need to require that
\beq\label{sources}
\frac{\delta J (z_i)}{\delta J(z_j)} =  \bar{\cal D}^2 \, \delta^{(8)} (z_i - z_j) \, , \qquad
\frac{\delta \bar{J} (z_i)}{\delta \bar{J}(z_j)} =  {\cal D}^2  \, \delta^{(8)} (z_i - z_j) \, ,
\eeq
where $ z \equiv (x^M, \theta^{\alpha}, \bar{\theta}^{\dot{\alpha}}) $ and $\delta^{(8)} (z_i - z_j) \equiv \delta^{(4)} (x_i - x_j) \delta^{(2)} (\theta_i - \theta_j)\delta^{(2)} (\bar{\theta}_i - \bar{\theta}_j)$. 
The constraints are responsible for the appearance of the covariant derivatives., which will play an important role in the manipulations of the perturbative computations.

The superfield formulation is not only useful to write automatically SUSY-invariant actions and find the SUSY variation of the fields, but also to perform the computation of quantum corrections.
In fact, it turns out that the Feynman diagrams can be collected and organized in a systematic way by associating Feynman rules directly to the superfields.
Only when performing the algebra of covariant derivatives (D-algebra), they reduce to ordinary graphs for the component fields.
The super-Feynman rules for the massless WZ model are \cite{Gates:1983nr} 
\begin{itemize}
\item
Superfield propagator  
\beq
\langle \Sigma (z_i) \bar{\Sigma} (z_j) \rangle =  \frac{1}{\square} \, \delta^{(8)} (z_i - z_j) \; \longrightarrow  \; 
\langle \Sigma (p) \bar{\Sigma} (-p) \rangle = - \frac{1}{p^2} \, \delta^{(4)} (\theta_i - \theta_j) \, .
 \label{relativistic superpropagators WZ model}
\eeq 

\item
Vertices. 
The are read directly from the interacting Lagrangian: the massless WZ model contains only cubic vertices with chiral or anti-chiral superfields.
The rule \eqref{sources} implies that we assign a factor of $  \bar{\cal D}^2 $ ($ {\cal D}^2 $) to every internal line exiting from a chiral (anti-chiral) vertex.
Moreover, we use one of these factors to complete a chiral (anti-chiral) integral at the vertex, in such a way that we only have  $ \int d^4 \theta $ integrals at each vertex. 

\end{itemize}

At this point, we perform the D-algebra: we get rid of the integration along the Grassmann variables using the spinorial delta functions in order to reach a result given by a local function of $(\theta, \bar{\theta})$ integrated in a single $d^4\theta$.  
In order to obtain such a local function, we need to take into account the residual covariant derivatives ${\cal D}$'s or $\bar{\cal D}$'s acting on the internal lines and use the identities (we call for convenience $ \delta_{ij} \equiv \delta^{(2)} (\theta_i - \theta_j) \, \delta^{(2)} (\bar{\theta}_i - \bar{\theta}_j ) $) 
\begin{align}
& \delta_{ij} \delta_{ij} = 0 \, , \qquad
\delta_{ij} {\cal D}^{\alpha} \delta_{ij} = 0 \, , \qquad
\delta_{ij} {\cal D}^{2} \delta_{ij} = 0 \, , \qquad
\delta_{ij} {\cal D}^{\alpha} \bar{\cal D}^{\dot{\alpha}} \delta_{ij} = 0 \, , \qquad \delta_{ij} {\cal D}^{\alpha} \bar{\cal D}^2 \delta_{ij} = 0 \, ,
 &  \nonumber \\
 &  \nonumber \\
 &   \delta_{ij} {\cal D}^{\alpha} \bar{\cal D}^2 {\cal D}^{\beta} \delta_{ij} = -\epsilon^{\alpha \beta} \delta_{ij} \, , \qquad   \delta_{ij} {\cal D}^{2} \bar{\cal D}^2  \delta_{ij} = \delta_{ij} \bar{\cal D}^2 {\cal D}^{2} \delta_{ij} =  \delta_{ij} \frac{ {\cal D}^{\alpha} \bar{\cal D}^2 \mathcal{D}_{\alpha}}{2}  \delta_{ij} =  \delta_{ij}  \, . &
 \label{rules covariant derivatives on delta functions}
\end{align}
A careful analysis reveals that the D-algebra method must be applied until we reach a configuration in which exactly two ${\cal D}$'s and two $\bar{\cal D}$'s survive in each loop. 
This amounts to integrate by parts spinorial derivatives at the vertices and trade products of them with space-time derivatives through commutation rules like
\beq
[{\cal D}^{\alpha}, \bar{\cal D}^2] = i \p^{\alpha \dot{\alpha}} \bar{\cal D}_{\dot{\alpha}} \, , \quad [\bar{\cal D}^{\dot{\alpha}}, {\cal D}^2] = -i \p^{\alpha \dot{\alpha}} {\cal D}_{\alpha} \, , \quad
{\cal D}^2 \bar{\cal D}^2 {\cal D}^2 = \square {\cal D}^2 \, , \quad  \bar{\cal D}^2 {\cal D}^2 \bar{\cal D} ^2 = \square \bar{\cal D}^2 \, .
\label{rules covariant derivatives giving momenta in supergraphs}
\eeq
Every configuration in which a loop ends up containing less than $2{\cal D}$'s $+ 2\bar{\cal D}$'s must be discarded, while the configurations with exactly two ${\cal D}$'s plus two $\bar{\cal D}$'s have a non-vanishing Grassmann integration and give a local expression in the spinorial coordinates. 

In this way, a supergraph is reduced to a sum of ordinary Feynman diagrams.
As usual, in momentum space these correspond to integrals over loop momenta, with momentum conservation at each vertex. In general UV and IR divergences arise, and suitable regularizations are required in order to perform the integrals. At the end of the calculation, going back to configuration space we obtain contributions that are given by local functions of the superspace coordinates integrated in $d^4x d^4\theta$. 

The WZ model is renormalizable by power counting.
The D-algebra method described above immediately tells us that the only quantum corrections of the theory arise in the form of non-chiral superspace integrals, which by construction can only contribute to the kinetic part of the effective action, but not to the superpotential term.
This is the perturbative proof of the non-renormalization theorem for the superpotential of the WZ model \cite{Grisaru:1979wc}. 

This fact does not mean that the coupling constant does not renormalize, instead it inherits a non-trivial behaviour only as a consequence of the wavefunction renormalization. 
Concretely, we have
\beq
\mathcal{L} = \int d^4 \theta \, (\bar{\Sigma} \Sigma) + \int d^2 \theta \, (\lambda \Sigma^3) \, \rightarrow
\mathcal{L}_{\rm ren} = \int d^4 \theta \, Z_{\Sigma} (\bar{\Sigma} \Sigma) + \int d^2 \theta \, Z_{\lambda} Z_{\Sigma}^{3/2} (\lambda \Sigma^3)  \, ,
\eeq
but the absence of chiral divergences implies 
\beq
Z_{\lambda} Z_{\Sigma}^{3/2} = 1 \; \; \Longrightarrow \; \;
Z_{\lambda} = Z_{\Sigma}^{-3/2}  \, .
\label{statement non-renormalization theorem on Z}
\eeq

\subsection{Renormalization in components}
\label{sect-Renormalization in components}

In view of the comparison with the non-relativistic case, where up to now most of the literature has used the components field formalism, it is instructive to show how the perturbative computations occur in components.

The Feynman rules can be read from the action \eqref{component action}, where the only difference with the usual QFT formulation is the presence of auxiliary fields with 2-point function $\langle F \bar{F} \rangle = 1.$
The scalar and fermionic propagator are the ordinary ones\footnote{The propagators can also be obtained by reducing the super-propagator \eqref{relativistic superpropagators WZ model} in components.}.
A direct inspection of the interacting action immediately gives the vertices, which are only cubic.
The study of ordinary Feynman diagrams gives a renormalized Lagrangian of the form
\beq
\mathcal{L}_{\mathrm{ren}} =  - Z \, \p^M \bar{\phi} \, \p_M \phi + i Z \bar{\psi} \bar{\sigma}^M \p_M \psi + Z \bar{F} F 
 + \left( 3 \lambda  Z_{\lambda} Z^{3/2} F \phi^2 + 3 \lambda   Z_{\lambda} Z^{3/2} \psi^{\alpha} \psi_{\alpha} \phi + \mathrm{h.c.}
\right)
\eeq
where we have used the SUSY condition $Z_{\phi} = Z_{\psi} = Z_F \equiv Z$. 
In this way, the non-renormalization theorem still leads to condition (\ref{statement non-renormalization theorem on Z}). In fact, since we have not eliminated the auxiliary field $F$, this is nothing but a trivial rephrasing of the superspace approach. 

On the other hand, a different way to proceed is to integrate out all the auxiliary fields and retain only the dynamical ones.
In this case we use the equations of motion of $F,$ which are algebraic relations, and we perform again the perturbative analysis.
Taking into acount the non-renormalization condition \eqref{statement non-renormalization theorem on Z}, it turns out that the renormalized action for the dynamical fields reads
\beq
\mathcal{L}_{\mathrm{ren}} = - Z \p^M \bar{\phi} \, \p_M \phi + i Z \bar{\psi} \bar{\sigma}^M \p_M \psi + \left( 3 \lambda  \psi^{\alpha} \psi_{\alpha} \phi - 9 Z^{-1} |\lambda|^2 |\phi|^4 + \mathrm{h.c.} \right)
\eeq
We observe that renormalization in component fields formulation is more subtle: while the cubic vertex is still non-renormalized, the quartic term instead renormalizes non-trivially, due the wavefunction renormalization.
Even if the non-renormalization theorem is still at work, this procedure shows that quantum corrections to the vertices may arise in the component field formulation containing only dynamical fields.

\subsection{The non-renormalization theorem}
\label{sect-The non-renormalization theorem}

The holomorphicity of the superpotential is a powerful constraint which forces all quantum corrections to vanish. At perturbative level, superspace techniques provide a straightforward proof, as we have just recalled. The non-perturbative derivation of this result comes instead from an argument due to Seiberg \cite{Seiberg:1993vc}, which we reviewed in the introduction \ref{int-Non-relativistic supersymmetry}.

This classical argument can be extended, following \cite{Weinberg:2000cr}, to the case of a generic superpotential $W$ giving the action
\beq
S= \int d^4 x \, d^4 \theta  \,  \bar{\Sigma}_a \Sigma_a + \int d^4 x \, d^2 \theta \,  
W(\Sigma_a) + \mathrm{h.c.} 
\label{WZ-generico}
\eeq
We introduce one extra chiral superfield $Y$, whose scalar part is set to $1$ to recover the original action, whereas the spinorial and auxiliary components vanish identically. 
A consistent assignment of R-charges for the superfields appearing in the action is $R(\Sigma_a)=0$ and $R(Y)=2$.
Upon quantization, the kinetic part of the action will be in general modified by the wave function renormalization, which we parametrize with the introduction of real superfields $Z_{ab}$ in the following way
\beq
\tilde{S}= \int d^4 x \, d^4 \theta  \, Z_{ab} \bar{\Sigma}_a \Sigma_b + \int d^4 x \, d^2 \theta \,  
Y \, W(\Sigma_a) + \mathrm{h.c.}
\label{WZ-generico2}
\eeq
Assuming that the regularization procedure does not spoil SUSY,  the Wilsonian effective action at a given scale $\lambda$
is of the form
\beq
\tilde{S}_\lambda= \int d^4 x \, d^4 \theta   \,  K(\bar{\Sigma}_a \Sigma_a,Z_{ab}, Y, \bar{Y}, \mathcal{D})
+\int d^4 x \, d^2 \theta \, W_\lambda (\Sigma_a, Y)  + \mathrm{h.c.}
\label{WZ-generico3}
\eeq
Then R-invariance and holomorphicity of the superpotential force $W_\lambda$ to be of the form
\beq
W_\lambda (\Sigma_a, Y) =Y \,  W_\lambda(\Sigma_a) 
\label{WZ-generico4}
\eeq
Taking the weak coupling limit $Y \rightarrow 0$, the only contribution to the superpotential is a tree-level vertex, and therefore we find $ W_\lambda(\Sigma_a) = W(\Sigma_a)$.


\section{The non-relativistic Wess-Zumino model}
\label{sect-The non-relativistic Wess-Zumino model}

We use the superfield formalism introduced in Section \ref{sect-Non relativistic superfields} to investigate the renormalization properties of the non-relativistic counterpart of the WZ model.
First of all, we find the Galilean-invariant version of the WZ model by applying the null reduction prescription for Berezin integration \eqref{Berezin integration null reduction} to the action in (\ref{classical rel WZ model}).

Using the decomposition of the superfields as $\Sigma = e^{imx^-} \Phi,$ we immediately observe that while the canonical Kahler potential survives the reduction being U(1) neutral, the holomorphic superpotential has charge 3 and is killed by the $x^-$ integration. 
The only way-out to obtain an interacting non-relativistic scalar model is then to introduce at least two species of superfields with different $m$ charges, and trigger them in such a way that also the superpotential turns out to be neutral.

The minimal non-trivial case is the null reduction of the $3+1$ dimensional relativistic WZ model with two massless fields described by the action
\beq
 S = \int d^4 xd^4 \theta    \, \le \bar{\Sigma}_1 \Sigma_1 + \bar{\Sigma}_2 \Sigma_2 \ri  + g  \int  d^4 x d^2 \theta    \, \Sigma_1^2 \Sigma_2 + \mathrm{h.c.}
 \label{WZM-before-null-red}
\eeq
with the decomposition
\beq
\Sigma_1 (x^M, \theta, \bar{\theta}) = \Phi_1 (x^{\mu}, \theta, \bar{\theta}) \, e^{imx^{-}} \, , \qquad \Sigma_2 (x^M, \theta, \bar{\theta}) = \Phi_2 (x^{\mu}, \theta, \bar{\theta}) \,  e^{-2imx^{-}} \, 
\label{superfields sector 1 and 2 non rel WZ model}
\eeq
which ensures that the superpotential is neutral under the $U(1)$ mass generator.
We will refer to these superfields as belonging to sector 1 and 2, respectively.

The null-reduced action simply becomes
\beq
\label{non-rel WZ action in superfield formalism} 
S = \int   d^3 x d^4 \theta \le \bar{\Phi}_1 \Phi_1 +  \bar{\Phi}_2 \Phi_2 \ri + g \int   d^3 x  d^2 \theta \,  \Phi_1^2 \Phi_2 + \mathrm{h.c.}
\eeq
Using the definitions in eq. \eqref{light-cone coordinates, 2nd appearance} and below, it is clear that in the non-relativistic case the 
time coordinate has twice the dimensions of the spatial ones.
In this way, the superfields have still mass dimension one and the coupling $g$ is dimensionless. Therefore the model shares classical scale invariance with its relativistic parent.
Moreover, the formulation with superfields obtained from the null reduction of $\mathcal{N}=1$ relativistic ones assures that the $2+1$ dimensional action is invariant under $\mathcal{N}=2$ non-relativistic SUSY.

\subsection{Expansion of the action in components}

In order to compare with the standard approach in the literature, we put the action to the component formalism.
We start with the null reduction of the kinetic part, since it reveals some non-trivial features.
Applying the prescription for the Berezin integration \eqref{Berezin integration null reduction} we can write 
\beq
S_{kin} = \int d^3 x \, \left[\bar{D}^2 \bar{\Sigma}_a \, D^2 \Sigma_a + \bar{D}_{\alpha} \bar{\Sigma}_a \; \bar{D}^{\alpha} D^2 \Sigma_a + \bar{\Sigma}_a \, \bar{D}^2 D^2 \Sigma_a  \right]  \, ,
\eeq
where the non-relativistic covariant derivatives are given in eq. \eqref{nonrelQD} and $  a=1,2 $ labels the two sectors of superfields.

We define the theta expansion of the Wess-Zumino chiral superfields as (here $\theta^1, \theta^2$ indicate components 1 and 2 of the $\theta^\alpha$ spinor)
\beq \label{sigma_components}
\begin{aligned}
& \Sigma_1 = \varphi_1 + \theta^1 \xi_1 + \theta^2 \, 2^{\frac14}\sqrt{m} \, \chi_1 - \frac12 \theta^\alpha \theta_\alpha F_1 \, , \\
& \Sigma_2 = \varphi_2 + \theta^1 \xi_2 + \theta^2 \, i 2^{\frac14} \sqrt{2m}  \, \chi_2 - \frac12 \theta^\alpha \theta_\alpha F_2 \, ,
\end{aligned}
\eeq
where a convenient rescaling of the grassmannian fields has been implemented in order to have the standard normalization of the kinetic terms, with $\varphi_a$ and $\chi_a$ sharing the same dimensions. Using these conventions we obtain
\beq
\begin{aligned}
S_{kin}  = & \int d^3 x \,  \left[ 2 i m \bar{\varphi}_1 (\p_t \varphi_1) + \bar{\varphi}_1 \p_i^2 \varphi_1 - 4 i m \bar{\varphi}_2 (\p_t \varphi_2) + \bar{\varphi}_2 \p_i^2 \varphi_2 + \bar{F}_1 F_1 + \bar{F}_2 F_2
\right. \\
& \left. + \sqrt{2} m \bar{\xi}_1 \xi_1 + 2 i m \bar{\chi}_1 (\p_t \chi_1) -  2^{\frac{1}{4}} i \sqrt{m} \, \bar{\xi}_1 (\p_1 - i \p_2) \chi_1  
  -  2^{\frac{1}{4}} i \sqrt{m} \bar{\chi}_1 (\p_1 + i \p_2) \xi_1 \right. 
\\ & \left. - 2 \sqrt{2} m \bar{\xi}_2 \xi_2 + 4 i m \bar{\chi}_2 (\p_t \chi_2) +  2^{\frac{1}{4}}  \sqrt{2m} \, \bar{\xi}_2 (\p_1 - i \p_2) \chi_2  
  -  2^{\frac{1}{4}}  \sqrt{2m} \bar{\chi}_2 (\p_1 + i \p_2) \xi_2 \right] \, .
\end{aligned}
\label{Azione modello WZ interagente al primo ordine}
\eeq
Integrating out the auxiliary fields $ F_{1,2} $ and $ \xi_{1,2} $ we find
\beq\label{actioncomponents}
\begin{aligned}
S_{kin}  =  \int d^3 x & \, \left[   2 i m \bar{\varphi}_1 \p_t \varphi_1 + \bar{\varphi}_1 \p_i^2 \varphi_1 
- 4 i m \bar{\varphi}_2 \p_t \varphi_2   + \bar{\varphi}_2 \p_i^2 \varphi_2   \right. \\
& \; \left. + 2 i m  \bar{\chi}_1 \p_t \chi_1 +  \bar{\chi}_1 \p_i^2 \chi_1 + 4 i m \bar{\chi}_2 \p_t \chi_2
 -  \bar{\chi}_2 \p_i^2 \chi_2 \right] \, ,
\end{aligned}
\eeq  
where $ \varphi_{1,2} $ and $ \chi_{1,2} $ are the dynamical non-relativistic scalar and fermion fields, respectively. 

If we Fourier-transform both the scalar and fermion fields
\beq
 \varphi (x^{\mu}) = \int \frac{ d\omega \, d^2 k }{(2 \pi)^3} \, a (\vec{k}) e^{-i ( \omega t - \vec{k} \cdot \vec{x}) } \, ,
\eeq
the free equations of motion lead to the following dispersion relations 
\beq
\omega_1 =   \frac{\vec{k_1}^2}{2 m} \, , \qquad \qquad \omega_2 = -  \frac{\vec{k_2}^2}{4 m} \, .
\eeq
The wrong sign for the energy of fields in sector 2 is due to U(1) invariance, which forces to assign a negative eigenvalue to the mass operator for $\Phi_2$ in the decomposition (\ref{superfields sector 1 and 2 non rel WZ model}). 

The problem can be avoided if we integrate by parts the derivatives in sector 2
\beq
\begin{aligned}
S_{kin}   = \int d^3 x & \, \left[   2 i m \bar{\varphi}_1 \p_t \varphi_1 + \bar{\varphi}_1 \p_i^2 \varphi_1 
+ 4 i m \varphi_2\p_t \bar{\varphi_2}    + \varphi_2\p_i^2 \bar{\varphi}_2     \right. \\
& \; \left. + 2 i m  \bar{\chi}_1 \p_t \chi_1 +  \bar{\chi}_1 \p_i^2 \chi_1 + 4 i m  \chi_2 \p_t\bar{\chi}_2   
 + \chi_2 \p_i^2 \bar{\chi}_2     \right] \, ,
\end{aligned}
\eeq 
and we interchange the roles of $\varphi_2$ with $\bar{\varphi}_2,$ and of  
$\chi_2$ with $\bar{\chi}_2$.
This operation is equivalent to reversing the role of creation and annihilation operators.
From the point of view of the superfield formulation, this is equivalent to interchanging all the components of $ \Phi_2$ with the components of $ \bar{\Phi}_2 $, without affecting the chirality property of the superfield itself (the Grassmann coordinates entering the superfield are not modified in this exchange).
From now on we will call $ \Phi_2$ the chiral superfield whose components\footnote{ For details about the null reduction of non-relativistic fermions, see Appendix \ref{app-conv}.} are $(\bar{\varphi}_2, \bar{\xi}_2, \bar{\chi}_2, \bar{F}_2)$, while the antichiral $\bar{\Phi}_2$ has components $(\varphi_2, \xi_2, \chi_2, F_2)$, and assign positive mass $2m$ to $\Phi_2$.

The component form of the interacting part of the action can be similarly obtained by means of the standard superspace manipulations combined with the Berezin integration \eqref{Berezin integration null reduction}.
After performing the redefinition of the fields and adding the contribution from the interacting part of the Lagrangian, we obtain 
\beq
\begin{aligned}
S   =   & \int d^3x  \; \Big[ 2 i m \bar{\varphi}_1 \p_t \varphi_1 + \bar{\varphi}_1 \p_i^2 \varphi_1 
+ 4 i m \bar{\varphi}_2 \p_t \varphi_2  + \bar{\varphi}_2 \p_i^2 \varphi_2  \\
& \quad  + 2 i m  \bar{\chi}_1 \p_t \chi_1 +  \bar{\chi}_1 \p_i^2 \chi_1  + 4 i m \bar{\chi}_2 \p_t \chi_2
 +  \bar{\chi}_2 \p_i^2 \chi_2 - 4 |g|^2 |\varphi_1\varphi_2|^2 -  |g|^2 |\varphi_1|^4   \\
&  -  i g \left( \sqrt{2} \varphi_1 \chi_1 (\p_1 - i \p_2) \bar{\chi}_2 -2 \bar{\varphi}_2 \chi_1 (\p_1 - i \p_2) \chi_1 + 2 \sqrt{2} \varphi_1 ((\p_1 - i \p_2)\chi_1) \bar{\chi}_2  \right) + {\rm h.c.} \\
 & + 2 |g|^2 \left( -|\varphi_1|^2 \bar{\chi}_1 \chi_1 - 4 |\varphi_1|^2 \bar{\chi}_2 \chi_2 + 2 |\varphi_2|^2 \bar{\chi}_1 \chi_1 + 2 \sqrt{2} \varphi_1 \varphi_2 \bar{\chi}_1 \bar{\chi}_2 + 2 \sqrt{2} \bar{\varphi}_1 \bar{\varphi}_2 \chi_2 \chi_1 \right)  \Big] \, .
\end{aligned}   
\label{Lagrangiana finale WZ interagente potenziale cubico}
\eeq
In component field formulation, there are standard quartic scalar interactions together with cubic derivative interactions, similar to the interacting part of a non-supersymmetric $1+1$ dimensional 
Galilean model recently studied in \cite{Yong:2017ubf}.

An alternative way to derive this action is to consider the null reduction of the component field formulation of the relativistic $3+1$ dimensional WZ model, where the auxiliary fields are integrated out.
In other words, the elimination of the non-dynamical fields from the action commute with the DLCQ.

\section{Quantum corrections in superspace}
\label{sect-Quantum corrections in superspace}

We study the renormalization properties of the model in eq.
 (\ref{non-rel WZ action in superfield formalism}) by means of the superfield formalism\footnote{The computation in component field formalism is performed in Appendix \ref{app_Non-relativistic Wess-Zumino model in components}.}.
As in the relativistic case, this procedure turns out to be very convenient to consider the contributions from various fields inside a unique supergraph.

First of all, we will collect the super-Feynman rules for the propagators and the vertices of the theory from the null reduction of the corresponding ones in the relativistic parent\footnote{As usual, these Feynman rules can be taken as well directly from the null reduced non-relativistic action (\ref{non-rel WZ action in superfield formalism}).} found in Section \ref{sect-Review of the relativistic Wess-Zumino model}.
Moreover, we will show how the non-relativistic properties of the theory strongly constrain the form of the supergraphs accounting for the quantum corrections of the theory.
A crucial role is played by the $U(1)$ symmetry associated to the mass central charge $M$, which implies that the particle number has to be conserved at each vertex and the only non-vanishing Green functions are the ones whose external particle numbers add up to zero.
The other fundamental property will be the causal structure of the inverse non-relativistic propagator, which is linear in the energy instead of quadratic as it happens in the relativistic case.


\subsection{Super-Feynman and selection rules}
\label{sect-Super-Feynman and selection rules}

\begin{figure}[h]
\begin{subfigure}[b]{0.5\linewidth}
\centering
\includegraphics[scale=1]{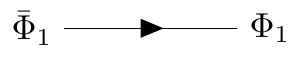}
\end{subfigure}
\begin{subfigure}[b]{0.5\linewidth}
\centering
\includegraphics[scale=1]{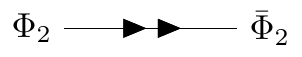}
\end{subfigure}
\begin{subfigure}[b]{0.5\linewidth}
\centering
\includegraphics[scale=0.9]{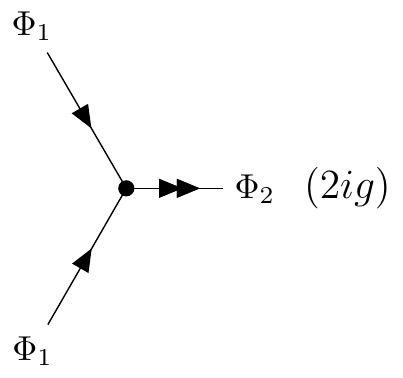}
\end{subfigure}
\begin{subfigure}[b]{0.5\linewidth}
\centering
\includegraphics[scale=0.9]{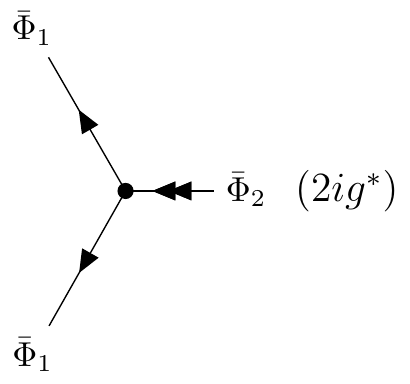}
\end{subfigure}
\caption{Propagators and vertices in superspace.}
\label{fig2-Superpropagators e vertices WZ model nonrel}
\end{figure}

The $U(1)$ particle conservation at each vertex can be implemented in a graphical way by associating to each propagator line a flow depicted with an arrow: a single one for the superfield  $\Phi_1$ with mass $m,$ and a double one for $\Phi_2$ with mass $2m.$
Consequently, the super-Feynman rules in the non-relativisitic ${\cal N}=2$ superspace are:

\begin{itemize}
\item
Superfield propagators. 
The null reduction simply acts as the replacement $\square \rightarrow 2 i M \, \p_t + \p_i^2 $, with $M = m$ or $ 2 m$ in the propagators of the relativistic parent (\ref{relativistic superpropagators WZ model}).
In momentum space, this is equivalent to send $ -p^2 \to 2 M \omega - \vec{p}^{\, 2}$. 
The result is
\beq
\langle \Phi_1 (\omega, \vec{p}) \bar{\Phi}_1 (-\omega, -\vec{p}) \rangle =  i\frac{ \,  \delta^{(4)} (\theta_1 - \theta_2)  }{2m \omega -\vec{p}^{\, 2} + i \varepsilon}  \, , \quad
\langle \bar{\Phi}_2 (\omega, \vec{p}) \Phi_2 (-\omega, -\vec{p}) \rangle =  i \frac{ \delta^{(4)} (\theta_1 - \theta_2)  }{4m \omega -\vec{p}^{\, 2} + i \varepsilon}  
\label{superpropagatori momentum space}
\eeq
The dimensional analysis of the denominator works correctly when remembering that the energy dimensions in the Galilean setting are taken to be
\beq
[\omega]= E^2 \, , \qquad [\vec{k}]= E \, , \qquad [m]=E^0 \, ,
\eeq
since we rescaled the $x^-$ direction as $x^- \rightarrow x^- /R$ and the time direction as $x^+ \rightarrow R x^+,$ being $R$ the radius of the circle along the compact null direction.
The propagators for both sectors have a retarded $ i \varepsilon $ prescription which follows the order of fields shown in fig. \ref{fig2-Superpropagators e vertices WZ model nonrel}, where the exchange of particles with anti-particles in sector 2 is manifest from  the reversed order of the fields with respect to sector 1. \footnote{In configuration space the $ i \varepsilon $ prescription translates into a retarded prescription for the propagator. In fact, the Fourier transform of \eqref{superpropagatori momentum space} reads ($M=m$ or $2m$)
\beq
D(\vec{x} , t)= \int \frac{d^2 p \,  d \omega}{(2 \pi)^3} 
i\frac{ \delta^{(4)} (\theta_1 - \theta_2)}{2 M \, \omega - \vec{p}^2 + i \varepsilon} \, e^{-i (\omega t - \vec{p} \cdot \vec{x})} =
 - \frac{i \, \Theta (t)}{4 \pi \,  t} e^{i\frac{M \vec{x}^2}{2t}} \,  \delta^{(4)} (\theta_1 - \theta_2)
\eeq
where $\Theta$ is the Heaviside function.
We will see that the appearance of the step function (a consequence of the linearity of the inverse propagator with respect to energy) will play a fundamental role in the determination of the renormalization properties of the model.}

\item 
Vertices. 
There are no subtleties in the non-relativistic limit and they are cubic vertices containing only chiral or anti-chiral superfields.
The particle number conservation at each vertex translates into the condition that the numbers of entering and exiting arrows have to match (see fig. \ref{fig2-Superpropagators e vertices WZ model nonrel}).
\end{itemize}

\vskip 10pt

What are the changes in the D-algebra procedure with respect to the relativistic case?
The null reduction does not affect the grassmannian part of the superspace, which means that the rules \eqref{sources} for the sources in the path integral still hold we have to count one extra $\bar{D}^2$ ($D^2$) for each chiral (anti-chiral) superfield entering or exiting a vertex, and use one of these factors at each vertex to complete the chiral integral to a $\int d^4 \theta$ one.
The only important difference is that in the present case the grassmannian derivatives are the non-relativistic ones in \eqref{nonrelQD}.
Every other step of the D-algebra that applies in the relativistic case can be performed as in Section \ref{sect-Renormalization in superspace} in order to reduce the supergraph to a combination of ordinary Feynman graphs for functions that are local in $(\theta, \bar{\theta})$. 
The identities \eqref{rules covariant derivatives on delta functions}, which immediately rule out all the diagrams without enough covariant derivatives running along the lines of the graph, still hold with the change $D, \bar{D} \rightarrow {\cal D}, \bar{\cal D} .$
The rules \eqref{rules covariant derivatives giving momenta in supergraphs} require instead the application of the DLCQ prescription and become (see eqs. (\ref{3d_derivatives}, \ref{useful_identities}) in Appendix \ref{app-conv} for more details)
\begin{align}
&[D^{\alpha}, \bar{D}^2] = \sqrt{2} M
\bar{D}_1 \delta^{\alpha}_{ 1}  + i (\bar{\sigma}^{\mu})^{\alpha \beta} \p_{\mu} \bar{D}_{\beta} 
 \, , \quad  [\bar{D}^{\alpha}, {D}^2] = -\sqrt{2}  M
D_1 \delta^{\a}_{ 1} - i (\bar{\sigma}^{\mu})^{\alpha \beta} \p_{\mu} D_{\beta}  \, ,  \nonumber & \\ 
& {D}^2 \bar{D}^2 {D}^2 = (2 i M \, \p_t + \p_i^2) D^2  \, , \qquad \qquad  \; \;  \bar{D}^2 {D}^2 \bar{ D} ^2 = (2 i M \, \p_t + \p_i^2) \bar{D}^2 
\label{rules covariant derivatives giving momenta in supergraphs 2}
\end{align}
where $  \mu \in \lbrace +,1,2 \rbrace$.

Since the only kind of interactions appearing in the superpotential are cubic as in the relativistic case, the possible topologies of supergraphs that can be built with the rules at our disposal are the same of the ordinary WZ model (in particular, they were studied \emph{e.g.} in \cite{Abbott:1980jk}, \cite{Sen:1981hk}).
In addition, there are certain selection rules which can be derived by the non-relativistic properties of the theory and that drastically reduce the number of non-vanishing diagrams.

We start analyzing the consequences of the retarded nature of the non-relativistic propagator.
The linearity in the energy $\omega$ implies that
\begin{srule} 
\label{srule2}
Arrows inside a Feynman diagram cannot form a closed loop.
\end{srule}

\begin{figure}[h]
\centering
\includegraphics[scale=1.4]{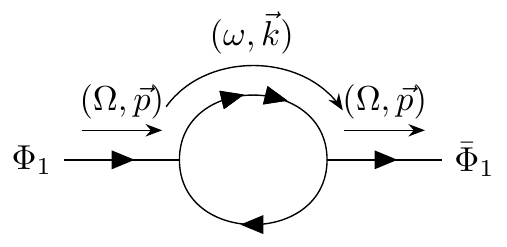}
\caption{One-loop correction to the self-energy of the $\Phi_1$ superfield.}
\label{fig3_rinormalizzione self-energy supercampo 1}
\end{figure}

This can be easily seen to be a consequence of the residue theorem in momentum space and is better illustrated with an example.
We consider the quantum correction to the self-energy of the superpropagator in sector 1, as depicted in fig. \ref{fig3_rinormalizzione self-energy supercampo 1}.

The contribution to the effective action from this diagram is\footnote{We denote the effective action with a subscript referring to the sector to which the superfield belongs, and with a superscript referring to the number of external lines to attach to the diagram.}
\beq
i \Gamma_1^{(2)}  (\Phi_1, \bar{\Phi}_1) =  4 |g|^2  \int d^4 \theta \, \frac{d\omega \, d^2k}{(2 \pi)^3}  \, \frac{ \Phi_1 (\Omega, \vec{p}, \theta) \bar{\Phi}_1 (\Omega, \vec{p}, \theta)}{\left[4 m \omega- \vec{k}^2 + i \varepsilon\right]\left[2 m (\omega-\Omega)- (\vec{k} -\vec{p})^2 + i \varepsilon \right]}  \, .
\eeq
The integral in $\omega$ is convergent and can be performed \emph{e.g.} by means of the residue theorem.
The poles of the integrand are located in
\beq
\omega^{(1)} = \frac{\vec{k}^2}{4 m} - i \varepsilon \, , \qquad
\omega^{(2)} = \Omega + \frac{(\vec{k}- \vec{p})^2}{2 m} - i \varepsilon \, , 
\eeq
and in particular are both in the lower-half complex plane, so that we can close the integration contour in the upper half-plane, obtaining
\beq
\Gamma_1^{(2)} (\Phi_1, \bar{\Phi}_1) = 0 \, .
\eeq
Analogously, in configuration space, 
the vanishing of the two-point function arises from the product of two 
Heaviside functions with opposite arguments, which would have support only in one point. 
By normal ordering, we choose to put this contribution to zero \cite{Bergman:1991hf}.

This argument can be generalized to the case where the analitic structure of the integrand in momentum space is given by a set of simple poles located from the same side of the complex plane in $\omega.$
In particular, this analitic structure corresponds from the diagrammatic point of view to a disposition of the arrows where they form a closed loop, which is the statement of the selection rule \ref{srule2}.
On the other hand, we need to be careful on the convergence conditions of the integral: the rule relies on the possibility to apply the residue theorem to perform the $\omega-$integration, which in turn requires the integrand to be sufficiently decreasing at infinity for applying Jordan's lemma.
The behaviour of the propagators should guarantee that this is always the case, but the $D$-algebra procedure can in principle spoil the convergence due to extra $\omega$ factors coming from the commutation rules \eqref{rules covariant derivatives giving momenta in supergraphs 2}.
As it will be discussed in Section \ref{sect-Renormalizability of the theory}, this never happens and then selection rule \ref{srule2} is true even before performing D-algebra.

\begin{figure}[H]
\centering
\begin{subfigure}[b]{0.4\linewidth}
\centering
\includegraphics[scale=0.5]{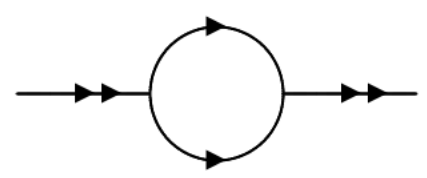}
\caption{}
\end{subfigure} 
\begin{subfigure}[b]{0.4\linewidth}
\centering
\includegraphics[scale=0.25]{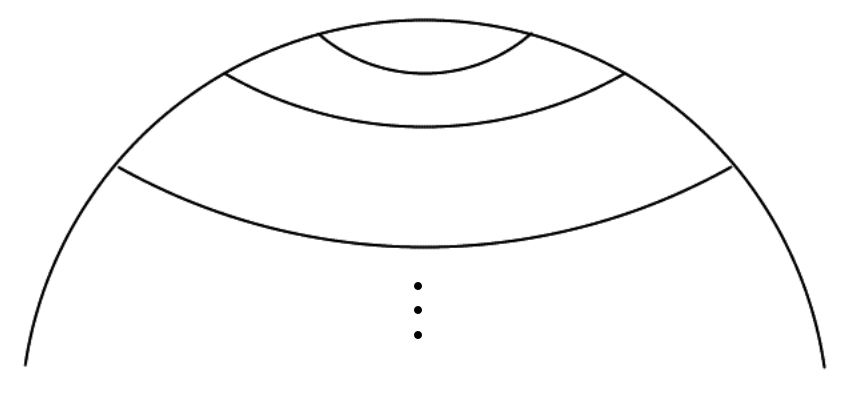}
\caption{}
\end{subfigure}
\caption{Configurations allowed (a) and forbidden (b) by selection rule \ref{srule2}.}
\label{fig6_selectionrule4}
\end{figure}

We stated that the supergraphs which we can build with the super-Feynman rules are in principle the same of the relativistic case, having the same building blocks (the propagators and the vertices are graphically are the same).
However, after imposing the stringent constraints provided by selection rule \ref{srule2}, the number of allowed diagrams drastically reduce.
For example, an immediate consequence is that at one loop, one-particle irreducible diagrams with two external lines admit only one non-vanishing configuration, given in fig. \ref{fig6_selectionrule4}(a). 
This rule is true also when the diagram is part of a bigger graph. As a consequence, the topology shown in fig. \ref{fig6_selectionrule4}(b) is always forbidden, when the number of horizontal lines is bigger than two.
 
\begin{figure}[h]
\centering
\begin{subfigure}[b]{0.4\linewidth}
\centering
\includegraphics[scale=0.4]{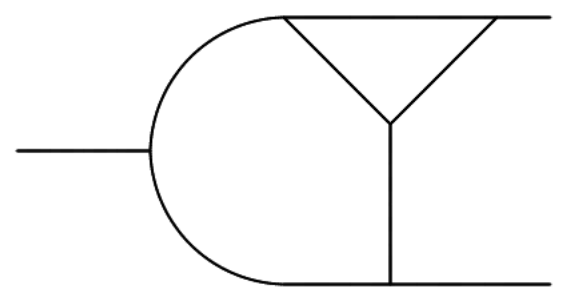}
\caption{}
\end{subfigure}
\begin{subfigure}[b]{0.4\linewidth}
\centering
\includegraphics[scale=0.4]{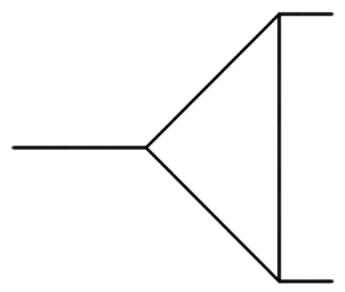}
\caption{}
\end{subfigure} \\
\begin{subfigure}[b]{0.3\linewidth}
\centering
\includegraphics[scale=0.25]{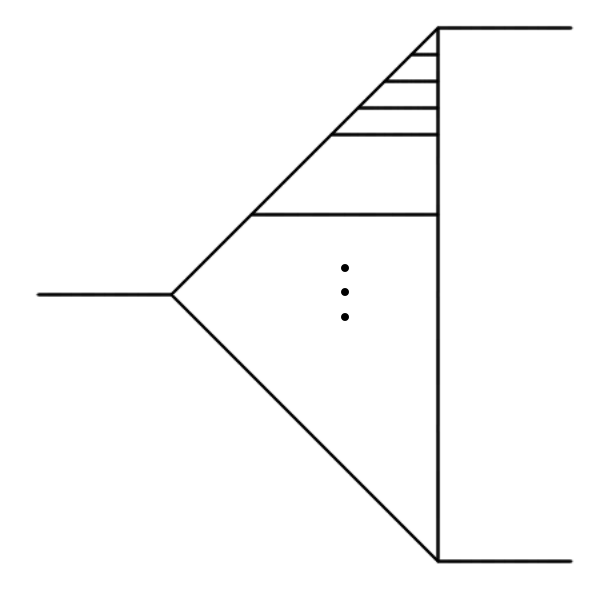}
\caption{}
\end{subfigure}
\begin{subfigure}[b]{0.3\linewidth}
\centering
\includegraphics[scale=0.25]{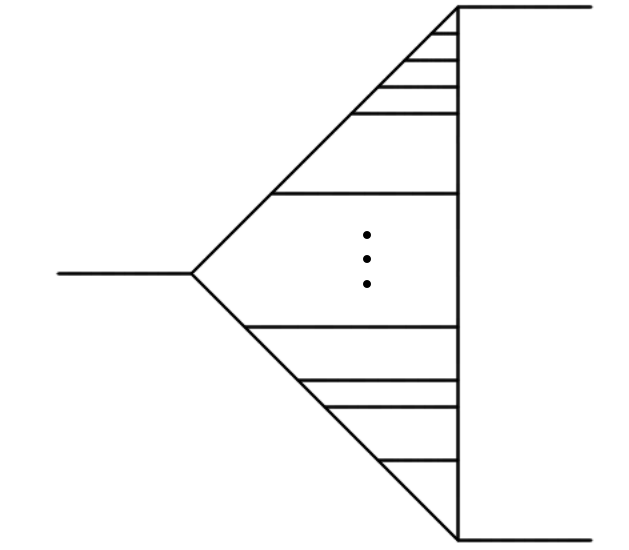}
\caption{}
\end{subfigure}
\begin{subfigure}[b]{0.3\linewidth}
\centering
\includegraphics[scale=0.25]{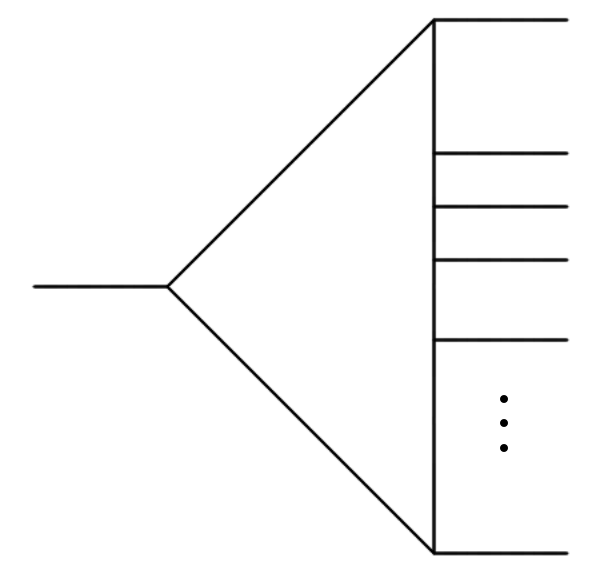}
\caption{}
\end{subfigure}
\caption{Set of vanishing (sub)diagrams due to particle number conservation.
In (e) the number of horizontal lines on the right side is required to be even.}
\label{fig4_selectionrule3}
\end{figure}

Further  selection rules can be obtained from the application of the particle number conservation:
\begin{srule}
\label{srule3}
The (sub)diagrams appearing in fig. \ref{fig4_selectionrule3} are forbidden by particle number conservation.
Configuration {\rm (e)} is forbidden only for an even number of horizontal lines on the right side of the vertical line.
\end{srule}

In order to prove the previous rule, the following procedure is applied:
\begin{itemize}
\item Consider a particular diagram obtained by using as building blocks propagators and 3-point vertices
\item Draw all the possible configurations of arrows that can be assigned to the lines
\item Check that particle number is conserved at every vertex (the number of entering and exiting arrows must be the same).
\end{itemize}
If there is no way to assign the arrows consistently at each vertex, then the diagram is forbidden and must be discarded.

\begin{figure}[h]
\centering
\begin{subfigure}[b]{0.4\linewidth}
\centering
\includegraphics[scale=0.4]{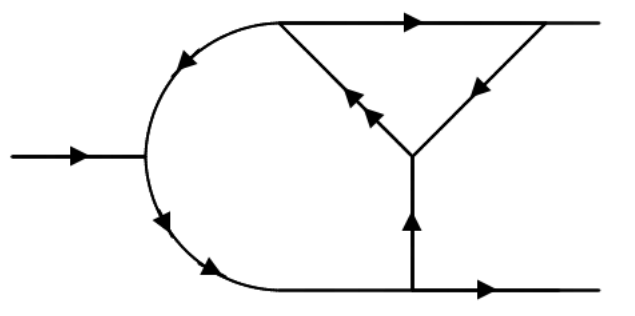}
\caption{}
\end{subfigure}
\begin{subfigure}[b]{0.4\linewidth}
\centering
\includegraphics[scale=0.4]{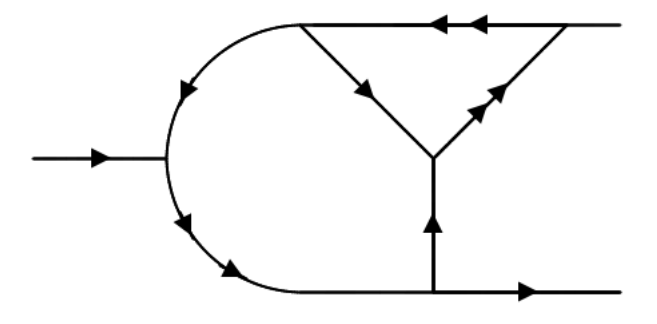}
\caption{}
\end{subfigure}
\begin{subfigure}[b]{0.4\linewidth}
\centering
\includegraphics[scale=0.4]{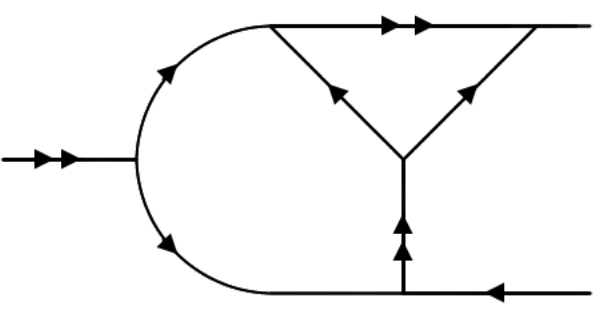}
\caption{}
\end{subfigure}
\begin{subfigure}[b]{0.4\linewidth}
\centering
\includegraphics[scale=0.4]{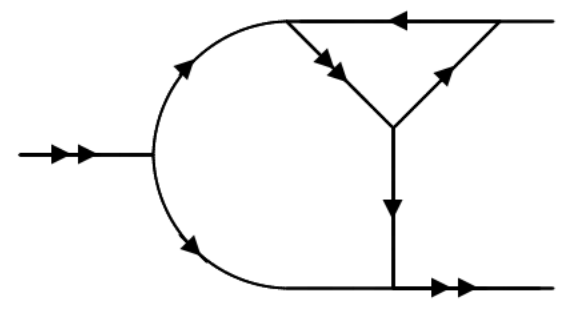}
\caption{}
\end{subfigure}
\begin{subfigure}[b]{0.4\linewidth}
\centering
\includegraphics[scale=0.4]{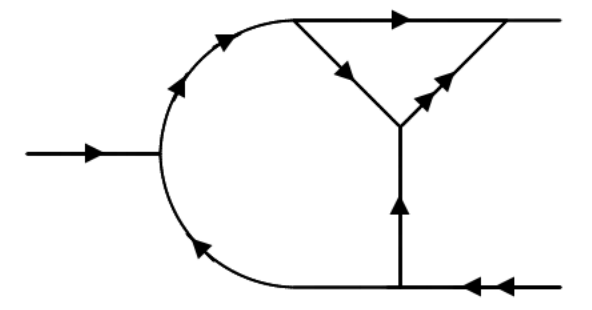}
\caption{}
\end{subfigure}
\begin{subfigure}[b]{0.4\linewidth}
\centering
\includegraphics[scale=0.4]{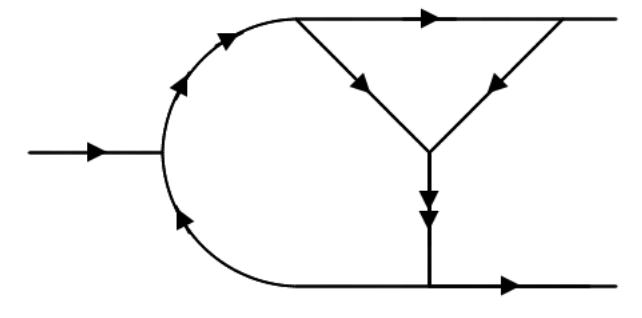}
\caption{}
\end{subfigure}
\caption{All possible configurations of arrows in an example of subdiagram. All of them are forbidden by particle number conservation at the upper right vertex. }
\label{fig5_example_selectionrule3}
\end{figure}

As an example, we consider diagram \ref{fig4_selectionrule3}(a) for which all  possible configurations of arrows are drawn in fig. \ref{fig5_example_selectionrule3}. It can be seen that in all the configurations we cannot consistently assign arrows in the top right vertex.

\subsection{Renormalizability of the theory}
\label{sect-Renormalizability of the theory}

We study the renormalizability of the model by considering the superficial degree of divergence of a generic supergraph with $L$ loops, $E=E_C+E_A$ external lines, $P$ internal propagators and $V=V_C+V_A$ vertices, where the $C$  and $A$ subscripts stand for chiral and anti-chiral, respectively.  

A connected graph satisfies the topological constraint
\beq
L= P-V+1  \, .
\label{relation between L,P,V}
\eeq
In addition, the fact that all the vertices of the theory are cubic implies that the relation $E + 2P = 3V$ also holds.
Combined with the previous constraint, this leads to
\beq
P= E + 3 L - 3 \, .
\label{relation with cubic vertices between P,E,L}
\eeq
Since the Galilean model comes from the null reduction of a massless WZ model, the propagator necessarily connects a chiral superfield with an anti-chiral one.
This requires the following relations to be true:
\beq
3 V_A = P + E_A \, , \qquad
3 V_C = P + E_C \, .
\label{relation on chiral and anti-chiral vertices}
\eeq
We consider the ingredients composing the integrand of a 1PI supergraph with these properties.
There is a product of $P$ super-propagators \eqref{superpropagatori momentum space} times factors of covariant derivatives $D, \bar{D}$ acting on the grassmannian delta functions. The precise counting is:
\begin{itemize}
\item One factor of $ (\bar{D}^2)^2 $ associated to each internal chiral vertex
\item One factor of $D^2$ associated to each internal anti-chiral vertex
\item One factor of $ \bar{D}^2 $ for each chiral vertex with an external line attached
\item One factor of $ D^2 $ for each anti-chiral vertex with an external line attached.
\end{itemize}
The total number of covariant derivatives is then
\beq
(D^2)^{2 V_A-E_A}  (\bar{D}^2)^{2 V_C-E_C} 
\eeq
On the other hand, D-algebra requires one factor $ D^2 \bar{D}^2 $ coming from each loop in order to contract the integral to a point in the $ (\theta, \bar{\theta}) $ space. This implies that in non-vanishing diagrams the remaining derivatives 
\beq
(D^2)^{2 V_A-E_A-L}  (\bar{D}^2)^{2 V_C-E_C-L}  
\eeq
are traded with powers of loop momenta, according to the D-algebra procedure explained in section \ref{sect-Renormalization in superspace}.

The constraints \eqref{relation on chiral and anti-chiral vertices} allow to express the total factor of covariant derivatives acting on the supergraph as
\beq
(D^2 \bar{D}^2)^{\frac{2}{3} P - L} \, (D^2)^{- \frac{E_A}{3}} \, (\bar{D}^2)^{-\frac{E_C}{3}}
\eeq
In addition to this factor, the diagram contains by assumption a set of $P$ propagators $1/\bigtriangleup$ with $ \bigtriangleup \equiv 2M \omega - \vec{k}^2 $, times $L$ integrations on the loop variables.
Looking at the superficial degree of divergence of the integral, the worst case occurs when identities \eqref{rules covariant derivatives giving momenta in supergraphs 2} can be used to trade $ D^2 \bar{D}^2$ with $ \bigtriangleup$, which then cancel internal propagators. The corresponding integral reads
\beq
\hspace{-0.4cm}
 \int d\omega_1 d^2 k_1 \dots d \omega_L d^2 k_L \, \frac{(D^2)^{- \frac{E_A}{3}} \, (\bar{D}^2)^{-\frac{E_C}{3}}}{\bigtriangleup^{L + \frac{P}{3}}} =  \int d\omega_1 d^2 k_1 \dots d \omega_L d^2 k_L  \, \frac{(D^2)^{- \frac{E_A}{3}} \, (\bar{D}^2)^{-\frac{E_C}{3}}}{\bigtriangleup^{2L + \frac{E}{3} -1}} \, ,
\eeq
where in the last step eq. \eqref{relation with cubic vertices between P,E,L} has been used.

The worst case for the convergence of the integral is a supergraph with 
$E_A = E_C = E/2$  where the remaining covariant derivatives also combine into inverse propagators. In this case the diagram schematically contributes as
\beq
  \int d\omega_1 d^2 k_1 \dots d \omega_L d^2 k_L \,  \frac{1}{\bigtriangleup^{2L + \frac{E}{2} -1}} 
\label{counting renormalizability amplitudes}
\eeq
The superficial degree of divergence is $\delta = 2 -E$. It is always negative for $ E \geq 3 $ and the corresponding integrals give finite contributions. 
For self-energy diagrams ($E=2$) logarithmic divergences arise, which can be subtracted by wave-function renormalization. 

The more problematic case for convergence is $E=1,$ which needs a careful treatment. 
The prototype for such case is a 1-loop diagram of the form
\beq
\int \frac{d \omega \, d^2 k}{2 M \omega - \vec{k}^2 + i \varepsilon}  
\eeq
After performing the $\omega$-integration, we can use dimensional regularization to compute the $\vec{k}$ integral. The result is zero since the integral is dimensionful and cannot  depend on any possible scale.
This shows that the non-relativistic WZ model is renormalizable.

While the check of renormalizability ensures that we can regularize the divergences coming from quantum corrections with a finite number of counterterms, in view of the application of the selection rule \ref{srule2}, the non-relativistic case requires a further property to prove: the loop integrals on $ \omega_1, \dots , \omega_L  $ are separately convergent. 
This can happen only if the integrand converges at least like $ 1/\omega^{2}_i $ for a given $\omega_i$-integration.

Let's then consider a generic loop $L_i$ inside the supergraph containing $ P_i $ propagators.
It is crucial that the inverse Galilean propagator is linear in the energy, because this information combined with the energy conservation at each vertex ensures that the $ P_i $ propagators provide a power $1/ \omega_i^{P_i}$.
Since in a loop we always have $P_i \geq 2$ (tadpoles are zero due to the previous argument), the convergence of the $\omega_i$-integral is guaranteed, as long as there are no $\omega_i$ powers at the numerator. 
Problems of convergence can arise if $\omega_i$ factors appear at the numerator fro D-algebra manipulations, but in the worst situation $D$-derivatives produce factors which cancel completely some propagators, contracting points in momentum space. In any case, this process leads to a loop with at least two propagators, which is sufficient to ensure the convergence of the integral.
Moreover we observe that adjacent loops which in the relativistic case would lead to overlapping divergences, have an even better convergence in $\omega$. 

In conclusion, all the energy integrals are convergent, they do not need to be regularized and we can compute them in the complex plane by using the residue theorem.
This property allows to apply selection rule \ref{srule2} to a given supergraph without worrying of D-algebra manipulations.


\subsection{Loop corrections to the self-energy}
\label{sect-Loop corrections to the self-energy}

Having at disposal the power of the selection rules coming from the properties of the non-relativistic model, we study the quantum corrections to the Galilean WZ model \eqref{non-rel WZ action in superfield formalism} by means of the supergraph formalism.
Due to the analysis in the previous section on the convergence of integrals along the energy variable, only integrals involving the spatial momentum need to be regularized and we choose the prescription of dimensional regularization with the minimal subtraction scheme.

In this way, we work in $d = 2- \epsilon$ dimensions and we introduce a mass scale $\mu$ to keep the coupling constant $g$ dimensionless. We define renormalized quantities \beq
\begin{cases}\label{ren_functions}
\Phi_a = Z_a^{-1/2} \,  \Phi_a^{(B)} = \le 1- \frac12 \delta_a \ri \Phi_a^{(B)} \qquad a=1,2\\
g = \mu ^{-\epsilon} Z_g^{-1} g^{(B)} = \mu^{-\epsilon}  (1 - \delta_g) g^{(B)} 
\end{cases}
\eeq
and determine counterterms proportional to $\delta_a$, $\delta_g$ 
\beq
\mathcal{L}_{\mathrm{ren}}   +
  \int d^4 \theta \, \le \delta_1 \bar{\Phi}_1 \Phi_1 + \delta_2 \bar{\Phi}_2 \Phi_2  \ri  + \int d^2 \theta \, \left[
\mu^{\epsilon} g \le \delta_g + \delta_1 + \frac12 \delta_2 \ri  \Phi_1^2 \Phi_2 \right] + \mathrm{h.c.} 
\eeq
to cancel UV divergences.  

In this section we investigate the quantum corrections to the self-energy: only 1PI diagrams are considered.
We follow this strategy:
\begin{itemize}
\item Draw all the possible topologies of 1PI supergraphs at a given loop order
\item Assign arrows to the lines of the diagram consistently with particle number conservation
\item Check if there is any part of the diagram where the arrows form a close loop, and discard the diagram if this happens
\item Apply D-algebra to find if the grassmannian nature of the amplitude forbids the diagram
\item Perform the remaining integral with the usual techniques, \emph{e.g} with dimensional regularization
\end{itemize}

\subsubsection*{One loop}

\begin{figure}[h]
\centering
\begin{subfigure}[b]{0.45\linewidth}
\includegraphics[scale=1.2]{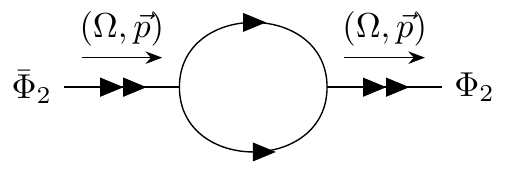}
\end{subfigure}
\begin{subfigure}[b]{0.45\linewidth}
\end{subfigure}
\caption{One-loop correction to the self-energy of $\Phi_2$.}
\label{fig7_rinormalizzione self-energy supercampo 2}
\end{figure}

We start with the 1PI diagrams contributing to the quantum corrections of the self-energy of the superfield in sector 1.
The selection rule \ref{srule2} applies, and immediately tells us that there are no possible configurations: the only admitted diagram would be the one depicted in Fig. \ref{fig3_rinormalizzione self-energy supercampo 1}, but we already showed that this vanishes by means of the residue theorem.
As a consequence,
\beq
\delta_1^{\rm (1loop)} = 0  
\label{wavefunction renormalization sector 1}
\eeq
For the one-loop self-energy in sector 2 we find, instead, that there is an allowed diagram, depicted in fig. \ref{fig7_rinormalizzione self-energy supercampo 2}.

After performing the D-algebra, the corresponding contribution reads
\beq
i \Gamma_2^{(2)} (\Phi_2, \bar{\Phi}_2) =  2 |g|^2  \int d^4 \theta \,   \frac{d\omega \, d^2 k}{(2 \pi)^3} \,  \frac{ \Phi_2 (\Omega, \vec{p}, \theta) \bar{\Phi}_2 (\Omega, \vec{p}, \theta) }{\left[2 m \omega- \vec{k}^2 + i \varepsilon \right] \left[ 2 m (\Omega-\omega)- (\vec{p} -\vec{k})^2 + i \varepsilon \right]}   \, .
\eeq
The crucial difference with the computations in sector 1 is the fact that the arrows of particle number and momenta are directed in the same way, and then the poles of the integrand are located in both sides of the complex plane
\beq
\omega^{(1)} = \frac{\vec{k}^2}{4 \tilde{m}} - i \varepsilon \, , \qquad
\omega^{(2)} = \Omega - \frac{(\vec{k}- \vec{p})^2}{2 \tilde{m}} + i \varepsilon \, .
\eeq
In this way the integral is non-vanishing and since the $ \omega $-integral is convergent, we can apply the residue theorem to obtain
\beq
 \Gamma_2^{(2)} (\Phi_2, \bar{\Phi}_2) = - \frac{|g|^2 }{m}  \int d^4 \theta  \, \Phi_2 (\Omega, \vec{p}, \theta) \bar{\Phi}_2 (\Omega, \vec{p}, \theta) \,  \int \frac{d^2 k}{(2 \pi)^2} \, \frac{1}{2 m \Omega - \vec{k}^2 - ( \vec{p} - \vec{k})^2 + i \varepsilon}   \, .
\eeq
The two-dimensional momentum integral can be now performed using dimensional regularization.
In generic spatial dimensions \emph{d} there exists a region in complex plane where the integral is convergent and we can translate the integration variable as
\beq
\vec{l} = \vec{k} - \frac{\vec{p}}{2} \, \Rightarrow \,
d \vec{l} = d \vec{k} \, ,
\eeq
giving
\beq
 \Gamma_2^{(2)} (\phi_2, \bar{\phi}_2) = \frac{2 |g|^2 }{(2 \pi)^2}  \int d^4 \theta  \, \phi_2 (\Omega, \vec{p}, \theta) \bar{\phi}_2 (\Omega, \vec{p}, \theta) \frac{2 \pi^{d/2}}{\Gamma(d/2)} \int_0^{\infty} dl \, \frac{l^{d-1}}{l^2 - m \Omega + \frac{\vec{p}^2}{4}}   \, .
\eeq
Focusing on its divergent part we obtain
\beq
\Gamma_2^{(2)} (\Phi_2, \bar{\Phi}_2) \to  \frac{|g|^2}{4 \pi m} \frac{1}{\varepsilon} \int d^4 \theta \, \Phi
_2 (\Omega, \vec{p}, \theta) \bar{\Phi}_2 (\Omega, \vec{p}, \theta)  \, .
\eeq
In the minimal subtraction scheme this leads to the counterterm  
\beq\label{superfield_ren}
\delta_2^{\rm (1loop)} = - \frac{|g|^2}{4 \pi m} \frac{1}{\varepsilon} \, .
\eeq

\subsubsection*{Two loops}

\begin{figure}[h]
\centering
\begin{subfigure}[b]{0.4\linewidth}
\includegraphics[scale=0.3]{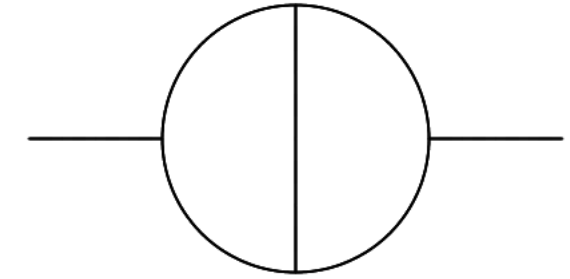}
\end{subfigure}
\begin{subfigure}[b]{0.4\linewidth}
\includegraphics[scale=0.3]{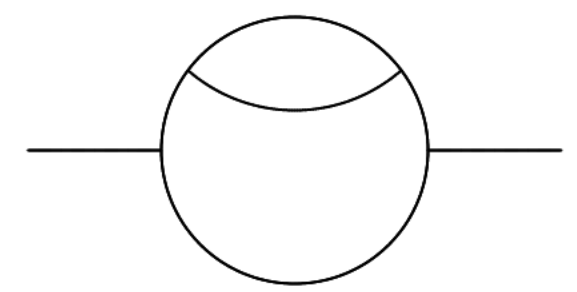}
\end{subfigure}
\caption{Topologies of possible two-loop corrections to the self-energies.}
\label{fig8_selfenergy a 2 loop}
\end{figure}

It turns out that the selection rules \ref{srule2} and \ref{srule3} are sufficient to rule out every two-loop corrections to the self-energy.
In particular, looking at the two possible two-loop topologies of diagrams depicted in fig. \ref{fig8_selfenergy a 2 loop}, it is easy to realize that no consistent assignments of arrows exist, or they vanish due to circulating arrows in a loop.

\subsubsection*{Three loops}
At three loops, the classification is richer and the possible cases in the relativistic setting are given in \cite{Abbott:1980jk}, where the three-loop $ \beta $ function was computed.

\begin{figure}[h]
\centering
\begin{subfigure}[b]{0.45\linewidth}
\centering
\includegraphics[scale=0.2]{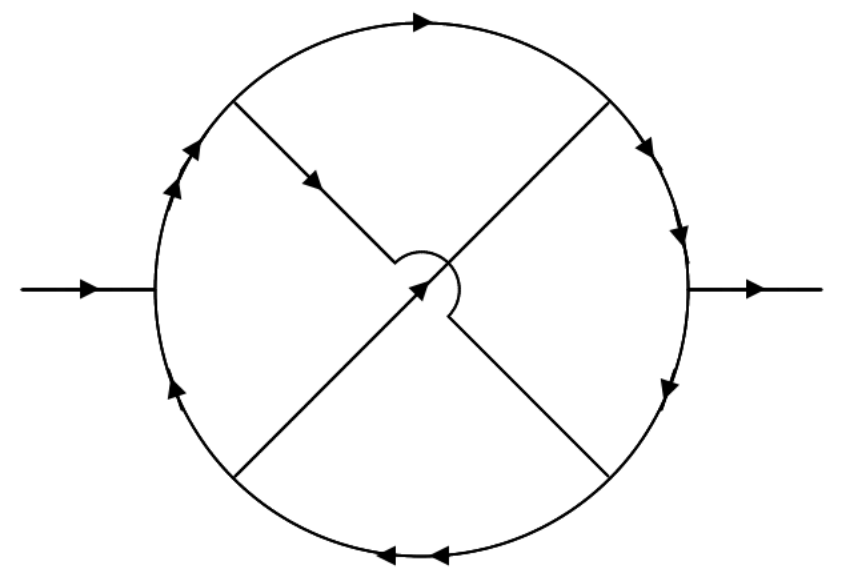}
\end{subfigure}
\begin{subfigure}[b]{0.45\linewidth}
\centering
\includegraphics[scale=0.2]{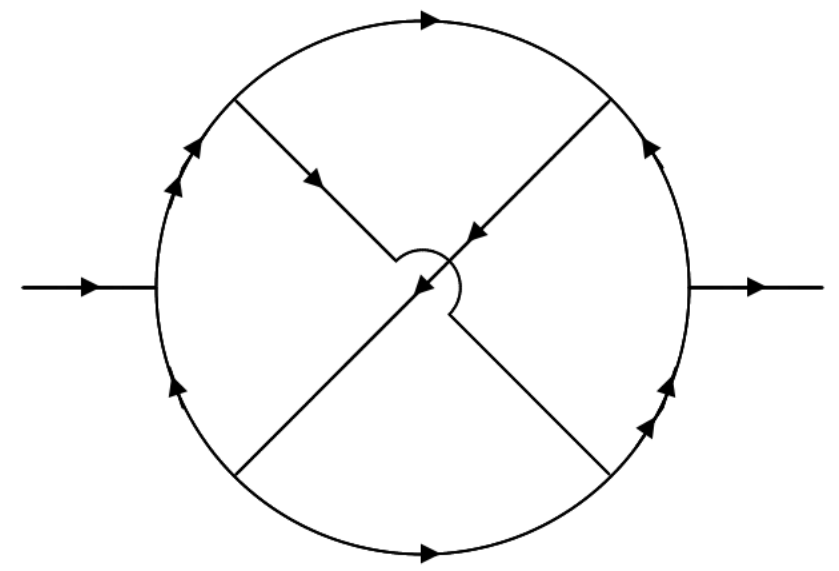}
\end{subfigure}
\begin{subfigure}[b]{0.45\linewidth}
\centering
\includegraphics[scale=0.26]{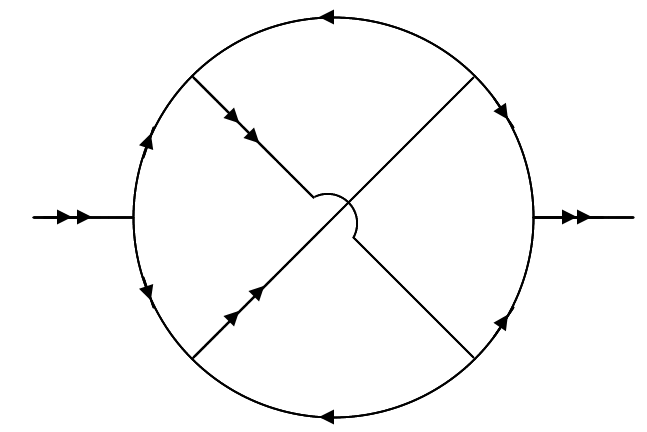}
\end{subfigure}
\begin{subfigure}[b]{0.45\linewidth}
\centering
\includegraphics[scale=0.2]{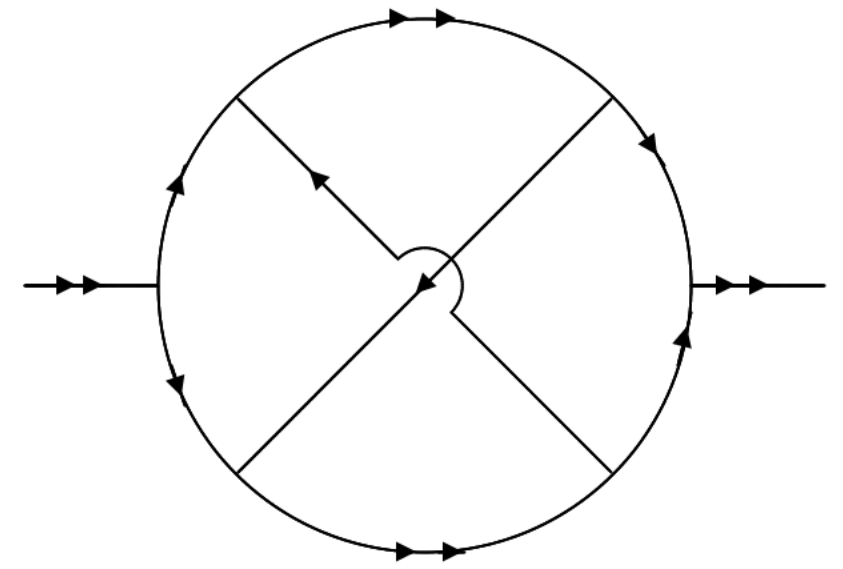}
\end{subfigure}
\caption{Non-trivial three-loop corrections to the self-energies. We depict all possible assignments of arrows in the lines.}
\label{fig8bis_esempio_selfenergy_2loop}
\end{figure}

In the non-relativistic case selection rules \ref{srule2} and \ref{srule3} discard almost all possible configurations, leading only to one non-trivial type of diagram, the non-planar one depicted in fig. \ref{fig8bis_esempio_selfenergy_2loop}.
However, looking at all possible assignments of arrows we conclude that there is always a circulating loop, which entails a vanishing result according to selection rule \ref{srule2}. Therefore, there are no three-loop corrections to the self-energies of both superfields.

\subsubsection*{Four loops}
At four loops we take the classification of all the topologies in the relativistic case from \cite{Sen:1981hk}, where the $\beta $-function for the relativistic WZ model was computed. 

\begin{figure}[h]
\centering
\begin{subfigure}[b]{0.32\linewidth}
\centering
\includegraphics[scale=0.18]{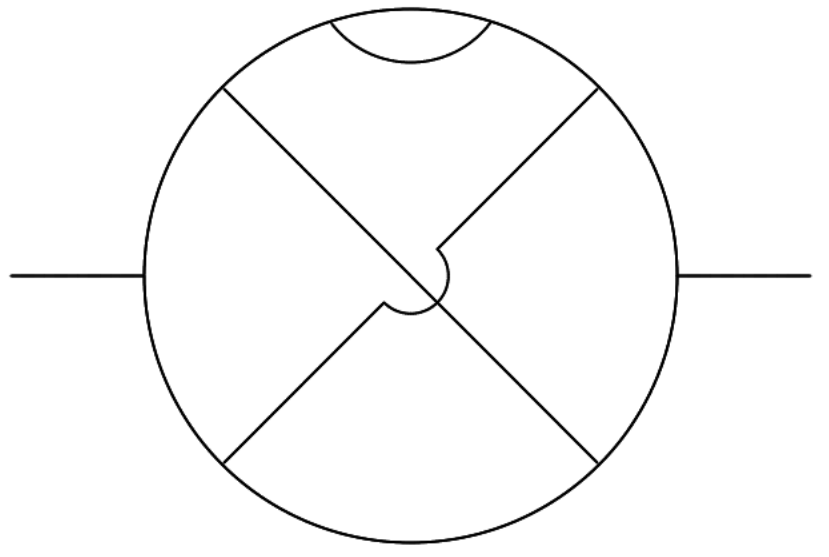}
\subcaption{}
\end{subfigure}
\begin{subfigure}[b]{0.32\linewidth}
\centering
\includegraphics[scale=0.18]{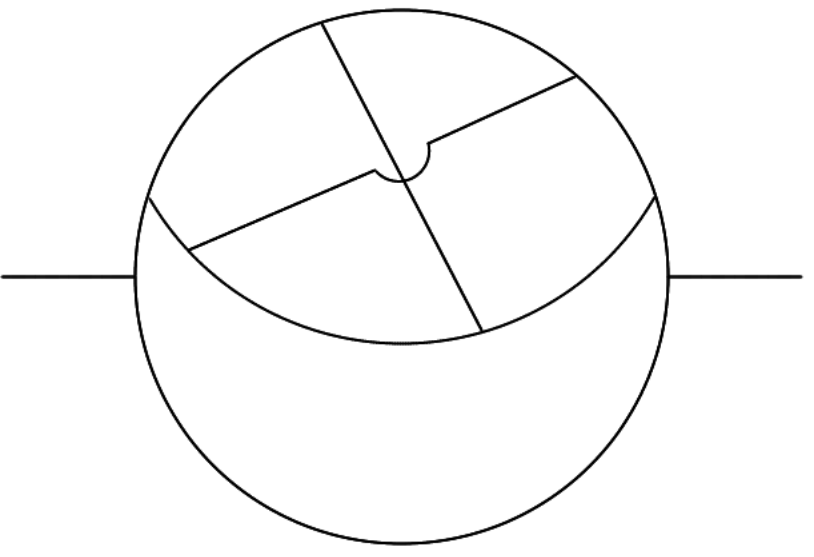}
\subcaption{}
\end{subfigure}
\begin{subfigure}[b]{0.32\linewidth}
\centering
\includegraphics[scale=0.18]{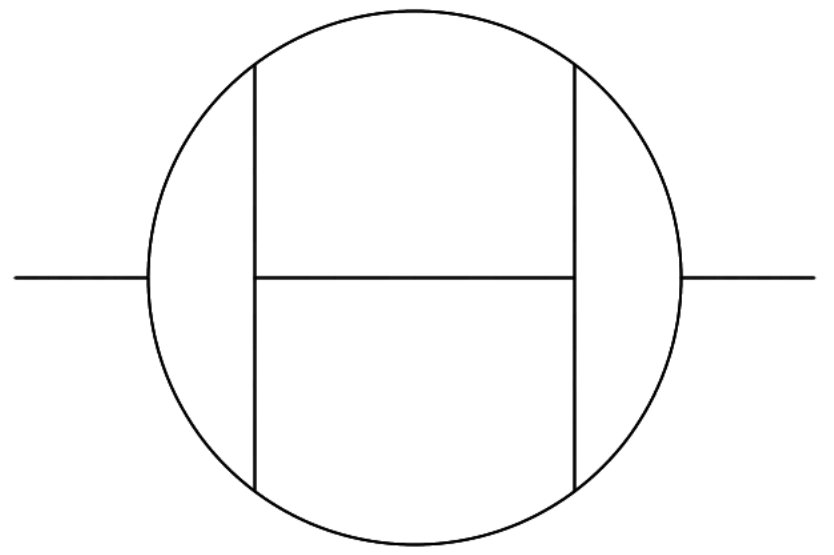}
\subcaption{}
\end{subfigure}
\caption{Non-trivial quantum corrections to the self-energy at four loops.}
\label{fig9_selfenergy_4_loops}
\end{figure}

\begin{figure}[h]
\centering
\begin{subfigure}[b]{0.35\linewidth}
\centering
\includegraphics[scale=0.18]{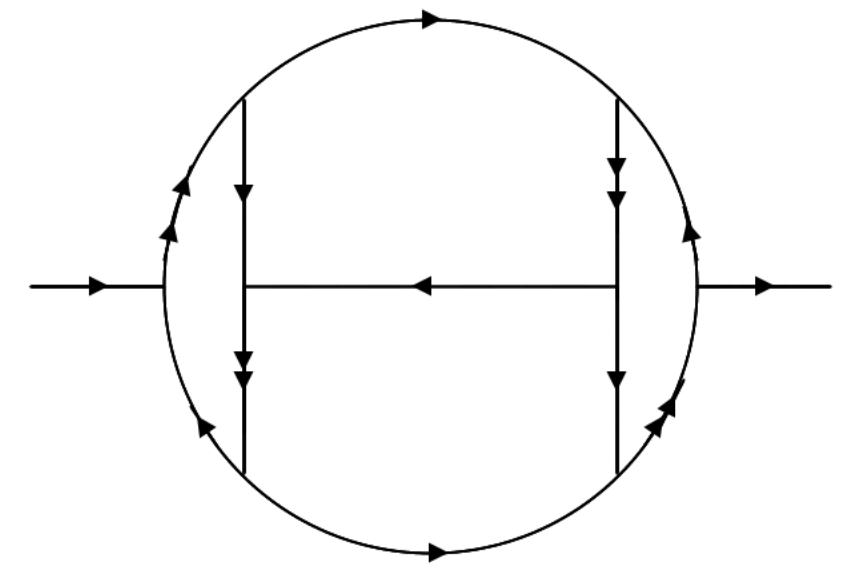}
\subcaption{}
\end{subfigure}
\begin{subfigure}[b]{0.35\linewidth}
\centering
\includegraphics[scale=0.18]{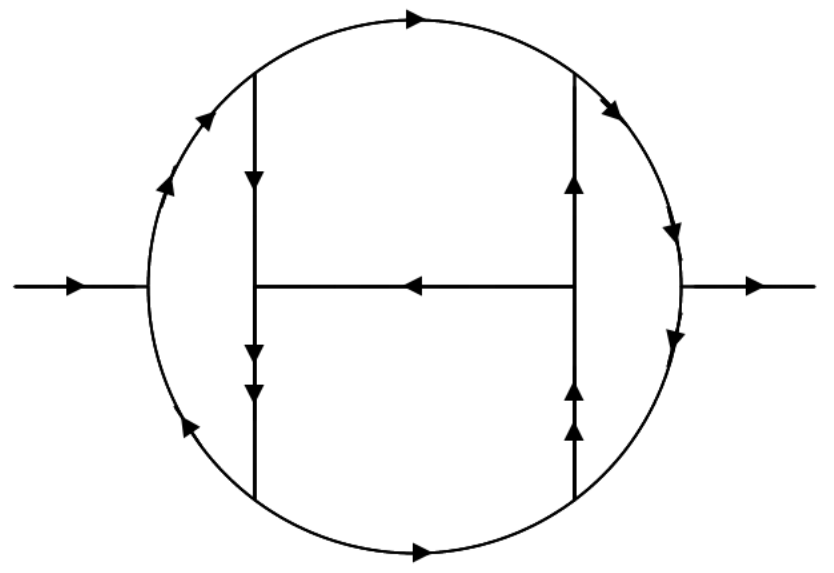}
\subcaption{}
\end{subfigure}
\begin{subfigure}[b]{0.35\linewidth}
\centering
\includegraphics[scale=0.18]{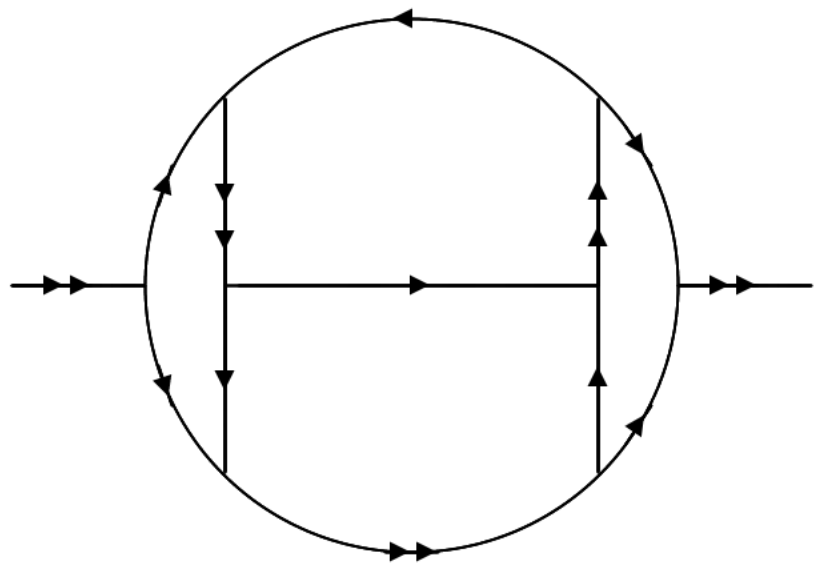}
\subcaption{}
\end{subfigure}
\begin{subfigure}[b]{0.35\linewidth}
\centering
\includegraphics[scale=0.18]{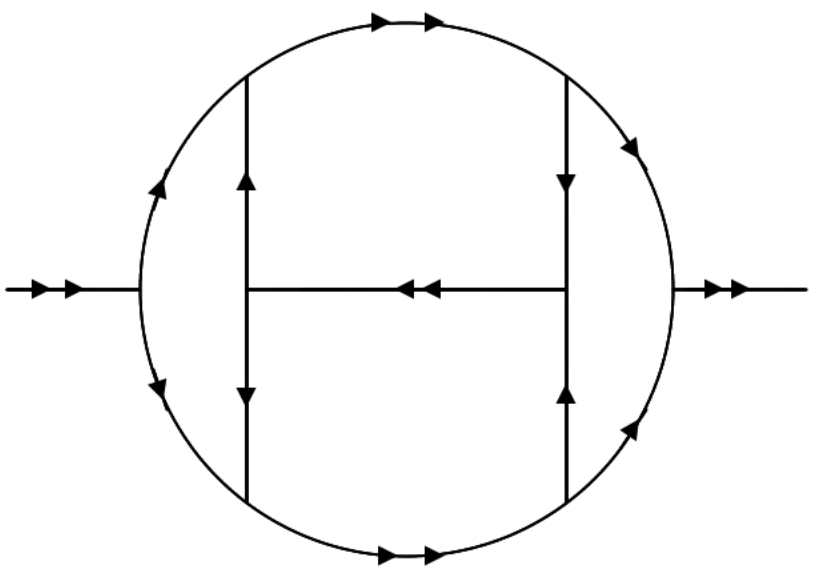}
\subcaption{}
\end{subfigure}
\caption{Allowed assignments of arrows to the lines of  diagram (c) in Fig. \ref{fig9_selfenergy_4_loops}.} 
\label{fig10_examples_selfenergy_4_loops}
\end{figure}

Following the same strategy of lower loops, we find few non-trivial diagrams listed in fig. \ref{fig9_selfenergy_4_loops}(a)-(c). 
The first two graphs contain as a subgraph the non-planar three-loop diagram already discussed, and then we discard them by similar arguments to the three-loop case.
The remaining case (c) allows for various configurations of arrows depicted in fig. \ref{fig10_examples_selfenergy_4_loops}(a)-(d), but all of them contain at least one subgraph where the arrows form a close loop.
This implies that the diagram does not give any quantum correction to the self-energy.


\subsubsection*{Higher loops}
Up to four loops we have found that non-vanishing quantum corrections to the self-energy appear only in sector 2 and only at one loop. 
Triggered by these results, the natural question which arises is whether the same pattern repeats at any loop order or we should expect non-vanishing contributions at higher loops. 

A deeper understanding of the problem can be found by applying the strategy followed at lower loops to find recursive configurations of diagrams that vanish, and possibly rule out every graph that it is possible to draw.
First of all, every diagram containing the stuctures in fig. \ref{fig4_selectionrule3} as a subgraph vanish by means of selection rule \ref{srule3}.

\begin{figure}[h]
\centering
\begin{subfigure}[b]{0.4\linewidth}
\centering
\includegraphics[scale=0.18]{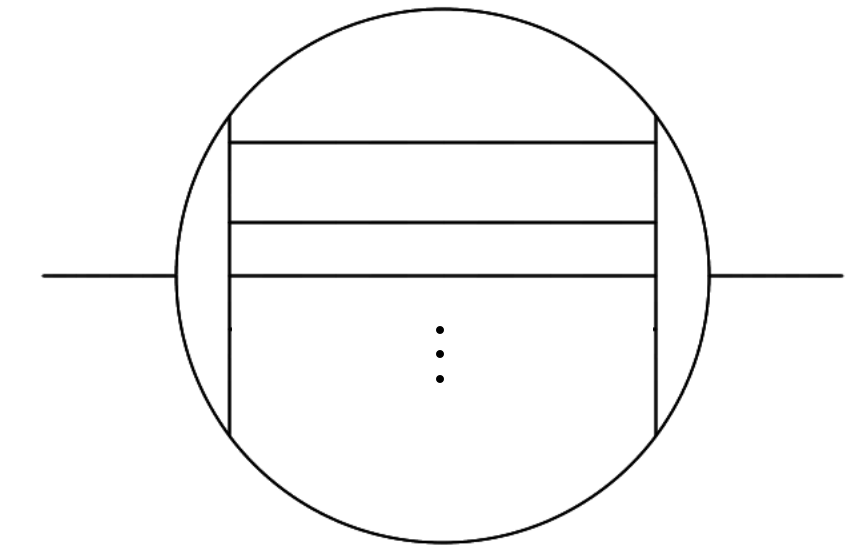}
\subcaption{}
\end{subfigure}
\begin{subfigure}[b]{0.4\linewidth}
\centering
\includegraphics[scale=0.18]{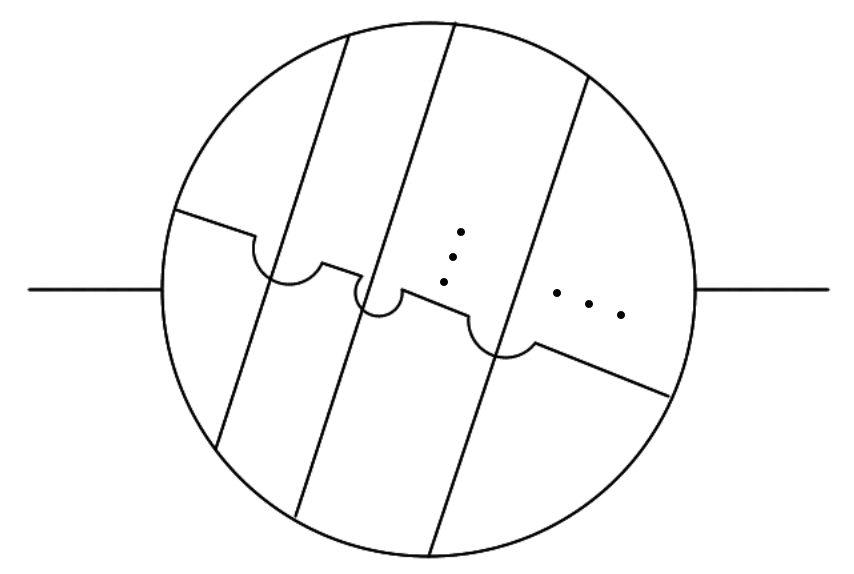}
\subcaption{}
\end{subfigure}
\begin{subfigure}[b]{0.4\linewidth}
\centering
\includegraphics[scale=0.18]{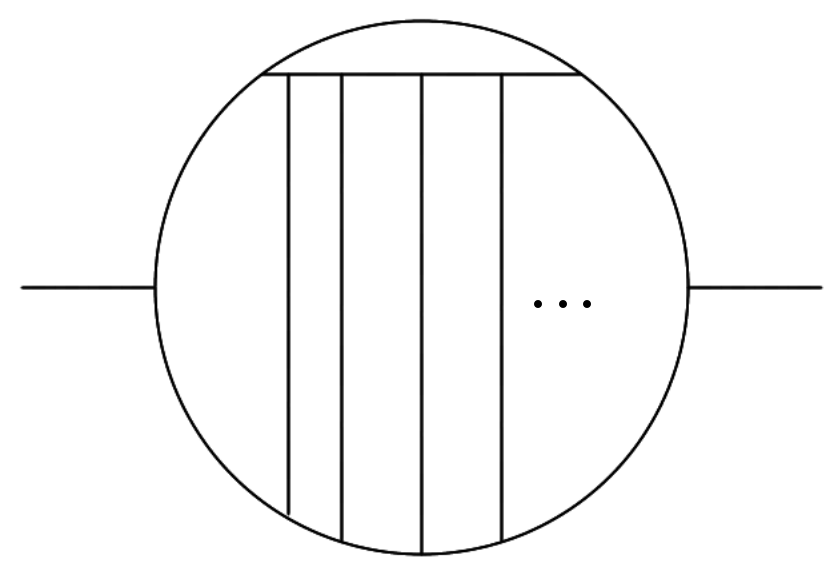}
\subcaption{}
\end{subfigure}
\begin{subfigure}[b]{0.4\linewidth}
\centering
\includegraphics[scale=0.18]{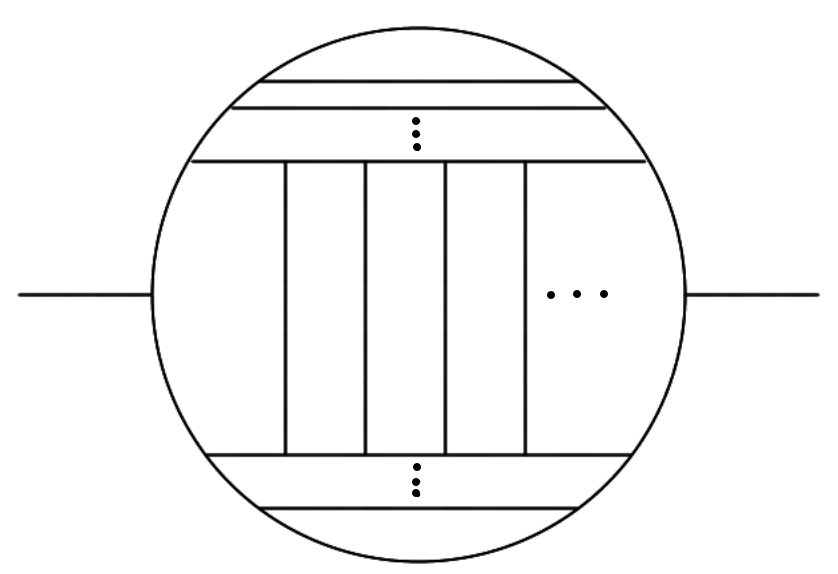}
\subcaption{}
\end{subfigure}
\begin{subfigure}[b]{0.4\linewidth}
\centering
\includegraphics[scale=0.18]{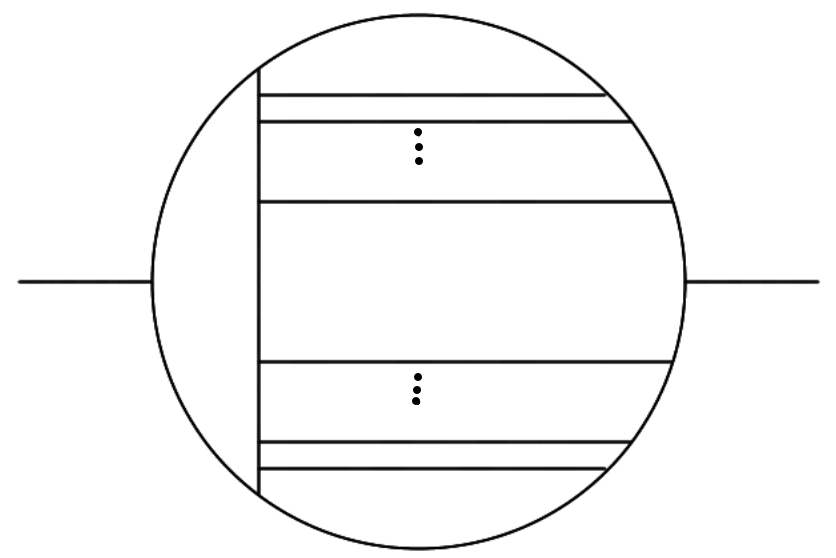}
\subcaption{}
\end{subfigure}
\caption{Non-trivial vanishing quantum corrections to self-energies at generic loop level.}
\label{fig11_candidates_nloop_prop}
\end{figure}

If we further apply selection rule \ref{srule2}, we can find among the set of allowed configurations other structures that vanish, depicted in fig. \ref{fig11_candidates_nloop_prop}. 
This is also true for all the diagrams that can be obtained by gluing different structures among the previous set.

\begin{figure}[h]
\centering
\begin{subfigure}[b]{0.4\linewidth}
\centering
\includegraphics[scale=0.18]{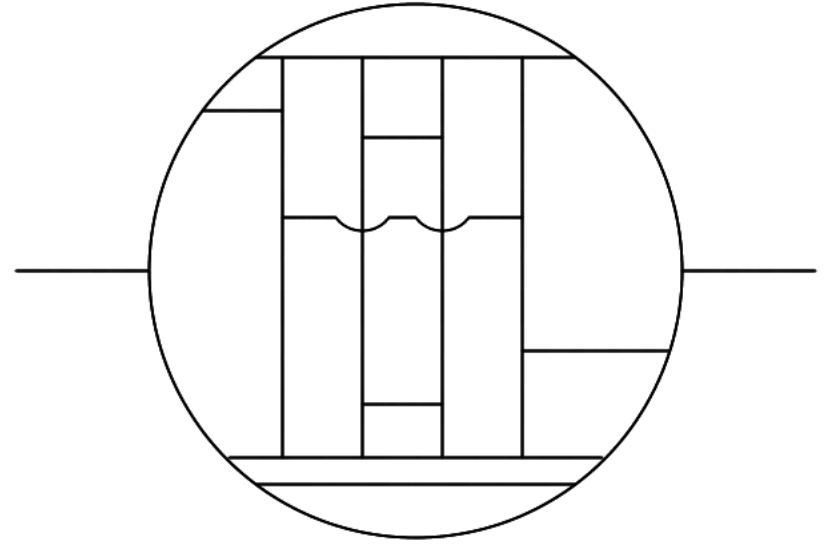}
\end{subfigure}
\caption{General self-energy diagram.}
\label{general_diagram}
\end{figure}

Although these topologies cover a vast number of diagrams, they are not exhaustive and in principle we cannot exclude the appearance of possible non-vanishing contributions from more general configurations, like the one in fig. \ref{general_diagram}. Nonetheless, based on the experience gained up to four loops, we expect that when the numbers of loops increases it becomes more and more difficult to realize configurations of arrows without closed loops. Therefore, we can quite safely conjecture that the self-energy of the $\Phi_1$ superfield is not corrected at quantum level, while the one for $\Phi_2$ is one-loop exact.

Independently of the validity of this conjecture, there are various things that we learned from the study of quantum corrections to the self-energy in the non-relativistic case.
First of all, it is evident that the computation is simpler than the relativistic parent ${\cal N}=1$ theory in 3+1 dimensions, because the selection rules greatly increase the number of vanishing contributions.
In fact, in the relativistic case the kinetic term acquires UV divergent corrections at any loop order, while in the non-relativistic case there are contributions coming only from the one-loop computation.
In particular, this shows that at quantum level the non-relativistic three-dimensional ${\cal N}=2$ WZ model cannot be obtained simply from null reduction of the four-dimensional relativistic model. 

\subsection{Loop corrections to the vertices}

The discussion in section \ref{sect-Renormalizability of the theory} shows that UV divergent contributions should not arise from any diagram with three or more external legs.
Moreover, the grassmannian nature of the superspace, not being affected by the null reduction, does not allow the production of any chiral integral at the perturbative level.
This means that the perturbartive non-renormalization theorem for the superpotential should still work, giving the constraint
\beq\label{ren_constraint}
\delta_g + \delta_1 + \frac12 \delta_2 = 0 \quad \Rightarrow \quad \delta_g^{\rm (1loop)} = \frac{|g|^2}{8\pi m } \frac{1}{\epsilon} \, .
\eeq

In this section we will study the case of the 1PI quantum corrections to the three-point vertex, both to investigate how the selection rules restrict the number of possible quantum corrections for configurations with three external fields, and to provide further evidence of the previous statements. 

As in the relativistic case, at one-loop there is no way to draw any three-point diagram as long as the model is massless. 

\begin{figure}[h]
\centering
\begin{subfigure}[b]{0.45\linewidth}
\centering
\includegraphics[scale=0.18]{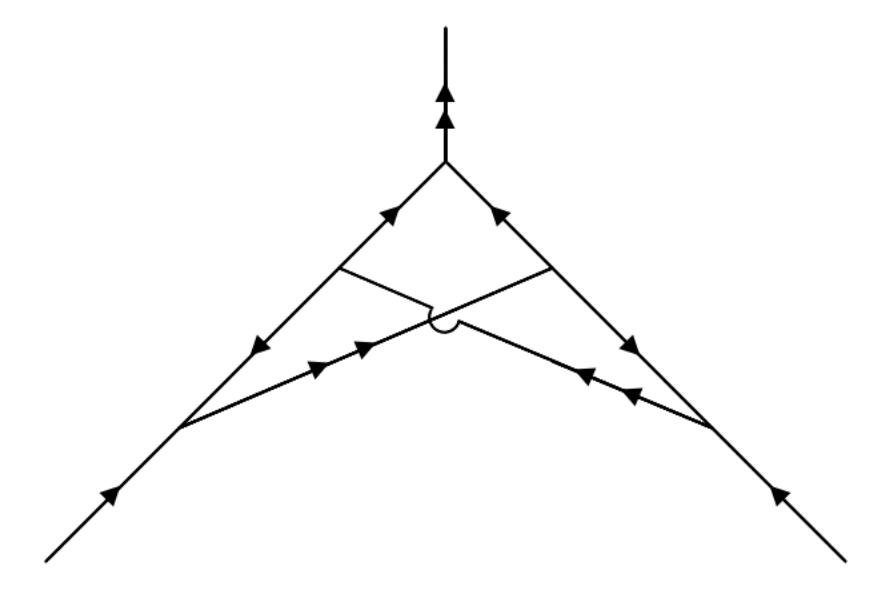}
\caption{}
\end{subfigure}
\begin{subfigure}[b]{0.45\linewidth}
\centering
\includegraphics[scale=0.18]{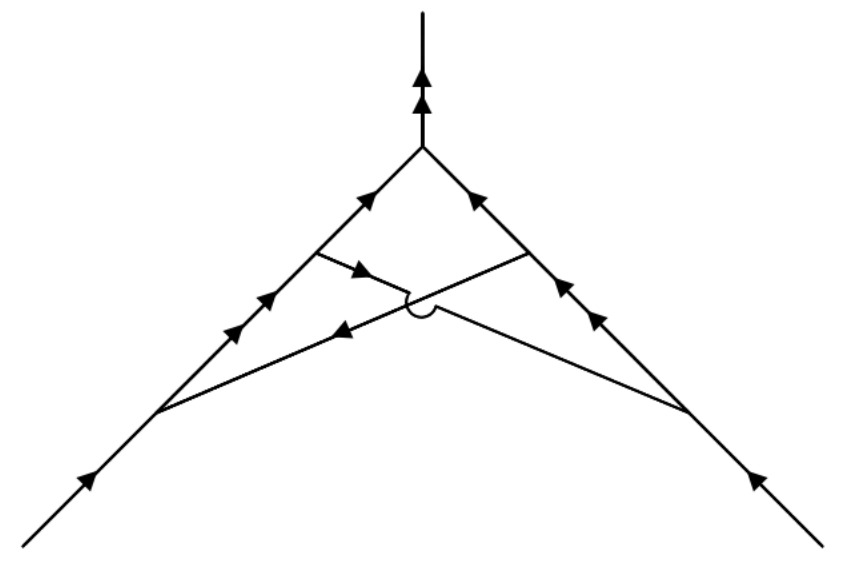}
\caption{}
\end{subfigure}
\begin{subfigure}[b]{0.45\linewidth}
\centering
\includegraphics[scale=0.18]{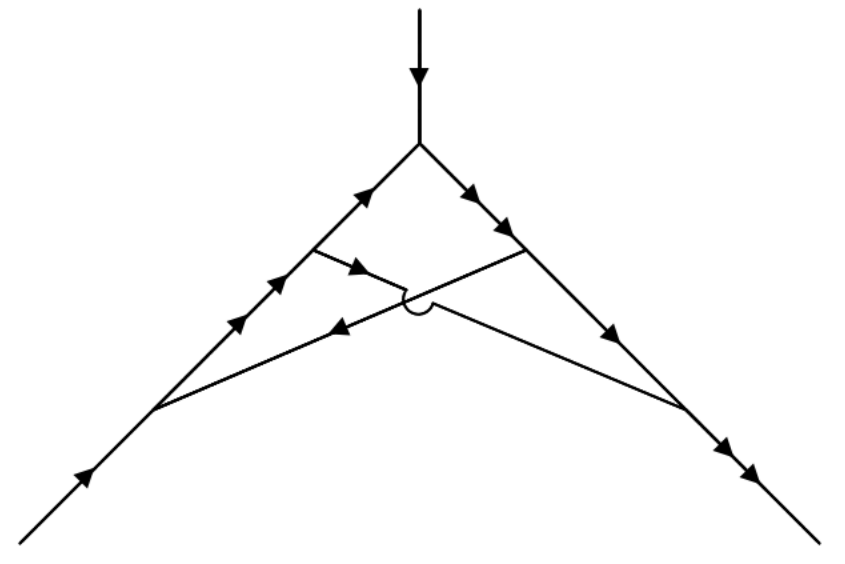}
\caption{}
\end{subfigure}
\begin{subfigure}[b]{0.45\linewidth}
\centering
\includegraphics[scale=0.18]{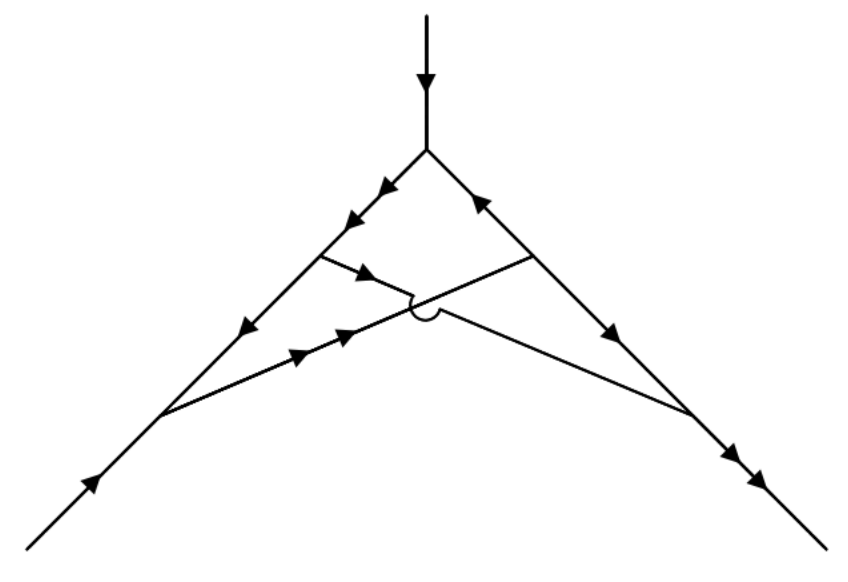}
\caption{}
\end{subfigure} 
\caption{Two-loop 1PI diagram for the three-point vertex. We depicted all the configurations of arrows associated to the lines.}
\label{fig14_2 loop supervertici a 3}
\end{figure}

At two loops the only supergraph allowed by particle number conservation is the one in fig. \ref{fig14_2 loop supervertici a 3}, where all possible configurations of arrows have been depicted. In all the configurations we see that a circulating loop of arrows appears, thus this diagram is ruled out by selection rule \ref{srule2}. 

\begin{figure}[h]
\centering 
\begin{subfigure}[b]{0.32\linewidth}
\centering
\includegraphics[scale=0.20]{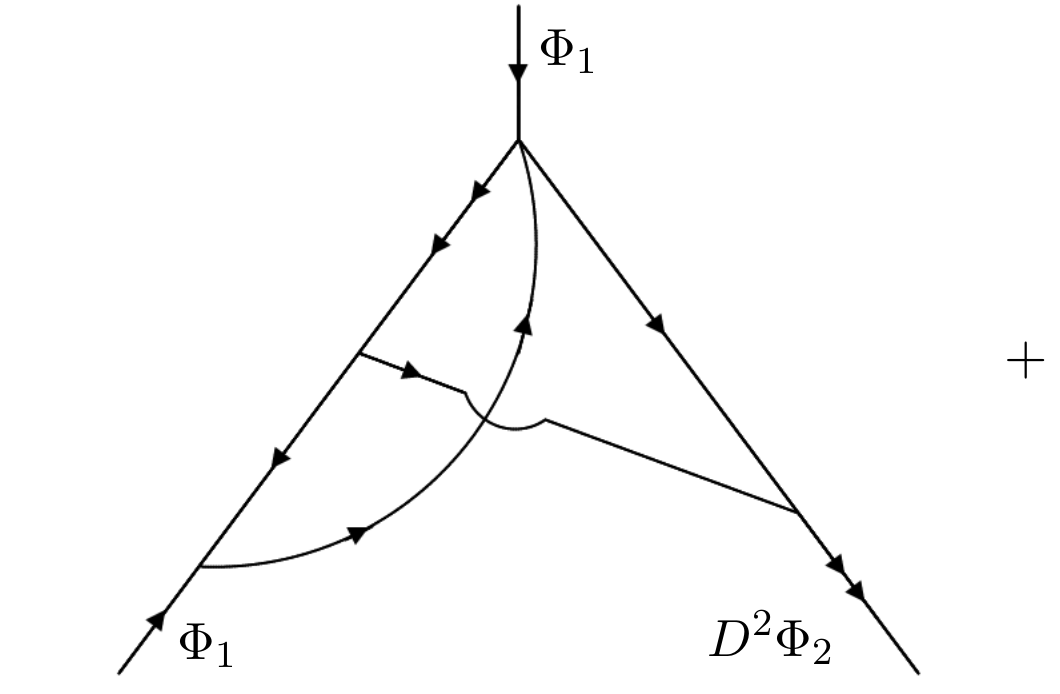}
\end{subfigure} 
\begin{subfigure}[b]{0.32\linewidth}
\centering
\includegraphics[scale=0.20]{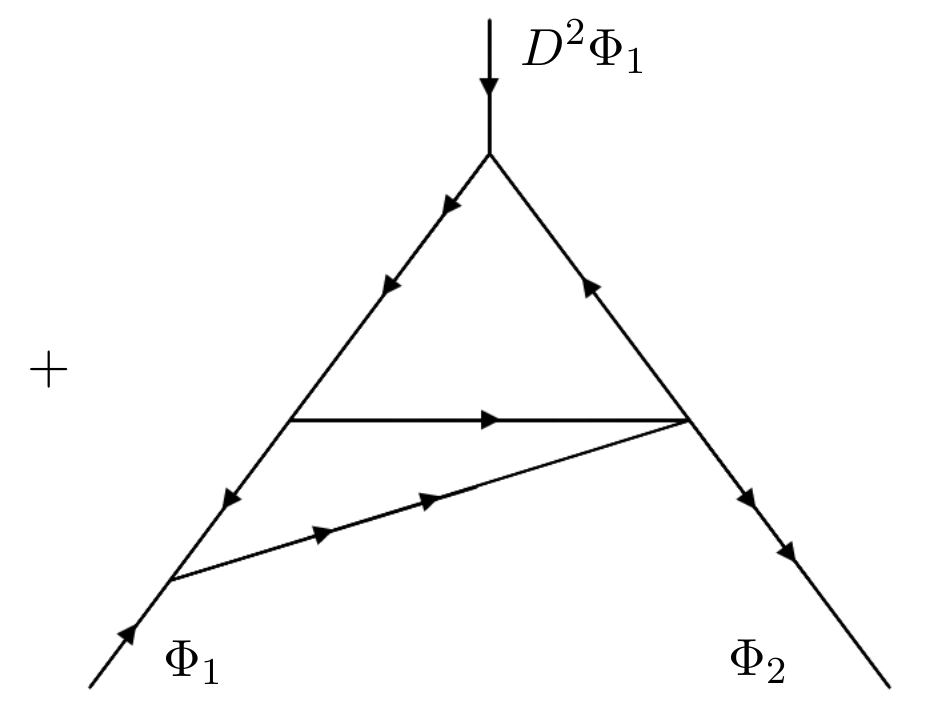}
\end{subfigure}
\centering
\begin{subfigure}[b]{0.32\linewidth}
\centering
\includegraphics[scale=0.20]{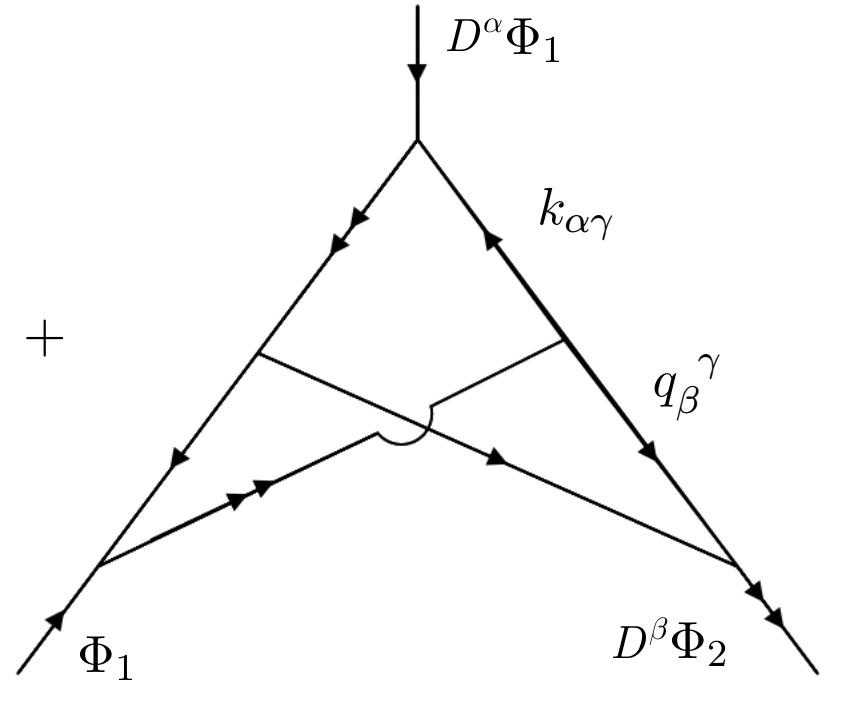}
\end{subfigure}
\caption{Diagrams resulting from the D-algebra reduction of diagram \ref{fig14_2 loop supervertici a 3}(d).}
\label{fig15_2 loop supervertici a 3_specific diagrams}
\end{figure}

Since this is a configuration where the number of chiral and anti-chiral vertices is different, the factors of  covariant derivatives are not only used to simplify propagators, but the application of the D-algebra \eqref{rules covariant derivatives giving momenta in supergraphs 2} gives additional powers of momenta at the numerator which might affect the convergence of the $\omega$ integrations.
In order to show that there is enough regularity to guarantee the convergence of the $\omega$ integral, we analyze the diagram in more details.
For instance, focusing on the arrow configuration \ref{fig14_2 loop supervertici a 3}(d), the result of D-algebra is given in fig. 
\ref{fig15_2 loop supervertici a 3_specific diagrams}.

In the first two diagrams the covariant derivatives act on the external fields or they are responsible for the simplification of some propagators. In fact, there are some effective 4-point vertices due to Dirac $ \delta $-functions arising in this way.
In both cases we are left with a loop containing three propagators whose arrows form a closed loop, and then there is enough regularity to apply the Jordan's lemma and conclude that they vanish.

Due to the structure of the external covariant derivatives, the only relevant contribution from the third diagram in fig. \ref{fig15_2 loop supervertici a 3_specific diagrams}  is proportional to
\beq
\begin{aligned}
 & \int \frac{d \omega_q d^2q}{(2 \pi)^3} \, \frac{d \omega_k d^2k}{(2 \pi)^3} \;
\epsilon_{\alpha \beta}\left( m(\omega_k + \omega_q) + \vec{k} \cdot \vec{q} \right)
  \, \frac{1}{2m \omega_k - \vec{k}^2 + i \varepsilon} \, \frac{1}{4m (\omega_{p_1} + \omega_k) - (\vec{p}_1 + \vec{k})^2 + i \varepsilon} \\
& \qquad  \qquad \qquad  \qquad\times  \frac{1}{2m (\omega_{k} + \omega_q - \omega_{p_2}) - (\vec{k} + \vec{q} - \vec{p}_2)^2 + i \varepsilon}
 \, \frac{1}{4m (\omega_{k} + \omega_q) - (\vec{k} + \vec{q})^2 + i \varepsilon}  \\
& \qquad  \qquad \qquad  \qquad\times  \frac{1}{2m \omega_{q}  -\vec{q}^2 + i \varepsilon}
 \, \frac{1}{2m (\omega_{p_1} + \omega_{p_2}- \omega_q) - (\vec{p}_1 + \vec{p}_2 - \vec{q})^2 + i \varepsilon}  
\end{aligned}
\eeq
where momenta $ (\omega_{p_a}, \vec{p}_a) $, $a=1,2$ refer to the external $\Phi_1, \Phi_2$ particles. 
At the numerator we have used the null reduction of the 4d expression $k_{\alpha \dot{\alpha}} \, q_{\beta}^{\,\,\, \dot{\alpha}} =  (\sigma^M)_{\alpha \dot{\alpha}} (\sigma^N)_{\beta}^{\,\,\, \dot{\alpha}} k_M q_N $.
 
If we now focus on the $\omega_k$ integration, we see that in the region of large $ \omega_k $ the worst integrand goes as $1/\omega_k^3$. This allows to apply Jordan's lemma and compute the integral by residue theorem. Since all the poles are on the same side of the complex plane the result is zero. 

The same pattern occurs for the other configurations of arrows in fig. \ref{fig14_2 loop supervertici a 3}(a)-(c). This provides a check of selection rule \ref{srule2} in this particular case.   

Extending the analysis of quantum corrections of the three-point vertex at higher loops, again we find that in the non-relativistic model the number of (finite) quantum contributions is drastically reduced compared to the relativistic case. 

\subsection{Non-relativistic non-renormalization theorem}
\label{sect-Non-relativistic non-renormalization theorem}

There is an ever stronger support to the absence of quantum corrections to the three-point vertex: we argue that it is possible to inherit the non-renormalization theorem from the relativistic parent theory.

We consider a generic Galilean WZ model for $n$ chiral superfields
\beq
S= \int d^3 x d^2 \theta d^2 \bar{\theta} \, \bar{\Phi}_a \Phi_a + g \int d^3 x d^2 \theta \, W(\Phi_a) + \mathrm{h.c.}
\label{WZ-generico-NR}
\eeq
obtained by null reduction of the relativistic one in eq. (\ref{WZ-generico}).

The same argument used in the relativistic case can be adapted here in order to
rule out quantum corrections to the F-term.  As in eq. (\ref{WZ-generico2})
we can introduce one extra chiral superfield $Y$ multiplying the superpotential,
which can be set equal to $1$ in order to reproduce eq. (\ref{WZ-generico-NR}), together with wave function renormalization superfields $Z_{ab}$
\beq
\tilde{S}= \int d^3 x \, d^4 \theta  \, Z_{ab} \bar{\Phi}_a \Phi_b + \int d^3 x \, d^2 \theta \,  
Y \, W(\Phi_a) + \mathrm{h.c.} 
\eeq
Since the non-relativistic limit via null reduction technique does not affect the grassmannian part of the superfields in the action, R-symmetry works in the same way as in the relativistic case. Therefore, as in the relativistic case, we assign R-charges  $R(\Phi_a)=0$ and  $R(Y)=2$. 

The regularization that we used, which corresponds to first performing the regular $\omega$-integrals and then the $\vec{k}$-integrals in dimensional regularization, preserves SUSY\footnote{The fact that our regularization scheme preserves SUSY can also be seen at the level of components fields, see Appendix \ref{app_Non-relativistic Wess-Zumino model in components}, where we find consistent results from the quantum corrections of the various fields.}. Therefore, the Wilsonian effective action at a given scale $\lambda$ will have the following general structure
\beq
\tilde{S}_\lambda= \int d^3 x \, d^4 \theta   \,  K(\bar{\Phi}_a \Phi_a,Z_{ab}, Y, \bar{Y}, D)
+\int d^3 x \, d^2 \theta \, W_\lambda (\Phi_a, Y) 
\eeq
R-invariance and holomorphicity of the superpotential, combined with the weak coupling limit, give as in the relativistic case
 $W_\lambda=Y \,  W(\Phi_a)$.


\section{Comments and discussion}

We have seen many interesting properties from the investigation of the null reduction of the $3+1$ dimensional WZ model.
The non-renormalization theorem is inherited because the DLCQ method does not affect the grassmannian part of the superfields and SUSY is preserved after quantization, which allow to import both the perturbative and the non-perturbative formulations of the theorem.

More surprisingly, the properties of the model coming from the Galilean invariance of the problem produce many interesting results.
Combining the retarded nature of the propagator with the mass conservation, we found a set of selection rules which eliminate from the study of quantum corrections a lot of supergraphs which are instead allowed in the relativistic parent theory.
As a result, the contributions to the self-energy vanish up to four loops, except for a one-loop diagram for the superfield $\Phi_2 .$

Extending the investigation to higher loops we have provided strong evidence that
the combination of the non-renormalization properties of the F-terms and the selection
rules, in particular the vanishing of loop diagrams whose arrows form a closed loop, sup-
presses Galilean UV divergences in a very efficient way and makes the model one-loop
exact.
This remarkable property is not shared by the relativistic parent theory, and then we have strong evidence that
the non-relativistic limit and the quantization of a theory do not commute: we cannot obtain the Galilean WZ model at quantum level simply by null-reducing the quantum relativistic one.
This observation points in the same direction of the fact that the non-relativistic trace anomaly studied in chapters \ref{chapt-Non-relativistic actions}, \ref{chapt-The heat kernal technique} is not simply the null reduction of the relativistic ones. 

The result we have found is reminiscent of relativistic gauge theories with extended SUSY, like for instance the relativistic $ \mathcal{N}=2 $ SYM in 3+1 dimensions. 
In that case extended supersymmetry constrains the corrections to the K\"ahler  potential to be related to the F-terms, which are protected by the non-renormalization theorem.
In the non-relativistic model discussed in this paper, instead the protection of the K\"ahler potential is related to the $U(1)$ charge conservation at each vertex, which in many
diagrammatic contributions constrains arrows to form a closed loop, so leading to a vanishing integral.
It would be interesting to investigate 
if a common hidden mechanism exists, which is responsible of the similar
mild UV behavior of these two rather different classes of theories.

The Galilean WZ model that we investigated is classically scale invariant, but not quantum mechanically.
In fact, choosing for simplicity the case where the coupling constant $g$ is real, we find from the non-renormalization theorem and the one-loop exactness of the self-energy that the beta function is given by
\beq
\beta_g=\frac{d g } {d \log \mu} =\frac{g^3}{ 4 \pi m}
\label{betafunction}
\eeq
and the theory is infrared free at low energies, like the model studied in \cite{Bergman:1991hf}.
It is interesting to observe that the result in ref. \cite{Bergman:1991hf} is also exact, but for different reasons: in that model there are only scalar fields with quartic interaction, and the non-relativistic invariance forbids any self-energy correction, while the contributions to the vertex can be resummed.


\part{Complexity}


\chapter{Complexity for warped AdS black holes}
\label{chapt-Complexity for warped AdS black holes}

\begin{center}
\emph{The work in this chapter has previously appeared in \cite{Auzzi:2018zdu,Auzzi:2018pbc}.}
\end{center}

In the first part of this thesis we focused on QFT aspects concerning the investigation of the trace anomaly and the renormalization properties of Galilean-invariant systems. 
In this context, there was recently much interest in the investigation of a particular kind of models with non-relativistic invariance, called Warped Conformal Field Theories (WCFTs) and studied in $1+1$ dimensions.
These are local field theories invariant under translations and chiral dilatations\footnote{We label the two coordinates as  $x^{\pm},$ which is the common choice in the literature. Despite the name, these are not light-cone coordinates.}  
\beq
x^{\pm} \rightarrow x^{\pm} + c^{\pm} \, , \qquad
x^- \rightarrow \lambda x^- \, ,
\eeq
where $ \lambda, c^{\pm}$ are constants.
If we require unitarity, locality and a spectrum bounded from below of the dilatation operator, this symmetry enhances to an infinite-dimensional group which can be either the conventional conformal group or a Virasoro times a Kac-Moody algebra, which give the warped conformal case \cite{Hofman:2011zj}.
The global subgroup in the latter case is $ SL(2,\mathbb{R}) \times U(1) .$ 

General classifications of quantum anomalies can be performed for this class of theories \cite{Jensen:2017tnb}: the computation requires to couple the system to an appropriate background, which turns out to be the NC geometry introduced in section \ref{sect-Newton-Cartan geometry}. 
On this background Weyl invariance can be defined and it is possible to interpret WCFTs as Lifshitz field theories with dynamical exponent $z=\infty.$

As for the trace anomaly, we can hope to compute the quantum anomalies for explicit models realizing the warped symmetry, but it turns out that there are very few examples of such models, and moreover they are all on the edge of non-locality: there is an infinite number of exactly marginal operators which are non-local along one of the coordinates in the 1+1 dimensional spacetime.
For these reasons, it may be convenient to investigate the quantum anomalies and other aspects of these theories from a different point of view, \emph{i.e.} using holography.
In fact, it was conjectured that WCFTs are dual to a non-trivial deformation of $ \mathrm{AdS}_3 $ which only preserves the isometries $ SL(2,\mathbb{R}) \times U(1) , $ called warped $\mathrm{AdS}_3$ spacetime \cite{Detournay:2012pc,Jensen:2017tnb,Anninos:2008fx,Anninos:2008qb,Hofman:2014loa}. 
First of all, the set of symmetries of this spacetime matches with the field theory side. Secondly, the entanglement entropy was studied in this context and an analog of the Cardy formula was found \cite{Detournay:2012pc}.
Entanglement entropy in this context has been studied by several authors \cite{Anninos:2013nja,Castro:2015csg,Azeyanagi:2018har,Song:2016pwx,Song:2016gtd}. 

There are various reasons to study WCFTs and their holographic dual.
First of all, from the theoretical point of view, it is interesting to consider a non-relativistic conformal field theory with an infinite dimensional symmetry group.
Since in the relativistic case the constraints given by the Virasoro algebra and the holomorphic structure of the symmetry group are so powerful to give remarkable results, we may hope to find similar phenomena in this case as well.
Secondly, it is thought that WCFT techniques can be applied in condensed matter systems like the Potts model, because they show an anisotropic scaling \cite{Cardy:1992tq}. 
The investigation of the field theory side is also useful to study higher spin theories.
In fact, these models require to consider all kind of conserved currents on equal footing, a treatment that can be obtained in the WCFT context.
This approach turns out to be useful to understand the modular properties of partition functions for such theories \cite{Kraus:2011ds, Gaberdiel:2012yb}.
From the point of view of holography, extremal rotating black holes have a near horizon which is topologically $\mathrm{AdS}_3 \times S^2$ in 3+1 dimensions.
This structure leads to a simplifications of low-energy S-matrix elements and is conjectured to be dual to a CFT (Kerr/CFT correspondence).
In particular, when going at fixed polar angle in this configuration, we precisely find a $\mathrm{WAdS}_3$ structure.
We then think that an analysis of WAdS/WCFT correspondence can also shed light on properties of extremal rotating black holes.

In this chapter we continue the investigation of non-local quantities from the gravity side by studying the computational complexity, which recently was proposed to describe the time evolution of the Einstein-Rosen Bridge (ERB) of a BH \cite{Stanford:2014jda, Brown:2015lvg}.
In the Complexity=Volume (CV) conjecture, complexity is proportional to the volume of a maximal codimension-one slice anchored at the boundary
\beq
C_V \sim   \frac{\textrm{Max}(V)}{G l} \, .
\eeq
In this expression $l$ is a length scale which depends from the holographic setting we are considering.
The precise proportionality factor also depends from the specific BH.

In the Complexity=Action (CA) conjecture, complexity is proportional to the gravitational action evaluated in the Wheeler-De Witt (WDW) patch, \emph{i.e.} the bulk domain of dependence of the extremal slice
\beq
C_A = \frac{I}{\pi \hbar} \, .
\eeq
Holographic complexity has been recently studied by many groups
in  various asymptotically AdS gravity backgrounds, see \emph{e.g.}
\cite{Lehner:2016vdi,Cai:2016xho,Chapman:2016hwi,Carmi:2017jqz,Chapman:2018bqj,Moosa:2017yvt,Moosa:2017yiz,Chapman:2018dem,Chapman:2018lsv,Barbon:2015ria,Bolognesi:2018ion,
Flory:2018akz,Flory:2019kah}.
In the context of warped AdS, previous studies were performed in \cite{Ghodrati:2017roz,Dimov:2019fxp}.

In order to test the holographic proposals by Susskind, we will study both the CV and the CA conjectures for BHs in Warped $ \mathrm{AdS}_3 $ seen as a solution of Einstein gravity coupled to electromagnetism with Chern-Simons term.

We point out that while this computation can provide important hints for the matching of complexity with the field theory side of the duality, the precise definition of complexity in QFT has yet to be completely understood.
Here we briefly review the state of the art of this problem\footnote{The reader interested only in the holographic approach to complexity can directly skip to section \ref{sect-Black holes in Warped AdS}.}.

A promising approach is based on Nielsen's geometric formalism, which involves the search for geodesics in the space of unitary evolutions \cite{Jefferson:2017sdb, Khan:2018rzm, Hackl:2018ptj, Chapman:2018hou, Caceres:2019pgf}.
The idea is to consider a quantum circuit model in order to obtain the target state $| \psi_T \rangle $ starting from a simple reference state $| \psi_R \rangle $ via the application of a unitary operator such that
\beq
| \psi_T \rangle = U |  \psi_R \rangle \, ,
\label{reference and target states, Nielsen}
\eeq
where $U$ is built with a set of simple elementary gates.
The unitary operator can be synthesized by means of an Hamiltonian such that
\beq
U = \vec{\mathcal{P}} \, \exp  \left[ \int_0^1 dt \, H(t) \right] \, ,
\eeq
where 
\beq
H(t)= \sum_I Y^I (t) M_I \, .
\eeq 
In the previous expressions $\vec{\mathcal{P}}$ is the path ordering such that the Hamiltonian at earlier times is applied to the first state (\emph{i.e.} the circuit is built starting from the right), $M_I$ is a set of generalized Pauli matrices and $Y^I$ are control functions which specify the tangent vector to a trajectory in the space of unitaries, given by
\beq
U(t) = \vec{\mathcal{P}} \, \exp  \left[ \int_0^t dt' \, H(t') \right] \, .
\eeq
The boundary conditions on such trajectories are $U(0)= \mathbf{1}$ and $U(1)=U,$ \emph{i.e.} we start from the reference state and we end up with the target state as in eq. (\ref{reference and target states, Nielsen}).
In order to give a measure of the difficulty to perform a path in the space of unitary, we have to define a cost function
\beq
\mathcal{D} (U(t)) = \int_0^1 dt \, F \le U(t) , \dot{U} (t) \ri \, ,
\label{Nielsen cost function}
\eeq
where the minimum corresponds to the optimal circuit, thus providing the definition of complexity.
The minimal requirements for a reasonable cost function imposed by Nielsen correspond to define (\ref{Nielsen cost function}) as a length functional for a Finsler manifold, which is a particular class of differentiable manifolds where a concept of distance can be introduced. 
In this context, the problem to find the optimal circuit corresponds to finding the length of geodesics in this particular geometry.

When considering free field theories, it is possible to regularize the theory by placing it on a lattice, which reduces the computation of complexity to the case of a set of harmonic oscillators.
Progress with this approach has been made in computing the complexity for a set of harmonic oscillators for the preparation of Gaussian states, while the computation is challenging for more general states.

Another approach from the field theory side is based on building a path in the space of states, where the distance is determined using the Fubini-Study metric on the space of the normalized states \cite{Chapman:2017rqy}.
In this case we define a path as
\beq
| \psi (\sigma) \rangle = U( \sigma) | \psi_R \rangle \, ,
\eeq
with 
\beq
U(\sigma) = \vec{\mathcal{P}} \, \exp \left[ - i \int_{s_i}^{\sigma} ds \, G(s)  \right] \, .
\eeq
In this context, $s$ parametrizes the progress along a path in the space of states starting from $s_i$ and ending at $s_f,$ with $\sigma \in [s_i, s_f],$ while $G(s)$ is a generator taken from an elementary set of hermitian operators.
The Fubini-Study line element is then  defined to be
\beq
ds^2 = d \sigma^2 \left( \left| \p_{\sigma} | \psi (\sigma) \rangle \right|^2 - \left| \langle \psi (\sigma) | \p_{\sigma} | \psi (\sigma) \rangle  \right|^2   \right) \, .
\eeq
The complexity is found by computing the geodesics corresponding to this geometry.
Even using this technique, most of the results are formulated for Gaussian states.
 
We conclude the review of field theory definitions of complexity with an approach slightly different from the previous ones, being based on a path integral optimization process \cite{Caputa:2017urj, Caputa:2017yrh, Bhattacharyya:2018wym}.
Consider a QFT defined in Euclidean spacetime $\mathbb{R}^d $ with coordinates denoted as $(x,z),$ being $x$ the vector referring to the spatial dimensions $\mathbb{R}^{d-1}$ and $z=-\tau$ the opposite of Eucidean time, which is interpreted to be the radial coordinate of $\mathrm{AdS}_{d+1}$ in the holographic picture.
The reference state can be considered to be the vacuum computed as a path integral over the spatial directions and $\varepsilon<z <\infty,$ where $\varepsilon$ is a UV cutoff:
\beq
\psi_{g_0, \lambda_0} [\phi(x)] = \int \prod_x \prod_{\varepsilon<z<\infty} \mathcal{D}\phi(x,z) \, e^{-S_{g_0, \lambda_0} [\phi]} \, \prod_x \delta \le \phi(x,\varepsilon) - \phi(x) \ri \, .
\label{vacuum reference state in path integral optimization}
\eeq 
In this definition $g_0$ is the flat space metric where the integration is performed
\beq
ds^2 = \frac{dx^2 + dz^2}{\varepsilon^2} \, ,
\eeq
while $\lambda_0$ is a general label for the coupling constants of the QFT with action $S_{g_0,\lambda_0} [\phi].$
The optimization process consists in letting the metric and the coupling constants to vary with the space coordinates $g(x,z),\lambda(x,z)$ with boundary conditions
\beq
g(x,\varepsilon) = g_0 \, , \qquad
\lambda(x, \varepsilon) = \lambda_0 \, .
\eeq
In this way, the general path integral defining a state is
\beq
\psi_{g(x,z), \lambda(x,z)} [\phi(x)] = \int \prod_x \prod_{\varepsilon<z<\infty} \mathcal{D}\phi(x,z) \, e^{-S_{g(x,z), \lambda(x,z)} [\phi]} \, \prod_x \delta \le \phi(x,\varepsilon) - \phi(x) \ri \, .
\eeq 
Of course this state differs from (\ref{vacuum reference state in path integral optimization}) in a non-trivial way, but it is possible to fine tune $g(x,z),\lambda(x,z)$ in order to find particular configurations such that
\beq
\psi_{g(x,z), \lambda(x,z)} [\phi(x)]  = e^{N[g,\lambda] - N[g_0, \lambda_0]} \, \psi_{g_0, \lambda_0} [\phi(x)] \, .
\eeq
If this is possible, it means that the wave-functions describe the same quantum state.
The optimization procedure consists in minimizing the functional $N[g,\lambda]$ appearing in the exponential, and this minimum value corresponds to complexity.
It turns out that most of the results obtained with this approach are limited to $d=2$ dimensions, where the functional can be related to the Liouville action.

Given the previous approaches to complexity from the field theory side, we understand how much is important to have computations from the gravitational side in order to have some feelings for the results to compare, and to understand which of the previous proposals should be taken to define complexity.
We start testing the holographic conjectures by Susskind for black holes in warped $\mathrm{AdS}_3$ spacetime.


\section{Black holes in Warped AdS}
\label{sect-Black holes in Warped AdS}

We consider BHs in a spacetime with Warped AdS$_3 $ asymptotic \cite{Moussa:2003fc,Bouchareb:2007yx,Anninos:2008fx}, which are interpreted to be dual to a boundary WCFT at finite temperature. The metric is given by
\beq
\label{BHole}
\frac{ds^2}{l^2} = dt^2 +\frac{ dr^2}{(\nu^2+3) (r-r_+)(r-r_-)}
+\le 2 \nu r -\sqrt{r_+ r_- (\nu^2+3)}  \ri dt d \theta +  \frac{r}{4}  \Psi d \theta^2 \, ,
\eeq
\beq
 \Psi(r)= 3 (\nu^2-1)  r +(\nu^2+3) (r_+ + r_-) - 4 \nu \sqrt{ r_+ r_- (\nu^2+3)} \, .
\eeq
We introduce the quantity $\tilde{r}_0$ as 
\beq
\tilde{r}_0= \max(0, \rho_0) \, , \qquad \rho_0=\frac{4 \nu \sqrt{r_+ r_- (\nu^2+3) }- (\nu^2+3)(r_++r_-)}{3 (\nu^2-1)} \, ,
\label{rtilde0}
\eeq
where $\Psi(\rho_0)=0.$ 
The range of coordinates is $r \in [\tilde{r}_0, \infty), t \in (-\infty, \infty)$ and $\theta \in [0,2 \pi]$ with the identification $\theta \sim \theta + 2 \pi .$
We denote with $r_-, r_+$ the inner and outer horizons, respectively. They satisfy $r_- \leq r_+$ and the particular case $r_-=r_+=0$ corresponds to empty warped $\mathrm{AdS}_3$ spacetime in Poincaré patch with timelike boundary parametrized by $(t,\theta).$
The parameter $\nu$ is related to the left and right central charges of the boundary WCFTs, which in Einstein gravity are \cite{Anninos:2008qb}
\beq
c_L=c_R=\frac{12 l \nu^2 }{G (\nu^2+3)^{3/2}} \, .
\eeq
Ordinary $\mathrm{AdS}_3$ spacetime can be seen as a fibration of the real line over $\mathrm{AdS}_2,$ but if a warping factor multiplies the fiber metric we obtain a spacetime whose asympthotic changes from $ SL(2, \mathbb{R})_L \times SL(2, \mathbb{R})_R$ to $ SL(2, \mathbb{R})_L \times U(1)_R$.
This warping is parametrized by $\nu$ and it is such that for $\nu=1$ the  Banados-Teitelboim-Zanelli (BTZ) black hole 
\cite{Banados:1992wn,Banados:1992gq} is recovered. In fact, in this case the change of coordinates 
\beq
r=\bar{r}^2 \, , \qquad t = \frac{\sqrt{r_+}-\sqrt{r_-}}{l^2} \bar{t} \, , \qquad 
\theta=\frac{l \bar{\theta} - \bar{t}}{l^2 (\sqrt{r_+}-\sqrt{r_-}) } \, , \qquad r_\pm=\bar{r}_\pm^2 
\label{change-BTZ}
\eeq
brings the metric to the standard BTZ form
\beq
ds^2   = - \frac{ \bar{r}^2 - \bar{r}^2_{+} - \bar{r}^2_{-}}{l^2} 
d \bar{t}^2 + \frac{ l^2 \bar{r}^2}{(\bar{r}^2-\bar{r}^2_{+})(\bar{r}^2- \bar{r}^2_{-})} d\bar{r}^2 - 2 \frac{\bar{r}_{+}\bar{r}_{-}}{l} d\bar{t} d\bar{\theta} + \bar{r}^2 d\bar{\theta}^2  \, .
\label{standardBTZ}
\eeq
For $\nu^2<1$ the solution is pathological because
it has closed time-like curves. For $\nu^2>1$ the solution is not sick and can
be realized as an exact vacuum
 solution of Topologically Massive Gravity (TMG)
 \cite{Moussa:2003fc,Bouchareb:2007yx},
New Massive Gravity (NMG) \cite{Clement:2009gq}
and also general linear combinations of the two mass terms \cite{Tonni:2010gb}.
We restrict our analysis to the case of positive $\nu$.
So  at the end we will consider just the case  $\nu \geq 1$.

Strictly speaking, the relation between area and entropy
 holds just in Einstein gravity: if we consider higher order
 corrections to the gravitational entropy, we have to 
 use the Wald entropy formula \cite{Wald:1993nt} instead of the geometrical area law.
So the CV conjecture should be
directly applicable just to Einstein gravity and should be
appropriately modified in order to take into account
higher order corrections in the gravitational action.
A proposal for such correction has been put forward in
\cite{Alishahiha:2015rta,Alishahiha:2017hwg}.
The CA conjecture can also be generalized to the case of
higher derivatives corrections to the gravitational action, see
e.g. \cite{Guo:2017rul,Ghodrati:2017roz,Qaemmaqami:2017lzs}.

As far as we know, there is no known non-pathological matter
content in field theory
supporting stretched warped BHs in Einstein gravity \cite{Anninos:2008qb}.
However, they can be obtained as solutions to a perfect
fluid stress tensor with spacelike quadrivelocity \cite{Gurses:1994bjn}. 
Alternatively they can arise as a solution of Chern-Simons-Maxwell electrodynamics 
coupled to Einstein gravity \cite{Banados:2005da,Barnich:2005kq}, 
but a wrong sign for the kinetic Maxwell term is required in order to have solutions with no closed time-like curves 
(which corresponds to $\nu^2 \geq 1$).
Moreover, warped BH can arise in string theory constructions, e.g. 
\cite{Compere:2008cw,Detournay:2012dz,Karndumri:2013dca}.
 In the following we take a pragmatical approach: we suppose that
a consistent  realization of stretched warped BHs in Einstein gravity exists,
and we investigate the CV conjecture.

We will use for concreteness the model studied in
 \cite{Banados:2005da,Barnich:2005kq}, which is 
 Chern-Simons-Maxwell electrodynamics 
coupled to Einstein gravity.
In order to have solutions without closed time-like curves,
a wrong sign for the kinetic Maxwell term is needed.
Solutions with positive Maxwell kinetic energy have $\nu^2<1$ and correspond
to G\"odel spacetimes.
We will see that the CA conjecture is so solid that can survive
to an unphysical action with ghosts.

\subsection{Conserved charges and thermodynamics}

In order to get a physical understanding of the computation of complexity, we need to find the conserved charges and the thermodynamics quantities of the BH.
Since we are studying the implications of taking the warped BH as a solution of Einstein gravity, the entropy is simply given by the area law
\beq
S=S_+=
\frac{l \pi}{4 G} (2 \nu r_+ - \sqrt{r_+ r_- (\nu^2+3)}) \, ,
\eeq
while the Hawking temperature\footnote{As in the standard $\mathrm{AdS}$ case, the Hawking temperature can be found by requiring the metric does not contain conical singularities after Wick rotating the $t$ coordinate.} and angular momentum are given by \cite{Anninos:2008fx}:
\beq
T= \frac{\nu^2+3}{4 \pi l } \,  \frac{r_+ -r_-}{2 \nu r_+ -\sqrt{(\nu^2+3) r_+ r_-} } \, ,
\qquad
\Omega=\frac{2}{(2 \nu r_+ -\sqrt{(\nu^2+3) r_+ r_-}) l } \, .
\label{T,Omega}
\eeq
At least formally, it is possible to associate an entropy via the area law to the surface corresponding to the inner horizon as
\beq
S_-=  \frac{ l \pi}{4 G} 
(  \sqrt{r_+ r_- (\nu^2+3)} - 2 \nu r_- ) \, .
\eeq
Following \cite{Castro:2012av,Giribet:2015lfa}, the existence of a holographic dual implies
 a quantization condition on the product of inner and outer entropies, 
 which in turn must be proportional to the conserved charges of the black hole which are quantized.
Since the angular momentum is the only quantized conserved charge, we obtain $J=S_- S_+ f(\nu)$,
where $f(\nu)$ is a so far arbitrary function which will be fixed by thermodynamics. 

Imposing that the resulting $dM$
is an exact differential, the  function $f(\nu)$ is fixed
and allows to solve for both the conserved charges:
\beq
M=\frac{1}{16 G} (\nu^2+3)
 \le  \le r_{-} + r_{+} \ri - \frac{\sqrt{r_{+}r_{-} (\nu^2 +3)}}{\nu} \ri \, ,
 \label{M guess}
\eeq
\beq
J=\frac{l}{32 G} (\nu^2+3) 
\le \frac{r_- r_+ (3+5 \nu^2)}{2 \nu}
 -(r_+ + r_-) \sqrt{(3+\nu^2) r_+ r_-} 
 \ri \, .
\label{J guess}
\eeq
Another approach to find these conserved charges is described in Appendix \ref{app-An explicit model for WAdS black holes}.

\subsection{Null coordinates and causal structure}
\label{sect-Null coordinates and causal structure}

In order to compute the CV and CA conjectures for this class of BHs, we need to know the causal structure of spacetime: this allows to depict extremal surfaces anchored at the boundaries and to build the WDW patch where the gravitational action will be computed.
In particular, the Penrose diagram is a useful tool to easily understand the causal properties of spacetime and to visualize these structures.

We start by finding null coordinates. The expression of the metric (\ref{BHole}) in Arnowitt-Deser-Misner (ADM) form is: 
 \beq
 \label{BHole-ADM}
 ds^2=-N^2 dt^2  +\frac{l^4 dr^2}{4 R^2 N^2}
 +l^2 R^2 (d \theta + N^\theta dt)^2
 \, , 
 \eeq
 where
 \beq
 R^2=\frac{r}{4} \Psi \, , \qquad
 N^2=\frac{l^2 (\nu^2+3) (r-r_+)(r-r_-)}{4 R^2} \, ,
 \qquad
 N^\theta= \frac{2 \nu r -\sqrt{r_+ r_- (\nu^2+3)} }{2 R^2} \, .
 \eeq
We consider a set of null geodesics
which satisfy $(d \theta + N^\theta dt)=0$; then 
 a positive-definite term in the metric (\ref{BHole-ADM})
 saturates to zero, and the null geodesics are given by the constant $u$ and $v$ trajectories \cite{Jugeau:2010nq}
 \beq
du = dt - \frac{l^2}{2 R N^2} dr \, , \qquad
dv = dt + \frac{l^2}{2 R N^2} dr  \,  .
 \eeq
These are the normal one-forms to the WDW null surfaces 
\beq
dv = v_\a dx^\a \, , \qquad du=u_\a dx^\a \, .
\label{null-normals}
\eeq
Moreover, the integral curves of $u^\a$ and $v^\a$ are 
null geodesics in the affine parameterization, \emph{i.e.}
\beq
u^\a D_\a u^\b=0 \, , \qquad v^\a D_\a v^\b=0 \, ,
\label{affine}
\eeq
where $D_\a$ is the covariant derivative.

Direct integration of them allows to define Eddington-Finkelstein coordinates as
\beq
u= t - r^* (r) \, , \qquad
v= t + r^* (r) \, ,  
\label{Kruskalcoordinates}
\eeq
where the tortoise coordinate $r^*$ is given by
\beq
r^*(r)= \int^r \frac{dr'}{f(r')} \, , \qquad    f(r)=\frac{2 R N^2}{l^2} 
=\frac{(\nu^2 +3)(r-r_-)(r-r_+)}{\sqrt{r \Psi(r)}} \, .
\label{derirstar}
\eeq 
Integrating eq. (\ref{derirstar}),  $r^*$ can be explicitly found  \cite{Jugeau:2010nq};
for $r_+ \neq r_-$ the explicit expression is
\beq
\begin{aligned}
r^{*}(r)= & \frac{\sqrt{3 \left(\nu^2-1\right)}}{\left(\nu^2+3\right)} 
\left\lbrace \frac{\sqrt{r_{+}(r_{+}-\rho_0)}}{r_+-r_-} \log \left(\frac{\left|r-r_{+}\right|}{\left(\sqrt{r} \sqrt{r_{+}-\rho_0}+\sqrt{r-\rho_0}\sqrt{r_+}\right)^2}\right) \right.\\
& \left. - \frac{\sqrt{r_{-}(r_{-}-\rho_0)}}{r_+-r_-} \log \left(\frac{\left|r-r_{-}\right|}{\left(\sqrt{r} \sqrt{r_{-}-\rho_0}+\sqrt{r-\rho_0}\sqrt{r_-}\right)^2}\right)  +2\log(\sqrt{r} +\sqrt{r-\rho_{0}}) \right\rbrace \, , 
\label{rstar}
\end{aligned}
\eeq
where $ \rho_0 $ was defined in eq. (\ref{rtilde0}).

The non-rotating case is defined by the condition $J=0$, 
and corresponds  to the following values:
\beq
r_-=0 \, , \qquad  \frac{r_+}{r_-}=\frac{4 \nu^2}{\nu^2+3} \, .
\label{non-rota}
\eeq
The two values in eq (\ref{non-rota}) can be mapped among each other by an isometry \cite{Jugeau:2010nq},
then we will always consider the case $ r_-=0, r_+ = r_h$ for  simplicity
when referring to the non-rotating case.
Curiously enough, in this case the Penrose diagram is the same as the ones for the Schwarzchild BH in 3+1 dimensions \cite{Jugeau:2010nq}. In the rotating case, for generic $(r_+,r_-)$, the Penrose diagram is  the same as the one of the Reissner-Nordstr\"om BH (see figs. 7 and 8 of  \cite{Jugeau:2010nq}).
The extremal limit corresponds to 
$r_+ = r_- $; in this case temperature is zero
and there is no thermofield double: the Penrose diagram 
has just one boundary.

In light of the conjectured WAdS/WCFT duality, it is puzzling that the spacetime has a Minkwoskian asymptotic, because it is not completely clear where the boundary theory should live.
We point out that in our computation we will always require the existence of a UV cutoff $\Lambda$ which induces a timelike boundary where we can think that the QFT lives. When taking the limit $\Lambda\rightarrow \infty,$ the timelike boundary goes to a single point in the Penrose diagram, where there is the future timelike infinity\footnote{We remind that the Penrose diagram for 2+1 dimensional WAdS contains at each point a factor of $S^1,$ so that the future timelike infinity is not actually a single point.}.
A similar issue arises when we will build the WDW patch for this black hole: we discuss in more details how we treat the problem in section \ref{subsect-Regularization of the WDW patch}.

\subsection{An explicit realization in Einstein gravity}
\label{sect-An explicit realization in Einstein gravity}

In view of the computation of the CA conjecture, we need to take a specific theory supporting warped BHs as a solution of Einstein gravity. For concreteness we will use a model introduced in \cite{Banados:2005da}, where the matter content is a Chern-Simons $U(1)$ gauge field. In order to find absence of closed time-like curves ($\nu^2 \geq 1$), a ghost-like kinetic Maxwell term is needed. 
We will see that the CA conjecture seems solid enough to survive to unphysical matter contents which include ghosts, giving a physical result consistent with expectations about complexity from quantum information.

We consider  Einstein gravity in $2+1$ dimensions with a negative cosmological constant, coupled to a $U(1)$ gauge field with both Maxwell and Chern-Simons terms 
\beq
I_{\mathcal{V}} = \frac{1}{16 \pi G} \int_{\mathcal{V}} d^3 x \, \left\lbrace \sqrt{-g} \left[ \le R + \frac{2}{L^2} \ri - \frac{\kappa}{4} F_{\mu \nu} F^{\mu\nu} \right] - \frac{\alpha}{2} \epsilon^{\mu\nu\rho} A_{\mu} F_{\nu \rho} \right\rbrace = \int_{\mathcal{V}} d^3 x \, \sqrt{-g} \, \mathcal{L} \, ,
\label{bulk action}
\eeq
where $\epsilon^{\mu \nu \rho} $ is the Levi-Civita tensorial density.
Here we put a coefficient $\kappa=\pm 1$ in front of  the Maxwell kinetic term

The equations of motion for the gauge field are
\beq
D_\mu F^{\a \mu} = -\frac{\a}{\kappa}  \frac{ \epsilon^{\a \nu \rho} }{\sqrt{g}} F_{\nu \rho} \, ,
\eeq
while the Einstein equations are
\beq
G_{\mu \nu} - \frac{1}{L^2} g_{\mu \nu} = \frac{\kappa}{2}  T_{\mu \nu} \, , \qquad
T_{\mu \nu} =   F_{\mu \a} F_{\nu}^{\,\,\, \a}-\frac14 g_{\mu \nu}
F^{\a \b} F_{\a \b} \, .
\eeq
We consider the set of coordinates $ (r,t, \theta) $ where the metric assumes the form (\ref{BHole}), and we choose a 
 gauge motivated by the ansatz from \cite{Banados:2005da}:
\beq
A= a dt+ (b+c r) d \theta \, , \qquad F = c \, dr \wedge d \theta \, ,
\label{ansatz per campo di gauge}
\eeq
where $ \lbrace a,b,c \rbrace $ is a set of constants. Thus, the Maxwell equations give:
\beq
\a = \kappa \frac{\nu}{l} \, .
\eeq
From the Einstein equations, we get, independently from $(r_+,r_-)$:
\beq
L = l \sqrt{\frac{2}{3-\nu^2} } \, , \qquad
c =\pm  l \sqrt{\frac{3}{2 } \frac{1-\nu^2}{\kappa}}
\, . 
\label{gaugau}
\eeq
The second equation shows that there is conflict between absence of closed time-like  curves
and the presence of ghosts ($\kappa=-1$).

The gauge parameter $a$ is not constrained by the equations of motion,
but the action depends explicitly on $a$ through the Chern-Simons term.
The value of $a$ is important to properly define 
the mass $M$ as a conserved charge  \cite{Barnich:2005kq}.
Formally, only for the value
\beq
a=
\frac{l}{\nu} \sqrt{\frac32} \sqrt{\nu^2-1}  \, .
\label{a-action}
\eeq
the mass is associated to the
Killing vector $\p/\p t$ and it does not depend
 on the $U(1)$ gauge transformations. 
For this value, the action density reads:
\beq
16 \pi G \, \sqrt{-g} \, \mathcal{L} = - \frac{l}{2} (\nu^2 +3) \equiv \mathcal{I} \, .
\label{densita}
\eeq
The comparison with the solution of  \cite{Banados:2005da} is
discussed in appendix \ref{app-An explicit model for WAdS black holes}.


\section{Complexity=Volume}
\label{sect-Complexity=Volume}

\subsection{Einstein-Rosen bridge}

\begin{figure}[h]
\begin{center}
\includegraphics[scale=0.5]{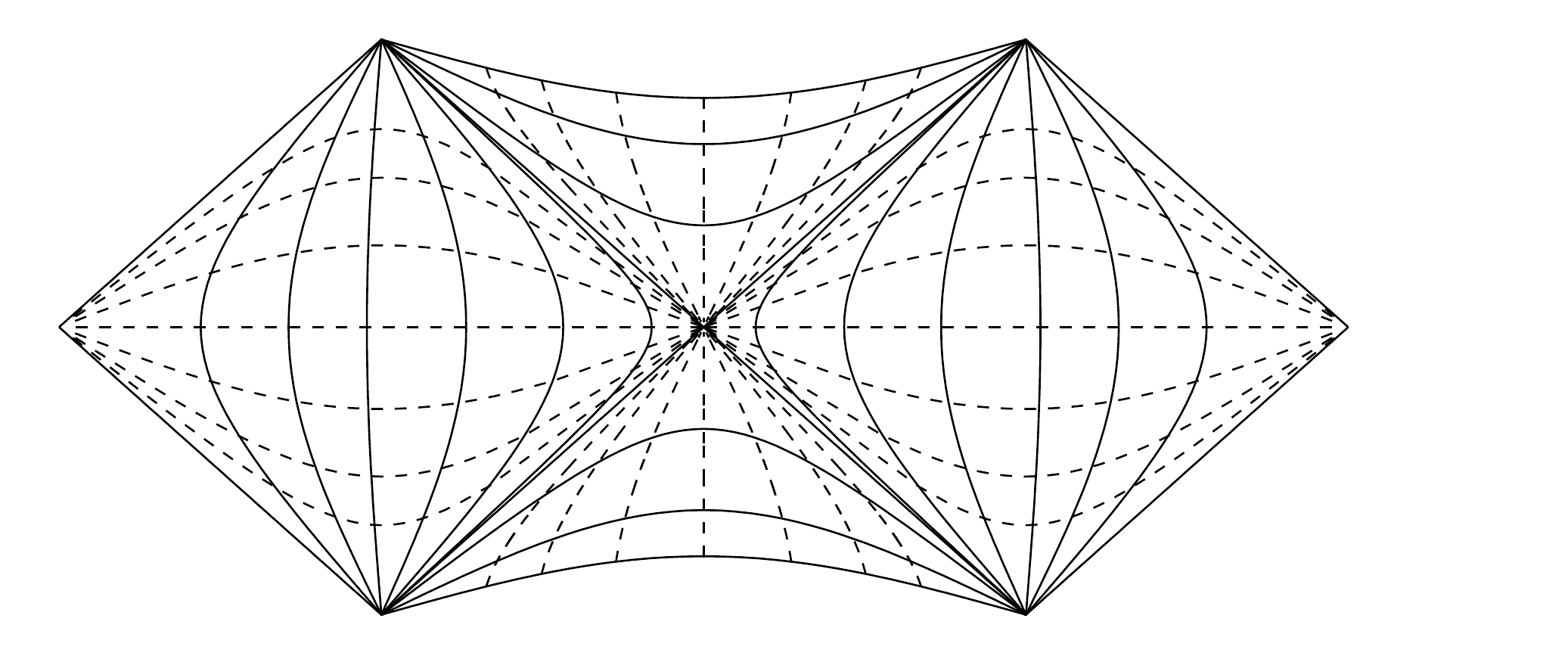}
\end{center}
\caption{Constant $r$ lines (solid) and constant $t$ lines (dashed) 
of the Penrose diagram in the non-rotating case.}
\label{fig-t}
\end{figure}

The Penrose diagram for the non-rotating case is shown in figure \ref{fig-t},
with some lines at constant $r$ and $t$. 
Both in the rotating and non-rotating cases,
for $r \rightarrow \infty$,  the asymptotic behavior of $r^*(r)$ is
\beq
r^* (r) \approx \frac{3 \sqrt{\nu^2-1}}{\nu^2+3} \log r \equiv C  \log r  \, .
\eeq
So we should first fix a cutoff surface at $r= \Lambda$ to make our calculations finite.
The WDW surface is bounded by lines with constant values of $v$ and $u$,
which in the Penrose diagram correspond to $45$ degrees lines.

On the left and right boundaries,
the time  coordinate $t$ diverges to $\pm \infty$ in the upper and lower sides, respectively.
From eqs. (\ref{Kruskalcoordinates}), a change of cutoff
from $\Lambda_1$ to $\Lambda_2$ implies a constant shift in the time coordinate
by $C \log \frac{\Lambda_2}{\Lambda_1}$.
For $\nu=1$ we recover the AdS asymptotic, $r^*(\infty)$ is finite and no shift is needed;
the Penrose diagram in this case is different and is the standard one of the BTZ black hole.

As done in \cite{Susskind:2014moa, Stanford:2014jda} for the AdS and the flat cases,
  we consider an extremal codimension-one bulk surface
extending between the left and the right side of the Kruskal diagram;
we denote the times at the left and right sides as $t_L, t_R$,  respectively. 
 The dual  thermofield  double  state has the following form:
 \beq
|\Psi_{TFD} \rangle   \propto   \sum_n  e^{-E_n \beta/2- i E_n (t_L + t_R)} | E_{n} \rangle_R  | E_{n} \rangle_L \, ,
\eeq
where $| E_{n} \rangle_{L,R}$ refer to the energy eigenstates of left and right boundary theories,
 $\beta$ is the inverse temperature.
The usual time translation symmetry in Schwarzschild coordinates
corresponds to a forward  time translation on the right side and a backward
translation on the left one \cite{Maldacena:2001kr}, \emph{i.e.}
\beq
 t_L \rightarrow t_L + \Delta t \, , \qquad t_R \rightarrow t_R - \Delta t \, .
 \eeq
 This corresponds to the invariance of the  thermofield  double state
  under the evolution described by the Hamiltonian 
  $ H=H_L- H_R $ in the associated couple of entangled WCFTs.
If instead we take time running forward on both the copies
of the boundaries, we introduce some genuine time dependence in the problem \cite{Hartman:2013qma}
and the volume of the maximal slice will depend on time \cite{Stanford:2014jda}.
We will then consider the symmetric case with equal boundary times 
\beq
 t_L=t_R=t_b/2 \, .
 \eeq
In order to regularize the divergences, the times at the left
and right boundaries  are evaluated
at the cutoff surface $r=\Lambda$.

\subsection{Null coordinates for the computation of the Volume}

In Section \ref{sect-Null coordinates and causal structure} we already introduced a set of null geodesics, but it turns out that there is a similar coordinate system that is convenient to study the volume of the Einstein-Rosen bridge anchored at the boundary.

We have the following conserved quantities along geodesics:
\bea
 K &=& 2 \dot{t} + \le 2 \nu r - \sqrt{r_{+}r_{-}(\nu^2 +3)} \ri \dot{\theta} \, ,  \nl
 P &=& \le 2 \nu r - \sqrt{r_{+}r_{-}(\nu^2 +3)} \ri \dot{t} + \frac{r}{2}  \Psi(r) \, ,
 \label{K,P conserved quantities}
\eea
where dots denote derivatives with respect to the geodesic affine parameter.

All the null geodesics that can be found in the spacetime are parametrized by these conserved quantities. By taking the particular choice $P=0$ with generic $K$ we find precisely the set of null coordinates defined in Section \ref{sect-Null coordinates and causal structure}.
In this case we take instead the special value $K=0$ with generic $P,$ getting a particular set of geodesics satisfying
\beq
\dot{t} = \frac{P \le 2 \nu r - \sqrt{r_{+}r_{-}(\nu^2 +3)} \ri  }{(\nu^2 +3)(r-r_{-})(r-r_{+})} \, , \qquad
\dot{\theta} =- \frac{2 P}{(\nu^2 +3)(r-r_{-})(r-r_{+})} \, , \qquad
\dot{r} = \pm P \, .
\eeq
These geodesics can be used to introduce null coordinates which are regular at the horizon.

The infalling geodesics correspond to
\beq
\frac{d\theta}{dr} = \frac{2}{(\nu^2 +3)(r-r_{-})(r-r_{+})} \, , \qquad
\frac{d t}{dr} = - \frac{  2 \nu r - \sqrt{r_{+}r_{-}(\nu^2 +3)} }{(\nu^2 +3)(r-r_{-})(r-r_{+})} \, ,
\eeq
and allow to define Eddington-Finkelstein coordinates $(w, \theta_w)$ such that
\beq
dw =dt + \frac{  2 \nu r - \sqrt{r_{+}r_{-}(\nu^2 +3)} }{(\nu^2 +3)(r-r_{-})(r-r_{+})}  dr \, , \qquad
d\theta_{w} = d\theta - \frac{2}{(\nu^2 +3)(r-r_{-})(r-r_{+})} dr \, .
\eeq
The finite expression for the coordinate change is\footnote{We called the null coordinate and the associated tortoise coordinate as $w, \tilde{r} (r)$ to distinguish them from the names $v, r^* (r)$ used for the quantities defined in section \ref{sect-Null coordinates and causal structure}. However, they are defined with the same spirit, the only difference being the particular choice of the parameters $K,P$ in the conserved quantities (\ref{K,P conserved quantities}) along a null geodesic.}
\beq
w= t + \tilde{r} (r) \, , \qquad
\theta_{w} = \theta - \frac{2}{(\nu^2 +3)(r_{+}-r_{-})} \log \left| \frac{r -r_{+}}{r-r_{-}} \right| \, ,
\label{TUR}
\eeq
where
\beq
\tilde{r} (r) =  \frac{ 2 \nu r_{+} - \sqrt{r_{+}r_{-}(\nu^2 +3)} }{(\nu^2 +3)(r_{+}-r_{-})} \log |r-r_{+}| - \frac{ 2 \nu r_{-} - \sqrt{r_{+}r_{-}(\nu^2 +3)} }{(\nu^2 +3)(r_{+}-r_{-})} \log |r-r_{-}| \, .
\label{RSTAR}
\eeq
In terms of these coordinates, the metric becomes
\beq
\frac{ds^2}{l^2} = dw^2 -dr d\theta_{w} + \le 2 \nu r - \sqrt{r_{+}r_{-}(\nu^2 +3)} \ri dw d\theta_{w}
+ \frac{r}{4}  \Psi(r)  d \theta_{w}^2 \, .
\label{metricEF}
\eeq

\subsection{Computation of the Volume in the non-rotating case}

In this section we will compute the volume of the ERB as a function of time \cite{Stanford:2014jda}.
We first study the non-rotating case, setting $r_+=r_h$ and $r_-=0$ in the metric with coordinates (\ref{metricEF}).

\begin{figure}[ht]  
\begin{center}
\includegraphics[scale=0.5]{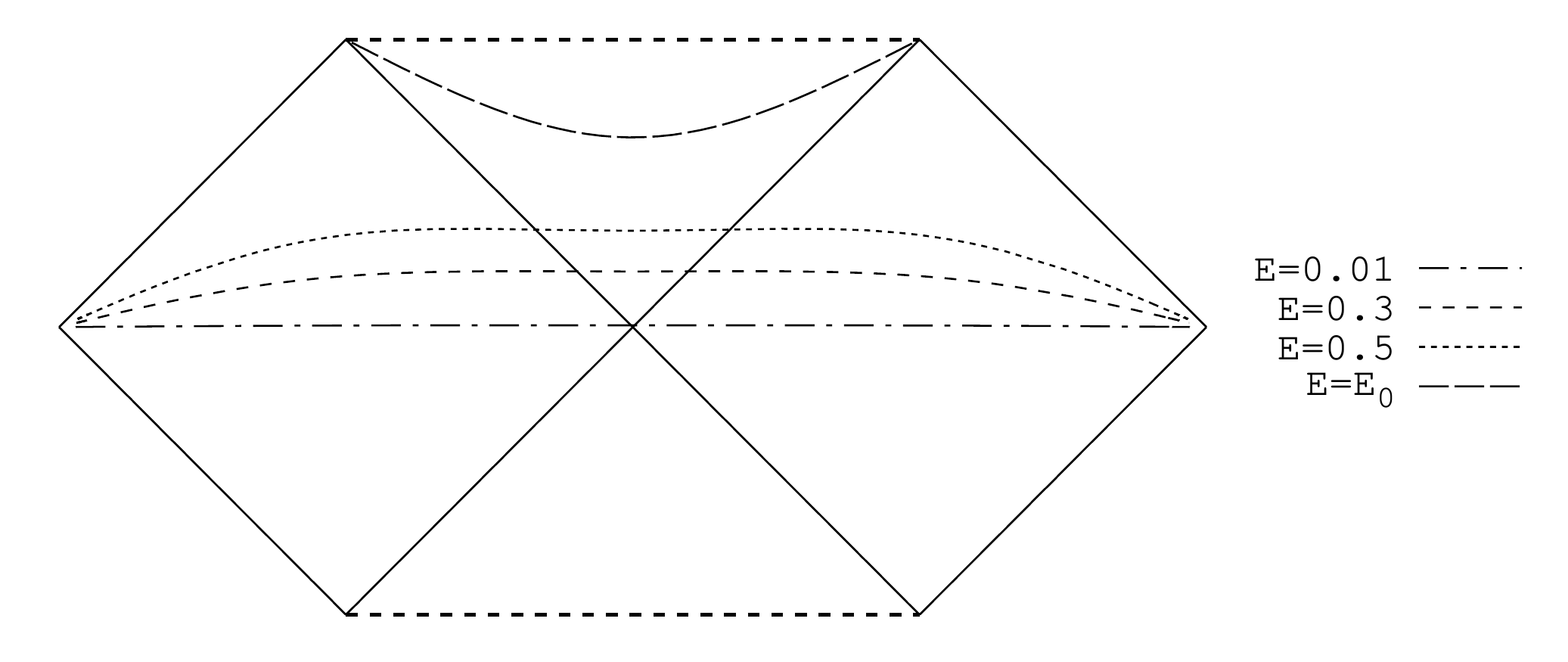}          
\end{center} 
\caption{Solutions to eqs. (\ref{shape}) for the non-rotating case,
 plotted in a Penrose diagram, for $\nu=2.5$ and $r_h=1$. The $E=E_0$ line,
 which sits at constant $r_{\rm min}=  \frac{r_h}{2}$, corresponds to the
 large $t_b$ limit.}
\label{figura0} 
\end{figure}

The minimal volume is chosen along the $0 \leq \theta_w \leq 2 \pi$ coordinate, and with
profile functions $w(\l)$, $r(\l)$, written in terms of a parameter $\l$. 
The volume integral will run from 
$\l_{\rm min}$ to $\l_{\rm max}$, with associated radii $r_{\rm min}$ and $r_{\rm max}$:
\beq
V=2 \cdot 2 \pi \int_{\l_{\rm min}}^{\l_{\rm max}}  d \l \,  l^2 \sqrt{ 
\frac{\dot{w}^2 r}{4} \left[ 3(\nu^2-1) r + (\nu^2+3) r_h \right]
 - \left(\dot{w} r \nu -\frac{\dot{r}}{2} \right)^2 } = 4 \pi  \int d\lambda \, \mathcal{L} (r,\dot{r},\dot{w})  \, .
\eeq
The factor $2$ takes into account the two sides of the Kruskal extension,
the $2 \pi$ is the result of the integration in $\theta_w$ and the
 dots denote derivatives with respect to $\l$.
The radius $r_{\rm max}$ plays the role of an ultraviolet cutoff;
we will take the limit $r_{\rm max} \rightarrow \infty$
at the end of the calculation.
The conserved quantity from translational invariance in $w$ gives
\beq
E=\frac{1}{l^2}  \frac{\p \mathcal{L}}{\p \dot{w}}=  \frac{\frac{ \nu^2+3 }{4} \dot{w} r (r_h-r)+ \frac{\nu r \dot{r} }{2}}
{\sqrt{ 
\frac{\dot{w}^2 r}{4} \left[ 3(\nu^2-1) r + (\nu^2+3) r_h \right]
 - \left(\dot{w} r \nu -\frac{\dot{r}}{2} \right)^2 }} \, .
 \label{conserved quantity}
\eeq
We use the reparametrization symmetry  for $\l$ in such a way that 
$V= 4 \pi l^2 \int d \l , $ which implies
\beq
\frac{\dot{w}^2 r}{4} \left[ 3(\nu^2-1) r + (\nu^2+3) r_h \right]
 - \left(\dot{w} r \nu -\frac{\dot{r}}{2} \right)^2  =1 \, ,  \, \qquad 
E=  \frac{ \nu^2+3 }{4} \dot{w} r (r_h-r)+ \frac{\nu r \dot{r} }{2} \, .
\eeq
We can then solve for $\dot{r},\dot{w}$:
\beq
\dot{r}=  2 \sqrt{
\frac{  4 E^2 +  \left(\nu ^2+3\right) r \left(r-r_h\right) }
{ r \left(3 \left(\nu ^2-1\right) r+\left(\nu ^2+3\right) r_h\right)
   } } \, ,
\qquad 
\dot{w}=
\frac{4}{(\nu^2 +3)(r_h-r)} \le \frac{E}{r}  - \frac{\nu}{2}  \dot{r} \ri \, ,
\label{shape}
\eeq
where we took $\l$ in the direction of increasing $r$.
These equations can be solved numerically; some example of solutions, plotted in the Penrose
diagram, are shown in figure \ref{figura0}.

The minimum radius $r_{\rm min}$ is a solution of $\dot{r}=0$:
\beq
 r_{\rm min}^2- r_h  r_{\rm min} + \frac{4 E^2}{(3+\nu^2)} = 0 \, , \qquad
 r_{\rm min}=\frac{r_h}{2} \le 1 \pm \sqrt{1-\frac{16 E^2}{ r_h^2 (3+ \nu^2) } }  \ri \, ,
 \label{mini}
\eeq
where the physical solution relevant for holographic complexity is the
one with the $+$ sign.  Conventionally,  $t_b=0$ corresponds to $E=0$ and $r_{\rm min}=r_h$.
The $t_b \rightarrow \infty$ limit, instead, corresponds to
coincident roots for $r_{\rm min}$ in eq.~(\ref{mini}), i.e.
 $E\rightarrow  \frac{ r_h}{4} \sqrt{\nu^2 + 3}$
 and $r_{\rm min}=  \frac{r_h}{2} $.
The minimal value of the radial coordinate is inside the black hole horizon 
$\frac{r_h}{2} \leq  r_{\rm min} \leq r_h$.

The volume can be written as an integral over tha radial coordinate
\beq 
V = 4 \pi l^2 \int \frac{dr}{\dot{r}}= 2 \pi l^2 \int_{r_{\rm min}}^{r_{\rm max}}  
\sqrt{ 
\frac
{ r \left(3 \left(\nu ^2-1\right) r+\left(\nu ^2+3\right) r_h\right)} 
{  4 E^2+\left(\nu ^2+3\right) r \left(r-r_h\right) }
} dr \, .
\eeq
Here we use a trick similar to the $\mathrm{AdS}$ case \cite{Carmi:2017jqz}.
We consider the difference of $w$ coordinates
\bea
\label{DELTAu}
& & w (r_{\rm max}) - w (r_{\rm min}) = \int_{r_{\rm min}}^{r_{\rm max}} dr \frac{\dot{w}}{\dot{r}} 
\nl
&=& \int_{r_{\rm min}}^{r_{\rm max}} dr \left[
\frac{2}{(\nu^2 +3)(r_h-r)} \le \frac{E}{ r }
\sqrt{ 
\frac
{ r \left(3 \left(\nu ^2-1\right) r+\left(\nu ^2+3\right) r_h\right)} 
{  4 E^2+  \left(\nu ^2+3\right) r \left(r-r_h\right) }}
  - \nu  \ri
\right] \, .
\eea
Note that this integral is not divergent for $r \rightarrow r_h$.
The volume can then be written as follows:
\bea
\frac{V}{4 \pi l^2} & = & E( w (r_{\rm max}) - w (r_{\rm min}) )
+ \int_{r_{\rm min}}^{r_{\rm max}} dr
\left\{
\frac{2 \nu E}{(\nu^2+3) (r_h-r)} \right.
\nl & &
\left.
-\frac{\sqrt{r \, 
[4 E^2 -  r (r_h-r)(\nu^2+3)] \,
[(\nu^2+3)r_h +3 r (\nu^2-1)]} }
{2 (\nu^2+3) r (r_h-r)}
\right\} \, .
\label{volume-u}
\eea
It is important to emphasize that
\beq
\lim_{r_{\rm max}\rightarrow \infty} w(r_{\rm max})- \tilde{r} (r_{\rm max})=t_R 
\eeq
is finite  and can be identified with the time on the right boundary.
In the limit $r_{\rm max} \rightarrow \infty$, we can use the explicit expression 
\beq
w(r_{\rm max}) - w (r_{\rm min})  = t_R + \tilde{r} (r_{\rm max}) - \tilde{r} (r_{\rm min}) \, ,
\label{DELTAu2}
\eeq
obtained specializing eq. (\ref{TUR}) with the values
\beq
w(r_{\rm max}) =t_R + \tilde{r} (r_{\rm max}) \, , \qquad
 w (r_{\rm min})=\tilde{r} (r_{\rm min}) \, .
\eeq
In fact, we have $t=0$ at $r=r_{\rm min}$ by symmetry considerations.

\begin{figure}[ht]  
\begin{center}
\includegraphics[scale=0.5]{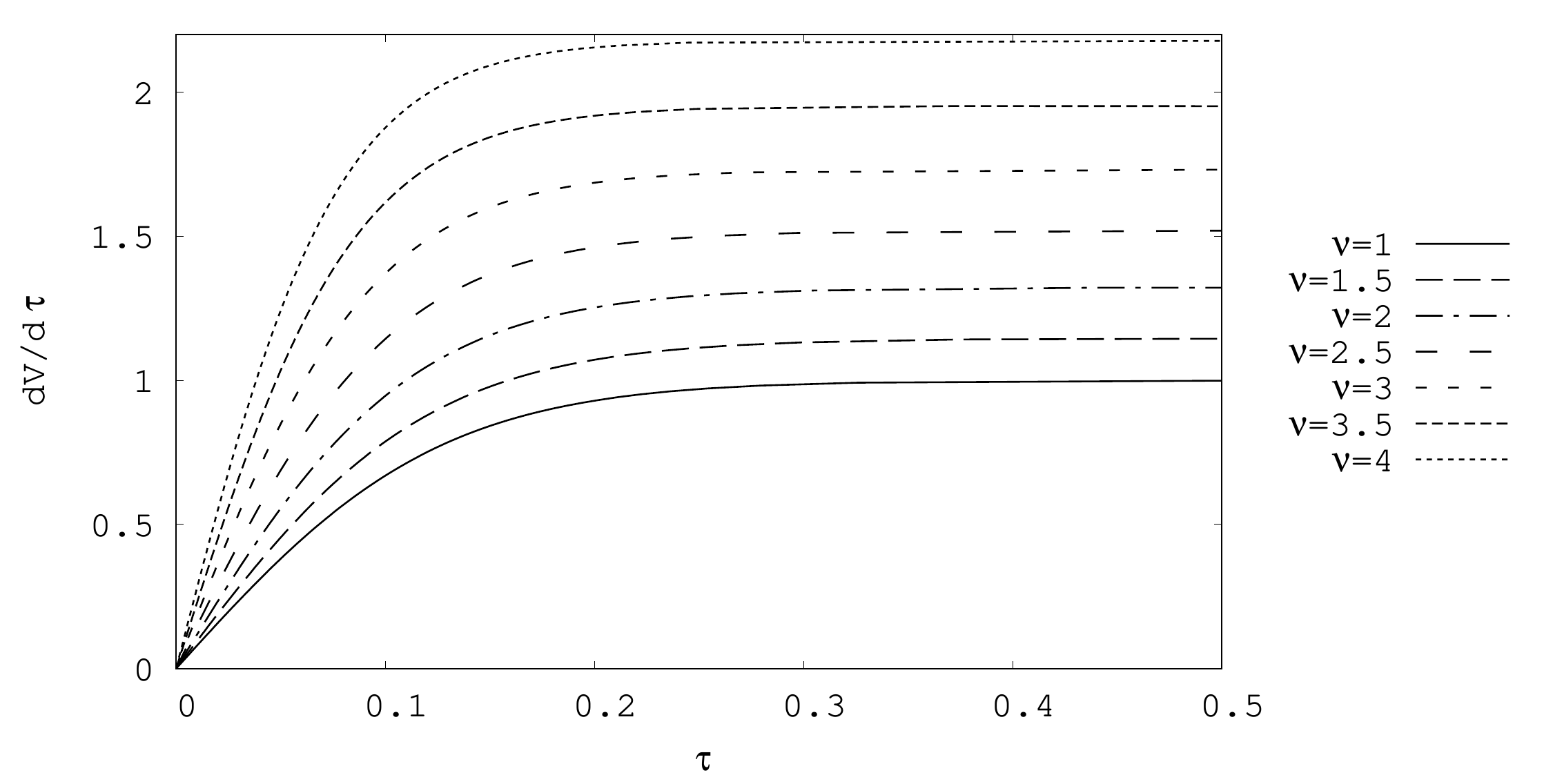}          
\end{center} 
\caption{Time dependence of $\frac{d V}{d \tau}$ in units of $ \pi l$,
for $r_h=1$ and various values of the warping parameter $\nu$.}
\label{figura1} 
\end{figure}

Taking into account that both $E$ and $r_{\rm min}$
depend on $t_R$ (see eq. (\ref{mini}) for the relation among $r_{\rm min}$ and $E$),
the time derivative of eq. (\ref{volume-u}) gives, 
after several cancellations among terms, the result
\beq
\frac{1}{2 l} \frac{d V}{d t_R}=\frac{d V}{d \tau} = 2 \pi l E \, ,
\eeq
where $\tau=l \, t_b=2 l \, t_R$.
At large $\tau$, $E$  approaches to the constant $E_0=\frac{ r_h}{4} \sqrt{\nu^2 + 3}$.
Computing the constant of motion \emph{E} in eq. (\ref{conserved quantity}) 
for the particular value $ r=r_{\rm min} $ shows that $ E>0$ for $\tau>0$ 
(corresponding to  $\dot{w}>0$) and $ E<0$ for $\tau<0$ (corresponding to $\dot{w}<0$).
 Numerical calculations with the full time dependence can be
 obtained by expressing $\tau$ in terms of $E$ using eqs. (\ref{DELTAu}-\ref{DELTAu2}),
  are shown in figure \ref{figura1}.
 For $\nu=1$ the results in \cite{Stanford:2014jda},
  \cite{Carmi:2017jqz} are recovered, under the change of variables
 in eq. (\ref{change-BTZ}).
 
\subsection{Computation of the Volume in the rotating case}
The procedure to compute the volume in the rotating case is similar to the non-rotating black hole, then we will be very schematic.
The volume functional using the metric in the coordinates (\ref{metricEF}) is written as
\beq
V= 4 \pi \int_{\l_{\rm min}}^{\l_{\rm max}} 
 d\lambda \, l^2 \sqrt{\frac{r \dot{w}^2}{4} \Psi
 - \le \frac{\dot{w}}{2} \le 2 \nu r - \sqrt{r_{+}r_{-}(\nu^2 +3)} \ri - \frac{\dot{r}}{2} \ri^2 } = 4 \pi \int d\lambda \, \mathcal{L} (r,\dot{r},\dot{w}) \, ,
\eeq
where we used the axial symmetry to put the codimension-one surface along the $\theta_w$ direction.
As before, there is a conserved quantity because the variable $w$ is cyclic
\beq
E = \frac{1}{l^2} \frac{\p \mathcal{L}}{\p \dot{w}} =  \frac{\frac{r \dot{w}}{4} \Psi
-\le \frac{\dot{w}}{2} \le 2 \nu r - \sqrt{r_{+}r_{-}(\nu^2 +3)} \ri - \frac{\dot{r}}{2} \ri \frac12  \le 2 \nu r - \sqrt{r_{+}r_{-}(\nu^2 +3)} \ri
}
{\sqrt{\frac{r \dot{w}^2}{4} \Psi
- \le \frac{\dot{w}}{2} \le 2 \nu r - \sqrt{r_{+}r_{-}(\nu^2 +3)} \ri - \frac{\dot{r}}{2} \ri^2 }} \, .
\eeq
The expression greatly simplifies choosing a parametrization for $ \lambda $ such that
$V= 4\pi l^2 \int d \lambda$, which corresponds to set
\beq
\frac{r \dot{w}^2}{4} \Psi
- \le \frac{\dot{w}}{2} \le 2 \nu r - \sqrt{r_{+}r_{-}(\nu^2 +3)} \ri - \frac{\dot{r}}{2} \ri^2  = 1 \, ,
\label{verde1}
\eeq
finding
\beq
E= -  \frac{\nu^2 +3}{4} \dot{w} (r-r_{-})(r-r_{+}) +  \frac{\dot{r}}{4} \le 2 \nu r - \sqrt{r_{+}r_{-}(\nu^2 +3)}  \ri \, .
\label{verde2}
\eeq
Solving eqs. (\ref{verde1}, \ref{verde2}), we obtain the inverse expressions 
\beq
\dot{r} =  2 \sqrt{\frac{4E^2 +  (\nu^2 +3)(r-r_{-})(r-r_{+})}{\le 2 \nu r - \sqrt{r_{+}r_{-}(\nu^2 +3)}  \ri^2 -  (\nu^2 +3)(r-r_{-})(r-r_{+})}} \, ,
\eeq
\bea
& \dot{w} &= \frac{2}{ (\nu^2 +3)(r-r_{-})(r-r_{+})} \nl 
& & \left[ \frac{\sqrt{4E^2 +  (\nu^2 +3)(r-r_{-})(r-r_{+})}
  \le 2 \nu r - \sqrt{r_{+}r_{-}(\nu^2 +3)}  \ri }{ \sqrt{\le 2 \nu r - \sqrt{r_{+}r_{-}(\nu^2 +3)}  \ri^2 -  (\nu^2 +3)(r-r_{-})(r-r_{+})}} 
 - 2 E \right] \, ,
\eea
which will be used later to conveniently express the volume functional in terms of the conserved quantity.
The minimum value $r_{\rm min}$  of the radial coordinate is obtained by solving $\dot{r}=0$:
\beq
r_{\rm min} = \frac{r_{+}+r_{-}}{2} \le 1 \pm \sqrt{1 - \frac{16E^2}{ (\nu^2 +3)(r_{+}+r_{-})^2}} \ri \, .
\eeq
As in the non-rotating case, the physical solution relevant for holographic complexity is the
one with the $+$ sign.  Conventionally, $ t_b=0 $ corresponds to $ E=0 $ and $ r_{\rm min}=r_{+}+r_{-}$. 
The $ t_b \rightarrow \infty $ limit corresponds to $ E \rightarrow  \frac{(r_{+}-r_{-})}{4} \sqrt{\nu^2 +3} $
 and $ r_{\rm min}=\frac{r_{+}+r_{-}}{2}$. 

\begin{figure}[ht]  
\begin{center}
\includegraphics[scale=0.5]{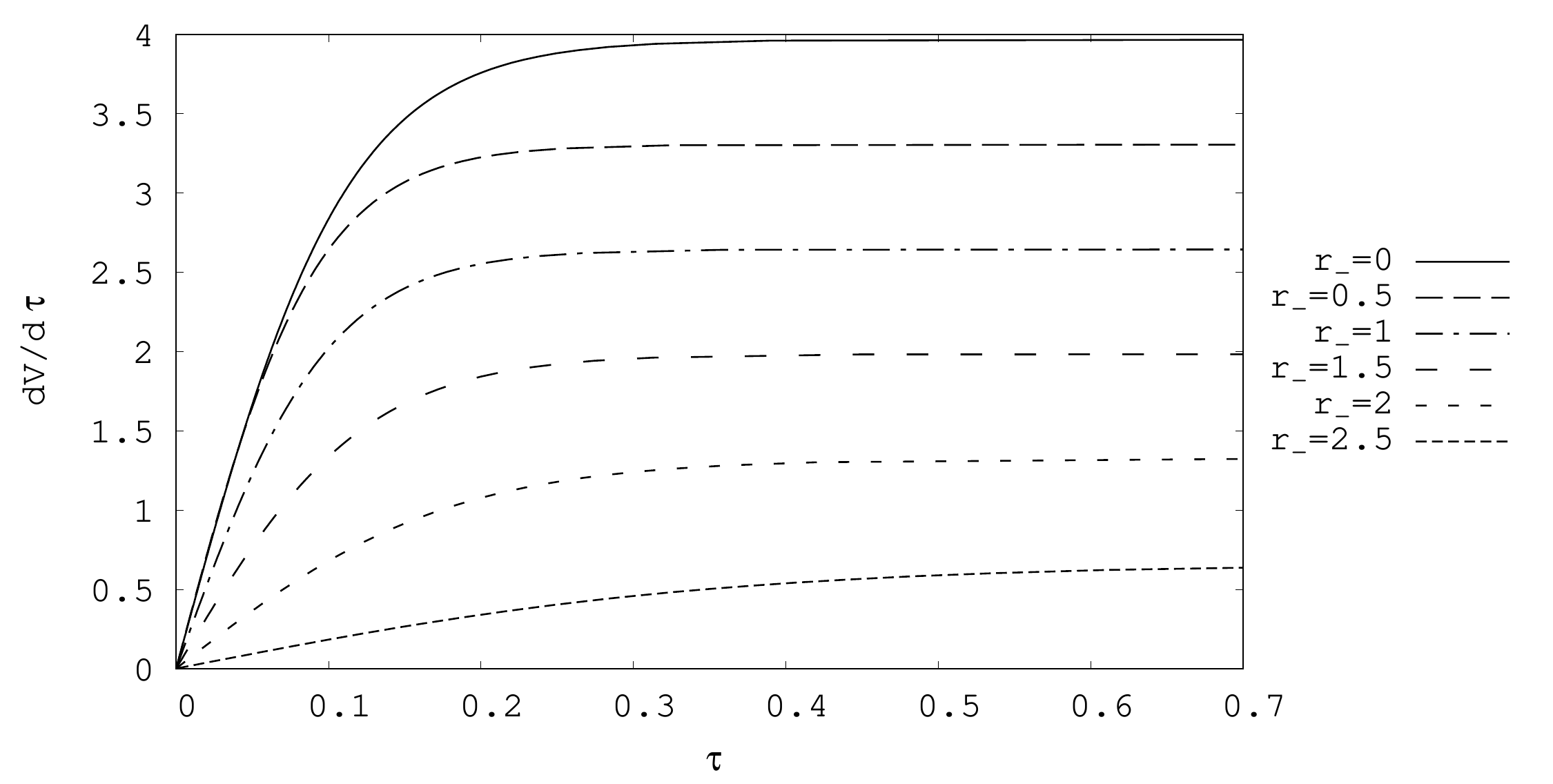}              
\end{center} 
\caption{Time dependence of $\frac{d V}{d \tau}$ in units of $ \pi l$, for $r_+=3$, $\nu=2$ and several values of $r_-$.
For other values of $\nu$ the plots are qualitatively similar.}
\label{figura2} 
\end{figure}

The volume expressed as an integral over the radial coordinate becomes
\beq
V= 2 \pi l^2 \int_{r_{\rm min}}^{r_{\rm max}} dr \sqrt{\frac{\le 2 \nu r - \sqrt{r_{+}r_{-}(\nu^2 +3)}  \ri^2 -  (\nu^2 +3)(r-r_{-})(r-r_{+})}{4E^2 + (\nu^2 +3)(r-r_{-})(r-r_{+})}} \, ,
\eeq
while the difference between the extermal values of the null coordinates is
\bea
&& w (r_{\rm max}) - w(r_{\rm min}) = \int_{r_{\rm min}}^{r_{\rm max}} dr \, 
\frac{1}{(\nu^2 +3)(r-r_{-})(r-r_{+})} \Biggl[\le 2 \nu r - \sqrt{r_{+}r_{-}(\nu^2 +3)}  \ri \Biggr.
\nl
&&
\Biggl.
- 2 E \sqrt{\frac{\le 2 \nu r - \sqrt{r_{+}r_{-}(\nu^2 +3)}  \ri^2 -
  (\nu^2 +3)(r-r_{-})(r-r_{+})}{4E^2 +  (\nu^2 +3)(r-r_{-})(r-r_{+})}} \,\,\, 
  \Biggr]
   \, .
\eea
As in the non-rotating case, the symmetry of the configuration sets $t=0$ at $r=r_{\rm min}$, giving the simple result
\beq
w (r_{\rm max}) - w(r_{\rm min}) = t_R + \tilde{r} (r_{\rm max}) - \tilde{r} (r_{\rm min}) \, .
\eeq
This will be used again to find the time derivative of the volume in terms of the conserved quantity along the surface.
In order to do that, we consider the relation (obtained by direct computation)
\bea
\frac{V}{4 \pi l^2}  & = & \int_{r_{\rm min}}^{r_{\rm max}} dr \, \left[ \frac{\sqrt{4E^2 +  (\nu^2 +3)(r-r_{-})(r-r_{+})}}
{2 (\nu^2 +3)(r-r_{-})(r-r_{+})}    \right.
\nl
& & \sqrt{ \le 2 \nu r - \sqrt{r_{+}r_{-}(\nu^2 +3)}  \ri^2 -  (\nu^2 +3)(r-r_{-})(r-r_{+})}
\nl & &
\left. - E \frac{2\nu r - \sqrt{r_{+}r_{-} (\nu^2 +3)}}{(\nu^2 +3)(r-r_{-})(r-r_{+})} \right] + E (w (r_{\rm max})- w(r_{\rm min})) \, .
\eea
Using the previous definitions and simplifying the expression, we formally obtain the same result of the non-rotating case
\beq
\frac{dV}{d \tau} = 2 \pi l E \, ,
\label{rateV}
\eeq
where $\tau=l \, t_b=2 l \, t_R$. 
At large $\tau$, $E$ approaches the constant value
\beq
E_0= \frac{(r_{+}-r_{-})}{4} \sqrt{\nu^2 +3} \, .
\eeq
Numerical calculation are shown in figure  \ref{figura2}.
As a consistency check, putting $ \nu=1 $ for the BTZ case, we obtain
\beq
\lim_{\tau \rightarrow \infty} \frac{dV}{d \tau} = \pi l (r_+ - r_-) \, ,
\eeq
which is the same result found in standard coordinates on the Poincar\'e
 patch when we perform the change of variables (\ref{change-BTZ}).

The late time limit of the maximal volume slices
 can be found also in a simpler way, as in \cite{Stanford:2014jda}.
In this limit, we expect that the maximal volume slice sits at constant $r$,
due to translation invariance in time\footnote{Besides time translation invariance, the configuration is also symmetric with respect to the angular direction, which means that the only coordinate remaining is the radial direction.}. We can then consider volume
slices at a constant $r=\hat{r}$. Extremizing the volume from the metric in eq. (\ref{BHole}),
we find that the only possible maximal constant-$r$ slice sits at
\beq
\hat{r}= \frac{r_{+}+r_{-}}{2} \, .
\eeq
Inserting this value back in the volume functional, we recover eq. (\ref{rateV})
with $E=E_0$.

The late time result can be written in terms of the Bekenstein-Hawking entropy and the Hawking temperature by observing that for the WAdS BH solutions in Einstein gravity the following identity is true
\beq
T S = \frac{ (r_+ -r_-)(3+\nu^2)}{16 G} \, .
\eeq
In this way the late time volume growth is
\beq
\lim_{\tau \rightarrow \infty} \frac{d V}{d \tau} = \frac{\pi l}{2}
(r_+ -r_-)  \sqrt{3+\nu^2}  = T S \frac{8 \pi G l }{\sqrt{3+\nu^2}} \, .
\label{late time volume rate}
\eeq


\section{Complexity=Action}
\label{sect-Complexity=Action}

The action of the WDW patch has several contributions, which can be
summarized as \cite{Lehner:2016vdi}
\beq
I = I_{\mathcal{V}} + I_{\mathcal{B}} + I_{\mathcal{J}} + I_{\rm ct} \, .
\label{composition of terms of the gravitational action}
\eeq
In this formula $ I_{\mathcal{V}} $ refers to the Einstein-Hilbert action in the bulk, $ I_{\mathcal{B}} $ to the codimension-one boundaries (timelike, spacelike or null) and $ I_{\mathcal{J}} $ to the codimension-two joints coming from intersections of other boundaries.
The contribution $ I_{\rm ct} $ is a counterterm to be added in order to ensure
the reparameterization invariance  of the action, which cancels the ambiguities
in the action arising from the normalization of null normals.

The bulk action integrand $\sqrt{g} \mathcal{L}$ in 
eq. (\ref{bulk action}) evaluated on the background defined in eqs. (\ref{BHole}) and (\ref{ansatz per campo di gauge})
 is constant and independent from the parameters $(r_+,r_-)$:
\beq
I_{\mathcal{V}} =\int dr dt d \theta \frac{\mathcal{I} }{16 \pi G} \, , \qquad
\mathcal{I} = 
- \frac{l}{2} (\nu^2 +3) + \frac{\kappa c^2}{l} - \alpha a c \, .
\eeq
In particular, the quantity $\mathcal{I}$ is the same introduced in eq. (\ref{densita}) , but without using the gauge choice $ a=
\frac{l}{\nu} \sqrt{\frac32} \sqrt{\nu^2-1}  $ which makes the conserved charges of the black hole well-defined.
We will perform this further simplification later.

The boundary terms contain two kind of contributions
\beq
I_{\mathcal{B}} = I_{\rm GHY} + I_{\mathcal{N}} \, ,
\eeq
where $I_{\rm GHY}$ refers to spacelike or timelike boundaries
(Gibbons-Hawking-York (GHY) term), while
 $ I_{\mathcal{N}} $ is the contribution for null boundaries.
The GHY term is given by
\beq
I_{\rm GHY} = \frac{  \varepsilon }{8 \pi G}  \int_{\mathcal{B}} d^2 x \, \sqrt{|h|} \, K \, ,
\label{GHY general}
\eeq
where $ \mathcal{B} $ is the appropriate boundary, \emph{h} the induced metric, \emph{K} the extrinsic curvature
and $ \varepsilon$ is equal to $+1$ if the boundary is timelike and $-1$ if it is spacelike.
It is well-known that the GHY term must be added to the Einstein-Hilber action in order to make the variational problem well-defined.

The new term appearing in the treatment of the action in the WDW patch refers to the null surface boundaries and is given by  \cite{Neiman:2012fx, Parattu:2015gga, Lehner:2016vdi}
\beq
I_{\mathcal{N}}= \frac{  1 }{8 \pi G}  \int_{\mathcal{B}} \tilde{\kappa} d \lambda d S \, ,
\eeq
where  $\lambda$ parameterizes the null direction of the surface,  $d S$
is the area element of the spatial cross-section orthogonal to the null direction
and $\tilde{\kappa}$ is the acceleration measuring the failure of $\lambda$ to be an affine parameter:
if we denote by $k^\a$ the null generator, $\tilde{\kappa}$  is defined by the relation:
$k^\mu D_\mu k^\a = \tilde{\kappa} k^\a$. 
 It turns out that the contribution to the action
$I_{\mathcal{N}}$ is not reparameterization-invariant \cite{Parattu:2015gga,Lehner:2016vdi}
 and it can be set to zero using an
affine parameterization for the null direction of the boundary
\cite{Lehner:2016vdi}. 

In the case of joints between spacelike and timelike surfaces, this contribution
was studied in \cite{Hayward:1993my}. 
The analysis for joints between a null surface and either a timelike, spacelike or another null surface were recently studied in
\cite{Lehner:2016vdi}. 
In the CA calculations done in the next sections, we will use these null joints
contributions several times:
\beq
I_{\mathcal{J}} = \frac{1}{8 \pi G} \int_{\Sigma} d \theta \sqrt{\sigma} \, \mathfrak{a} \, ,
\label{jjoints}
\eeq
where $ \sigma_{ab} $ is the induced metric over the joint (in this case, it is 1-dimensional)
and $ \mathfrak{a} $  depends on the kind of joint, but it is chosen in such a way that the action is additive when inserting new boundaries inside a given region of spacetime.
Let us denote with $k^\a$ the future directed null normal to a null surface
(which is also tangent to the surface), $n_\a$ the normal to a spacelike surface 
and $s_\a$ the normal to a timelike surface, both directed outwards the volume of interest. 
In the case of the intersection between two null surfaces with normals  $k^\a_1$ and $k^\a_2$ the integrand is given by
\beq
\mathfrak{a} = \eta  \log \left| \frac{k_1 \cdot k_2}{2} \right| \, ,
\label{jnn}
\eeq
while in the case of intersection of a null surface with normal $k^\a$
and a spacelike surface with normal $n_\a$ (or a timelike surface with normal $s_\a$):
\beq
\mathfrak{a} = \eta  \log \left|  k \cdot n \right| \, , \qquad
\mathfrak{a} = \eta \log \left|  k \cdot s \right| \, .
\label{jns}
\eeq
In eqs. (\ref{jnn}-\ref{jns}) we set $ \eta= + 1 $ 
if the joint lies in past of the spacetime volume of interest, 
and $ \eta=-1 $ if the joint lies in the future of the relevant region.
Note that  eqs.  (\ref{jnn}) and (\ref{jns})  are ambiguous because of 
the normalization of the null normal $k^\a .$ 
This ambiguity is related to the boundary term on the null surfaces
and does not affect the late-time limit of the complexity, but just
the finite-time behavior\footnote{These ambiguities could be related to various ambiguities 
of the dual circuit complexity of the quantum state,
such as the choice of the reference state,
the specific set of elementary gates and the amount of tolerance 
that one introduces to describe the accuracy with which the final state should
be constructed. }.
As discussed in \cite{Carmi:2017jqz},
we will partially fix this ambiguity by requiring that the null vector $k^\mu$
have constant scalar product with the boundary time killing vector $\p/\p t$.

The following counterterm  \cite{Lehner:2016vdi} must be added to the 
boundary term of null boundaries, in order to make the action 
invariant under reparameterization
\beq
I_{\rm ct}=\frac{1}{8\pi G}\int d\theta \, d\lambda \, \sqrt{\sigma} \, \Theta\log{|\tilde{L}\Theta|} \, ,
\label{counterterm}
\eeq
where  $\lambda$ is the affine parameter of the null geodesics which delimit the boundary, and
\beq
\Theta = D_{\alpha} k^{\alpha} =\frac{1}{\sqrt{\sigma}}\frac{\partial\sqrt{\sigma}}{\partial \lambda} 
\eeq
is the expansion of the congruence of null geodesics on the hypersurface.
The parameter $\tilde{L}$ appearing in eq. (\ref{counterterm}) is an arbitrary length scale
which is needed for dimensional reasons, whose physical meaning is so far obscure.

\subsection{Computation of the action in the non-rotating case}

\begin{figure}[h]
\begin{center}
\includegraphics[scale=0.7]{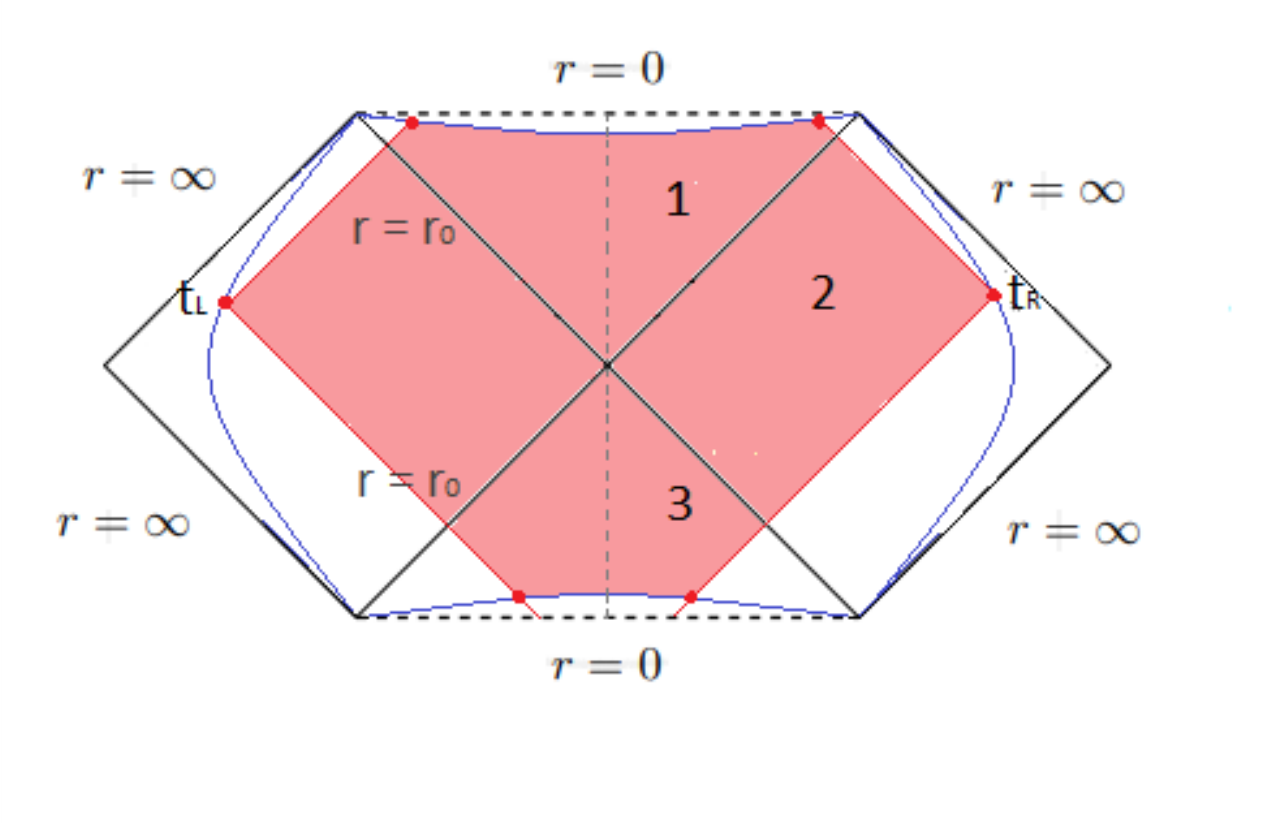}
\end{center}
\caption{Penrose diagram  for the non-rotating BH, with the WDW patch  for $t_b<t_C $. In this picture we called the horizon radius $r_0 .$}
\label{fig1}
\end{figure}

The Penrose diagrams with the WDW patch associated to the boundary region for the non-rotating case are shown in figures
\ref{fig1} and \ref{fig2}. 
We choose without loss of generality the symmetric configuration $t_L = t_R = \frac{t_b}{2} .$

A difference between the volume and action conjectures is that the extremal surface goes very deep inside the horizon but stays away from curvature singularities, while the WDW patch reaches the singularities of the black hole.
For this reason, we need both a IR cutoff $\varepsilon_0$ and a UV cutoff $\Lambda.$
We will see that the time derivative of the action is independent from both of them, and moreover the curvature singularities do not give any problem or divergence during the computation.

The structure of the WDW patch in the non-rotating case changes with time:
at early times it looks like in figure \ref{fig1}, while at late times like in
figure \ref{fig2}. In particular, there exists a critical time $ t_C $
 such that the bottom vertex of the patch touches the past singularity.
This is given by
\beq
t_C = 2(r^*_{\Lambda} - r^* (0) )\, ,
\eeq
where $r^*_{\Lambda} \equiv r^*(\Lambda)$.
We will separate the calculation of the action in  two cases.
At the end we will express the results
in terms of 
\beq
\tau=l (t_b - t_C) \, ,
\eeq 
which is the boundary time rescaled with the $\mathrm{AdS}$ radius $l$
for dimensional purposes and with the origin translated 
at the critical time $t_C$.

\subsubsection{Initial times $ t_b<t_C $}

{\bf{ Bulk contributions:}}
We decompose the WDW patch into three regions and we use the symmetry of the configuration to write the bulk action as
\beq
I_{\mathcal{V}}  = 2 \le I^{1}_{\mathcal{V}} + I^{2}_{\mathcal{V}}  + I^{3}_{\mathcal{V}}  \ri \, ,
\eeq
where 
\bea
I^{1}_{\mathcal{V}}  &=& \frac{\mathcal{I}}{16 \pi G} \int_0^{2\pi} d \theta \int_{\varepsilon_0}^{r_h} dr  \int_0^{v - r^* (r)} dt = \frac{\mathcal{I}}{8 G} \int_{\varepsilon_0}^{r_h} dr \le \frac{t_b}{2} + r^*_{\Lambda} - r^* (r) \ri \, ,
\nl 
I^{2}_{\mathcal{V}}  &=& \frac{\mathcal{I}}{16 \pi G} \int_0^{2\pi} d \theta \int_{r_h}^{\Lambda} dr  \int_{u + r^* (r)}^{v - r^* (r)} dt = \frac{ \mathcal{I}}{4 G} \int_{r_h}^{\Lambda} dr \le  r^*_{\Lambda} - r^* (r) \ri \, ,
\label{bulk action CA conjecture} \\
I^{3}_{\mathcal{V}}  &=&   \frac{\mathcal{I}}{16 \pi G} \int_0^{2\pi} d \theta \int_{\varepsilon_0}^{r_h} dr  \int_{u + r^* (r)}^0 dt = \frac{ \mathcal{I}}{8  G} \int_{\varepsilon_0}^{r_h} dr \le - \frac{t_b}{2} + r^*_{\Lambda} - r^* (r) \ri \, .
\nonumber
\eea
Summing all the contributions, we get the result
\beq
I_{\mathcal{V}}  =
\frac{\mathcal{I}}{2 G} \int_{\varepsilon_0}^{\Lambda} dr \le  r^*_{\Lambda} - r^* (r) \ri \equiv
I^0_{\mathcal{V}}  \, .
\eeq
This contribution is time-independent.

{\bf GHY surface  contributions:}
The constant $r$ surface, inside the horizon, is a spacelike surface 
whose induced metric in the $x^i=(t,\theta)$ coordinates reads: 
\beq
h_{ij} = l^2 \begin{pmatrix}
1 & \nu r \\
\nu r & \frac{r}{4} \Psi(r)
\end{pmatrix} \, , \qquad
\sqrt{h} = \frac{l^2}{2} \sqrt{(\nu^2 +3)r(r_h-r)} \, .
\eeq
The normal vector to these slices is
\beq
n^{\mu} = \le 0 \, , - \frac{1}{l} \sqrt{(\nu^2 +3)r(r_h-r)} \, , 0 \ri \, ,
\qquad n^\a n_\a = -1 \, ,
\label{normal-constant-r}
\eeq
and the extrinsic curvature is
\beq
K = \frac{1}{2 l} \sqrt{\nu^2 +3} \frac{2r - r_h}{\sqrt{r(r_h-r)} } \, .
\eeq
In the GHY term we then set $\varepsilon=-1$ because the surface is spacelike.
We now have all the ingredients to compute the two contributions to the GHY term coming from the regions near the past and future singularities:
\beq
I^1_{\rm GHY} =
- \frac{ (\nu^2 +3) l}{16  G} \left[ \le 2r - r_h \ri \le \frac{t_b}{2} +r^*_{\Lambda} - r^*(r) \ri  \right]_{r=\varepsilon_0} \, ,
\eeq
\beq
I^2_{\rm GHY}  =
- \frac{ (\nu^2 +3) l}{16 G} \left[ \le 2r - r_h \ri \le - \frac{t_b}{2} +r^*_{\Lambda} - r^*(r) \ri  \right]_{r=\varepsilon_0} \, .
\eeq
Consequently, the total GHY contribution is
\beq
I_{\rm GHY} = 2 \le I^1_{\rm GHY}+ I^2_{\rm GHY} \ri =
- \frac{ (\nu^2 +3) l}{4 G} \left[ \le 2r - r_h \ri \le  r^*_{\Lambda} - r^*(r) \ri  \right]_{r=\varepsilon_0} \equiv I^0_{\rm GHY} \, ,
\eeq
which is time-independent.

{\bf Joint  contributions:}
There are four joints between null and 
spacelike surfaces at $r=\varepsilon_0$ (nearby  the future and past  singularities)
and two joints at $r=\Lambda$.
The normal to the constant $r$ spacelike surfaces is 
$n^\a$ given by eq. (\ref{normal-constant-r}), while the normal
to the lightlike surfaces are $u^\a$, $v^\a$ from eq. (\ref{null-normals}).
From eq. (\ref{jns}), the four joint contributions nearby the singularities vanish in the limit $\varepsilon_0 \rightarrow 0,$
while the two joint contributions nearby the UV cutoff are time-independent
(see eq. \ref{jnn}).

{\bf Total:}
Summing all the terms coming from the bulk, the boundary and the joint contributions, we find
that the action of the WDW patch is time-independent.

\subsubsection{Later times $ t_b> t_C $}

\begin{figure}[h]
\begin{center}
\includegraphics[scale=0.7]{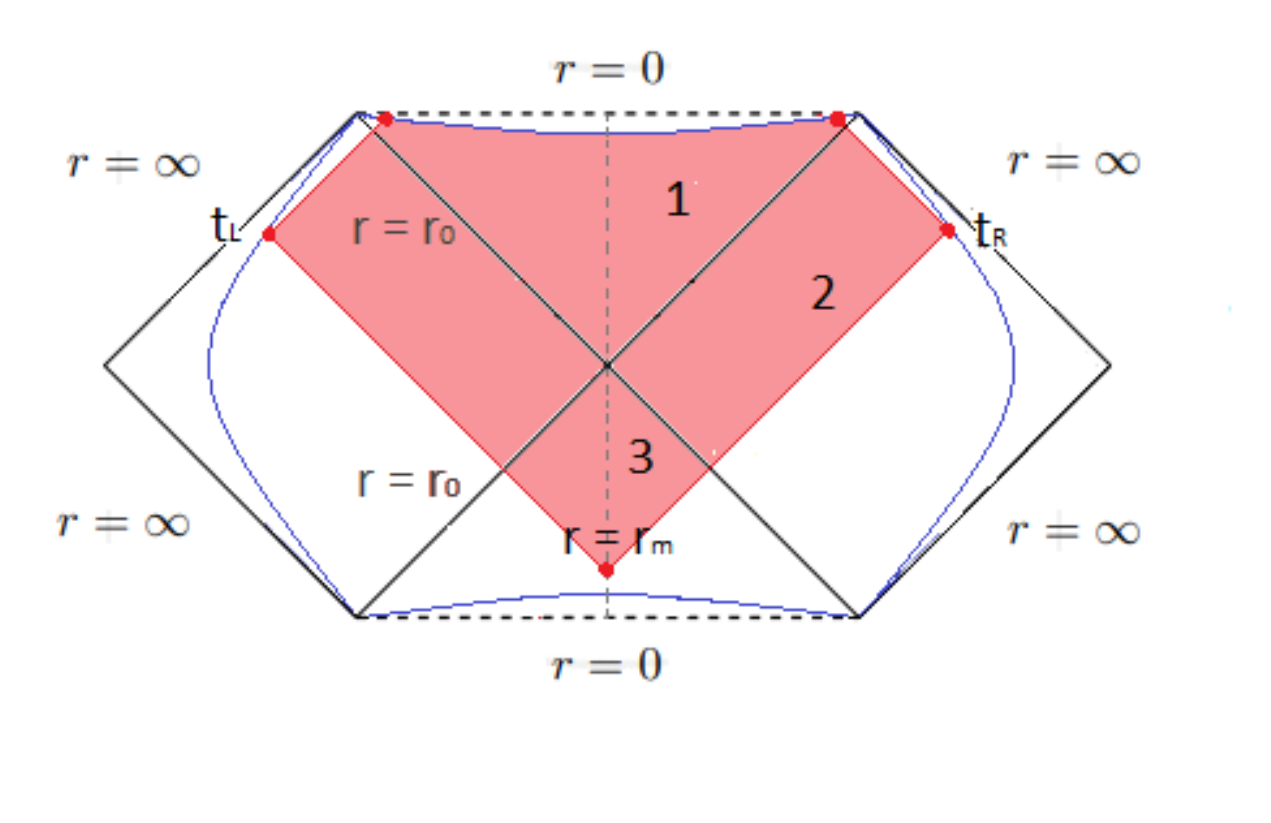}
\end{center}
\caption{Penrose diagram  for the non-rotating BH, with the WDW patch  for $t_b>t_C $. In this picture we called the horizon radius $r_0 .$}
\label{fig2}
\end{figure}

After the critical time $ t_C , $ the WDW patch moves and the 
lower vertex of the diagram does not reach the past singularity (see figure \ref{fig2}).
This vertex is defined via the relation
\beq
\frac{t_b}{2} - r^*_{\Lambda} + r^* (r_m ) = 0 \, .
\label{definizione vertice basso}
\eeq
The evaluation of the null joint contributions will require the computation of the time derivative of the tortoise coordinate,
which is done by differentiating eq. (\ref{definizione vertice basso}):
\beq
\frac{d r_m}{dt_b} =
-\frac12 
\le \frac{dr^*(r_m)}{d r_m} \ri^{-1} \, .
\label{time-derivative}
\eeq

{\bf Bulk contributions:}
The bulk action is the same of the case $ t_b <t_C , $ apart from the last contribution which becomes
\beq
I^{3}_{\mathcal{V}}  (t_b >t_C) =
   \frac{\mathcal{I}}{16 \pi G} \int_0^{2\pi} d \theta \int_{r_m}^{r_h} dr  \int_{u + r^* (r)}^0 dt 
   = \frac{ \mathcal{I}}{8 G} \int_{r_m}^{r_h} dr \le - \frac{t_b}{2} + r^*_{\Lambda} - r^* (r) \ri \, .
\eeq
We can re-write this contribution in the following way:
\beq
 I^{3}_{\mathcal{V}}  (t_b >t_C)  = I^{3}_{\mathcal{V}}  (t_b <t_C) +
\frac{ \mathcal{I}}{8 G} \int^{r_m}_{\varepsilon_0} dr \le  \frac{t_b}{2} - r^*_{\Lambda} + r^* (r) \ri \, .
\eeq
Since the other contributions to the bulk action are unchanged, the total result is
\beq
I_{\mathcal{V}}  (t_b >t_C)  =
I^0_{\mathcal{V}}  + \frac{ \mathcal{I}}{4 G} \int^{r_m}_{\varepsilon_0} dr \le  \frac{t_b}{2} - r^*_{\Lambda} + r^* (r) \ri \, ,
\eeq
the first term being time-independent.
Then the time derivative of the bulk action is
\beq
\frac{dI_{\mathcal{V}} }{d t_b} (t_b >t_C)  =
 \frac{ \mathcal{I}}{8 G} \, r_m =
 \frac{1}{8 G} \left[ - \frac{l}{2} (\nu^2 +3) + \frac{\kappa c^2}{l} - \alpha a c \right] r_m \, ,
 \label{bulk-late-time-non-rot}
\eeq
where the defining relation (\ref{definizione vertice basso}) is used in order to obtain a vanishing
contribution from the upper integration extreme.

{\bf GHY surface  contributions:}
After the critical time $  t_C $ we only have a contribution from the future singularity,
 because the lower part of the WDW patch does not reach the past singularity.
We are only left with
\beq
I_{\rm GHY}  = 2 I^1_{\rm GHY}  =
- \frac{(\nu^2 +3) l}{8 G} \left[ \le 2r - r_h \ri \le \frac{t_b}{2} +r^*_{\Lambda} - r^*(r) \ri  \right]_{r=\varepsilon_0} \, ,
\eeq
which is time-dependent.
The time derivative of this term gives
\beq
\lim_{\varepsilon_0 \rightarrow 0} \frac{dI_{\rm GHY}}{d t_b} (t_b >t_C)  =
\frac{(\nu^2 +3)l}{16 G} \, r_h \, .
\label{GHY-late-time}
\eeq

{\bf Joint  contributions:}
Following the same procedure of the case $ t_b <t_C$,  we find that the null joints 
at the UV cutoff give time-independent contributions, 
while the joint at the future singularity gives a vanishing result.
The contribution from the remaining  null-null  joint between $u^\a$ and  $v^\a$
at $ r=r_m$ is instead time-dependent, because $r_m$ is function of time (see eq. (\ref{time-derivative})).
We find that this contribution to the action is given by
eq. (\ref{jjoints}), with $\mathfrak{a}$ defined in eq. (\ref{jnn}):
\beq
\mathfrak{a}= \log \left| A^2 \frac{ u^\a v_\a}{2}
 \right| = \log \left| A^2 \frac{1}{l^2} \frac{ \Psi(r) }{(\nu^2 +3) (r-r_h)} \right| \, .
 \label{null-joint}
\eeq
The normalization factor $A^2$ corresponds to an ambiguity
in the contribution to the action due to the null joint \cite{Lehner:2016vdi},
because the normalization of the two null normals $u^\a$ and  $v^\a$
which delimitate the WDW patch is in principle not fixed by the metric. 
The action contribution from eq. (\ref{null-joint}), evaluated for $ r=r_m$, gives
\beq
I_{\mathcal{J}} = - \frac{ l}{4 G} \sqrt{\frac{r_m}{4}
\Psi(r_m) } 
\log \left| \frac{l^2}{A^2} \frac{(\nu^2 +3)(r_m-r_h)}
{\Psi(r_m)}  \right| \, ,
\eeq
whose time derivatives is
\beq
\begin{aligned}
 \frac{dI_{\mathcal{J}}}{d t_b}  = & - \frac{ l}{16 G}   \frac{d r_m}{dt_b}  
\frac{6(\nu^2 -1) r_m+ (\nu^2 +3) r_h }{\sqrt{ r_m \left[ 3 (\nu^2 -1) r_m + (\nu^2 +3) r_h \right]}}
\log \left| \frac{l^2}{A^2} \frac{(\nu^2 +3)(r_m-r_h)}{\Psi(r_m)}  \right|
+ \\
& - \frac{ l}{8 G}  \frac{d r_m}{dt_b}   
\frac{4 \nu^2  r_h \sqrt{r_m \left[ 3 (\nu^2 -1) r_m + (\nu^2 +3) r_h \right] }}{(r_m-r_h)\le 3 r_m (\nu^2 -1) + (\nu^2 +3) r_h \ri}  \, .
\end{aligned}
\eeq
Inserting eq. (\ref{time-derivative}) we obtain a further simplification:
\beq
\begin{aligned}
 \frac{dI_{\mathcal{J}}}{d t_b}  = &  \frac{l}{32 G}  
\frac{(\nu^2 +3)(r_m-r_h) \le 6(\nu^2 -1) r_m + (\nu^2 +3) r_h \ri}{ 3 (\nu^2 -1) r_m + (\nu^2 +3) r_h }
\log \left| \frac{l^2}{A^2} \frac{(\nu^2 +3)(r_m-r_h)}{\Psi(r_m)}  \right|
+ \\
& + \frac{l}{16 G}  \frac{4 \nu^2 (\nu^2 +3) r_m r_h }{ 3 r_m (\nu^2 -1) + (\nu^2 +3) r_h}  \, .
\end{aligned}
\eeq

{\bf Total:}
The total time derivative of the action is finally given by
\beq
\begin{aligned}
\frac{d I}{d t_b} & = \frac{1}{8 G} \left[ - \frac{l}{2} (\nu^2 +3) + \frac{\kappa c^2}{l} 
- \a a c \right] r_m + \frac{(\nu^2 +3)l}{16 G} \, r_h  + \frac{l}{16 G}
  \frac{4 \nu^2 (\nu^2 +3) r_m r_h }{ 3 r_m (\nu^2 -1) + (\nu^2 +3) r_h}  \\
& + \frac{l}{32 G}  
\frac{(\nu^2 +3)(r_m-r_h) \le 6(\nu^2 -1) r_m + (\nu^2 +3) r_h \ri}{ 3 (\nu^2 -1) r_m + (\nu^2 +3) r_h }
\log \left| \frac{l^2}{A^2} \frac{(\nu^2 +3)(r_m-r_h)}{\Psi(r_m)}  \right| \, .
\end{aligned}
\label{gran-totale1}
\eeq

We can now perform the late time limit of the previous rate. 
In this limit $ r_m \rightarrow r_h , $ which implies that the term in the second line vanishes and we find:
\beq
\lim_{t_b \rightarrow \infty} \frac{d I}{d t_b} 
= \frac{(\nu^2 +3)l}{16 G} \, r_h + \frac{1}{8 G} \le \frac{\kappa}{l} c^2 - \alpha  a c \ri r_h  \, .
\eeq
Note that the general result (\ref{gran-totale1}) depends on $A^2$, 
while its late time limit does not.
Using the value of $a$ given in eq. (\ref{a-action}),
we can now evaluate the combination appearing in the rate of the action
\beq
\frac{\kappa}{l} c^2 - \alpha a c = 0 \, ,
\eeq
finding
\beq
\lim_{t_b \rightarrow \infty} \frac{1}{l} \frac{d I}{d t_b} 
= \lim_{\tau \rightarrow \infty} \frac{d I}{d \tau} = \frac{\nu^2 +3}{16 G} \, r_h = M = TS \, .
\label{non-rotating late time action growth}
\eeq
This late-time results can also be recovered using the approach
by  \cite{Brown:2015lvg} (see Appendix \ref{another-way} for details).

\begin{figure}[h]
\begin{center}
\includegraphics[scale=0.5]{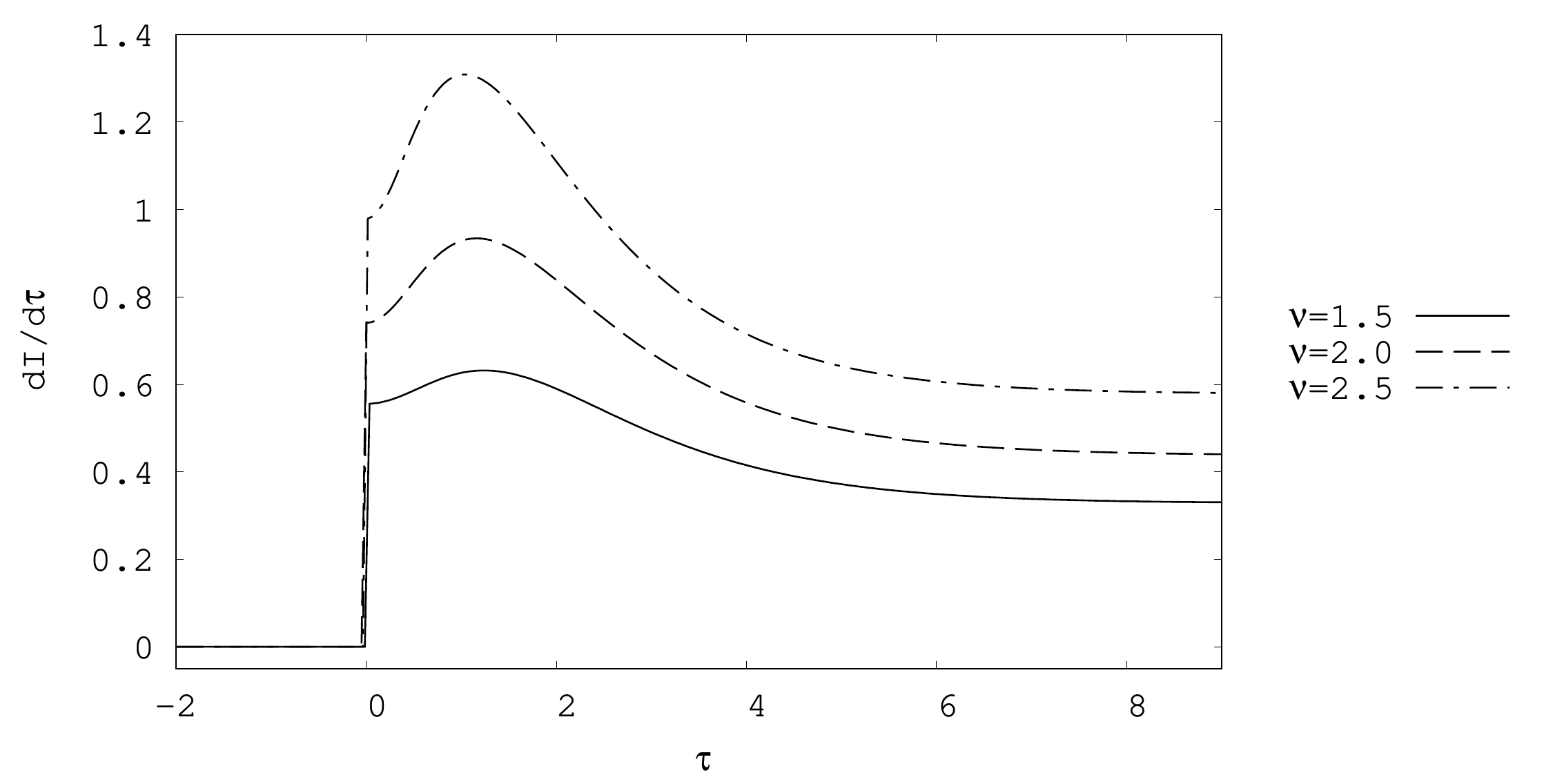}
\end{center}
\caption{Time dependence of the WDW action in the non-rotating case
for different values of $\nu$. We set $G=1$, $l=1$, $r_h=1$ 
and $A=2$. The critical time $t_C$ corresponds to $\tau=0$. }
\label{CA-time-dep1}
\end{figure}
\begin{figure}[h]
\begin{center}
\includegraphics[scale=0.5]{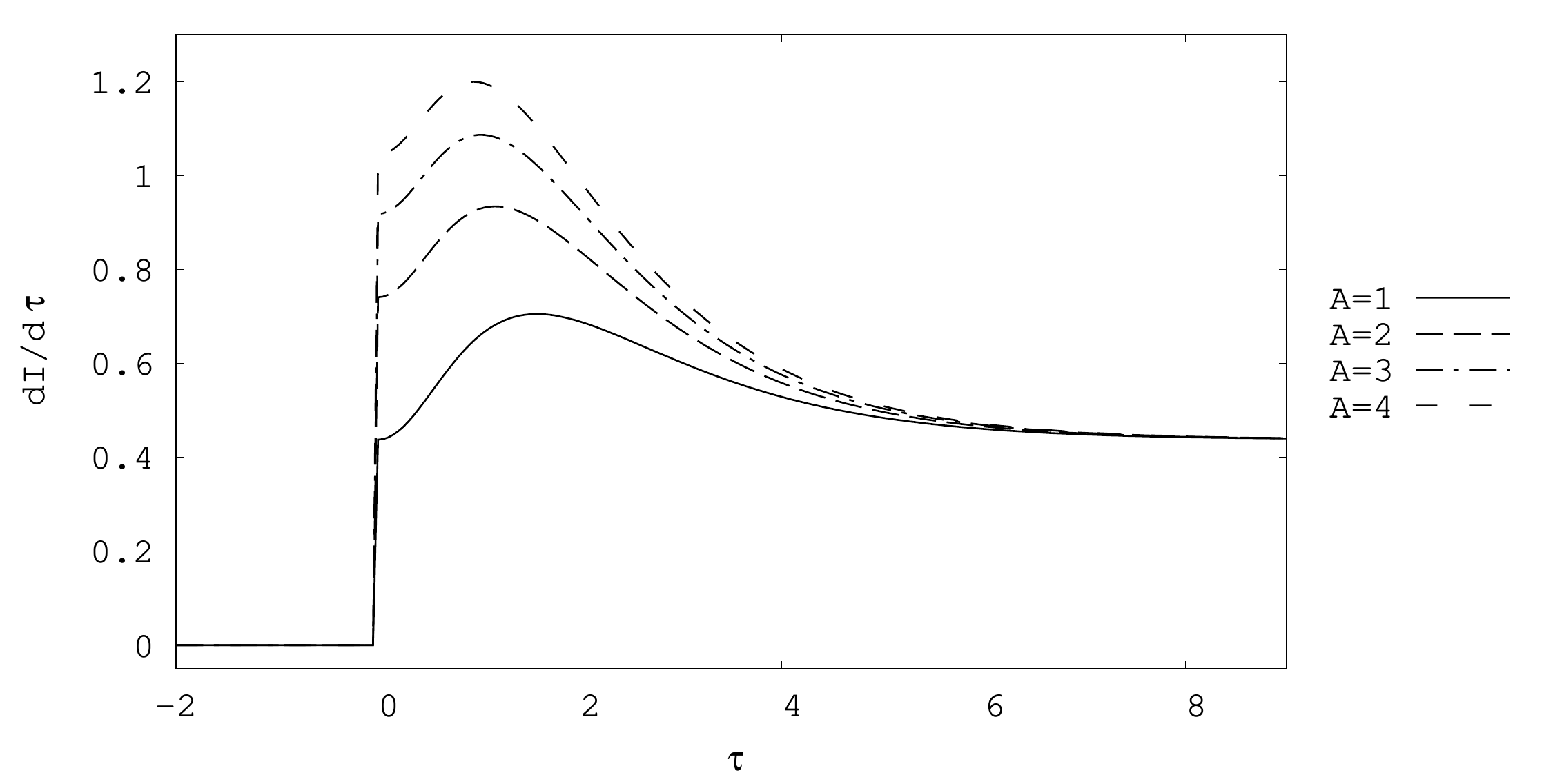}
\end{center}
\caption{Time dependence of the WDW action in the non-rotating case for different values of 
the parameter $A$. We set $G=1$, $l=1$, $r_h=1$ 
and $\nu=2$.}
\label{CA-time-dep-2}
\end{figure}
Numerical plots of the time dependence of the action rate (\ref{gran-totale1})
 for different values of $\nu$ 
are shown in figure \ref{CA-time-dep1}.
The same qualitative structure as for the AdS case \cite{Carmi:2017jqz} is found;
in particular the growth rate of the action is a decreasing function at late times.
As in \cite{Carmi:2017jqz}, 
the late-time limit then overshoots the asymptotic rate, which 
was previously believed \cite{Brown:2015lvg}
to be associated to an universal
upper bound, conjectured by Lloyd \cite{Lloyd}. 
There is some dependence
at finite time on the parameter $A$, see figure \ref{CA-time-dep-2};
this is a feature also of the AdS case \cite{Lehner:2016vdi,Chapman:2016hwi,Carmi:2017jqz}. The late-time limit is instead independent from $A$.

\subsection{Computation of the action in the rotating case}
\label{sect-Computation of the action in the rotating case}

\begin{figure}[h]
\begin{center}
\includegraphics[scale=0.6]{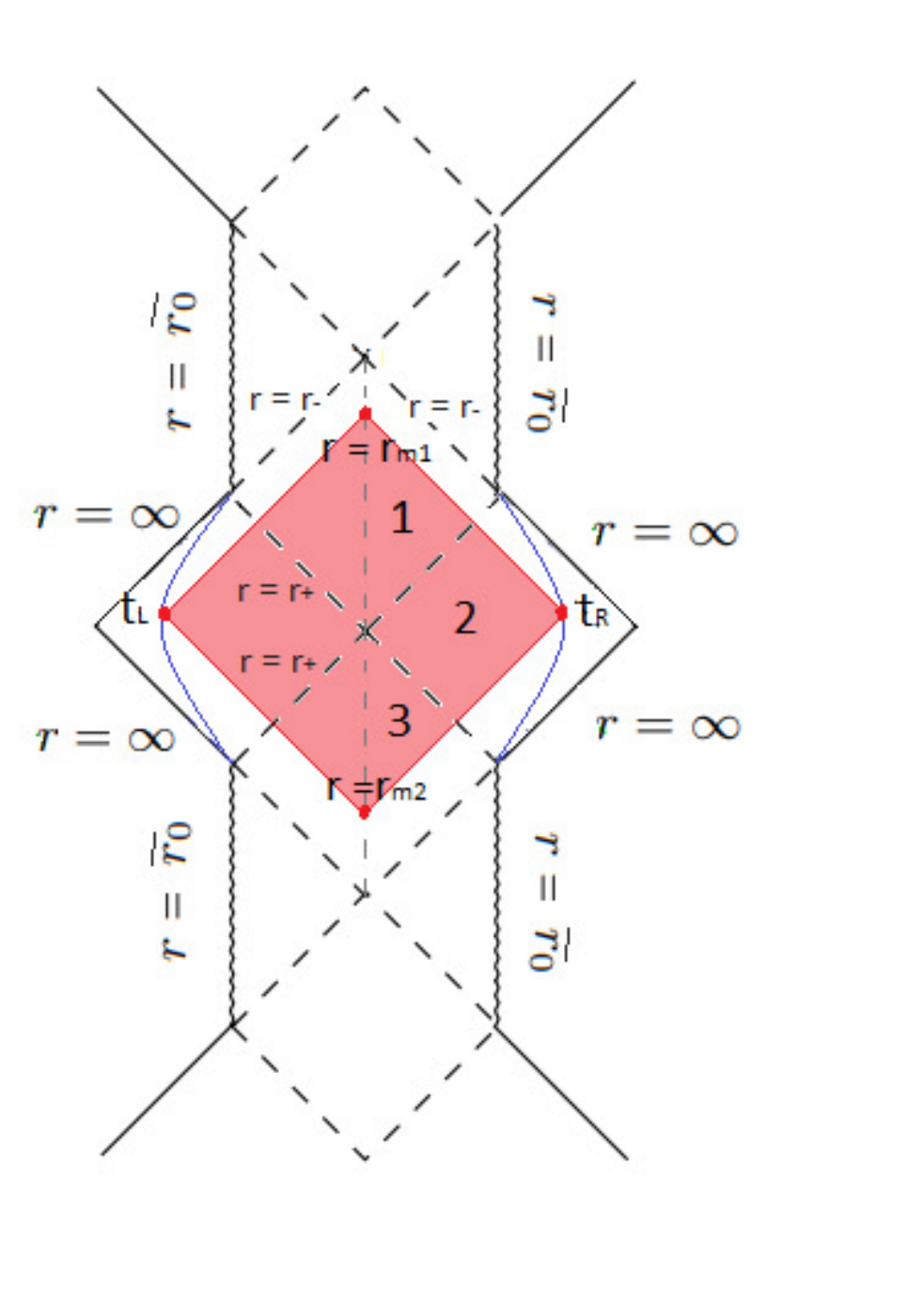}
\end{center}
\caption{Penrose diagram for the WDW patch in the rotating case.}
\label{fig3}
\end{figure}

In the rotating case (see figure \ref{fig3}) we
 do not need to distinguish between initial and later times, 
 because in this case the form 
of the WDW patch is the same at any time and the complexity is already non-vanishing at initial times. We define $\tau=l \, t_b$.
We call $  r_{m1}, r_{m2} $ the null joints referring respectively to the top and bottom vertices of the spacetime region of interest.
Due to the structure of the Penrose diagram in the rotating case (similar to the 3+1 dimensional diagram 
for a Reissner-Nordstrom black hole), we do not have boundaries contributing to the GHY term.

The definition of the null joints in terms of the tortoise coordinates are:
\beq
\frac{t_b}{2} + r^*_{\Lambda} - r^* (r_{m1} ) = 0 \, , \qquad
\frac{t_b}{2} - r^*_{\Lambda} + r^* (r_{m2} ) = 0 \, .
\label{definizione dei joint}
\eeq
It will be useful to differentiate with respect to time these expressions to find
\beq
\frac{d r_{m1}}{dt_b} =
\frac12 \le \frac{dr^*}{d r_{m1}} \ri^{-1} \, , \qquad
\frac{d r_{m2}}{dt_b} =
-\frac12 \le \frac{dr^*}{d r_{m2}} \ri^{-1} \, .
\label{derivate joint}
\eeq

{\bf Bulk contributions:}
We can still split the WDW patch into three regions covering only the right half of the diagram, which contribute as
\bea
I^{1}_{\mathcal{V}}  &=&  \frac{ \mathcal{I}}{8 G} \int_{r_{m1}}^{r_+} dr \le \frac{t_b}{2} + r^*_{\Lambda} - r^* (r) \ri \, ,
\qquad
I^{2}_{\mathcal{V}}  =  \frac{ \mathcal{I}}{4 G} \int_{r_+}^{\Lambda} dr \le  r^*_{\Lambda} - r^* (r) \ri \, , \nl 
I^{3}_{\mathcal{V}}  &=&   \frac{ \mathcal{I}}{8 G} \int_{r_{m2}}^{r_+} dr \le - \frac{t_b}{2} + r^*_{\Lambda} - r^* (r) \ri \, .
\label{bulk action rotating case}
\eea
The whole bulk contribution then amounts to
\bea
I_{\mathcal{V}}  &=& \frac{ \mathcal{I}}{2 G} \int_{r_+}^{\Lambda} dr \le  r^*_{\Lambda} - r^* (r) \ri  + \nl
&+& \frac{ \mathcal{I}}{4 G} \left[ \int_{r_{m1}}^{r_+} dr \le \frac{t_b}{2} + r^*_{\Lambda} - r^* (r) \ri 
+ \int^{r_{m2}}_{r_+} dr \le \frac{t_b}{2} - r^*_{\Lambda} + r^* (r) \ri \right] \, .
\eea
The rate of the bulk action is
\beq
\frac{dI_{\mathcal{V}} }{d t_b}  =
 \frac{ \mathcal{I}}{8 G} (r_{m2}- r_{m1} ) \, ,
\eeq
where the relations (\ref{definizione dei joint}) are used to obtain a vanishing result when differentiating the endpoints of integration.
The result simplifies when performing the late time limit, when
$ r_{m1} \rightarrow r_+$ and  $ r_{m2} \rightarrow r_-$,
and the bulk action time-derivative becomes
\beq
\lim_{t_b \rightarrow \infty} \frac{d I_{\mathcal{V}} }{d t_b} = - \frac{(\nu^2 +3)l}{16 G} \, (r_+ - r_-) + \frac{1}{8 G} \le \frac{\kappa}{l} c^2 - \alpha  a c \ri (r_+ - r_-) \, .
\label{bulk-late-time-rot}
\eeq

{\bf{Null joint contributions}}:
As in the non-rotating case,  the joints at $ r= \Lambda $ give a time-independent contribution, 
and then they are not of interest to find the rate of complexity.
We have two time-dependent  contributions coming from the top and bottom joints.

As a function of $r$, these contributions are proportional to:
\beq
\mathfrak{a} = \eta \log \left| A^2 \frac12 u^\a v_\a \right|  = 
\eta\log \left| \frac{A^2}{l^2} \frac{r \Psi (r)}{(\nu^2 +3)(r-r_-)(r-r_+)} \right|
 \, .
\eeq
For  both $ r=r_{m1}$ and  $ r=r_{m2}$  we have to insert $ \eta_1=-1. $ 

The action of each joint is
\beq
I_{\mathcal{J}}^{k} = - \frac{1}{4 G} \sqrt{\frac{r_{k}}{4} \Psi(r_{k})}
 \log \left| \frac{l^2}{A^2} F(r_k) \right| \, ,
 \qquad 
 F(r_k) \equiv \frac{(\nu^2 +3)(r_{k}-r_-)(r_{k}-r_+)}{r_{k} \Psi (r_{k})}  \, ,
\eeq
and the corresponding time derivative is
\bea
& & \frac{dI_{\mathcal{J}}^k}{d t_b}  =
   - \frac{ l}{8 G}   \frac{d r_{k}}{dt_b}   \left\{  
    \sqrt{r_{k} \Psi (r_{k})} 
\frac{d}{d r_{k}} \le 
\log \left| \frac{l^2}{A^2}   F(r_{k}) 
  \right| \ri 
+  \right. \nl
& & \left.   
 + \frac12 \frac{ 6(\nu^2 -1) r_{k} + (\nu^2 +3) (r_+ + r_-) -4 \nu \sqrt{(\nu^2 +3) r_+ r_-} }{\sqrt{ r_{k} \Psi(r_{k})}} 
\log \left| \frac{l^2}{A^2} F(r_{k})    \right|
   \right\} \, .
\eea
Using eqs. (\ref{derivate joint}) in the previous expression,
 it is possible to find the complete time dependence of the null contributions.
 Since the expression is rather cumbersome, we only write the late-time limit
\beq
\lim_{t_b \rightarrow \infty} \frac{d I_{\mathcal{J}}^k}{d t_b} =  \frac{(\nu^2 +3)l}{16 G} \, (r_+ - r_-) \, ,
\qquad
k=1,2 \, .
\eeq

{\bf Total:}
Summing all the previous asymptotic expressions, the late-time limit of the action growth is given by
\beq
\lim_{t_b \rightarrow \infty} \frac{d I}{d t_b} = \frac{(\nu^2 +3)l}{16 G} \, (r_+ - r_-) - \frac{1}{8 G} \le \frac{\kappa}{l} c^2 - \alpha  a c \ri (r_+ - r_-) \, .
\eeq
Taking into account eq. (\ref{a-action}) we finally find
\beq
\lim_{t_b \rightarrow \infty} \frac{1}{l} \frac{d I}{d t_b} = \lim_{\tau \rightarrow \infty} \frac{d I}{d \tau} = \frac{(\nu^2 +3)}{16 G} \, (r_+ - r_-)  = TS \, .
\label{late time action rate}
\eeq
The late-time limit can be recovered also with the
methods introduced in  \cite{Brown:2015lvg} and the results agree; 
details of the explicit calculation  can be found  in appendix \ref{another-way}.


\subsection{Adding the counterterm}

In eq. (\ref{composition of terms of the gravitational action}) we included in the gravitational action a counterterm accounting for the reparametrization invariance of the result, but so far we did not include it in the computation.
The reason is that it does not play an important role in this specific case. In fact, we observe from eq. (\ref{counterterm}) that the expression contains an arbitrary length scale $\tilde{L}$ which introduces another ambiguity in the action!
While adding the counterterm eliminates the parameter $A$ related to the ambiguity in giving a normalization for the null normals, and renders the action invariant under reparametrizations, on the other hand there is no way to fix a priori the value of the scale $\tilde{L}.$
In this way, the various graphs depicted in fig. \ref{CA-time-dep-2} are simply substituted by analogous graphs where the free parameter is the value associated to the counterterm length $\tilde{L}.$
On the other hand, the late time result was independent from the ambiguity in the normalization of null normals, and the same can be proven to be true about the dependence from the length scale $\tilde{L}$ after adding the counterterm.

From this discussion it seems that adding or not the counterterm does not change anything meaniningful in the action, and then its role appears to be insignificant.
However, we will see in chapters \ref{chapt-Subregion complexity for warped AdS black holes} and \ref{chapt-Subregion action complexity of the BTZ black hole} that it will play a role in determining the sub/super-additivity properties of the subregion action, and that will introduce only the length scale $\tilde{L}$ in exchange of all the ambiguous parameters related to null normals when multiple null surfaces are considered in the geometric set-up.
Moreover the role of the counterterm was found to play an important role in time-dependent configurations \cite{Chapman:2018dem, Chapman:2018lsv}.

\subsection{Comments and discussion}

There are some expectations about computational complexity which we can check with the late time volume and action rates in eqs. (\ref{late time volume rate}) and (\ref{late time action rate}), which we report here for convenience
\beq
 \lim_{\tau \rightarrow \infty} \frac{d V}{d \tau} = T S \frac{8 \pi G l }{\sqrt{3+\nu^2}} \, , \qquad
  \lim_{\tau \rightarrow \infty} \frac{d I}{d \tau}= TS \, ,
\eeq
\beq 
T S = \frac{ (r_+ -r_-)(3+\nu^2)}{16 G} \, .
\eeq
In  \cite{Stanford:2014jda} it has been proposed that the asymptotic rate of increase of complexity should be proportional to the product of temperature times entropy
\beq
\frac{dC}{d\tau} \simeq T S \, .
\label{stimaTS}
\eeq
The main motivation comes from the fact that complexity growth rate is an extensive
quantity which has the dimensions of an energy, and which should vanish
for a static object as an extremal BH.
Moreover, the authors of \cite{Cai:2016xho} proposed the following bound for the complexity growth rate:
\beq
\frac{dC}{d\tau} \lesssim
  \left[ (M- \Omega J - \Phi Q)_{+} - (M-\Omega J - \Phi Q)_{-} \right] \, ,
\label{bound2}
\eeq
where $ \pm $ indicate that the corresponding values of the quantities are computed at the outer and inner horizons.
With suitable units for complexity, the bound (\ref{bound2}) seems to be saturated in several cases.

For WAdS BHs, the angular velocities computed on the inner and outer horizons are:
\beq
\Omega_+  =\frac{2}{l(2 \nu r_+ -\sqrt{(\nu^2+3) r_+ r_-} )} \, ,
\qquad
\Omega_-  = \frac{2}{l(2 \nu r_- -\sqrt{(\nu^2+3) r_+ r_-} )} \, .
\eeq
If we use the values of mass and angular momentum in eqs. (\ref{M guess})-(\ref{J guess}), we find that
\beq
  (M- \Omega_+ J ) - (M-\Omega_- J )= \frac{ (r_+ -r_-)(3+\nu^2)}{16 G} = TS \, .
\label{TS-MJ}
\eeq
For the purpose of  the case studied in this paper, the saturation of the bound in 
eq.  (\ref{bound2}) is equivalent to eq. (\ref{stimaTS}).
Since both the volume and the action rate for late times are proportional to the product $TS,$ we satisfy all the previous requirements: in particular, they both vanishes in the extremal case $r_+ = r_- .$

In asymptotically AdS$_D$ spacetime, we have that the coefficient of proportionality between
 complexity and volume \cite{Susskind:2014moa} is usually taken as
\beq
 C = (D-1) \frac{V}{G l} \, ,
\eeq
and the late-time rate of growth of the volume is
\beq
\lim_{\tau \rightarrow \infty} \frac{d V}{d \tau} =\frac{8 \pi  G l}{D-1}  T S  \, .
\eeq
For comparison, in the case of flat spacetime BHs, 
\beq 
\lim_{\tau \rightarrow \infty} \frac{d V}{d \tau} \approx \frac{G r_h}{D-3}  T S \, ,
\eeq
where $r_h$ is the horizon radius ($\approx$ refer to a neglected order one prefactor \cite{Susskind:2014moa}).
Consequently,  the proportionality coefficient between the late time rate of growth
of the volume and $TS$ depends on the kind of asymptotic of the spacetime.

In order to compare with the AdS$_3$ case, we can write the rate of growth of the volume in WAdS as
\beq
\frac{d V}{d \tau} \rightarrow 
S T \,  4 \pi G l \, \eta \,  , \qquad \eta= \frac{2}{\sqrt{3+\nu^2}} \, .
\eeq
We may interpret the details of this result in distinct ways, depending 
on the exact holographic dictionary that we may conjecture between volume and complexity.
For example, it could be that complexity approaches at 
 late time  to $\eta \, TS$ (note
that $\eta \leq 1$  if we impose $\nu^2 \geq 1$);
if this is true, warping would make complexity rate decreases.
On the other hand, it could also  be that in spaces with WAdS asymptotic
 the holographic dictionary between
complexity and volume is changed by some non-trivial function of the warping parameter $\nu$;
for example, if we would have that
\beq
C=  \frac{2 }{G l \eta} V  \, ,
\eeq
the asymptotic complexity increase rate would be still $TS$ for every $\nu$.
The investigation of Complexity=Action conjecture suggests that the latter possibility may be preferrable.

We notice that in the case of the action computation, the only terms which contribute are the bulk and the joints,
while in the non-rotating case there is also a surface GHY contribution.
Although the details of the calculation are 
 quite different,  the final result is a continuous function of the parameters
 of the solution $(r_+,r_-)$.
A curious feature of  the non-rotating case is that there exists an initial time period
($t<t_c$)  in which complexity is constant; this is the same as in the AdS 
case \cite{Carmi:2017jqz}.

\chapter{Subregion complexity for warped AdS black holes}
\label{chapt-Subregion complexity for warped AdS black holes}

\begin{center}
\emph{The work in this chapter has previously appeared in \cite{Auzzi:2019fnp}.}
\end{center}

By analogy with entanglement entropy, an interesting further extension of the CV and CA conjectures consists in the case where the physical state is mixed, \emph{i.e.} we consider a subregion of the full boundary.
In this case, it is well known that the information properties of the subregion on the boundary are encoded in some extremal slices in the bulk, \emph{i.e.} the Ryu-Takayanagi (RT) and the Hubeny-Rangamani-Takayanagi (HRT) surfaces for the static and time-dependent cases, respectively.
Moreover, a bulk region which naturally encodes all the informations coming from a subsystem on the boundary is the entanglement wedge, defined as the bulk domain of dependence of the spacetime region bounded by the RT surface and the subregion on the boundary \cite{Headrick:2014cta}.

There are two proposals which naturally generalize the CV and CA conjectures for mixed states \cite{Carmi:2016wjl}:
\begin{itemize}
\item The CV requires to compute the spacetime volume of an extremal codimension-one surface anchored at the boundary and bounded by the RT (HRT) surface for a static (time-dependent) configuration. We will denote as $\mathcal{C}_V$ such quantity.
\item  The CA requires to compute the gravitational action in the domain given by the intersection between the WDW patch and the entanglement wedge. We will denote this quantity as $\mathcal{C}_A .$
\end{itemize}
Subregion complexity has been recently 
studied by many authors, \emph{e.g.}
\cite{Ben-Ami:2016qex,Abt:2017pmf,Abt:2018ywl,Agon:2018zso,Alishahiha:2018lfv,Caceres:2018blh,Roy:2017kha,Roy:2017uar,Bakhshaei:2017qud,Bhattacharya:2019zkb,Chen:2018mcc,Auzzi:2019mah}.  

\begin{figure}[h]
\centering
\includegraphics[scale=0.4]{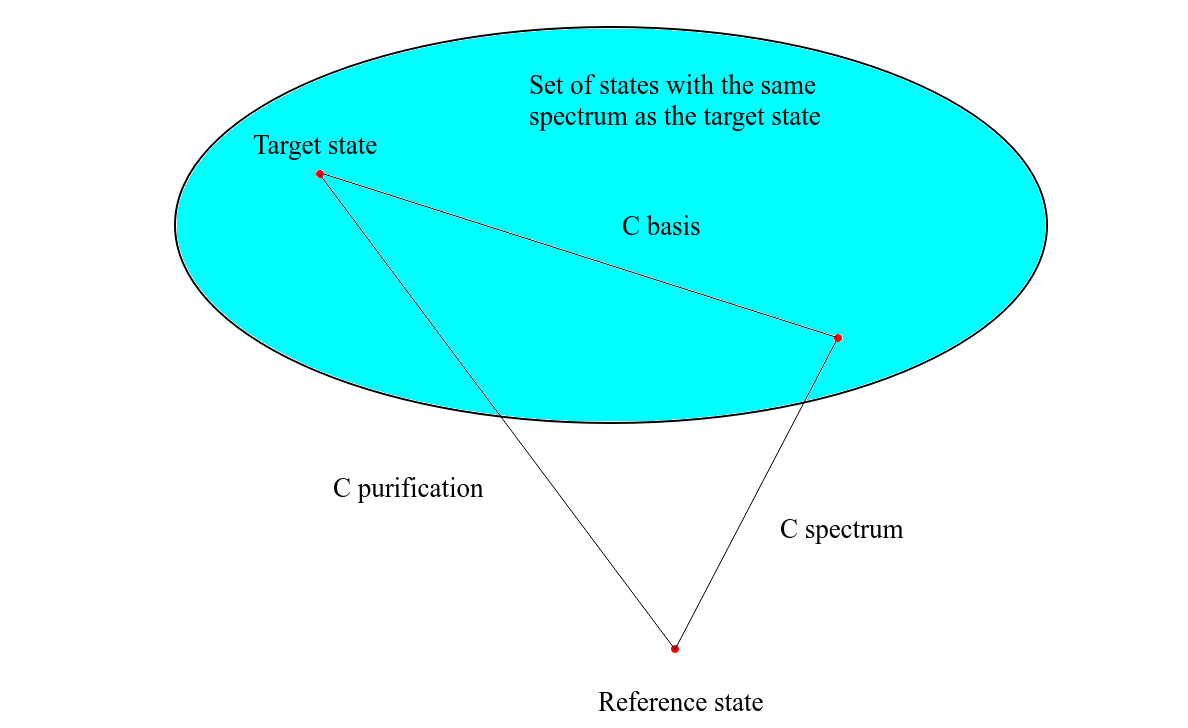}
\caption{Various definitions of subsystem complexity: basis, spectrum and purification.
The coloured area represents the states with the same spectrum as the target state $\rho,$ in particular the one which is reached with the least number of operations is $\rho_{spec}$ and gives the spectrum complexity.}
\label{fig-Subsystem_complexity_QFT}
\end{figure}

The investigation of the two conjectures for mixed states can give additional hints on which of them is preferrable, and may also help to identify a correct quantity to match from the field theory side.
Various notions of complexity exist from an analysis of tensor networks \cite{Agon:2018zso}:
 \begin{itemize}
 \item  Purification complexity $\mathcal{C}_P$, which 
 can be defined as the minimal number of gates needed to transform the initial
 pure state (plus some ancillary external qubits) into a purification of the  
 mixed state $\rho.$
\item Spectrum complexity $\mathcal{C}_S$, which can be defined as the minimal number
of operations needed to prepare a mixed state $\rho_{spec}$ with the same spectrum as $\rho.$
\item Basis complexity $\mathcal{C}_B$, which can be defined as the minimum number of gates
needed to prepare   $\rho$ from $\rho_{spec}$.
 \end{itemize}
The previous definitions are pictorially represented in fig. \ref{fig-Subsystem_complexity_QFT}.
The spectrum complexity does not reduce to 
complexity when computed on pure states, and so it is not a good candidate
 as a field theory dual of CV or CA. Instead both $\mathcal{C}_P$ and $\mathcal{C}_B$ might be
 in principle  reasonable candidates as duals of holographic complexities.
These issues were recently investigated by
 \cite{Agon:2018zso,Alishahiha:2018lfv,Caceres:2018blh,Ghodrati:2019hnn}.
 
There are some interesting properties that we can investigate for the subregion complexity: structure of UV divergencies, sub/superadditivity and temperature dependence.
The first aspect can be useful for the matching from the field theory side.
In this context, it is also interesting to understand if logarithmic or constant pieces arise, which usually contain universal informations in the case of the entanglement entropy.

The sub/superadditivity properties can be matched with the various definitions of computational complexity proposed from the tensor network analysis.
In particular, it was conjectured in  \cite{Agon:2018zso} that $\mathcal{C}_P$  should be subadditive for the left $L$
and right $R$ factors of the thermofield double state $TD$.
An analog guess was made about superadditivity of $\mathcal{C}_B$.

The volume complexity $\mathcal{C}_V$ is in general superadditive
because the volume is always a positive-definite quantity: given two regions A,B and their union, we have
\beq
 \mathcal{C}_V(A \cup B)  \geq  \mathcal{C}_V( A) +  \mathcal{C}_V( B)  \, .
\eeq
In the special case where the theormofield (TD) double state is considered at vanishing boundary time $t_b$ and the subregions are chosen to be the left (L) and right (R) boundaries separately, this inequality saturates:
\beq
 \mathcal{C}_V( TD, t_b=0) =  \mathcal{C}_V( L) +  \mathcal{C}_V( R) \, .
 \label{saturation}
\eeq
The situation is less clear when the action is considered, because it is not positive-definite.
 An interesting technical point which arises  in the CA conjecture
 is due to an arbitrary length  scale $\tilde{L}$ which appears in
the counterterm needed to make the action reparameterization invariant \cite{Lehner:2016vdi}.
Depending on the choice of $\tilde{L}$, for the AdS neutral black hole,
one can get either \cite{Alishahiha:2018lfv}
that CA is superadditive or subadditive for the $L,R$ sides of the thermofield  double. 

The investigation of the temperature dependence leads to the conclusion that $\mathcal{C}_B$ decreases with temperature $T$ and approaches zero for $T \rightarrow \infty$, while $\mathcal{C}_P$ should not have strong dependence on $T$.
As studied in \cite{Agon:2018zso,Alishahiha:2018lfv},  for the AdS neutral black hole
the behaviour of subsystem action complexity as a function of temperature 
also depends on $\tilde{L}$.  

Following the analysis started in chapter \ref{chapt-Complexity for warped AdS black holes}, here we compute the divergences of subregion complexity for the left and right factors of the thermofield double state,
 in the case of black holes  in asymptotically warped $ \mathrm{AdS}_3 $ spacetimes.
We  investigate the temperature dependence of subregion complexity in each of the conjectures
and the sub/superadditivity properties of the CA conjecture.

The configuration that we consider is a the particular subregion identified by only one of the two boundaries of the spacetime (left L or right R).
In this case the computations can be performed analitically and the RT surface degenerates to the bifurcation surface, which implies that the associated subregion complexity refers only to the part of the volume (action) external to the horizon.
The study of sub/superadditivity in this set up coincides with the investigation of the sign of the complexity internal to the horizon, because we have to study the condition (\emph{e.g.} in the case of superadditivity)
\beq
\mathcal{C} (TD, t_b=0) - \le  \mathcal{C} (L) + \mathcal{C} (R) \ri \geq 0 \, \Rightarrow \,
\mathcal{C}_{\rm ext} - \mathcal{C}_{\rm out} = \mathcal{C}_{\rm int} \geq 0 \, .
\eeq


\section{Subregion Complexity=Volume}
\label{sect-Subregion Complexity=Volume}

In this section we compute the divergences of the volume complexity at $t_b=0$ for the generic rotating black hole\footnote{We have seen during the computation of the Complexity=Volume conjecture in section \ref{sect-Complexity=Volume} that the limit $r_- \rightarrow 0$ is smooth, despite the radical change of the Penrose diagram in the two cases. For this reason, we will focus immediately on the general rotating situation, and we put the non-rotating limit in Appendix \ref{appe-volume} as a check that everything works well.}.
The time dependence of the volume studied in section \ref{sect-Complexity=Volume} tells us that the complexity is a monotonically increasing quantity, and then its minimum is obtained in the case $t_b=0 .$

In this configuration, the extremal surface is a constant $t=0$ bulk slice, connecting
the two $t_L=0$ and $t_R=0$ regions on the left and right boundaries.
The RT surface is a line at a constant value of the radial coordinate $ r=r_+$. 
We  denote by $ V(L) $ the volume of the codimension-one extremal surface anchored at the entire left boundary of the spacetime,
 and by $ V(R) $ the corresponding volume for the right boundary.
The symmetry of the problem implies that the subregion complexity on the two boundaries separately is the same, and then
\beq
V_{\rm ext} =  V(L) + V(R) = 2 V(L) \, .
\eeq

The volume can be computed directly from the determinant of the induced metric on the $t=0$ slice
\bea
V(L)  &=& 2 \pi l^2  \int_{r_+}^{\Lambda} dr \, G(r) \, , \nl
G(r) &=&  
 \sqrt{\frac{r \le 3(\nu^2 -1)r +(\nu^2 +3)(r_+ + r_-) -4 \nu \sqrt{r_+ r_- (\nu^2 +3)} \ri}{4 (\nu^2 +3)(r-r_-)(r-r_+)}} \, ,
\eea
where $\Lambda$ is an UV cutoff.
We investigate the possible divergences of the integral, which are near the outer horizon or near to the cutoff surface.
When $ r \rightarrow r_+ $, the function $ G(r) $ can be approximated as
\beq
G(r)  =\frac{g}{\sqrt{r-r_+}} + \mathcal{O} \le  \sqrt{r-r_+} \ri  \, , \quad
g=
 \sqrt{\frac{r_+ \le 4 \nu^2 r_+ + (\nu^2 +3)r_- -4 \nu \sqrt{r_+ r_- (\nu^2 +3)} \ri}{4 (\nu^2 +3)(r_+ - r_-)}} \, .
\eeq
Then the contribution to the volume coming from the region nearby the outer horizon is not divergent, because we obtain
\beq
2 \pi l^2 \int_{r_+}^{r_+ + \varepsilon} dr \, G(r) \approx
2 \pi l^2 \int_{r_+}^{r_+ + \varepsilon} dr \, \frac{g}{\sqrt{r-r_+}} \approx
4 \pi l^2  g \sqrt{\epsilon}  \, .
\eeq
At $r \rightarrow \infty$, the function  $G(r)$ can be expanded as
\beq
G(r) = \sqrt{\frac{3(\nu^2 -1)}{4 (\nu^2 +3)}} + \frac{\nu \le \nu (r_+ + r_-) - \sqrt{r_+ r_- (\nu^2 +3)} \ri}{\sqrt{3(\nu^2 -1)(\nu^2 +3)}} \frac{1}{r} + \mathcal{O} \le \frac{1}{r^2} \ri \, .
\eeq
Upon integration, the first two terms give rise to a linear
 and a logarithmic divergences. Consequently, the divergence of the volume is
\beq
V(L) = \pi l^{2} \sqrt{ \frac{3 (\nu^{2} -1)}{\nu^{2}+3} } \Lambda 
+ \frac{32 \pi G l^{2} \nu^{2}}{(\nu^{2}+3)^{3/2} \sqrt{3 (\nu^{2}-1)}} M \log \Lambda 
+ \mathcal{O} \left( \Lambda^{0} \right) \, .
\label{divergenza-volume}
\eeq
Interestingly, the logarithmically divergent term is proportional to the mass $M$.

\section{Subregion Complexity=Action}
\label{sect-Subregion Complexity=Action}

\begin{figure}[h]
\centering
\begin{subfigure}[c]{0.4\linewidth}
\includegraphics[scale=0.53]{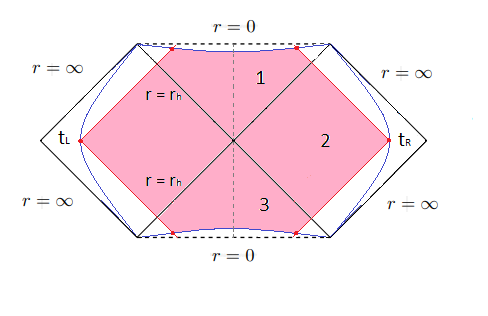}
\end{subfigure} \hspace{15mm}
\begin{subfigure}[c]{0.4\linewidth}
\includegraphics[scale=0.53]{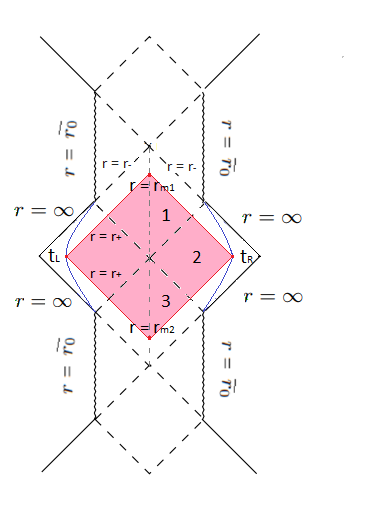}
\end{subfigure}
\caption{Penrose diagram and WDW patch at $t_b=0$ for the non-rotating (left)
and rotating (right) black holes.}
\label{WDW}
\end{figure}

The Penrose diagram and the WDW patch corresponding to the case
\beq
t_b= t_L = t_R = 0 \, ,
\eeq
which by symmetry argument corresponds to the minimum of the action, are depicted in fig. \ref{WDW}, where we have also drawn the surfaces at constant radius $r=\Lambda$ taken as the UV cutoff.
We call $  r_{m1}, r_{m2} $ the null joints referring respectively to the top and bottom vertices of the spacetime region of interest.
The definition of the null joints in terms of the tortoise coordinates are
\beq
\frac{t_b}{2} + r^*_{\Lambda} - r^* (r_{m1} ) = 0 \, , \qquad
\frac{t_b}{2} - r^*_{\Lambda} + r^* (r_{m2} ) = 0 \, ,
\eeq
where $r^*_{\Lambda} \equiv r^*(\Lambda)$. 
At  $ t_b =0$,  we get
\beq
r^*_{\Lambda} = r^* (r_{m1} ) = r^* (r_{m2} ) \equiv r^* (r_m) \, ,
\label{top and bottom joints at t=0}
\eeq
and the configuration is symmetric,  so the future and past interior actions are the same.

In this section we compute the divergences of the total action
 of the WDW patch at $t_b=0$ in the rotating case.
The calculation for the non-rotating case involves slightly different details
which are sketched in appendix \ref{appe-total};
as expected, the  result reproduces the $r_- \rightarrow 0$ limit of the rotating case.

The conditions \eqref{definizione dei joint}  cannot be solved analitically; we then consider a series expansion to obtain a closed form for $r_m$.
Both at $ r =\Lambda\rightarrow \infty$ and at  $ r \rightarrow r_-$ the function $r^*(r)$ diverges to $+\infty$,
so we study the behaviour around these points:
\begin{itemize}
\item
Nearby $r \approx r_-$,  we find
\beq
r^* (r_m)  = - \frac{\sqrt{3(\nu^2 -1)}}{\nu^2 +3} \tilde{A} \log \left| r_m - r_- \right| + \tilde{B} + \mathcal{O}(r_m-r_-) \, ,
\label{series expansion rm arond r-}
\eeq
where $\tilde{B} $ is a constant and 
\beq
\tilde{A}=\frac{\sqrt{r_- \Psi(r_-)}}{(r_+-r_-) \sqrt{3 (\nu^2-1)}} > 0 \, .
\eeq
\item Around $ r= \Lambda $,
\beq
r^*_{\Lambda} = \frac{\sqrt{3(\nu^2 -1)}}{\nu^2 +3} \log \Lambda + \tilde{C} + \mathcal{O}(\Lambda^{-1}) \, ,
\label{espa-lambda}
\eeq
where $ \tilde{C} $ is the finite piece of order $ \Lambda^0$ . 
\end{itemize}
Consequently, eq. \eqref{top and bottom joints at t=0} gives: 
\beq
r_m - r_-  \approx \, \Lambda^{- 1/\tilde{A}} \, 
  \exp\left[ \frac{(\tilde{B}-\tilde{C}) (\nu^2 +3)}{\tilde{A} \sqrt{3(\nu^2 -1)}} \right]  \, .
\label{rmin approssimato}
\eeq
We study the contributions to the gravitational action given in eq. (\ref{composition of terms of the gravitational action}), computed in the relevant spacetime region corresponding to this case, focusing on the divergent terms.

{\bf Interior bulk term:}
The interior bulk term can be obtained from eq. (\ref{bulk action rotating case}) and it is written as
 \beq
I^{\rm int}_{\mathcal{V}}  = 2(I_{\mathcal{V}}^1 + I_{\mathcal{V}}^3) 
 = - \frac{l}{4G} (\nu^2 +3) \left[ (r_+ - r_m) r^*_{\Lambda} - \int_{r_m}^{r_+} dr \, r^* (r) \right] \, .
 \label{inte-aaa}
\eeq
The last integral in eq. (\ref{inte-aaa}) is finite, because the function $ r^* (r) $
has integrable singularities around $r = r_-, r_+$.
The divergent part of the internal bulk action therefore comes only from the divergence of the tortoise coordinate near the boundary and the result is
\beq
I^{\rm int}_{\mathcal{V}}  = - \frac{l}{4G} \sqrt{3(\nu^2 -1)} (r_+ - r_-)  \log \Lambda +\mathcal{O}(\Lambda^0) \, .
\label{internal-bulk-action}
\eeq

{\bf External bulk term:}
The external part of the bulk term, taken from eq. (\ref{bulk action rotating case}), is given by
\beq
I^{\rm ext}_{\mathcal{V}} = 2 \, I_{\mathcal{V}}^2 = - \frac{l}{4G} (\nu^2 +3) \int_{r_+}^{\Lambda} dr \, \le r^*_{\Lambda} - r^*(r) \ri \, .
\label{general formula external bulk rotating case}
\eeq
In this case the divergence structure is richer: there are some contributions coming from the constant term $r^*_{\Lambda}$ multiplied by the integration range, and in addition we have another contribution from the integral of the tortoise coordinate near infinity.
In both cases, we need to consider the behaviour of $r^* (r)$ near infinity
\beq
r^* (r) = \alpha \log (4 r) + \beta + \frac{\gamma}{r} + \mathcal{O} (r^{-2}) \, ,
\eeq
where 
\bea
\beta &=&
-2 \frac{ \sqrt{r_+ \Psi(r_+)} \log \le \sqrt{r_+} + \sqrt{r_+ - \rho_0} \ri - 
 \sqrt{r_- \Psi(r_-)} \log \le \sqrt{r_-} + \sqrt{r_- - \rho_0} \ri }{(\nu^2 +3)(r_+ - r_-)} \, ,
 \nl
\alpha&=&\frac{\sqrt{3(\nu^2-1)}}{\nu^2 +3}  \, , \qquad
\gamma=\frac{\sqrt{3(\nu^2-1)}}{2(\nu^2+3)} \, (\rho_0-2 r_+-2 r_-)  \, .
\eea
The divergences of  (\ref{general formula external bulk rotating case}) then are
\bea
I^{\rm ext}_{\mathcal{V}} &=&  \frac{l}{4G} (\nu^2 +3) \left[- \alpha  \Lambda + \le \alpha r_+ + \gamma \ri \log \Lambda
\right] + \mathcal{O} (\Lambda^{0}) 
\nl
&=&  -  \Lambda \frac{l}{4G}  \sqrt{3(\nu^2 -1)}+
\frac{l}{8 G}  \sqrt{3(\nu^2 -1)} (\rho_0-2 r_-) (\log \Lambda) 
 + \mathcal{O} (\Lambda^0) \, .
\label{external-bulk-action}
\eea

{\bf Joint terms:}
The action evaluated on the WDW patch has four joint contributions: 
two on the cutoff surface $r = \Lambda$ and two in the region inside the black and white hole, coming from the top and bottom vertices. 
They can all be directly evaluated from  eq. (\ref{jjoints}).
The joint inside the black hole, located at $r = r_{m}$, gives the following contribution:
\beq
\begin{aligned}
I^{r_{m}}_{\mathcal{J}} & = - \, \frac{l}{8 G} \sqrt{r_m \Psi (r_m)} \log \left| \frac{l^2}{A^2} \frac{\le \nu^2 +3 \ri \le r_m - r_- \ri \le r_m - r_+ \ri}{r_m \Psi (r_m)} \right| = \\
& = \frac{l}{8 G} \sqrt{3 \le \nu^2 -1 \ri} \le r_+ - r_- \ri \log \Lambda + \mathcal{O} \le \Lambda^0 \ri \, .
\end{aligned}
\eeq
The joint nearby the cutoff surface gives:
\beq
\begin{aligned}
I^{\Lambda}_{\mathcal{J}} & = \frac{l}{8 G} \sqrt{\Lambda \Psi \le \Lambda \ri} \log \left| \frac{l^2}{A^2} \frac{\le \nu^2 +3 \ri \le \Lambda - r_- \ri \le \Lambda - r_+ \ri}{\Lambda \Psi \le \Lambda \ri} \right| = \\
& =\Lambda  \frac{l}{8 G} \sqrt{3 \le \nu^2 -1 \ri} \log \left| \frac{l^2}{A^2} \frac{\nu^2 +3}{3 \le \nu^2 -1 \ri} \right|  + \mathcal{O} \le \Lambda^0 \ri \, .
\end{aligned}
\eeq
Summing  the contributions of the four joints\footnote{Here we put a symmetry factor of 2 in the previous computations of the joints due to the symmetry of the configuration.}, we find
\beq
I^{\rm tot}_{\mathcal{J}} =\Lambda \frac{l}{4 G} \sqrt{3 \le \nu^2 -1 \ri} \log \left| \frac{l^2}{A^2} \frac{\nu^2 +3}{3 \le \nu^2 -1 \ri} \right|  + \frac{l}{4 G} \sqrt{3 \le \nu^2 -1 \ri} \le r_+ - r_- \ri \log \Lambda + \mathcal{O} \le \Lambda^0 \ri \, .
\eeq

{\bf Counterterm:}
The WDW patch is bounded by four codimension-one null surfaces;
they are all the same by symmetry, and so from (\ref{counterterm}) we find
\beq
I_{\rm ct} = \frac{l}{4 G} \int_{r_m}^{\Lambda} dr  \, \frac{\p_r (r \Psi (r) )}{\sqrt{r \, \Psi (r)}} \,
 \log \left| \frac{2 A \tilde{L}}{l^2} \frac{\p_r (r \, \Psi (r))}
{4 \sqrt{r \Psi (r)}} \right| \, .
\eeq
Since $\Psi(r)$ is linear in $r$,
the only divergence  comes from the region near $r = \Lambda,$ giving
\beq
I_{\rm ct} = \Lambda  \frac{l}{4 G} \sqrt{3 \le \nu^2 -1 \ri} \log \left| \frac{\tilde{L}^2 A^2}{l^4} \, 3 \le \nu^2 -1 \ri \right| 
+ \mathcal{O} \le \Lambda^0 \ri \, .
\eeq

{\bf Total action:}
Summing all the contributions, the divergences of the total action are
\beq
I_{\rm tot}= \frac{l}{4G} \sqrt{3 (\nu^2-1)} \Lambda \left( \log \left( \frac{\tilde{L}^2}{l^2} (\nu^2+3)  \right) -1 \right)
+(\log \Lambda) \frac{l}{4 G}  \sqrt{3(\nu^2-1)} \left( \frac{\rho_0}{2}-r_- \right) \, ,
\label{azione-totale}
\eeq
where $\rho_0$ was defined in eq. (\ref{rtilde0}).
As expected, the divergent contribution in the counterterm cancels the dependence on the ambiguous normalization constant $A$ appearing in the divergent contribution of the joints.


\subsection{Action of internal region and subregion complexity}

Now we focus on the divergence structure of the external action, which in the subregion action prescription corresponds to the complexity associated to one of the boundaries of the spacetime.
We will see that this investigation will not only provide a classification of the UV divergences, but will also be sufficient to find the sub/superadditivity properties of the action, giving also an interesting temperature behaviour.
The external bulk term was already identified in eq. (\ref{external-bulk-action}).

{\bf Joint terms:} 
In the interior of the black hole, there are four contributions of the form (\ref{jjoints}),
which are all in principle divergent because the function $f(r)$ defined in eq. (\ref{derirstar}) satisfies $f(r_+)=f(r_-)=0$.
There are other four joints inside the white hole.
As in the AdS case \cite{Agon:2018zso},  due to the signs $\eta=\pm 1$ of each joint,
these divergences will partially cancel each other.

It is useful to introduce the Kruskal coordinates $ (U,V) $ defined for $r>r_-$ as in \cite{Jugeau:2010nq}
\bea
U&=& \sgn (r-r_+) \, e^{b_{*} (r^* (r)-t)}=  \sgn (r-r_+) \,  e^{- b_* u} \, ,  \nl
 V&=&e^{b_{*}(r^* (r)+t)}= e^{b_* v} \, ,
\eea
where 
\beq
b_* = \frac{f'(r_+)}{2} =\frac{(\nu^2 +3)(r_+ - r_-)}{2 \sqrt{r_+ \Psi(r_+)}} \, .
\eeq
These coordinates satisfy the relation
\beq
\log |UV|  =   2b_* r^* (r) =f'(r_+) r^*(r)
\label{log UV coordinates}
\eeq 
which is useful to simplify expressions involving the joints.
Note that, since $r_* \rightarrow - \infty$ when $r \rightarrow r_+$,
the external horizon  corresponds to $U=0$ (black hole horizon for the right boundary)
and $V=0$ (white hole horizon for the right boundary).

\begin{figure}[h]
\centering
\includegraphics[scale=0.7]{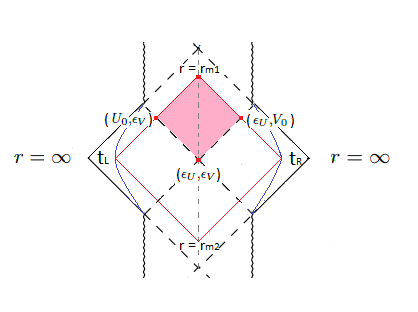}
\caption{Joint terms needed for the action of the black hole interior.}
\label{horizon-joint}
\end{figure}

Let us consider a contribution coming from sums of joints nearby the horizon.
We follow the prescription given in \cite{Agon:2018zso}, introducing the regulators
 $ \varepsilon_U , \varepsilon_V $ to move the joints off the horizon 
 by an infinitesimal quantity.   For instance, if we evaluate the sum of the contributions
  of two terms with the same $V=\varepsilon_V$,
 from eq. (\ref{jjoints}), we find a term proportional to
\beq
\begin{aligned}
& \log \left| \frac{l^2}{A^2} \frac{f(r_{U_1, \varepsilon_V})}{2R(r_+)} \right| - \log \left| \frac{l^2}{A^2} \frac{f(r_{U_2, \varepsilon_V})}{2R(r_+)} \right|  =  \int_{r_{U_2, \varepsilon_V}}^{r_{U_1, \varepsilon_V}} \frac{dr}{f(r)} f'(r) \approx  f'(r_+) \int_{r_{U_2, \varepsilon_V}}^{r_{U_1, \varepsilon_V}} \frac{dr}{f(r)} = \\
& = f' (r_+) \left[ r^* (r_{U_1, \varepsilon_V}) - r^* (r_{U_2, \varepsilon_V}) \right] =
\log \left| U_1 \varepsilon_V \right| - \log \left|U_2 \varepsilon_V  \right| = \log \left| \frac{U_1}{U_2} \right| \, ,
\end{aligned}
\label{difference of logarithms at horizon}
\eeq
where in the last steps we simplified the result by means of eq. (\ref{log UV coordinates}).

This expression tells us that in the limit $ \varepsilon_V \rightarrow 0, $ the difference of joints at
 the horizon is regular and the divergences coming from each term separately cancel.
We could perform the same trick  exchanging the $ U \leftrightarrow V $ coordinates, 
since the previous manipulations are symmetric under this transformation.
Combining these two results, one can conclude that
\beq
\log \left| \frac{l^2}{A^2} \frac{f(r_{U,V)}}{2 R(r_+)} \right| = \log \left| U V \right| + F(r_+) \, ,
\label{logarithm at the horizon}
\eeq
where the function $ F(r) $ is regular at the horizon and is given by
\beq
F(r) = \log \left| \frac{l^2}{A^2} \frac{f(r)}{2 R(r)} \right| - f'(r_+) r^* (r)   \, .
\label{BIGF}
\eeq

There are four joint contributions inside the black hole and four inside the white hole;
by symmetry they are the same and the total contribution is twice the ones of the black hole:
\beq
\begin{aligned}
I^{\mathrm{int}}_{\mathcal{J}} = & \, - 2 \times  \frac{l}{4 G}  \sqrt{\frac{r_+}{4} \Psi(r_+)} \left[ \log \left|\frac{l^2}{A^2} \frac{f(r_{\epsilon_U,\epsilon_V})}{2 R(r_+)}  \right|-\log \left|\frac{l^2}{A^2} \frac{f(r_{U_0,\epsilon_V})}{2 R(r_+)}  \right|-\log \left|\frac{l^2}{A^2} \frac{f(r_{\epsilon_U,V_0})}{2 R(r_+)}  \right| \right]  
\\ & - 2 \times \frac{l}{4 G} \sqrt{\frac{r_m}{4} \Psi(r_m)} \log \left| \frac{l^2}{A^2} \frac{f(r_m)}{2 R(r_m)} \right|  \, .
\label{joint tot rotating case}
\end{aligned}
\eeq
Thus, using the relations (\ref{log UV coordinates}) and (\ref{logarithm at the horizon}), this expression simplifies to
\bea
I^{\mathrm{int}}_{\mathcal{J}} &=& 
  \frac{l}{4 G} \sqrt{r_+ \Psi(r_+)} \left[ 2 b_* r^*_{\Lambda} 
  + F(r_+) \right]  -  \frac{l}{4 G} \sqrt{r_m \Psi(r_m)} \log \left| \frac{l^2}{A^2} \frac{f(r_{m})}{2 R(r_m)} \right| 
\nl
&=& \frac{l}{2 G} \sqrt{3(\nu^2-1)} (r_+- r_-) \log \Lambda +\mathcal{O}(\Lambda^0) \, .
\eea

{\bf Counterterms:}
The possible dependences from the UV cutoff 
can arise only from the $ r=r_m $ endpoint of integration.
However, putting the expansion (\ref{rmin approssimato}) inside the counterterm, 
we find that no  divergent pieces appear.

{\bf Internal and external action:}
Putting together all the terms contributing to the interior action in the rotating case, we find that
the divergent part is
\beq
I^{\rm int} = \frac{l}{4G} \sqrt{3(\nu^2 -1)} (r_+ - r_-) \log \Lambda + \mathcal{O} (\Lambda^0) \, .
\eeq
Subtracting this expression from eq. (\ref{azione-totale}), we find
the divergences of the external action, which correspond to the subsystem complexity:
\beq
I^{\rm ext}= \frac{l}{4G} \sqrt{3 (\nu^2-1)} \Lambda \left( \log \left( \frac{\tilde{L}^2}{l^2} (\nu^2+3)  \right) -1 \right)
+(\log \Lambda) \frac{l}{4 G}  \sqrt{3(\nu^2-1)} \left( \frac{\rho_0}{2}-r_+ \right) \, .
\label{external-action}
\eeq


\section{Comments and discussion}


\subsection{Regularization of the WDW patch}
\label{subsect-Regularization of the WDW patch}

\begin{figure}[h]
\centering
\includegraphics[scale=0.7]{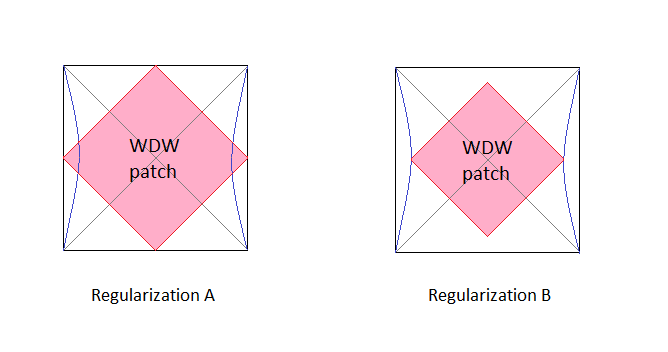}
\caption{Two different regularizations can be
chosen for the action of the WDW patch for black holes in AdS
(here for illustrative purpose we show
the case of non-rotating black hole in asymptotically AdS spacetime).}
\label{2-regs}
\end{figure}

In AdS one can consider two different regularizations \cite{Carmi:2016wjl}
for the CA conjecture (see figure \ref{2-regs}): 
\begin{itemize}
\item  The edge
of the WDW can end on the asymptotic AdS boundary 
(regularization $A$).
\item 
The edge of the WDW patch can end on the regulator surface
(regularization $B$).
\end{itemize}
The two regularizations give the same complexity rate
at large times. In asymptotically AdS spaces,
if one introduces appropriate counterterms
in regularization $A$ one can reproduce the same results
as in regularization $B$ \cite{Kim:2017lrw,Akhavan:2019zax}.

\begin{figure}[h]
\centering
\includegraphics[scale=0.7]{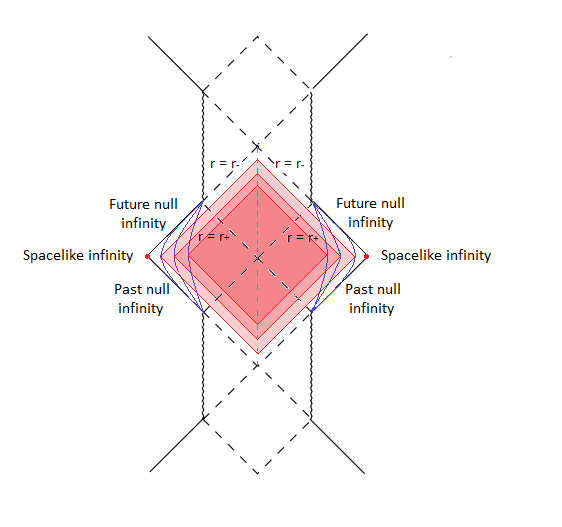}
\caption{In WAdS the causal structure resemble the one
of asymptotically Minkowski spacetime. Regularization $A$
would give a WDW patch with a corner which is located
at the spacelike infinity, and so would give a time-independent complexity.
Moreover, the WDW patch in regularization $B$  covers the entire black hole interior
in the limit of infinite cutoff $\Lambda$.}
\label{cutoff-var}
\end{figure}

In WAdS the structure of the Penrose diagram is radically different from AdS,
and it resembles instead the one of asymptotically Minkowski space:
the right corner of the Penrose diagram corresponds to
$r \rightarrow \infty$ and arbitrary $t$ (spacelike infinity).
The $45$ degrees boundaries correspond to the 
future null infinity and past null infinity (see figure \ref{cutoff-var}).

In all the previous works on the CA conjectures in WAdS 
\cite{Ghodrati:2017roz, Dimov:2019fxp},
 regularization $B$ was implicitly used.  This approach gives 
 the expected result for the complexity rate at late time
 $\dot{\mathcal{C}}_A \propto TS$ in the case of Einstein gravity, see eq. (\ref{late time action rate}).
 We used as well this regularization in the previous section
  to compute subregion action complexity.

It is not straightforward to generalise regularization $A$
to the case of WAdS, because then the corner of the WDW patch
would be located at the spacelike infinity point for all values of the time.
This would give the unphysical result that complexity is time-independent.

If in regularization $B$
we  sent the UV cutoff to infinity,
 we would find that the WDW patch
includes all the interior of the black hole.
This is the same part of the Penrose diagram which gives the 
linear growth of complexity at large time; so sending the cutoff
to infinity is equivalent to sending the time to infinity with finite cutoff, 
which gives a divergent internal action.
This explains why the action of the internal part of 
the WDW is UV divergent in WAdS, while it is finite in AdS.


\subsection{Role of the counterterm}

Following  \cite{Lehner:2016vdi}, 
we inserted in the gravitational action a counterterm of kind (\ref{counterterm})
which is needed in order to maintain reparameterization invariance
in presence of null boundaries. 
This term is not necessarily unique.

Let us borrow some notation from  \cite{Lehner:2016vdi}.
We consider a null hypersurface defined by the function $\Phi(x^\a)=0$.
The hypersurface can be described by parametric equations $x^\mu(\lambda,\theta^A)$,
where $\lambda$ is the affine null parameter and $\theta^A$ is constant on each null 
generator on the surface. 
The vectors
\beq
k^\mu=\frac{\p x^\mu}{\p \lambda} \, , \qquad
e^\mu_A= \frac{\p x^\mu}{\p \theta^A} 
\eeq
are tangent to the surface, while $k^\a$ is the null normal to the surface.
Let us denote by
\beq
\sigma_{AB}= g_{\a \b} e^\a_A e^\b_B
\eeq
the induced metric on the transverse directions $\theta_A$.
Also, one can introduce the following tensor
\beq
B_{AB}= e^\a_A e^\b_B D_\a k_\b \, ,
\eeq
which describes the behaviour of the congruence of null generators.


In principle, as discussed in the Appendix B of \cite{Lehner:2016vdi}, 
in presence of null boundaries
we can also allow for Lagrangians depending on combinations of the Riemann tensor
$\hat{R}_{ABCD}$ computed from the transverse induced metric $ \sigma_{AB} $.
Moreover,  contributions
containing the tensor $ B_{AB}$
are also allowed. A priori  we could have a counterterm of the type
\beq
\mathcal{L}_{\rm ct} (\hat{R}, \hat{R}_{AB}, \hat{R}_{ABCD}, B_{AB}, \Theta) \, ,
\eeq
where we should require that the total action is
 reparametrization-invariant.
Dramatic restrictions arise from the fact that 
we are working in 3 dimensions, which means that the null surfaces are 2-dimensional 
and that the induced metric $ \sigma_{AB} $ is 1-dimensional.
This implies that 
\beq
\hat{R}_{ABCD}=0 \, , \qquad
\hat{R}_{AB} = 0 \, , \qquad
\hat{R} = 0 \, , \qquad
B_{AB} = \frac12 \Theta \, \sigma_{AB} \, ,
\eeq
and then there is no space for curvature terms other than the geodesic expansion parameter
 $ \Theta$, which we already considered for the counterterm (\ref{counterterm}).

\subsection{Structure of divergences}
For the BTZ black hole, the only divergence in the holographic subregion
complexity is linear in the cutoff $\Lambda$.
In WAdS$_3$, we found that the two versions of holographic subregion complexity
have all a linear and a logarithmic divergence in $\Lambda$.
The coefficient of the linear divergence, as in the BTZ case, can be positive or negative
depending on  the counterterm parameter $\tilde{L}$.
The coefficient of the logarithmic divergence is  independent of $\tilde{L}$;
it is instead a function of  the black
holes parameters $(r_+,r_-)$, or equivalently of $(T,J)$.
In each of the two versions, the logarithmic divergence of the subregion complexity
is proportional to a different quantity:
\begin{itemize}
\item In the CA conjecture, eq. (\ref{external-action}) gives a result proportional to
$K_A=   \frac{\rho_0}{2}-r_+ $, with a positive coefficient.
\item In the CV conjecture, eq. (\ref{divergenza-volume}) gives a term proportional to the mass $M$,
with a positive coefficent.
\end{itemize}

\subsection{Sub/superadditivity}
In AdS black holes the internal action  $I^{\rm int}$ at $t_b=0$
 is finite \cite{Alishahiha:2018lfv,Agon:2018zso} and has
a sign which depends on the choice of the counterterm parameter $\tilde{L}$.
In turn, depending from the sign of $I^{\rm int}$, the action subregion complexity
can be sub/superadditive.
Instead, in WAdS$_3$ the interior action $I^{\rm int}$ is always positive 
and independent of the counterterm length scale;
as a consequence, $\mathcal{C}_A$ subregion complexity of the left and right side
of the thermofield double is superadditive.
Moreover, $I^{\rm int}$ is proportional to the product of
 temperature and entropy of the black hole:
\beq
I^{\rm int} = \frac{4 \sqrt{3(\nu^2 -1)}}{\nu^2 +3} \, l \,  TS \, \log \Lambda + \mathcal{O} (\Lambda^0) \, .
\eeq
By construction,  $\mathcal{C}_V$
is superadditive and saturates superadditivity
(\ref{saturation}) for the left and right side of thermofield double at $t_b=0 .$

\subsection{Temperature behaviour}
\label{sect-Temperature behaviour}

For neutral black holes in AdS, subregion $\mathcal{C}_A$ has  different properties 
depending on  the regularization parameter $\tilde{L}$.  For $\tilde{L} \ll l$,
$C_A$ increases with temperature,
whereas,  for $\tilde{L} \gg l$, $\mathcal{C}_A$ decreases with temperature. 
Instead, for neutral black holes in AdS, subregion $\mathcal{C}_V$ is an increasing powerlike
function of temperature \cite{Chapman:2016hwi} (for AdS$_3$, actually, it is independent of temperature).
 
\begin{figure}[h]
\centering
\includegraphics[scale=0.8]{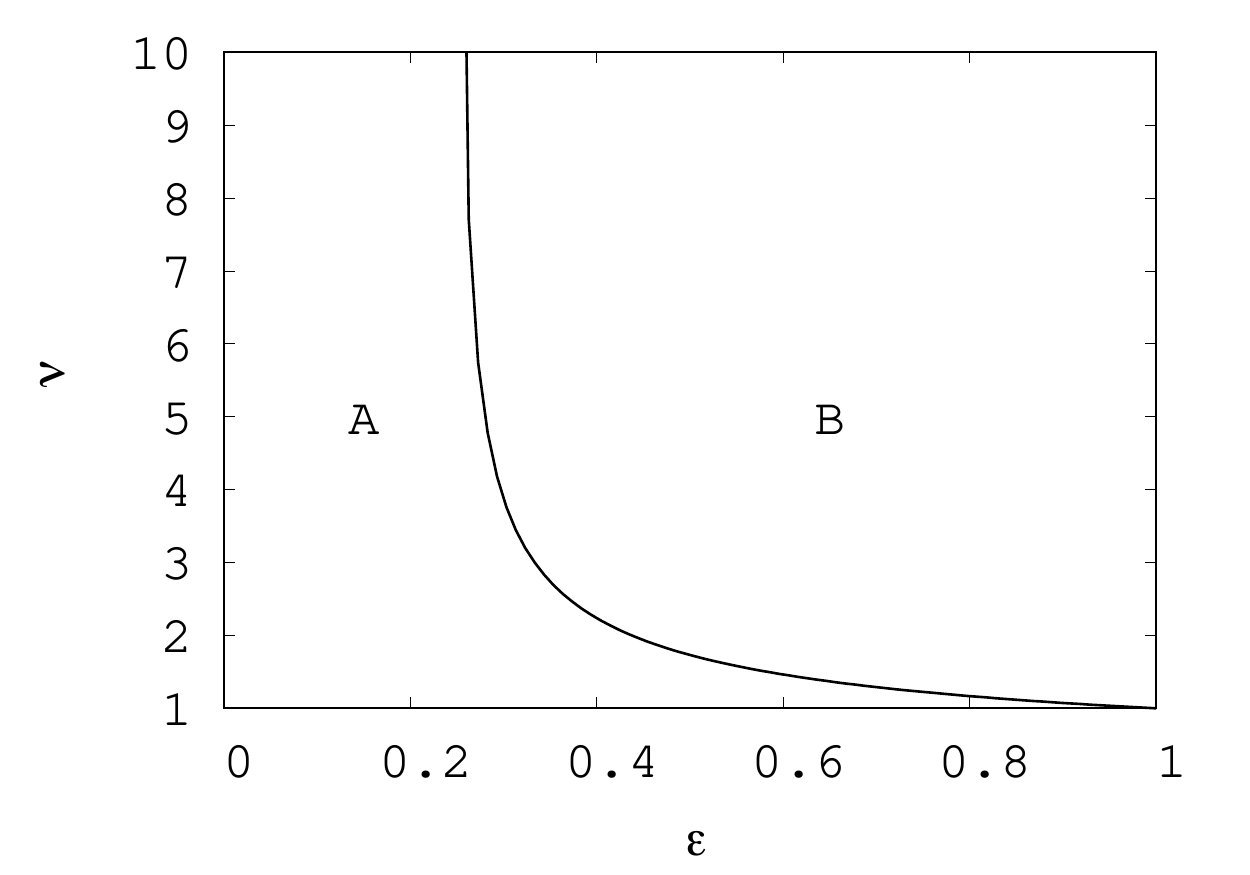}
\caption{Regions A and B in $(\epsilon, \nu)$ plane.}
\label{regionsAB}
\end{figure}

In WAdS$_3$, the leading dependence on temperature of the subsystem complexity
is in the $\log \Lambda $ terms. To this purpose we introduce 
\beq
C_J=\frac{\p M}{\p T} \Big|_J \, , \qquad C_A=\frac{\p K_A}{\p T} \Big|_J \, , 
\label{tante-C}
\eeq
which are explicitly computed in Appendix \ref{appe-temperature}. $C_J$
is the specific heat at constant $J$.
We note that the scale $r_+$ factorises from the quantities (\ref{tante-C}), then it is convenient to introduce
\beq
\epsilon=r_-/r_+  , \qquad
 0 \leq \epsilon < 1 \, ,
\eeq
and to study the sign of (\ref{tante-C}) as a function of $(\epsilon,\nu)$.
Let us define (see figure \ref{regionsAB})
\bea
{\rm Region \,\,\, A:} &&  \qquad  0< \epsilon< \frac{\nu^2+3}{4 \nu^2} \\
{\rm Region \,\,\, B:} &&  \qquad   \frac{\nu^2+3}{4 \nu^2} < \epsilon <1 \, .
\eea

The angular momentum $J$ is negative in region A and positive in region B,
while it vanishes along the two curves
\beq 
\epsilon=0 \,  \qquad {\rm and} \qquad \epsilon= \frac{\nu^2+3}{4 \nu^2} \, .
\eeq
It is interesting that the two quantities $(C_J,C_A)$
change sign in regions A,B: 
\begin{itemize}
\item $C_J$ is positive in region B and negative in region A.
\item $C_A$ is negative in region B and positive in region A.
\end{itemize}
As a consequence, in the region where $C_J>0$,
subregion $\mathcal{C}_V$  at constant $J$ increases with temperature 
while $\mathcal{C}_A$ decreases. In the thermodynamically unstable region
where $C_J<0$,  subregion $\mathcal{C}_V$ decreases with temperature
while $\mathcal{C}_A$ increases.


\chapter{Subregion action complexity of the BTZ black hole}
\label{chapt-Subregion action complexity of the BTZ black hole}

\begin{center}
\emph{The work in this chapter has previously appeared in \cite{Auzzi:2019vyh}.}
\end{center}

In chapters \ref{chapt-Complexity for warped AdS black holes} and \ref{chapt-Subregion complexity for warped AdS black holes} we investigated various aspects of complexity: we obtained the time evolution of complexity for both the volume and action conjectures, we studied their divergence structure, we observed that they are both superadditive and we studied the temperature behaviour.
The task of these computations was to obtain some informations for the putative field theory side of $\mathrm{WAdS/WCFT}$ correspondence, but we also had in mind to compare the warped case with the traditional $\mathrm{AdS}$ case, in order to find some universal aspects of the conjectures.

While in the general case volume and action differ from the transient behaviour of complexity for intermediate times, but they agree at late times, we observed that subregion complexity helps in finding situations where the two conjectures lead to different results.
Motivated by this observation, we find important to search for other examples where the holographic computation discriminates between the two proposals.
Some of them were already considered in the literature, \emph{e.g.} by studying defects \cite{Chapman:2018bqj} or time-dependent backgrounds \cite{Chapman:2018dem, Chapman:2018lsv}.

In this chapter we consider the subregion complexity for asymptotically $\mathrm{AdS}_3$ spacetime with a generic subsystem on the boundary, \emph{i.e.} without restricting to the particular case where the subregion is taken to be one of the disconnected boundaries of the eternal black hole, which was studied in \cite{Agon:2018zso}.
The volume case was considered in previous works \cite{Ben-Ami:2016qex, Abt:2018ywl}, and will be reviewed for comparison with the action computation later.

In this chapter we will find the following analytic  result for the subregion complexity of a segment 
of length $l$ in the BTZ black hole background:
\beq
\label{CABTZ}
 \mathcal{C}_{A}^{\rm BTZ} = \frac{l}{\varepsilon} \frac{c}{6 \pi^2} \log \left(\frac{\tilde{L}}{L} \right)
- \log  \left(\frac{2\tilde{L}}{L} \right) \frac{S^{\rm BTZ}}{\pi^2} + \frac{1}{24} c \, ,
\eeq
where $\tilde{L}$ is the counterterm length scale, $\varepsilon$
is the UV cutoff, $c$ the CFT central charge and $S^{\rm BTZ}$ the Ryu-Takayanagi (RT)
entanglement entropy of the segment subregion. 
This shows a direct connection at equilibrium, in the case of the one segment subregion,
between action complexity and entropy.
This expression is  also valid for the particular case of  AdS$_3$,
which was previously studied in \cite{Carmi:2016wjl,Chapman:2018bqj}.

One may wonder if such a simple connection between subregion complexity and 
entanglement entropy is valid also for more general subsystems.
For this reason, we compute action complexity in the case of a 
two segments subregion in AdS$_3$. This quantity has again
a linear divergence proportional to the total size of the region and a logarithmic
divergence proportional to the divergent part of the entropy.
However, if the separation between the two disjoint segments is small,
there is no straightforward relation between the finite part of complexity and entropy,
as we will derive in eq. (\ref{CA2}).


\section{Subregion complexity for a segment in AdS$_3$}
\label{sect:AdS}

 It is useful to review the AdS$_{3}$ calculation \cite{Carmi:2016wjl,Chapman:2018bqj,Caceres:2019pgf}
to set up the notation and the procedure, and as a warm-up for the more complicated BTZ case.
 We consider the Einstein-Hilbert action with negative cosmological constant in $2+1$ dimensions
\beq
S=\frac{1}{16 \pi G} \int \le R+\frac{2}{L^2} \ri   \sqrt{-g} \, d^3 x  \, ,
\label{HIlagrangian}
\eeq
which has as a solution AdS$_3$ spacetime,
whose metric in Poincar\'e coordinates reads
\beq
ds^2 = \frac{L^2}{z^2} \le -dt^2 + dz^2 + dx^2 \ri \, .
\eeq
The AdS curvature is $R=-6/L^2$ and $L$ is the AdS length.
The central charge of the dual conformal field theory is related to the bulk quantities via the expression
\beq
c=\frac{3 L}{2 G} \, .
\eeq

Two common regularizations \cite{Carmi:2016wjl} are used in the CA conjecture 
(see figure \ref{2-regs}): 
\begin{itemize}
\item Regularization $A$:
 the WDW patch is built 
starting from the boundary $z=0$ of the spacetime and a cutoff is then introduced
at $z=\varepsilon$.
\item Regularization B: the WDW patch is built from the surface $z=\varepsilon$.
\end{itemize}
As for the warped case, we will use in the main text the regularization $B,$ while the comparison with regularization $A$
is discussed in Appendix \ref{app:other-reg}.

\begin{figure}[h]
\center
\begin{tabular}{cc}
\includegraphics[scale=0.5]{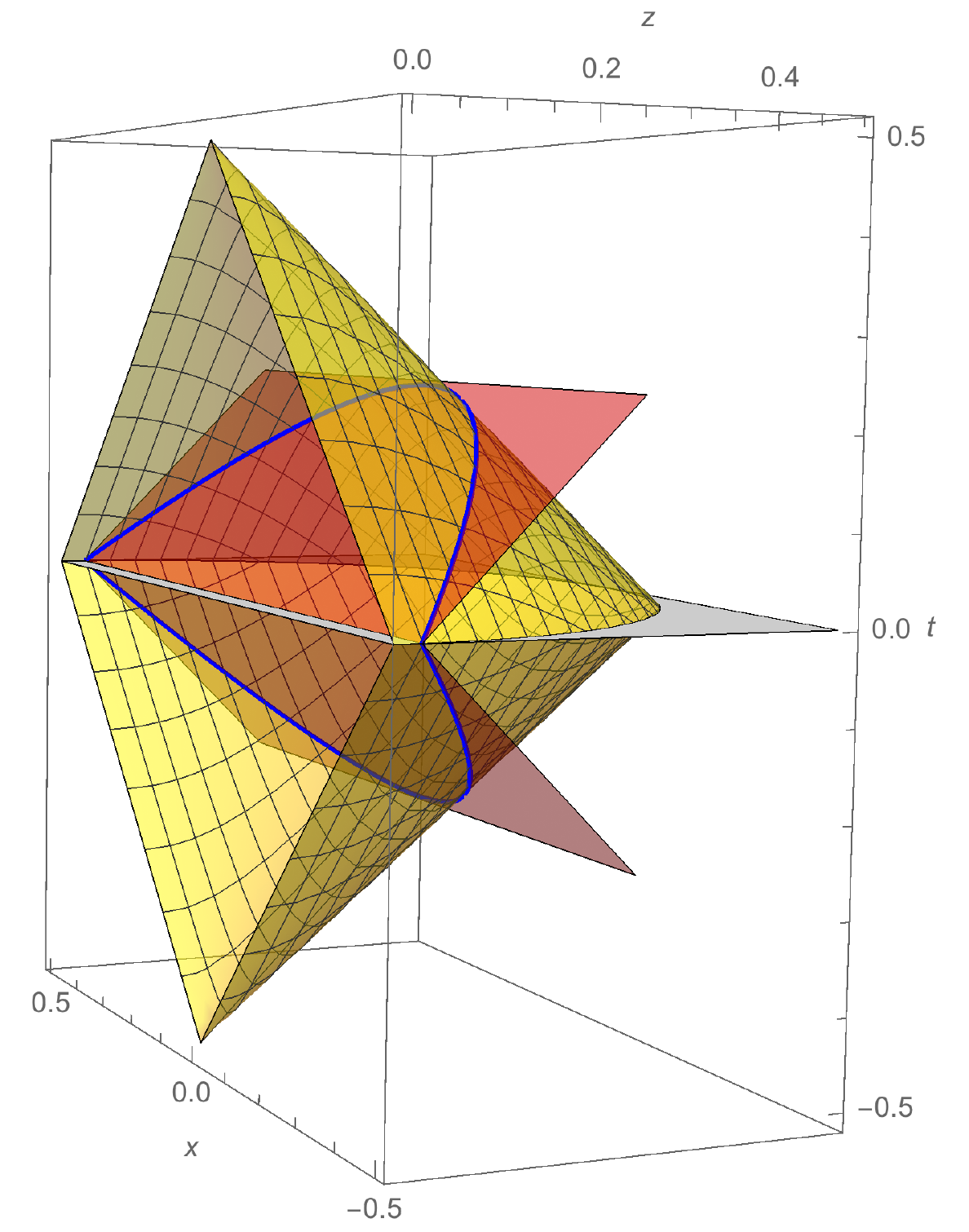} & \includegraphics[scale=0.55]{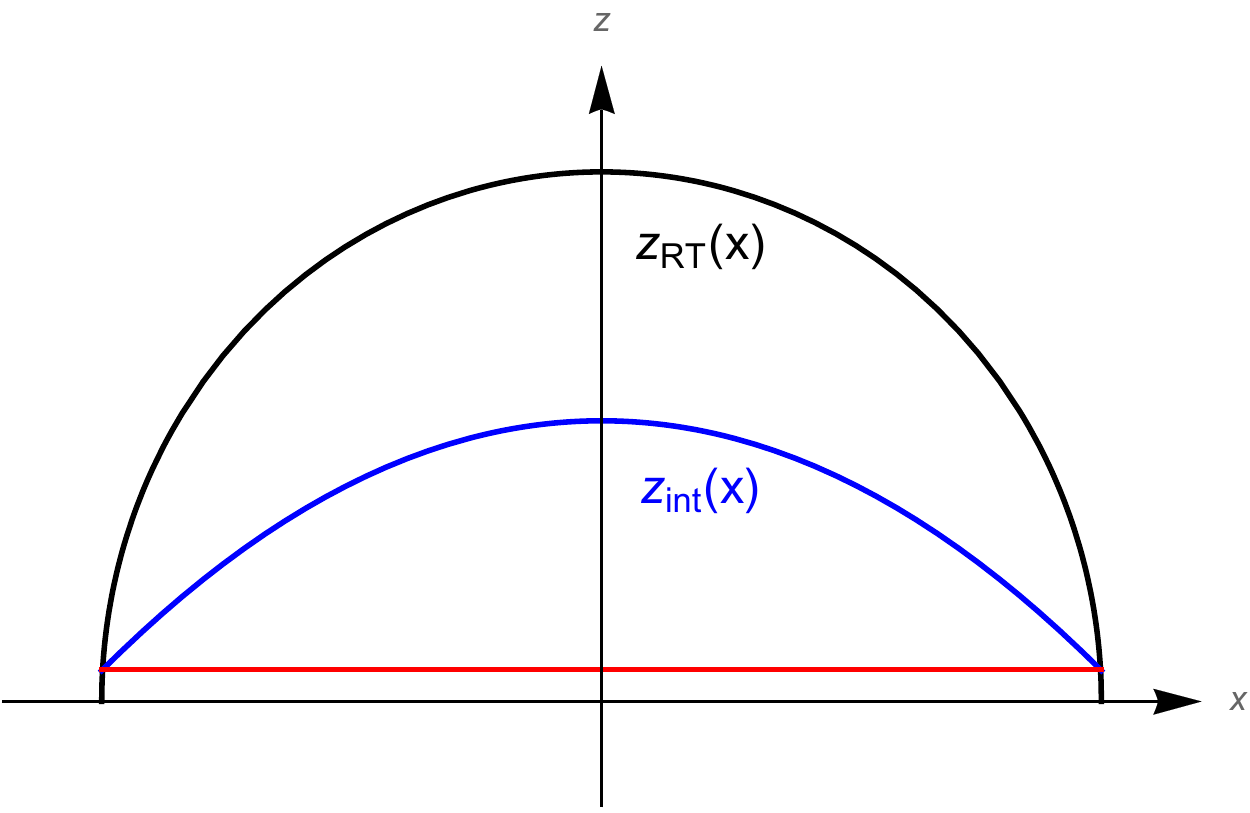}
\end{tabular}
\caption{Left: Intersection of WDW patch  with entanglement wedge in the $(x,z,t)$ space.
The boundary of the entanglement wedge is in yellow,
while the boundary of the WDW patch is in red.
Right: intersections in the $(x,z)$ plane, with $z_{RT}$ in black, $z_{int}$ in blue
and the cutoff $z=\varepsilon$ in red.}
\label{ads}
\end{figure}

We consider a subregion on the boundary given by a strip of length $l$ and
for convenience we take $x \in \left[ - \frac{l}{2}, \frac{l}{2} \right]$,
  at the constant time slice $t=0$.
This choice is possible due to the translation invariance of the set-up along these directions.  
The geometry relevant to the computation of  action complexity 
is the  intersection between  the entanglement wedge  \cite{Headrick:2014cta}  of the subregion
with the WDW patch \cite{Brown:2015bva,Brown:2015lvg}, see figure \ref{ads}.
Notice that among the various boundary terms involving the null surfaces, there is a codimension-three joint coming from the intersection between the WDW patch, the entanglement wedge and the boundary at $z=0$, $x=\pm l/2.$ 
This kind of joint exists only in regularization $B$ and can a priori give a non-vanishing contribution.
Since we will check that regularization $A$ gives a similar result for the subregion action in Appendix \ref{app:other-reg}, we believe that this joint at most shifts the action by an overall constant.

We use regularization $B$ with a cutoff a $z=\varepsilon$ and we consider all the ingredients describing the boundaries of the spacetime region of interest.
The RT surface is given by the space-like geodesic
\beq
t=0 \, , \qquad
z^2 + x^2 = \le  \frac{l}{2} \ri^2 \, ,
\eeq
which is a circle of radius $ l/2$.
A useful alternative parametrization of the curve is
\beq
z_{RT}=\sqrt{\left( \frac{l}{2} \right)^2 -x^2} \, .
\eeq
The entanglement wedge is a cone whose null boundaries are parameterized by
\beq
\label{twedge-ads}
t_{\rm EW} = \pm \le \frac{l}{2} - \sqrt{z^2 + x^2}   \ri \, .
\eeq
The boundaries of the WDW patch, which are attached to the regulator surface, are described by the equations
\beq
t_{\rm WDW} = \pm \le z- \varepsilon  \ri \, .
\eeq
In order to split the bulk term and to compute the joint contributions, we need to find the intersection curve between the null boundary of the WDW patch and the one of the entanglement wedge, parametrized by
\beq
z_{\rm int}= \frac{(l + 2 \varepsilon)^2 -4 x^2 }{4 (l +2 \varepsilon) } \, 
\qquad {\rm or} \qquad
x_{\rm int} = \frac{1}{2} \sqrt{(l+2 \varepsilon ) (l-4 z+2 \varepsilon )} \, .
\label{intersez}
\eeq
The introduction of the UV cutoff surface for the radial coordinate z induces a constraint on the values that the transverse coordinate can assume. In particular, the maximal one corresponds to the x coordinate of the RT surface evaluated at the cutoff, \emph{i.e.}
\beq
x_{\rm max} \equiv x_{\rm RT} (z=\varepsilon) = \sqrt{\le \frac{l}{2} \ri^2 - \varepsilon^2}  \, .
\label{x max AdS}
\eeq
This shift from $x=l/2$ is necessary for a correct regularization
of the on-shell action.
We remind the contributions to the action 
\beq
I=I_{\cal V}+I_{\cal N}+I_{\rm ct}+ I_{\mathcal{J}} \, ,
\eeq
where $I_{\cal V}$ is the bulk term (see eq. \ref{HIlagrangian}), 
$I_{\cal N}$ the null boundary term (see (\ref{null-bou})), 
$I_{\rm ct}$ the counterterm (\ref{countertermlabelbis}) and
$I_{\mathcal{J}}$ the null joint contribution (\ref{joint}).

In the following computations there will be various symmetry factors induced by the fact that we consider for most of the terms only the region $x>0, t>0.$
We will include all of these factors when giving the total bulk, joint and counterterm expressions.
In all the computations of this chapter it will be understood that the results are given up to the finite term, namely we omit $\mathcal{O}(\varepsilon)$ contributions, which vanish in the limit $\varepsilon \rightarrow 0 .$

\subsection{Bulk term}

The curvature is constant and so the Einstein-Hilbert term (\ref{HIlagrangian})
is proportional to the spacetime volume.
We can split the bulk contribution in two parts, based on the intersection between 
the WDW patch and the entanglement wedge, which we parametrize with the function $ z_{int} (x) . $
 In the first region the WDW patch is subtended by the entanglement wedge. 
 Consequently, we integrate along time $ 0 \leq t \leq t_{\rm WDW} (z)$, 
  then the radial direction along $ \varepsilon \leq z \leq z_{\rm int} (x)$,  and finally along the coordinate $ 0 \leq x \leq x_{\rm max}$:
 \beq
 I^1_{\cal V}  = - \frac{L}{4 \pi G} \int_{0}^{x_{max}} dx \int_{\varepsilon}^{z_{int}} dz \int_{0}^{t_{WDW}} dt \, \frac{1}{z^{3}} 
 \label{integralozzo1}
 \eeq
In the second region the entanglement wedge is under the WDW patch, 
 then the integration involves the endpoints $ 0 \leq t \leq t_{\rm EW}(z,x), z_{\rm int} (x) \leq z \leq z_{\rm RT} (x) $
 and finally $ 0 \leq x \leq x_{\rm max}$ :
  \beq
I^{2}_{\cal V}  = - \frac{L}{4 \pi G} 
\int_{0}^{x_{max}} dx \int_{z_{int}}^{z_{RT}} dz \int_{0}^{t_{EW}} dt \, \frac{1}{z^{3}} 
 \label{integralozzo2}
 \eeq
A direct evaluation of the integrals gives: 
   \bea
I^1_{\cal V} &=&  - \frac{L}{16 \pi G} \frac{l}{\varepsilon} - \frac{L}{4 \pi G} \log \le  \frac{\varepsilon}{l} \ri
- \frac{L}{8 \pi  G} \, .
\nl
I^{2}_{\cal V} &=&  \frac{L}{8 \pi G} \log \le \frac{\varepsilon}{l} \ri
+ \frac{L( \pi ^2 +8)}{64 \pi  G} \, .
\eea
The total result of the bulk action is: 
\beq
 I_{\cal V}^{\rm AdS} = 4 (I_{\cal V}^1 + I_{\cal V}^2) 
 = - \frac{L}{4 \pi G} \frac{l}{\varepsilon} + \frac{L}{2 \pi G} \log \le \frac{l}{\varepsilon} \ri + \frac{L \pi}{16 G}
 \, . \label{total-bulk-ads}
\eeq

\subsection{Null boundary counterterms}

A hypersurface described by the scalar equation $\Phi(x^a)=0$ has a normal vector $k_a= - \p_a \Phi$.
If the hypersurface is null, $k_a k^a=0$ and then it can be shown \cite{Poisson:2009pwt}
 that the hypersurface is generated by null geodesics,
which have $k^\a$ as a tangent vector.

In correspondence of a null boundary,
the following term appears in the action
\beq
I_{\cal N}= \int dS \, d\lambda \,  \sqrt{\sigma}
\kappa \, ,
\label{null-bou}
\eeq
where $\lambda$ is the geodesic parameter, $S$ the transverse spatial directions,
$\sigma$ is the determinant of the induced metric on $S$
and $\kappa$ is defined by the geodesic equation
\beq
k^\mu D_\mu k^\a =\kappa \,  k^\a \, .
\eeq
In our case, the null normals to the WDW patch and the entanglement wedge
 are given respectively by the following one-forms:
\beq
\mathbf{k}^{\pm}= \alpha \le \pm dt -dz  \ri \, , \qquad
\mathbf{w}^{\pm} = \beta \le \pm dt + \frac{z dz}{\sqrt{z^2 + x^2}} + \frac{x dx}{\sqrt{z^2 + x^2}} \ri \, ,
\eeq
where $\a,\b$ are arbitrary constants that will cancel in the final result.
We denote by $({k}^{\pm})^\mu$ and $({w}^{\pm})^\mu$ the corresponding vectors.
It can be checked that they correspond to an affine
 parametrization of their null surfaces, i.e.
 \beq
(k^\pm)^\mu  D_\mu (k^\pm)^\a =0 \, , \qquad
(w^\pm)^\mu  D_\mu (w^\pm)^\a =0 \, .
\label{geo-geo}
 \eeq
The term (\ref{null-bou}) vanishes in our calculation because we used
an affine parameterization, see eq. (\ref{geo-geo}).

We still need to include the contribution from the counterterm, 
which ensures the reparameterization invariance of the action:
\begin{equation}
\label{countertermlabelbis}
I_{\rm ct} = \frac{1}{8 \pi G} \int d\lambda \, dS \, \sqrt{\sigma} \, \Theta \, \log \left| \tilde{L} \, \Theta \right| \, ,
\end{equation}
where $\Theta$ is the expansion scalar of the boundary geodesics
and $\tilde{L}$ is an arbitrary scale.
If an affine parameterization is used, the following result holds \cite{Poisson:2009pwt}:
\begin{equation}
\label{expansion}
\Theta
= D_{\mu} k^{\mu} \, .
\end{equation}
We can then evaluate eq. (\ref{countertermlabelbis}) on each boundary:
\begin{itemize}
\item
The counterterm on the entanglement wedge boundary
vanishes because  $\Theta=0$. 
This agrees with the calculations in \cite{Headrick:2014cta}.

\item
For the boundary of the WDW patch we obtain:
\bea
I_{\rm ct}^{\rm WDW}  &= &  - \frac{L}{2 \pi G} \int_0^{x_{\rm max}} dx \int_{\varepsilon}^{z_{\rm int}} \frac{dz}{z^2} \log \left| \alpha
\frac{\tilde{L} z}{L^2} \right|  \nl
& = & \frac{L}{4 \pi G} \frac{l}{\varepsilon} \left[ 1 + \log \le \alpha \frac{\tilde{L}\varepsilon}{L^2} \ri \right]  
+ \frac{L}{4 \pi G} \log \le \frac{\varepsilon}{l} \ri \log \le \alpha^2 \frac{\varepsilon l \tilde{L}^2}{L^4} \ri 
\nl
&+& \frac{L}{2 \pi G} \log \le \frac{\varepsilon}{l} \ri
+ \frac{L \pi}{12 G}  \, .
\label{controt}
\eea

\end{itemize}

\subsection{Joint terms}

The  contribution to the gravitational action 
coming from a codimension-two joint, 
given by intersection of two codimension-one null surfaces \cite{Lehner:2016vdi}, is 
\begin{equation}
\label{joint}
I_{\mathcal{J}} = \frac{\eta}{8 \pi G} \int dx \sqrt{\sigma} 
\log \left| \frac{\mathbf{a_{1}} \cdot \mathbf{a_{2}}}{2} \right|
\end{equation} 
where $\sigma$ is the induced metric determinant on the codimension-two surface, 
$\mathbf{a_{1}}$ and $\mathbf{a_{2}}$ are the null normals
 to the two intersecting codimension-one null surfaces and $\eta=\pm 1$ depending from the orientation of the normals to the surface.
The four joints give the following contributions:
\begin{itemize}
\item
The joint at the UV cutoff  $ z= \varepsilon$ is characterized by the data
\beq
\sqrt{\sigma} = \frac{L}{\varepsilon} 
\, , \qquad
\log \left| \frac{\mathbf{k}^- \cdot \mathbf{k}^+}{2} \right| = \log \left| \alpha^2 \frac{\varepsilon^2}{L^2} \right| \, ,
\eeq
and then from the general expression (\ref{joint}) we find
\beq
I^{\rm cutoff}_{\mathcal{J}} 
= - \frac{L}{4 \pi G} \frac{l}{\varepsilon} \log \le \alpha \frac{\varepsilon}{L} \ri \, .
\eeq

\item
The joint coming from the intersection of the regions with $t>0$ and $t<0$ of the entanglement wedge corresponds to the RT surface and is described by
\beq
\sqrt{\sigma}= \frac{2l L}{l^2 - 4 x^2 } \, , \qquad
\log \left| \frac{\mathbf{w}^+ \cdot \mathbf{w}^-}{2} \right| = \log \left| \beta^2 \frac{l^2 - 4x^2}{4 L^2} \right| \, ,
\eeq
which gives
\beq
I_{\mathcal{J}}^{\rm RT}  
= \frac{L}{4 \pi G} \log \le \frac{\varepsilon}{l} \ri \log \le \frac{\beta^2 \varepsilon l}{L^2} \ri
+ \frac{L \pi}{48 G}  \, .
\eeq

\item
The last two joint terms come from the intersections between the 
null boundaries of the WDW patch and the ones of the entanglement wedge:
\beq
\sqrt{\sigma} =  \frac{4L (l+2 \varepsilon)}{(l-2x+2 \varepsilon)(l+2x + 2 \varepsilon)}  \, ,  
\eeq
\beq
\log \left| \frac{\mathbf{k}^+ \cdot \mathbf{w}^+}{2} \right| = \log \left|  \frac{(l-2x+2\varepsilon)(l+2x+2\varepsilon)}{4L(4x^2 + (l+2\varepsilon)^2)} \right|^2 \, .
\eeq
Therefore they evaluate to
\beq
I^{\rm int}_{\mathcal{J}} 
 = - \frac{L}{2 \pi G} \log \le \frac{\varepsilon}{l} \ri  \log \le \frac{\alpha \beta}{2} 
 \frac{\varepsilon l}{L^2} \ri - \frac{5 \pi L}{48  G} \, .
\eeq
\end{itemize}
Summing all the joint contributions we find
\beq
I_{\mathcal{J}}^{\rm tot} = - \frac{L}{4 \pi G} \frac{l}{\varepsilon} \log \le \alpha \frac{\varepsilon}{L} \ri 
+ \frac{L}{4 \pi G} \log \le \frac{\varepsilon}{l} \ri \log \le \frac{4 L^2}{\alpha^2 \varepsilon l} \ri
- \frac{\pi L}{12 G}  \, . 
\label{jj-tot}
\eeq
Note that the dependence on the normalization constant 
$\beta$ of the normals cancels in (\ref{jj-tot}); this is due to the fact that the null surfaces 
which have the RT surface as boundaries have vanishing expansion parameter $\Theta$.
Also, when summing the joint term (\ref{jj-tot}) with the counterterm contribution (\ref{controt})
 the double logarithmic terms cancel and the dependence on $\a$ disappears.

\subsection{Complexity}
Summing all the contributions, the action complexity is:
\beq
\mathcal{C}_A^{\rm AdS}=\frac{ I_{\rm tot}^{\rm AdS}}{\pi}=
\frac{c}{3 \pi^2 } \left\{
\frac{l}{2 \varepsilon} \log \le \frac{\tilde{L}}{L} \ri 
- \log \le \frac{2 \tilde{L}}{L} \ri \log \le \frac{l}{\varepsilon} \ri + \frac{\pi^2}{8} \right\} \, .
\label{AdS-CA}
\eeq
The calculations is in agreement with \cite{Caceres:2019pgf}.
In the expression for the complexity we recognize
a term proportional to the entanglement entropy of the segment:
\beq
S^{\rm AdS}=\frac{c}{3} \log \left(
\frac{l }{\varepsilon}  \right)  \, .
\eeq
This suggests that the complexity for a single interval 
has a leading divergence  proportional to the length of the subregion on the boundary, 
a subleading divergence proportional to the entanglement entropy and a constant finite piece.
We test this expression for the BTZ case in the next section.


\section{Subregion complexity for a segment in the BTZ black hole}
\label{sect:BTZ}

\begin{figure}[h]
\center
\begin{tabular}{cc}
\includegraphics[scale=0.5]{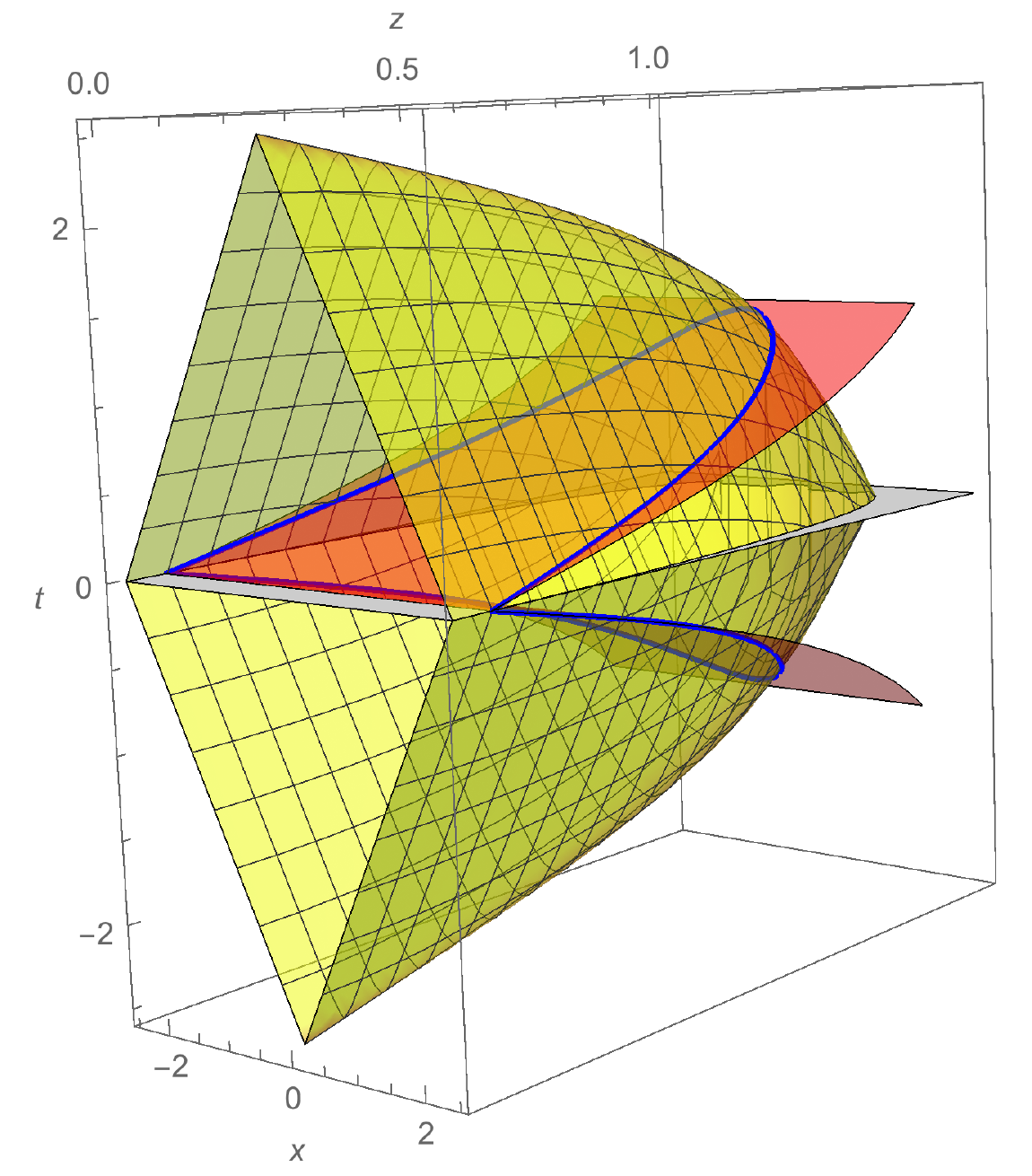} & \includegraphics[scale=0.55]{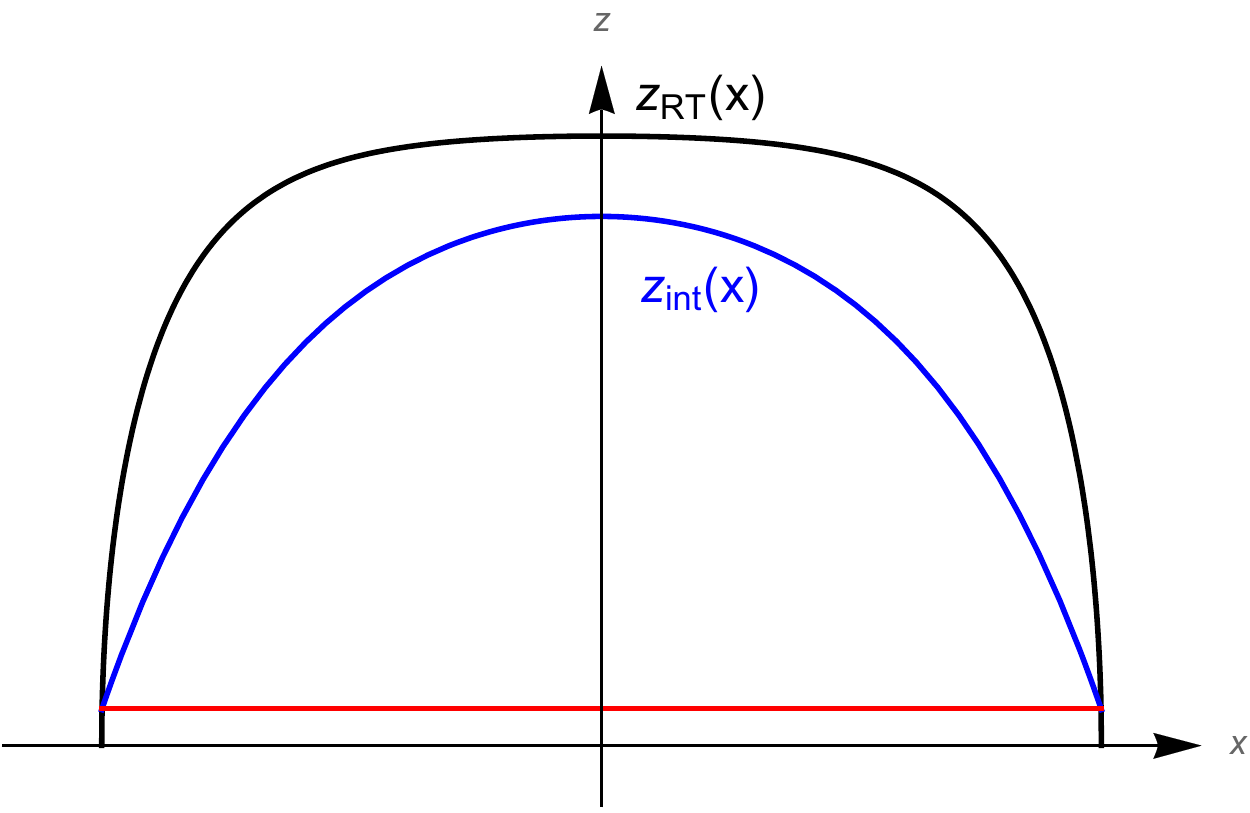}
\end{tabular}
\caption{ Region relevant to the action computation for a segment in the BTZ case, for $l=5$.
Left: Intersection of WDW patch  with entanglement wedge in the $(x,z,t)$ space.
The boundary of the entanglement wedge is in yellow,
while the boundary of the WDW patch is in red.
Right: intersections in the $(x,z)$ plane, with $z_{RT}$ in black, $z_{int}$ in blue
and the cutoff $z=\varepsilon$ in red.}
\label{fig-BTZ}
\end{figure}

We consider the metric of the planar BTZ black hole in $2+1$ dimensions with non-compact coordinates $(t,z,x)$ 
\beq
\label{metric}
ds^2 = \frac{L^2}{z^2} \le - f dt^2 + \frac{dz^2}{f} + dx^2 \ri \, ,
\qquad f = 1- \le \frac{z}{z_h} \ri^2 \, ,
\eeq
where $L$ is the $\mathrm{AdS}$ radius and $z_h$ is the position of the horizon.
The mass, temperature and entropy are:
\beq
M = \frac{L^2}{8 G z_h^2} \, , \qquad
T = \frac{1}{2 \pi z_h} \, , \qquad
S= \frac{\pi L^2}{2 G z_h} \, .
\eeq
The geometry needed to evaluate the subregion complexity for a segment is shown in figure
\ref{fig-BTZ}

The RT surface
 is a spacelike geodesic which lies on a constant time slice $t=0$ and which is
  anchored at the edges of the boundary subregion \cite{Balasubramanian:2011ur}:
\begin{equation}
\label{RT surface}
x_{\pm}(z)= \frac{1}{4} z_h \left[ \log \le \frac{J+1}{J-1} \ri^2 
+  \log \le \frac{z_h^2 -J z^2 \pm \sqrt{z_h^4 - \left( 1+J^{2}  \right) z_h^2 z^{2}+J^{2} z^{4}}}
 {z_h^2 + J z^2 \pm \sqrt{z_h^4 - \left( 1+J^{2}  \right) z_h^2 z^{2}+J^{2} z^{4}}} \ri^2 \right] \, ,
\end{equation}
where
\begin{equation}
J = \coth \left( \frac{l}{2 z_h} \right) \, .
\end{equation}
The turning point of the geodesic is at $x_{\pm} (z_{*})=0$, where
\begin{equation}
z_{*}= z_h \tanh \left( \frac{l}{2 z_h} \right) \, .
\end{equation}
Since $z_{*} < z_h $ for every value of the boundary subregion size $l$,
 the geodesic never penetrates inside the event horizon of the black hole\footnote{In a certain sense, this is opposite in spirit with respect to the motivations for introducing complexity: a quantity measuring the evolution of the Einstein-Rosen bridge connecting the interior of black holes, in a context where the WDW reaches the singularities. In time-dependent configurations, the RT can go deep inside the horizon, while in the static case we observe that it does not happen. Anyway, we will be able to find some important conclusions and a relation with the entanglement entropy.}. 
It is convenient to invert eq. (\ref{RT surface}) for the following computations:
\begin{equation}
\label{zgeod}
 z_{RT} =
 z_h \sqrt{\frac{\cosh \left( \frac{l}{z_h} \right) - 
 \cosh \left( \frac{2x}{z_h} \right)}{\cosh \left( \frac{l}{z_h} \right) +1}} \, .
\end{equation}
In our static case, the entanglement wedge coincides with the causal wedge  \cite{Czech:2012bh,Hubeny:2012wa,Wall:2012uf}, 
which can be  constructed by sending null geodesics from
 the causal diamond on the boundary into the bulk. 
 The explicit expressions of such geodesics 
 are \cite{Hubeny:2012wa}
\bea
 \tilde{x}_{\rm EW} (z,j) &=&  \frac{z_h}{2} \log  \le \frac{\sqrt{z_h^2 + j^2 (z^2 - z_h^2)}+ j z}{\sqrt{z_h^2 + j^2 (z^2 - z_h^2)}- j z} \ri \, ,  \nl
 \tilde{t}_{\rm EW} (z,j)&=& \pm \left[ \frac{l}{2} + \frac{z_h}{2} \log   \le \frac{\sqrt{z_h^2 + j^2 (z^2 - z_h^2)}- z}{\sqrt{z_h^2 + j^2 (z^2 - z_h^2)}+z} \ri \right] \, .
\label{t,x wedge}
\eea
We obtain an analytical expression for the boundary of the entanglement 
wedge in terms of a unique explicit relation between $(t,z,x)$ by determining $j=j(z,x)$ 
from the first equation in  (\ref{t,x wedge})
and then inserting it 
 into the second equation of
   (\ref{t,x wedge}). The result
   can be written as
\begin{equation}
\label{twedge}
t_{\rm EW} =\pm \left[
 \frac{l}{2} - z_h \, \mathrm{arccoth}  \le \frac{\sqrt{2} z_h \cosh \le \frac{x}{z_h} \ri}{\sqrt{2z^2 + z_h^2 \cosh \le \frac{2x}{z_h} \ri - z_h^2}}  \ri \right]
 \, .
\end{equation}
The WDW patch is delimited by the radial null geodesics
\beq
t_{\mathrm{WDW}} =\pm \frac{z_h}{4} 
\log \le \frac{z_h+z}{z_h-z} \frac{z_h-\varepsilon}{z_h+\varepsilon}
\ri^2 \, .
\eeq
We will need again the intersection between the boundaries of the WDW patch and of the entanglement wedge 
\beq
\label{xintpos}
t_{\rm int}=t_{\mathrm{WDW}} \, ,  \qquad
 z_{\rm int} = z_h \frac{\cosh \left[ \frac{l}{2 z_h} + \mathrm{arctanh} \le \frac{\varepsilon}{z_h} \ri  \right] - \cosh \le \frac{x}{z_h} \ri}{\sinh \left[ \frac{l}{2 z_h} + \mathrm{arctanh} \le \frac{\varepsilon}{z_h} \ri   \right]} \, . 
 \eeq
This curve is plotted in fig. \ref{fig-BTZ}. 
As in the AdS case, we denote by $ x_{\rm max} $ the maximum value of the transverse coordinate, 
which is reached when we evaluate the RT surface at $ z= \varepsilon$:
\beq
x_{\rm max} \equiv x_{\rm RT} (z= \varepsilon) = 
z_h \, \mathrm{arccosh}  \left[ \sqrt{1- \frac{\varepsilon^2}{z_h^2}} \, \cosh \le \frac{l}{2 z_h} \ri \right] \, .
\label{xmax}
\eeq

\subsection{Bulk contribution}
\label{sect-Bulk contribution 2nd regularization}

We split the integration region as in the AdS case, see
eqs. (\ref{integralozzo1},\ref{integralozzo2}).
The total bulk action is given by $I_{\cal V} = 4 ( I^1_{\cal V}  + I^2_{\cal V}), $ which combine into the expression 
\beq
\begin{aligned}
I_{\cal V}   =  \frac{L}{8 \pi G z_h} & \int_{0}^{x_{max}(\varepsilon)} dx \, \left\lbrace \frac{4 \sinh \left[ \frac{l}{2 z_h} + \mathrm{arctanh}  \le \frac{\varepsilon}{z_h} \ri  \right]}{\cosh \le \frac{l}{2 z_h} + \mathrm{arctanh} \le \frac{\varepsilon}{z_h} \ri \ri - \cosh \le \frac{x}{z_h} \ri }  -\frac{4 z_h}{ \varepsilon }  \right. \\
& \left.  + 2 \coth \le \frac{x}{z_h} \ri \log \left| \frac{\sinh \le \frac{l-2x}{2 z_h} \ri  \sinh^2 \left[ \frac{l+2x+2z_h \, \mathrm{arctanh} (\varepsilon/z_h)}{4 z_h} \right]}{\sinh \le \frac{l+2x}{2 z_h} \ri  \sinh^2 \left[ \frac{l-2x+2z_h \, \mathrm{arctanh} (\varepsilon/z_h)}{4 z_h} \right]}  \right| \right\rbrace \, .
\end{aligned}
\label{final integrand bulk}
\eeq
This integral can be computed analytically, giving
\beq
I_{\cal V} = - \frac{L}{4 \pi G} \frac{l}{\varepsilon} - \frac{L}{2 \pi G}  \log \le \frac{\varepsilon}{l} \ri + \frac{L}{2 \pi G}  \log \left[ \frac{2 z_h}{l} \sinh \le \frac{l}{2 z_h} \ri \right] + \frac{\pi L}{16 G} \, . 
\eeq

\subsection{Null normals}
In order to compute the counterterms due to the null surfaces and
the joint contributions, the null normals are needed.
It is convenient to use an affine parameterization, which can
be found using the following Lagrangian description of geodesics:
\begin{equation}
\label{lagrangian}
\mathcal{L} = \frac{L^2}{z^2} \le - f(z) \, \dot{t}^{2} + \frac{\dot{z}^{2}}{f(z)} + \dot{x}^{2} \ri
\end{equation}
where the dot represents the derivative with respect to the affine parameter $\lambda$. 
Since the Lagrangian does not depend on $t$ and $x$, we have two constants of motion
\begin{equation}
\label{motion}
E = - \frac12 \frac{\p \mathcal{L}}{\p \dot{t}} = \frac{L^2}{z^2} f(z) \, \dot{t} \, , \qquad J = \frac12 \frac{\p \mathcal{L}}{\p \dot{x}} =\frac{L^2}{z^{2}} \dot{x} \, .
\end{equation}
Imposing the null condition $\mathcal{L} = 0$ and making use of eq. (\ref{motion}) leads to
\begin{equation}
\label{zdot}
\dot{z} = \pm \, \frac{z^2}{L^2} \, \sqrt{E^{2} - J^{2} f(z)} \, .
\end{equation}
Therefore, from eqs. (\ref{motion}) and (\ref{zdot}), the tangent vector to the null geodesic is
\begin{equation}
\label{nullvector}
V^{\mu} = \left( \dot{t}, \, \dot{z}, \, \dot{x} \right) = \left(\frac{z^2}{L^2 f(z)} E , \, \pm \, \frac{z^2}{L^2} \, \sqrt{E^{2} - J^{2} f(z)} , \, \frac{z^2}{L^2} \, J \right)  \, .
\end{equation}
Lowering the contravariant index with the metric tensor, we get the normal one-form to the null geodesic
\begin{equation}
\label{nullform}
\mathbf{V} = V_{\mu} dx^{\mu} = - E \, dt \, \pm \, \frac{\sqrt{E^{2} - J^{2} f(z)}}{f(z)} \, dz + J \, dx \, .
\end{equation}

The null geodesics which bound the WDW patch are $x$-constant curves,
and then they correspond to the choice $J=0$. This gives the normals
\begin{equation}
\label{wdwnormals}
\mathbf{k^{+}}= k_{\mu}^{+} dx^{\mu} = \alpha \left( dt - \frac{dz}{f(z)} \right) \, ,
 \qquad \mathbf{k^{-}}= k_{\mu}^{-} dx^{\mu} = \alpha \left( - \, dt - \frac{dz}{f(z)} \right) \, ,
\end{equation}
where $\a$ is an arbitrary constant.

The null geodesics that bound the entanglement wedge are normal to the RT surface,
\emph{i.e.}
\begin{equation}
V_{\mu} \, \frac{dX^{\mu}_{RT} (x)}{dx} = 0 \, , \qquad 
X^{\mu}_{\rm RT} (x) = \left( 0, z_{RT}, \, x \right) \, ,
\label{rtvector}
\end{equation}
where $z_{RT}$ is given in eq. (\ref{zgeod}). 
With this condition and eqs. (\ref{nullform}) and (\ref{rtvector}), 
we find a relation between the two constants of motion $E$ and $J$ which  gives (for $t>0$ and $t<0$ respectively)
\beq
\mathbf{w^{\pm}} = w_{\mu}^{\pm} dx^{\mu} =
\beta \left( \pm dt + a \, dz +b \, dx \right)  \, , 
\eeq
where
\beq
a = \frac{e^{-\frac{x}{z_h}} \left( e^{\frac{2x}{z_h}}+1 \right) z z_h^2}{\left(z_h^2-z^{2} \right) \sqrt{4 z^2 + e^{- \frac{2x}{z_h}} \le e^{\frac{2x}{z_h}} -1 \ri^2 z_h^2}} \, , \qquad
b=\frac{e^{-\frac{x}{z_h}} \left( e^{\frac{2x}{z_h}}-1 \right) z_h}{\sqrt{4 z^2+ e^{- \frac{2x}{z_h}} \le e^{\frac{2x}{z_h}} -1 \ri^2 z_h^2}} \, .
\eeq

\subsection{Null boundaries and counterterms}
\label{sect-BTZ-countertems}

The term in eq. (\ref{null-bou}) vanishes because we used an affine parameterization.
The counterterm in eq. (\ref{countertermlabelbis}) gives:
\begin{itemize}
\item For the null normals of the boundary of the entanglement wedge, 
this contribution vanishes because $\Theta=D_{\mu} (w^{\pm})^{\mu} = 0$.
\item For the null normals of the boundary of the WDW patch, a direct calculation gives $\Theta = \frac{\alpha z}{L^2}$ and:
\beq
\begin{aligned}
I_{\rm ct}^{\rm WDW} & = - \frac{L}{2 \pi G} \int_0^{x_{\rm max}} dx \int_{\varepsilon}^{z_{\rm int} (x)} dz \, \frac{1}{z^2} \log \left| \frac{\tilde{L}}{L^2} \, \alpha z \right| = \\
& =  \frac{L}{2 \pi G} \int_0^{x_{\rm max}} dx  \, \left\lbrace \frac{1 + \log \left| \frac{\tilde{L}}{L^2} \, \alpha \varepsilon \right|}{\varepsilon} + \frac{\sinh \le \frac{l}{2 z_h} + \mathrm{arctanh} \le \frac{\varepsilon}{z_h} \ri \ri}{z_h \left[ \cosh \le \frac{x}{z_h} \ri - \cosh \le \frac{l}{2 z_h} \ri + \mathrm{arctanh} \le \frac{\varepsilon}{z_h} \ri  \right]}  \times \right. \\
& \left. \times \le 1 + \log \left| \frac{\tilde{L} z_h \alpha}{L^2} \frac{\cosh \le \frac{l}{2 z_h} + \mathrm{arctanh} \le \frac{\varepsilon}{z_h} \ri \ri - \cosh \le \frac{x}{z_h} \ri}{\cosh \le \frac{l}{2 z_h} + \mathrm{arctanh} \le \frac{\varepsilon}{z_h} \ri \ri} \right|  \ri  \right\rbrace \, .
\end{aligned}
\eeq
\end{itemize}

\subsection{Joint contributions}
\label{sect-BTZ-joints}

We evaluate the joint terms in eq (\ref{joint}):
\begin{itemize}
\item The joint at the cutoff surface:
\begin{equation}
I_{\mathcal{J}}^{\rm cutoff} = - \, \frac{L}{4 \pi G} \, \int_{0}^{x_{\rm max} } \frac{dx}{\varepsilon} \left| \frac{\alpha^{2} \, z_h^2 \, \varepsilon^{2}}{L^2 (z_h^2-\varepsilon^{2})} \right| \, .
\end{equation}
\item The joint at the RT surface:
\begin{equation}
I_{\mathcal{J}}^{\rm RT} = - \, \frac{L}{4 \pi G z_h} \, \int_{0}^{x_{\rm max} } dx \, \frac{\sinh \le \frac{l}{z_h} \ri}{\cosh \le \frac{l}{z_h} \ri - \cosh \le \frac{2x}{z_h} \ri} \log \left| \frac{\beta^2 z_h^2}{2 L^2} \frac{\cosh \le \frac{l}{z_h} \ri - \cosh \le \frac{2x}{z_h} \ri}{\cosh^2 \le \frac{x}{z_h} \ri} \right| \, .
\label{joint-BTZ-RT}
\end{equation}
\item The two joints coming from the intersection between the null boundaries of the 
WDW patch and the ones of the entanglement wedge:
\begin{equation}
\begin{aligned}
I_{\mathcal{J}}^{\rm int} & =  \frac{L}{2 \pi G z_h} \int_{0}^{x_{\rm max}} dx \, 
\frac{\sinh \le \frac{l}{2 z_h} + \mathrm{arctanh} \le \frac{\varepsilon}{z_h} \ri \ri}{\cosh \le \frac{l}{2 z_h} + \mathrm{arctanh} \le \frac{\varepsilon}{z_h} \ri \ri - \cosh \le \frac{x}{z_h} \ri} \times \\
& \times \log \left| \frac{e^{x/z_h} \alpha \beta z_h^2}{L^2} \frac{\left[ \cosh \le \frac{l}{2 z_h} + \mathrm{arctanh} \le \frac{\varepsilon}{z_h} \ri \ri - \cosh \le \frac{x}{z_h} \ri \right]^2}{ 1 + e^{2x/z_h} \cosh \le \frac{l}{2 z_h} + \mathrm{arctanh} \le \frac{\varepsilon}{z_h} \ri \ri - 2 e^{x/z_h} } \right| \, .
\end{aligned}
\label{joint-BTZ-EW}
\end{equation}
\end{itemize}
All the joints contributions and the counterterm are correctly regularized by the prescription induced from the UV cutoff at $z=\varepsilon$.

\subsection{Complexity}

We performed all the integrals analytically and we
 further simplified the result using various dilogarithm identities, including the relation
\bea
  8 \, \mathrm{Re} \left[  \mathrm{Li}_2 \le \frac{1 + i e^{\frac{y}{2}}}{1+ e^{\frac{y}{2 }}} \ri 
 - \mathrm{Li}_2 \le \frac{1}{1 + e^{\frac{y}{2 }}} \ri - \mathrm{Li}_2 \le 1 + i e^{\frac{y}{2 }} \ri
   - \mathrm{Li}_2 \le \frac{e^{\frac{y}{2 }} -i}{1+ e^{\frac{y}{2 }}} \ri
 \right] =
\nl
 =  -  \frac{7 \pi^2 }{6 }
+  4 \le \log \le 1 + e^{\frac{y}{2}} \ri \ri^2 
+ \log 2 \left[  2 y - 4  \log \le \frac{e^{y} -1}{y}  \ri  
 + 4 \log \left( \frac{2 }{y} \sinh \frac{y}{2 }  \right) \right] \, ,
\eea
which can be proved by taking a derivative of both side of
the equation with respect to $y$.
In this way the action subregion complexity is given by
\beq
 \mathcal{C}_{A}^{\rm BTZ} = \frac{c}{3 \pi^2 } 
 \left\{ \frac{l}{2 \varepsilon} \log \le \frac{\tilde{L}}{L} \ri -
 \log \le \frac{2 \tilde{L}}{L} \ri
  \log \left( \frac{2 z_h}{\varepsilon} \sinh \le \frac{l}{2 z_h} \ri \right) 
 +\frac{\pi^2}{8} \right\} \, .
\label{total action BTZ simplified}
\eeq
Introducing the entanglement entropy of an interval
\beq
S^{\rm BTZ}=\frac{c}{3} \log \left(
\frac{2 z_h }{\varepsilon} \sinh \left( \frac{l}{2 z_h} \right)
\right)  \, ,
\label{entanglement entropy BTZ}
\eeq
we can then write it in the form
\beq
 \mathcal{C}_{A}^{\rm BTZ} = \frac{l}{\varepsilon} \frac{c}{6 \pi^2} \log \left(\frac{\tilde{L}}{L} \right)
- \log  \left(\frac{2\tilde{L}}{L} \right) \frac{S^{\rm BTZ}}{\pi^2} + \frac{1}{24} c \, .
\label{ACTION-C}
\eeq
The divergencies of eqs. (\ref{ACTION-C}) are the same 
as in the AdS case eqs. (\ref{AdS-CA}), which is recovered for $z_h=0$.

A non-trivial cross-check can be done in the 
 $l  \gg z_h$ limit. Keeping just the terms linear in $l$
 in eq. (\ref{total action BTZ simplified}), we find agreement
 with the subregion complexity $\mathcal{C}_A^{\rm{BTZ},R}$ computed for one side of the Kruskal diagram,
 see \cite{Agon:2018zso,Alishahiha:2018lfv}:
 \beq
\mathcal{C}_A^{\rm{BTZ},R}= \frac{ c}{6 }
\frac{l  }{ \pi^2 } \left[
 \frac{1}{\varepsilon} \log \le \frac{\tilde{L}}{L} \ri
 - \frac{1}{ z_h}  \log \le \frac{2 \tilde{L}}{L} \ri \right] \, .
\label{external total action BTZ}
\eeq
Note that in this limit the $\log \varepsilon$ divergence disappears because
it is suppressed by the segment length $l$.

For comparison, the volume complexity  of an interval for the BTZ
\cite{Alishahiha:2015rta,Abt:2017pmf} is:
\beq
\mathcal{C}_V^{\rm BTZ} =\frac{2 \, c}{3} \le  \frac{l}{\varepsilon} - \pi \, \ri ,
\eeq
and it is non-trivially independent on temperature.
Subregion $\mathcal{C}_V$ at equilibrium is a topologically protected quantity:
for multiple intervals, the authors of \cite{Abt:2017pmf} found the following result
using the the Gauss-Bonnet theorem
\beq
\mathcal{C}_V^{\rm AdS}=  \mathcal{C}_V^{\rm BTZ} = \frac{2 \, c}{3} \le \frac{l_{tot}}{\varepsilon} + \kappa \ri \, ,
\label{CCVA}
\eeq
where $l_{tot}$ is the total length of all the segments and $\kappa$
is the finite part, that depends on topology
\beq
\kappa=- 2 \pi \chi + \frac{\pi}{2} \, m \, ,
\label{CCVB}
\eeq
where $\chi$ is the Euler characteristic  of the extremal surface
(which is equal to $1$ for a disk)
and $m$ is the number of ninety degrees junctions between RT 
surface and boundary segments. 
It would be interesting to see if 
 a similar result could be established for the  CA conjecture.
This motivates us to study the two segment case in the next section.



\section{Subregion complexity for two segments in AdS$_3$}
\label{sect:AdS-2seg}

\begin{figure}
\begin{center}
\includegraphics[scale=0.6]{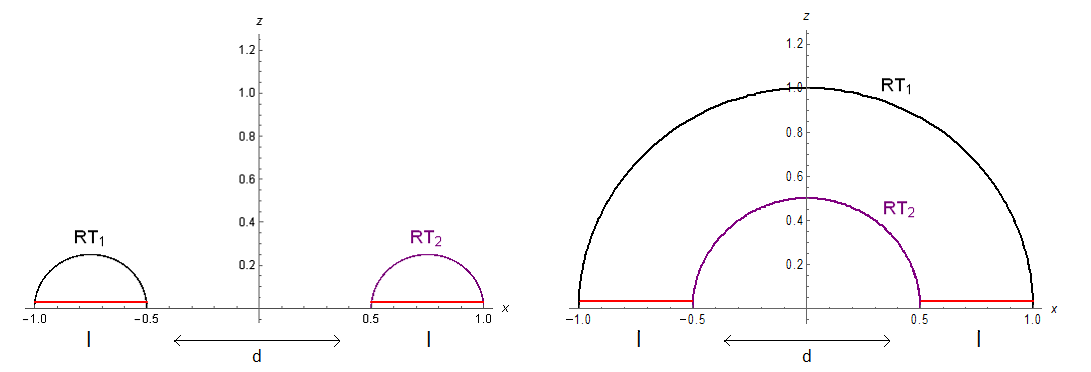}
\caption{The possible RT surfaces for disjoint subregions of length $l=0.5$ with a separation $d=1$, 
on the slice $t=0$. }
\label{RT-disjoint}
\end{center}
\end{figure}

In this section we  evaluate the holographic subregion action complexity 
for a  disjoint subregion on the boundary of $\mathrm{AdS}_{3}$ spacetime. 
We consider two segments of size $l$ with a separation equal to $d$,
 located on the constant time slice $t=0$. 
 For simplicity, we work with a symmetric configuration, 
 in which the two boundary subregions are respectively given by 
 $x \in \left[  -l - d/2 , -d/2 \right]$ and $x \in \left[ d/2 ,l + d/2 \right]$.  
According to the values of the subregions size $l$ and of the separation $d$,
 there are two possible extremal surfaces anchored at the boundary
  at the edges of the two subregions \cite{{Rangamani:2016dms,Headrick:2019eth}}:
\begin{itemize}
 \item The extremal surface (which in this number of dimension is a geodesic)
 is given by the union of the RT surfaces for the individual subregions.
 This is the minimal surface for $d>d_0$, where $d_0$ is a critical distance.
 \item The extremal surface connects the two subregions. 
 This configuration is minimal for $d< d_0$.
\end{itemize} 
The two cases are shown in Fig. \ref{RT-disjoint}. 
The geodesic with the minimal area provides the holographic entanglement entropy 
for the union of the disjoint subregions. 
The critical distance corresponds to the distance for which both the extremal surfaces
have the same  length, \emph{i.e.}
\beq
d_0=(\sqrt{2}-1) l \, .
\eeq

\begin{figure}[h]
\center
\begin{tabular}{ccc}
\includegraphics[scale=0.5]{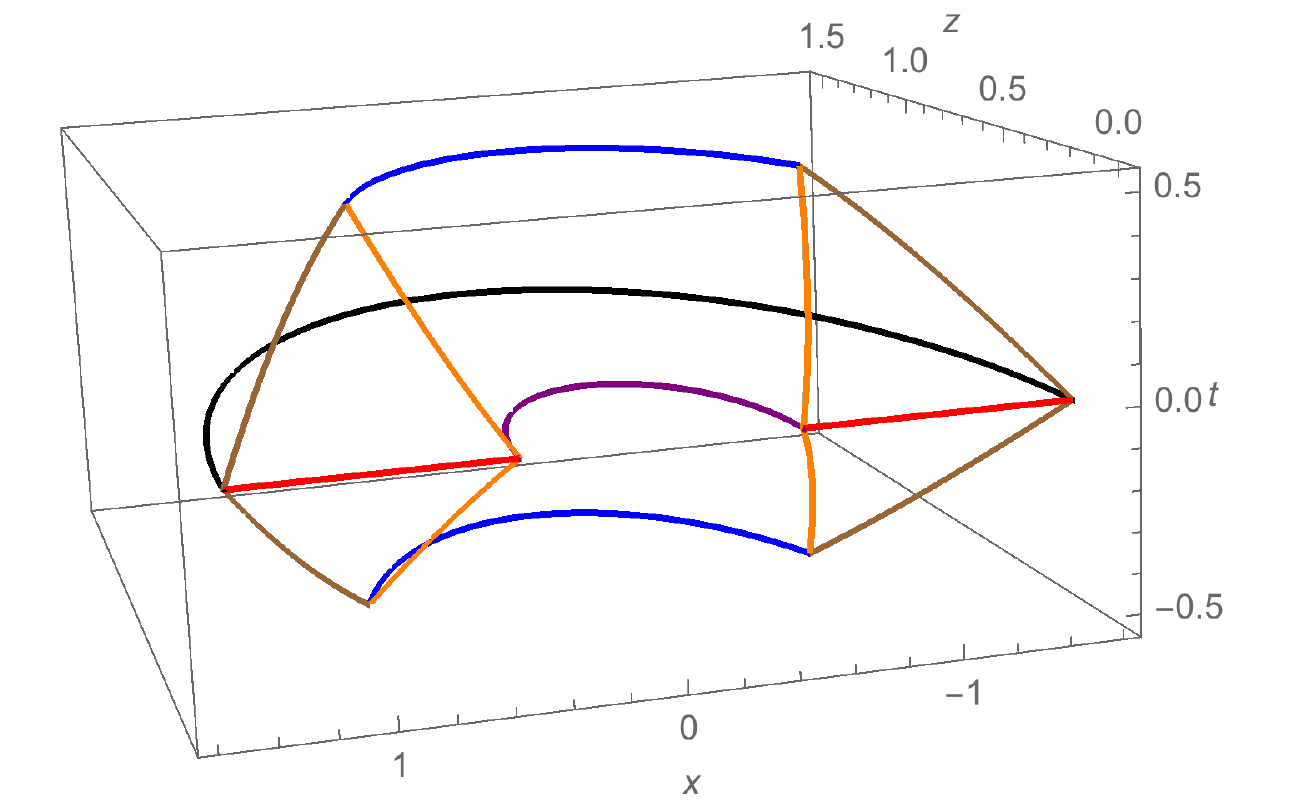} & \includegraphics[scale=0.5]{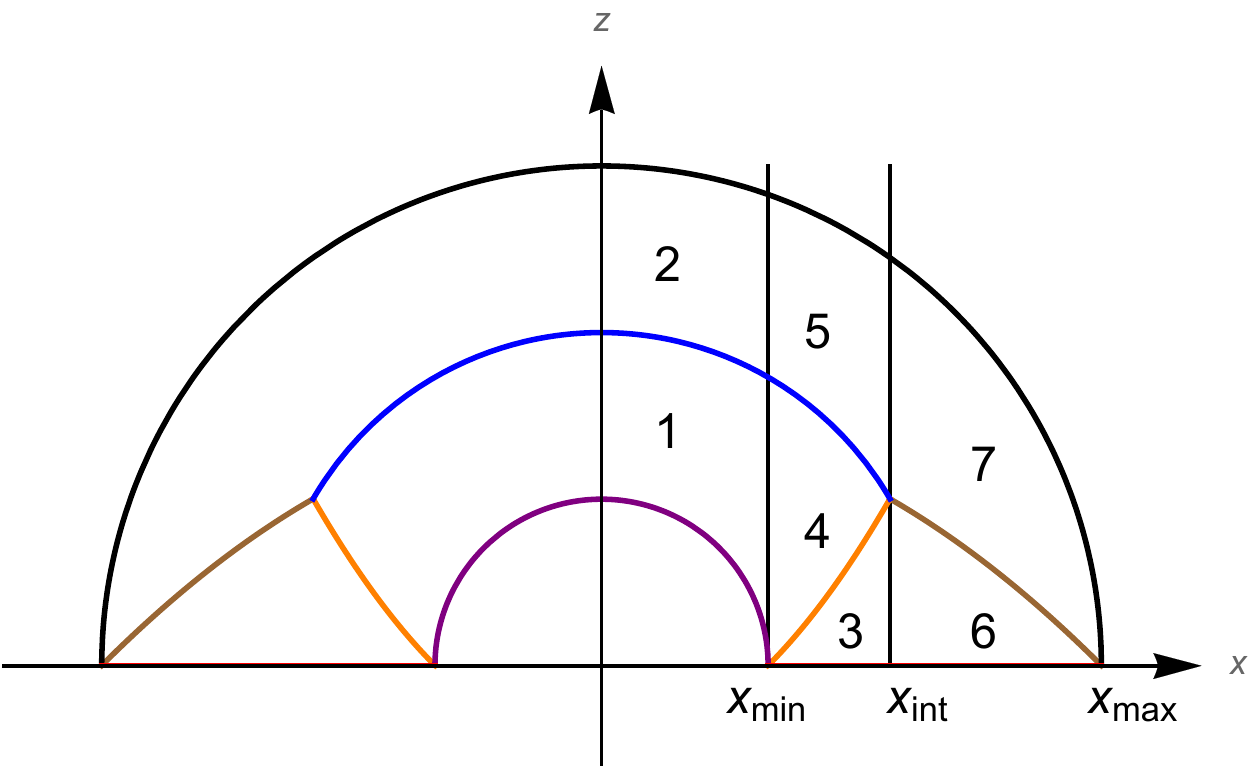} &
 \includegraphics[scale=0.5]{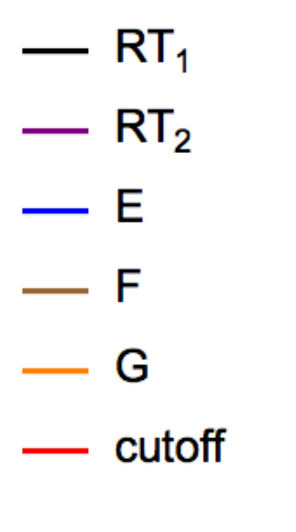}
\end{tabular}
\caption{Left: Bulk region relevant to the action subregion calculation for 
two segments in AdS.
 Right: projection in the $(x,z)$ plane. The regions in which the bulk integral 
 is splitted are numbered. }
\label{geometry-disjoint}
\end{figure}

In the first configuration (see left in Fig. \ref{RT-disjoint}), 
we have two non-intersecting entanglement wedges
and then
\begin{equation}
\label{disjoint-complexity-case1}
\mathcal{C}_A^1=2 \,  \mathcal{C}_A^{\rm AdS}  \, .
\end{equation}
For the second configuration (right in Fig. \ref{RT-disjoint}),
 we must perform a new computation. The spacetime region of interest is 
 symmetric both with respect to the $x=0$ slice and to the $t=0$ one. 
As a consequence, we can evaluate the action on the region with $t>0$ and $x>0$
 and introduce opportune symmetry factors. A schematic representation is shown in fig.
 \ref{geometry-disjoint}.

The RT surface is the union of the spacelike geodesics anchored 
at the edges of the region $x \in \left[ -l-d/2 , l+d/2 \right]$ and
 $x \in \left[ -d/2 , d/2 \right]$.
 We will denote such geodesics as RT$_1$ and RT$_2,$ respectively:
\begin{equation}
\label{zRT-small-big}
z_{RT_1} (x) = \sqrt{\left( \frac{2l+d}{2} \right)^{2} -x^{2}} \, , 
\qquad \qquad  z_{RT_2} (x) = \sqrt{\left( \frac{d}{2} \right)^{2} -x^{2}} \, . 
\end{equation}
With the introduction of the cutoff surface at $z= \varepsilon$, 
RT$_2$ is truncated at $x=x_{min} $ 
and RT$_1$ at $x= x_{max}$, defined by
\beq
x_{min} \equiv x_{\rm RT_2} (z=\varepsilon) = \sqrt{\le \frac{d}{2} \ri^2 - \varepsilon^2} \, ,
\qquad
x_{max} \equiv x_{\rm RT_1} (z=\varepsilon) = \sqrt{\le \frac{d+ 2 l}{2} \ri^2 - \varepsilon^2} \, .
\eeq
 The null boundaries of the entanglement wedge can be built by 
 sending null geodesics from  RT$_1$ and RT$_2$:
\begin{equation}
\label{tEW}
t_{EW_1}  = \frac{2l+d}{2} - \sqrt{z^{2} + x^{2}} \, ,
\qquad
t_{EW_2} = - \frac{d}{2} + \sqrt{z^{2} + x^{2}} \, .
\end{equation}
The WDW patch, anchored at the cutoff in the present regularization, is bounded by the null surface
\begin{equation}
\label{wdw-ads}
t_{WDW} = z - \varepsilon \, .
\end{equation} 
The intersection curve $E$ between the null boundaries of the entanglement wedge,
(built from  RT$_1$ and RT$_2$, see eq. (\ref{tEW})) is 
\begin{equation}
\label{zEW-EW}
t_{E} = \frac{l}{2} \, , \qquad \qquad z_{E} = \frac{1}{2} \sqrt{\left( d+l \right)^{2} -4 x^{2}} \, .
\end{equation}
The intersection $F$ between the boundary of the
WDW patch eq. (\ref{wdw-ads})
and the null surface anchored at RT$_1$ is:
\beq
\label{zWDW-EWbig}
t_{F} = \frac{1}{4} \left[ d + 2 \left( l - \varepsilon \right) - \frac{4 x^{2}}{d+ 2 \left( l + \varepsilon \right)} \right] \, , 
\qquad
 z_{F}  =t_{F}  + \varepsilon \, .
\eeq 
The  intersection $G$
 between the WDW patch eq. (\ref{wdw-ads})
  and the null surface anchored at RT$_2$ gives
\begin{equation}
\label{zWDW-EWsmall}
t_{G}  = - \frac{d}{4} + \frac{x^{2}}{d - 2 \varepsilon} - \frac{\varepsilon}{2} \, , 
\qquad z_{G}  = t_{G}  + \varepsilon\, .
\end{equation}
 The intersection among the three curves described above 
 (obtained solving the condition $z_{E}= z_{F}= z_{G}$) gives
\beq
x_{int} (\varepsilon) = \frac{\sqrt{\le d-2 \varepsilon \ri \left[ d+2 \le l+ \varepsilon \ri \right]}}{2} \, .
\eeq

\subsection{Bulk contribution}

As shown in Fig. \ref{geometry-disjoint}, the total bulk contribution can be divided into 7 parts for computational reasons:
\beq
I_{\cal V}= 4 \sum_{i=1}^{7} I_{bulk}^{i} \, ,
\eeq
where
\bea
I_{\cal V}^{1} &=& - \frac{L}{4 \pi G} \int_{0}^{x_{min} } dx \int_{z_{RT_2} }^{z_{E} } dz \int_{0}^{t_{EW_2} } \frac{dt}{z^3} \nl
I_{\cal V}^{2}  &=&- \frac{L}{4 \pi G} \int_{0}^{x_{min} } dx \int_{z_{E} }^{z_{RT_1} } dz \int_{0}^{t_{EW_1} } \frac{dt}{z^3}  \nl
I_{\cal V}^{3} &=& - \frac{L}{4 \pi G} \int_{x_{min} }^{x_{int} } dx \int_{\varepsilon}^{z_{G} } dz \int_{0}^{t_{WDW}} \frac{dt}{z^3} \nl
I_{\cal V}^{4} & = &- \frac{L}{4 \pi G} \int_{x_{min} }^{x_{int} } dx  \int_{z_{G} }^{z_{E} } dz \int_{0}^{t_{EW_2} } \frac{dt}{z^3}  \nl
I_{\cal V}^{5}  & = & - \frac{L}{4 \pi G} \int_{x_{min} }^{x_{int} } dx \int_{z_{E} }^{z_{RT_1} } dz \int_{0}^{t_{EW_1}} \frac{dt}{z^3} \nl
I_{\cal V}^{6}  &=& - \frac{L}{4 \pi G} \int_{x_{int} }^{x_{max} } dx \int_{\varepsilon}^{z_{F} } dz \int_{0}^{t_{WDW} } \frac{dt}{z^3} \nl
I_{\cal V}^{7} & = & - \frac{L}{4 \pi G} \int_{x_{int} }^{x_{max} } dx \int_{z_{F} }^{z_{RT_1} } dz \int_{0}^{t_{EW_1} } \frac{dt}{z^3} \, . 
\eea
All the integrals can be evaluated analytically. Since the expressions are rather 
cumbersome, we only write just the total expression 
\bea
\label{I-tot-bulk-disjoint-1}
I_{\cal V} &=& - \frac{c}{6 \pi} \left\{ \frac{2 l}{\varepsilon} - 2 \log \frac{d(d+2l)}{\varepsilon^2}
 + \frac{ \pi^2}{2 }   + 8 \, \mathrm{arctanh} \, \sqrt{\frac{d}{d+2l}}  \right. \nl
&& \,\,\,\,\,\, \left.
- 2  \left[ {\rm Li}_{2} \le \frac{\sqrt{d (d+2l)}}{d+l} \ri - {\rm Li}_{2} \le - \frac{\sqrt{d (d+2l)}}{d+l} \ri \right]  \right\} \, .
\eea
We already expressed the result in terms of the central charge for later convenience.

\subsection{Counterterms}

The counterterms for the null boundaries of the entanglement wedge vanish as usual.
We can separate the counterterm for the null boundaries of the WDW patch in two contributions:
\bea
 I_{ct,I} &=& \frac{L}{2 \pi G} \int_{x_{min} }^{x_{int} } dx \int_{\varepsilon}^{z_{G} } \frac{dz}{z^2} \, \log \le \frac{\tilde{L} \, \alpha \, z}{L^2} \ri  \, , \nl 
 I_{ct,II}&=&  \frac{L}{2 \pi G} \int_{x_{int} }^{x_{max} } dx \int_{\varepsilon}^{z_{F} } \frac{dz}{z^2} \, \log \le \frac{\tilde{L} \, \alpha \, z}{L^2} \ri \, .
\eea

\subsection{Joint contributions}

We have to include several joint contributions to the action:
\begin{itemize}
\item Joints on the cutoff at $z=\varepsilon$. The null normals are
\beq
\label{normals-wdw-disjoint}
\mathbf{k^{\pm}} = \alpha \le \pm dt - dz \ri \, ,
\eeq
and the contribution is:
\beq
I_{\varepsilon} = - \frac{L}{4 \pi G} \int_{x_{min} }^{x_{max}} dx \, \frac{\log \le \frac{\alpha^2 \varepsilon^2}{L^2} \ri }{\varepsilon} 
= - \frac{L}{2 \pi G} \frac{l \, \log \le \frac{\alpha \, \varepsilon}{L} \ri}{\varepsilon} \, .
\eeq

\item Joint on RT$_1$. The null normals to such surfaces are
\beq
\label{normals-wedge-big}
\mathbf{w_{1}^{\pm}} = \beta \le \pm dt + \frac{z}{\sqrt{z^2 + x^2}} dz + \frac{x}{\sqrt{z^2 + x^2}} dx \ri \, ,
\eeq
which gives
\beq
\begin{aligned}
I_{RT_1} & = - \frac{L}{2 \pi G} \int_{0}^{x_{max} } dx \, \frac{d+2 l}{\le d+ 2 l \ri^2 -4 x^2}
 \log  \frac{\beta^2  \left[ \le d+2 l \ri^2 - 4 x^2 \right]}{4 L^2} = \\
& = \frac{L}{4 \pi G} \log \le \varepsilon \ri \log \le \frac{\beta^2 \varepsilon}{L^2} \ri - \frac{L}{4 \pi G} 
\log \le d+ 2 l \ri \log \frac{\le d+2 l \ri \beta^2}{L^2} + \frac{L \pi}{48 G}  \, .
\end{aligned}
\eeq

\item Joint on RT$_2$. The null normals to these surfaces are
\beq
\label{normals-wedge-small}
\mathbf{w_{2}^{\pm}} = \gamma \le \pm dt - \frac{z}{\sqrt{z^2 + x^2}} dz - \frac{x}{\sqrt{z^2 + x^2}} dx \ri \, ,
\eeq
and the action is:
\beq
\begin{aligned}
I_{RT_2} & = - \frac{L}{2 \pi G} \int_{0}^{x_{min} } dx \, \frac{d}{d^ 2 -4 x^2} \log  \frac{\gamma^2  \le d^2 - 4 x^2 \ri}{4 L^2} = \\
& = \frac{L}{4 \pi G} \log \le \varepsilon \ri \log \le \frac{\gamma^2 \, \varepsilon}{L^2} \ri 
- \frac{L}{4 \pi G} \log \le d \ri \log \frac{d \, \gamma^2}{L^2} + \frac{L \pi}{48 G} \, .
\end{aligned}
\eeq

\item Joints between the two null boundaries of the entanglement wedge, curve $E$.
The normals are $ \mathbf{w_{1}^{+}} $ and $\mathbf{w_{2}^{+}}$.
The contribution gives 
\beq
I_{E}  = \frac{L}{\pi G} \int_{0}^{x_{int} } dx \, \frac{d+l}{\le d+l \ri^2 -4 x^2} \log \left[ \frac{\beta \, \gamma \left[ \le d+l \ri^2 -4 x^2 \right]}{4 L^2} \right] \, .
\eeq

\item Joint between  the null boundary of the WDW patch and the null boundary 
of the entanglement wedge anchored at RT$_1$, curve $F$.
The  normals are $\mathbf{k^{+}}$ and $\mathbf{w_{1}^{+}} $.
The term gives
\beq
I_{F}  = \frac{2 L}{\pi G} \int_{x_{int} }^{x_{max} } dx \, \frac{d+ 2 \le l + \varepsilon \ri}{\le d+ 2 \le l + \varepsilon \ri \ri^2 - 4 x^2} \log \left[ \frac{\alpha \, \beta \left[ \le d + 2 \le l + \varepsilon \ri \ri^2 - 4 x^2 \right]^2}{16 L^2 \left[ \le d + 2 \le l + \varepsilon \ri \ri^2 + 4 x^2 \right]} \right] \, .
\eeq

\item Joint between
 the null boundary of the WDW patch and the null boundary of the entanglement wedge anchored at RT$_2$ (curve $G$)
 with normals $\mathbf{k^{+}} $ and $ \mathbf{w_{2}^{+}}$.
 The contribution gives
\beq
I_{G}  = \frac{2 L}{\pi G} \int_{x_{min} }^{x_{int} } dx \, \frac{d-2 \varepsilon}{4 x^2 - \le d - 2 \varepsilon \ri^2 } \log \left[ \frac{\alpha \, \gamma \le d -2 \varepsilon +2 x \ri^2 \le d-2 \varepsilon -2 x \ri^2}{16 L^2 \left[ 4 x^2 + \le d-2 \varepsilon \ri^2 \right]} \right] \, .
\eeq

\end{itemize}

\subsection{Complexity}
Adding up all the contributions and using polylog identities, we find:
\beq
\label{CA2}
\begin{aligned}
\mathcal{C}_A^2 & =  \frac{c}{3\pi^2 }  \left\{ \log \le \frac{\tilde{L}}{L} \ri \frac{l}{\varepsilon} 
 -  \log \le \frac{2 \tilde{L}}{L} \ri \log \le \frac{d(d+2l)}{\varepsilon^2} \ri - \frac{\pi^2}{4}  \right.  \\
& + \left[  \log \le \frac{\tilde{L}}{L} \ri 
+ \log \le \frac{2(d+l)}{\sqrt{d(d+2l)}} \ri \right]
 \log \le \frac{(d+l+\sqrt{d(d+2l)})^2}{l^2} \ri  \\
& \left. + \mathrm{Li}_{2} \le \frac{\sqrt{d (d+2l)}}{d+l} \ri - \mathrm{Li}_{2} \le - \frac{\sqrt{d (d+2l)}}{d+l} \ri \right\} \, .
\end{aligned}
\eeq
The divergences of (\ref{CA2}) are the same as in eq. (\ref{disjoint-complexity-case1});
in particular, the subleading divergences are still poportional to the entanglement entropy
\beq
S = \frac{c}{3} \log \frac{d(d+2l)}{\varepsilon^2} \, .
\eeq
The  finite part is instead a more complicated function of $d,l$ compared to the single interval case.

\section{Mutual complexity}
\label{sec:mutual}

Consider a physical system which is splitted into two sets $A,B$.
The mutual information is defined as
\beq
I(A|B) = S (A) + S (B) - S (A \cup B) \, .
\eeq
Since the entanglement entropy is shown to exhibit a subadditivity behaviour, 
\emph{i.e.} the entanglement entropy of the full system is less than the sum of the entropies
 related to the two subsystems, the mutual information is a positive quantity.

Another quantity which measures the correlations between
 two physical subsystems was defined in \cite{Alishahiha:2018lfv,Caceres:2019pgf}
  and called \emph{mutual complexity}:
\beq
\Delta \mathcal{C}  =
\mathcal{C}(\hat{\rho}_A)+\mathcal{C}(\hat{\rho}_B) - \mathcal{C}(\hat{\rho}_{A \cup B}) \, .
\label{mutual1}
\eeq
where $\hat{\rho}_A$, $\hat{\rho}_B$ are the reduced density matrices 
in the Hilbert spaces localised in  $A$ and $B$.
If $\Delta \mathcal{C} $ is always positive, complexity is subadditive;
if it is always negative, complexity  is superadditive.
By construction, in the CV conjecture complexity is always 
superadditive, i.e.  $\Delta \mathcal{C} \leq 0$. Instead, in the $CA$ conjecture,
 no general argument is known which fixes the sign of $\Delta \mathcal{C}$.
$\Delta \mathcal{C}$ is a finite quantity in all the holographic
conjectures. Moreover,  $\Delta \mathcal{C} =0$ for $d>d_0$
 because in this case  the RT surface is  disconnected and then 
$\mathcal{C}(\hat{\rho}_A)+\mathcal{C}(\hat{\rho}_B) = \mathcal{C}(\hat{\rho}_{A \cup B}) $.
We will check that this quantity  is generically discontinuous
at $d=d_0$.  

\begin{figure}[h]
\center
\begin{tabular}{ccc}
\includegraphics[scale=0.5]{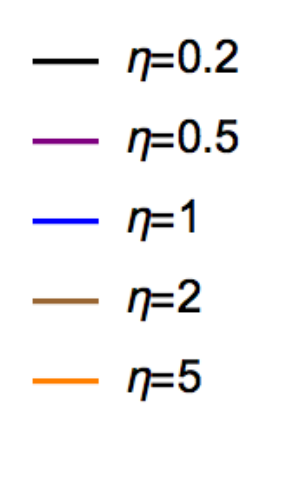} &
\includegraphics[scale=0.5]{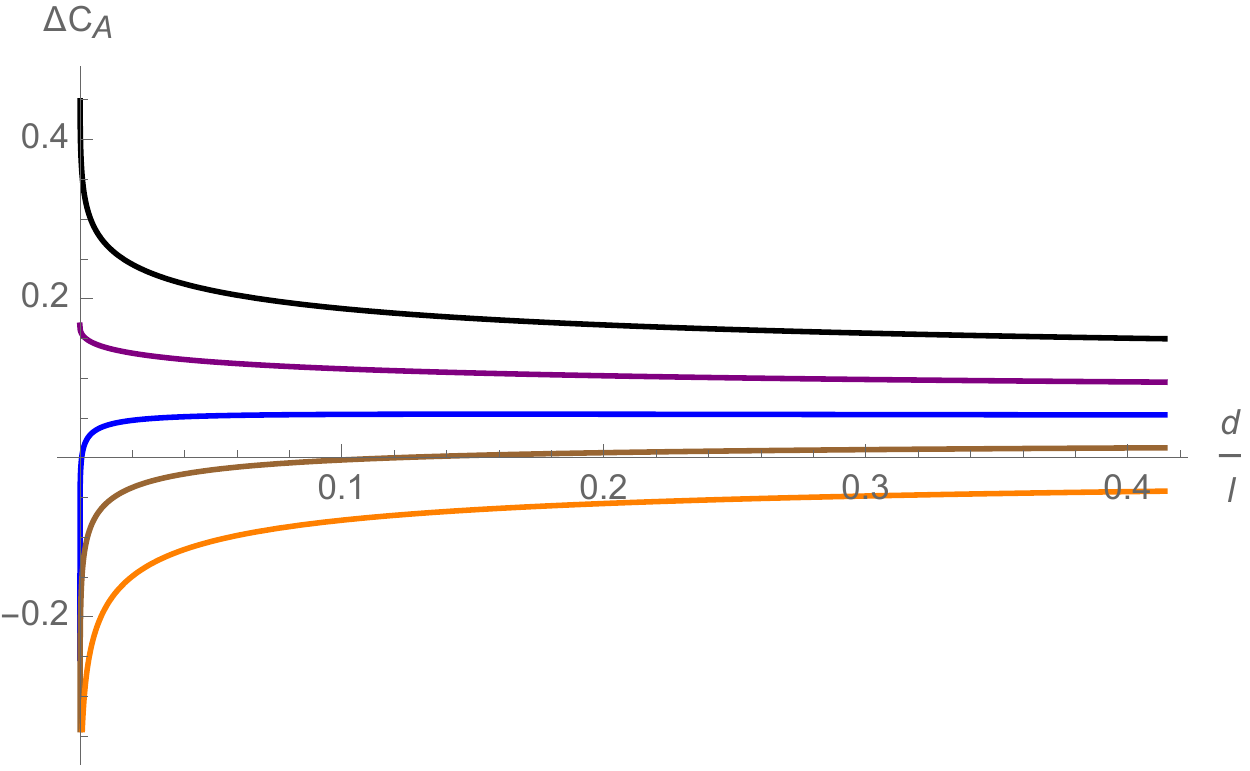} 
\end{tabular}
\caption{Mutual complexity $\Delta \mathcal{C}_A$ for
 several values of $\eta=\tilde{L}/L$  as a function of $\frac{d}{l} \in [0,\frac{d_0}{l}=\sqrt{2}-1]$.}
\label{mutual}
\end{figure}

In the case of two disjoint intervals, from eq. (\ref{CA2}) we find that
the action mutual complexity is:
\beq
\begin{aligned}
\Delta \mathcal{C}_A & =  \mathcal{C}_A^1 - \mathcal{C}_A^2 
 =  \frac{c}{3 \pi^2 } \left\{  \log \le \frac{2 \tilde{L}}{L} \ri \log \le \frac{d(d+2l)}{l^2} \ri + \frac{\pi^2}{2} \right.   \\
& 
- \left[  \log \le \frac{\tilde{L}}{L} \ri 
+ \log \le \frac{2(d+l)}{\sqrt{d(d+2l)}} \ri \right]
 \log \le \frac{(d+l+\sqrt{d(d+2l)})^2}{l^2} \ri  \\
& \left. - \mathrm{Li}_{2} \le \frac{\sqrt{d (d+2l)}}{d+l} \ri + \mathrm{Li}_{2} \le - \frac{\sqrt{d (d+2l)}}{d+l} \ri \right\} \, .
\end{aligned}
\eeq
The function $\Delta \mathcal{C}_A$ 
is plotted in figure \ref{mutual} for various $\eta=\tilde{L}/L$. 
From the figure, we see that this quantity can be either positive or negative.
At small $d$, the behavior of  $\Delta \mathcal{C}_A$ is:
\beq
\Delta \mathcal{C}_A \approx \frac{c}{3 \pi^2} 
\log \le \frac{2 \tilde{L}}{L} \ri \log \le  \frac{2 d}{l} \ri \, .
\eeq
For the value $\tilde{L}/L=1/2$, the behaviour of $\Delta \mathcal{C}_A$ at $d \rightarrow 0$
switches from $-\infty$ to $\infty$.

If $\eta \leq 1/2$, CA is subadditive for all values of $d/l$.
For $\eta > 1/2, $ it is always possible to find small enough distances giving a superadditive behaviour.
Moreover, there is a critical $\eta_0 \approx 2.465$ in such a way that complexity 
of two disjoint  intervals is always superadditive if $\eta>\eta_0$.
In order to have a positive definite subregion complexity,
we should require $\eta>1$. So it seems that it is not possible to achieve
an universally subadditive complexity in a physically consistent setting.

A similar behaviour of subregion CA is found
in the thermofield double state where the subsystems are taken as
 the two disconnected boundaries of spacetime.
This case was investigated for asymptotically $\mathrm{AdS}$ black holes in $D$ dimensions 
\cite{Agon:2018zso,Alishahiha:2018lfv}, 
showing that the complexity=action is subadditive when $\eta<\hat{\eta}_D$ and
superadditive for $\eta>\hat{\eta}_D$.
The value of $\hat{\eta}_D$ is given by the zero of $g_D(\eta)$ \cite{Agon:2018zso}:
\beq
g_D(\eta)= \log( (D-2) \eta)+\frac12 \le \psi_0(1)-\psi_0\le \frac{1}{D-1} \ri \ri +\frac{D-2}{D-1} \pi \, ,
\eeq
where $\psi_0(z)=\Gamma'(z)/\Gamma(z)$ is the digamma function.
For $D=3$,  $\hat{\eta}_3 \approx 0.1$. 

In the CV conjecture, we can use eqs.  (\ref{CCVA}) and  (\ref{CCVB}) from \cite{Abt:2017pmf} 
to determine mutual complexity.
Considering the case of a double segment,
we find that for $d<d_0$ the mutual complexity is constant:
\beq
\Delta \mathcal{C}_{V}= -\frac{4 \, c}{3} \pi \, .
\eeq

\subsection{Strong super/subaddivity for overlapping segments}

Given two generically overlapping regions $A$ and $B$,
entanglement entropy satisfies the strong subadditivity property:
\beq
\tilde{\Delta} S = S (A) + S(B) -S(A \cup B) - S(A \cap B) \geq 0 \, .
\label{gene}
\eeq
Inspired by this relation, we can define \cite{Caceres:2019pgf}
by analogy a generalization of the mutual complexity as:
\beq
\tilde{\Delta} \mathcal{C}(A,B) =
\mathcal{C}(\hat{\rho}_A)+ \mathcal{C}(\hat{\rho}_B) - \mathcal{C}(\hat{\rho}_{A \cup B}) - \mathcal{C}(\hat{\rho}_{A \cap B}) \, .
\eeq
This definition generalizes eq. (\ref{mutual1}) to the case where $A \cap B \ne \emptyset$.
We can investigate the sign of this quantity in the case of two overlapping segments.
 
 Suppose that we consider the regions given by 
 two intervals of lengths $a,b$ which intersect 
in a segment of length $c$. The union of these intervals 
is a segment of total length $a+b-c$. From eq. (\ref{ACTION-C}) we find
\beq
\tilde{\Delta} \mathcal{C}_A^{\rm BTZ}=-\log \le \frac{2 \tilde{L}}{L}\ri \tilde{\Delta} S^{\rm BTZ} \, , 
\eeq
where $\tilde{\Delta} S^{\rm BTZ}$ is the quantity defined in (\ref{gene}), computed
 for the two overlapping intervals in the BTZ background.
Then $ \mathcal{C}_A$ is strongly subadditive for $\tilde{L}/L < 1/2$
and strongly superadditive for $\tilde{L}/L > 1/2$.


\section{Comments and discussion}

We studied the CA  subregion complexity
conjecture in AdS$_3$ and in the BTZ background.
The main results of this chapter are:
\begin{itemize}
\item
In the case of one segment, we find that subregion complexity for AdS$_3$ and for the BTZ
 can be directly related to the entanglement entropy, see eqs. (\ref{CABTZ}).
\item
 In the case of a two segments subregion,
complexity in AdS$_3$ is a more complicated function of 
the lengths and the relative separation of the segments, see eqs. (\ref{CA2}).
Subregion complexity carries a 
different  amount of information compared to the entanglement entropy. In particular,
for two disjoint segments the mutual complexity (defined in eq. (\ref{mutual1})) is not
proportional to mutual information. 
\end{itemize}

We find that the sign of action mutual complexity $\Delta \mathcal{C}_A$
 of a two disjoint segments subregion
depends drastically on $\eta=\tilde{L}/L$ (see figure \ref{mutual}): 
\begin{itemize}
\item For $ \eta \geq \eta_0 \approx 2.465 $, $\Delta \mathcal{C}_A$ is always negative, and so
$\mathcal{C}_A$ is superadditive as $\mathcal{C}_V$ and  $\mathcal{C}_{V 2.0}$. 
\item For $\frac12 < \eta < \eta_0 $,  $\Delta \mathcal{C}_A$ is negative at small 
$d$ and positive at large $d$. This region should be partially unphysical, because 
 in order to obtain a positive-definite $\mathcal{C}_A$,
we have to  require $\tilde{L} > L$ and so $\eta >1$.
\item In the unphysical region $0< \eta \leq 1/2$, action complexity is subadditive.
\end{itemize}


\chapter{Conclusions and outlook}
\label{chapt-Conclusions and outlook}

In this thesis we studied various aspects of non-relativistic theories, both from the point of view of quantum field theory and of gravity computations in the context of general relativity.
We briefly summarize the results and we give some hints for further directions to follow for future investigations.

\section*{Trace anomaly}

The terms entering the trace anomaly for a $2+1$ dimensional Galilean-invariant field theory coupled to a Newton-Cartan background split into an infinite number of sectors which are invariant under Weyl transformations
\beq
\langle T^i_{\,\,\, i} - 2 T^0_{\,\,\, 0} \rangle = \sum_{k=0}^{\infty} \mathcal{A}_k \, .
\eeq
Each sector is distinguished only from the number of appearances\footnote{In the previous formula, the index $k$ indeed counts the number of appearances of $n.$} of the one-form $n$ giving the local time direction in Newton-cartan geometry.
In particular, there exists one sector which can be written as the dimensional reduction along a null direction of the $3+1$ dimensional relativistic trace anomaly
\beq
\mathcal{A}_0 = a E_4 - c W^2_{MNPQ} + \mathcal{A}_{\rm ct} \, ,
\eeq
while the next-to minimal subsector with one appearance of $n$ has vanishing trace anomaly $\mathcal{A}_1 = 0 .$
The classification for the other sectors in not known, apart from the existence of an infinite tower of type B anomalies which can be built using the Weyl tensor.

Studying the trace anomaly in the specific cases of a free scalar and a free fermion minimally coupled to Newton-Cartan gravity, we found that the coefficients of the minimal sector $\mathcal{A}_0$ of the trace anomaly are the same of the $3+1$ dimensional relativistic case, apart from an overall normalization $1/m$ involving the mass of the non-relativistic particle.
This result suggests the existence of an $a-$theorem for the coefficient of the Euler density as in the relativistic parent theory, whose physical interpretation could be the fact that bound states tend to be broken when adding energy to the system.

We also investigated what happens when a source for the particle number is turned on, finding a surprising violation of the invariance under gauge transformations and Milne boosts, the local version of Galilean boosts.
On the other hand, we found that no gravitational anomaly arises, even when adding the particle number source.

There are several open questions to answer. We propose the following ones:
\begin{itemize} 
\item The relation between the anomaly coefficients and the 
correlation functions of the energy-momentum tensor multiplet should be
clarified \cite{Osborn:1989td,Osborn:1991gm}. In the case of vacuum correlation function, these  correlators 
 have support just at coincident points.
It would be interesting to check if the anomaly coefficients can be related
to the form of the finite-density correlators evaluated at separated points.
\item It would be interesting to attempt a perturbative proof
using Osborn's local renormalization group  approach; 
this was initiated in  \cite{Auzzi:2016lrq}. The main 
missing ingredient to the proof is to control the positivity of
some anomaly coefficients whose relativistic analog turn out
to be proportional to the Zamolodchikov metric.
Local renormalization group for Lifshitz theories was studied in
\cite{Pal:2016rpz}.
\item The relation between the anomaly and the dilaton effective action
should be investigated; in the relativistic case, 
this leads to a proof of the $a$-theorem \cite{Komargodski:2011vj}. 
The study of non-relativistic dilaton was initiated in \cite{Arav:2017plg}.
  \item The anomaly coefficients for anyons
coupled to Newton-Cartan backgrounds should be computed. This may be interesting for 
condensed matter applications, as the quantum Hall effect.
\item An analysis of the  Wess-Zumino consistency conditions
for trace anomalies in presence
of gauge and Milne boost violations would clarify the nature
 of the anomalies and their possible relevance for the properties of the Renormalization Group flow.
Due to the large number of terms involved,
this seems a rather challenging task. 
\end{itemize}

\section*{Supersymmetry}

We considered a $\mathcal{N}=2$ Galilean-invariant Wess-Zumino model in $2+1$ dimensions obtained as the null reduction of the relativistic parent in one higher dimension.
Using the null reduction technique, we build a non-relativistic formulation of the superspace and we obtain a supergraph formalism with D-algebra rules which is very similar to the relativistic one.
As a result, the non-renormalization theorem can be easily imported and then we find that no quantum corrections arise in the superpotential.

Furthermore, the properties of the model to conserve the particle number and to have a retarded propagator, coming from the non-relativistic invariance, allow to produce a set of selection rules which greatly constrain the quantume corrections of the model.
This not only allows to find more simplifications than in the relativistic case, but we also find that the self-energy is one-loop exact.

This study can be considered as the starting point for many other developments:
\begin{itemize}
\item The result we have found is reminiscent of relativistic gauge theories with extended supersymmetry, like for instance the relativistic $ \mathcal{N}=2 $ SYM in 3+1 dimensions \cite{Bilal:2001nv}. 
In that case extended supersymmetry constrains the corrections to the K\"ahler  potential to be related to the F-terms, which are protected by the non-renormalization theorem.
In the non-relativistic model discussed in this paper, instead the protection of the K\"ahler potential is related to the $U(1)$ charge conservation at each vertex, which in many
diagrammatic contributions constrains arrows to form a closed loop, so leading to a vanishing integral.
It would be interesting to investigate 
if a common hidden mechanism exists, which is responsible of the similar
mild UV behavior of these two rather different classes of theories.
\item As a continuation of the study of non-relativistic conformal anomalies
in curved Newton-Cartan background
\cite{Jensen:2014hqa,Arav:2016xjc,Auzzi:2015fgg,Auzzi:2016lxb,Pal:2017ntk,Pal:2016rpz,Auzzi:2016lrq},
it would be interesting to study superconformal anomalies
of a Galilean supersymmetric theory in the presence of a classical Newton-Cartan supergravity source.
 \item A non-relativistic theory of a chiral superfield coupled 
to a Chern-Simons gauge field, which   is invariant under the 
conformal extension of the $ \mathcal{S}_2 \mathcal{G}$
algebra, was constructed in \cite{Leblanc:1992wu}.
We expect that further examples  of
 $ \mathcal{S}_2 \mathcal{G}$ theories may be constructed by coupling 
 the F-term interacting theory discussed in this paper to a Chern-Simons gauge field.
 These examples should contain trilinear derivative couplings
 between scalars and fermions and then
 should be different from existing constructions of non-relativistic
 SUSY Chern Simons theory built from the $c \rightarrow \infty$
 non relativistic limit (see e.g. \cite{Nakayama:2008qz,Nakayama:2008td,Lee:2009mm,Lopez-Arcos:2015cqa,Doroud:2015fsz}).
 In analogy to the non-SUSY example studied in \cite{Bergman:1993kq},
 we expect that for special values of the F-term coupling $g$ the
 resulting theory is conformal.   We leave this as a topic for further work.
 These extensions may provide a useful theoretical SUSY setting  
 for studying  non-abelian anyons \cite{Doroud:2015fsz,Doroud:2016mfv}
 and non-relativistic particle-vortex dualities \cite{Turner:2017rki}.
 \item The opposite ultra-relativistic limit $c \rightarrow 0$ gives the Carroll group and can be found from a particular choice of the parameters in the Bargmann algebra \cite{Zhang:2019gdm}.
 It would be interesting to build supersymmetric extensions of the Carroll group starting from the supersymmetric extensions of the Bargamann algebra.
\end{itemize}

\section*{Complexity}

We studied the holographic Complexity=Volume and Complexity=Action conjectures for black holes in asymptotically warped $\mathrm{AdS}_3$ spacetime.
The volume rate is a monotonically increasing function of time which goes to a constant for late times, while the action rate is always positive, but reaches a maximum and then decreases to a constant for late times.
In the action case, we also observe that when the black hole is non-rotating, there exists a critical time under which the complexity rate vanishes.
In both cases, the value reached for $t \rightarrow \infty$ is proportional to the product $TS$ of temperature and entropy, and vanishes in the extremal case.
All these properties are qualitatively the same as for the BTZ black hole.

We also investigated the complexity conjectures in the warped case when the state on the boundary is mixed, in particular when only one of the two boundaries is taken as a subregion.
In this case, we found that there is a richer structure of divergences than in the BTZ case, containing both a linear and a logarithmic divergence in the UV cutoff.
In the volume case the behaviour of subregion complexity is superadditive; in the action case, the leading linear divergence depends from a length scale $\tilde{L}$ appearing in a counterterm needed to have reparametrization invariance of the action.
While this behaviour is the same of the $\mathrm{AdS}$ case, the presence of an additional logarithmic divergence is responsible for the fact that the action is also superadditive, contrarily to the BTZ black hole, where the property depends from the counterterm scale.
Furthermore, the coefficient of the logarithmic divergence is temperature-dependent and the behaviour is different between the volume and the action: this allows to discern between the two conjectures.

Finally, we studied the subregion complexity=action in the BTZ background for a generic segment on the boundary.
We learnt from the analytic computation that the result for a single interval is very elegant: the result is a sum of a divergent term in the UV cutoff which is linear in the length of the region on the boundary, a term proportional to the entaglement entropy of the configuration, and a constant contribution which seems to be a topological term.
The analysis of the case with two disjoint intervals reveals that the result still holds for the divergent part of complexity, but we have an additional finite term given by a function of the distance between the two intervals and their length.
This tells us that the mutual complexity carries different information with respect to the mutual information, which is the analog quantity computed with the entanglement entropy instead of complexity.

There are some possible future developments that we can be intersted to study:
\begin{itemize}
\item Warped black holes can be realized also as solutions 
of Topological Massive Gravity and New Massive Gravity.
It would be interesting to study both Action and Volume in these examples, in order
to get control on both the conjectures in the case of higher derivatives terms in
the gravity action.  The Action conjecture for higher derivatives gravity
was already  studied by several authors in
in \cite{Alishahiha:2017hwg,Guo:2017rul,Ghodrati:2017roz,Qaemmaqami:2017lzs},
but always in the late-time limit.  In particular, ref. \cite{Ghodrati:2017roz} studied the late-time limit
of Action conjecture for warped black holes in Topological Massive Gravity; the asymptotic growth of the action
is not proportional to $TS$.
\item It would be interesting to study complexity from the field theory side.  A proper definition of complexity in quantum field theory has several subtleties, including
the choice of the reference state and the allowed set of elementary quantum gates.
Recently, concrete calculations have been performed in the case of free field theories 
and an approach based on tensor networks in connection with the Liouville action was proposed in \cite{Caputa:2017yrh}.
\item In the CV conjecture, subregion complexity for multiple intervals
 in the BTZ background is independent of
temperature and can be computed using topology from the Gauss-Bonnet theorem,
 see  \cite{Abt:2017pmf}. It would be interesting to investigate if a similar
 relation with topology holds also for CA.
 The complicated structure of the finite terms in eqs.
 (\ref{CA2}) suggests that 
 such relation, if exists, is more intricated than in CV.
 \item One of the obscure aspects of the CA conjecture is the physical meaning
of the scale $\tilde{L}$ appearing in the action counterterm eq. (\ref{counterterm})
on the null boundaries. A deeper understanding of the role of this parameter
is desirable. In particular, its relation with the field theory side of the correspondence
remains completely unclear.
\end{itemize}

\begin{spacing}{0.9}


\begin{appendices} 

\addtocontents{toc}{\protect\setcounter{tocdepth}{1}}
\appendix

\chapter{Conventions}
\label{app-conv}

In this Appendix we collect SUSY conventions in $3+1$ and $2+1$ dimensions. For conventions in four dimensions we primarily refer to \cite{Martin:1997ns}.

\subsection*{Spinors}
In $3+1$ dimensional Minkowski space-time we take the metric 
$\eta^{MN} = \mathrm{diag} (-1,1, 1 , 1) $
and denote left-handed Weyl spinors as $ \psi_{\alpha}$, while right-handed ones as $ \bar{\chi}^{\dot{\alpha}}$.

Spinorial indices are raised and lowered as
\beq\label{raise&lower}
\psi^{\alpha} = \epsilon^{\alpha \beta} \psi_{\beta} \, , \qquad
\bar{\chi}_{\dot{\alpha}} = \epsilon_{\dot{\alpha} \dot{\beta}} \bar{\chi}^{\dot{\beta}} 
\eeq
where the Levi-Civita symbol is chosen to be
\beq
\epsilon^{\alpha \beta} = \epsilon^{{\dot{\alpha}}{\dot{\beta}} } = - \epsilon_{\alpha \beta} = - \epsilon_{{\dot{\alpha}}{\dot{\beta}} }   = \begin{pmatrix}
0 & 1 \\
-1 & 0
\end{pmatrix} 
\eeq
Contractions of spinorial quantities are given by
\beq
 \chi \cdot \psi = \chi^{\alpha} \psi_{\alpha} 
 = \psi \cdot \chi \, ,  \qquad
 \bar{\chi} \cdot \bar{\psi} = \bar{\chi}_{\dot{\alpha}} \bar{\psi}^{\dot{\alpha}}
  = \bar{\psi} \cdot \bar{\chi}  
  \eeq
Complex conjugation changes the chirality of spinors. The prescription for the signs is
\beq
(\psi^{\alpha})^{\dagger} =  \bar{\psi}^{\dot{\alpha}} \, , \qquad
(\psi_{\alpha})^{\dagger} =  \bar{\psi}_{\dot{\alpha}} \, , \qquad
(\bar{\chi}^{\dot{\alpha}})^{\dagger} =  \chi^{\alpha} \, , \qquad
(\bar{\chi}_{\dot{\alpha}})^{\dagger} =  \chi_{\alpha} 
\label{complex conjugation 3+1 spinors}
\eeq

We use sigma matrices
\beq
\sigma^{M} = (\mathbf{1}, \sigma^i) \, ,
\, \qquad
\bar{\sigma}^{M} = (\mathbf{1},  -\sigma^i)  
\eeq
where we have defined $(\bar{\sigma}^M)^{\dot{\alpha} \alpha} = \epsilon^{\dot{\alpha} \dot{\beta}}\epsilon^{\alpha \beta} (\sigma^M)_{\beta \dot{\beta}} $.
They satisfy the following set of useful identities 
\begin{align}
\nonumber
& (\sigma^M)_{\alpha \dot{\alpha}} (\bar{\sigma}_M)^{\dot{\beta} \beta} = - 2 \delta_{\alpha}^{\,\,\, \beta} \delta_{\dot{\alpha}}^{\,\,\, \dot{\beta}} \, , \quad
  (\sigma^M)_{\alpha \dot{\alpha}} (\sigma_M)_{ \beta  \dot{\beta}} = - 2 \epsilon_{\alpha \beta} \epsilon_{\dot{\alpha} \dot{\beta}} \, , \quad 
  \mathrm{Tr} (\sigma^M \bar{\sigma}^N) = - 2 \eta^{MN} \, , &  \\
  &   \le \sigma^M \bar{\sigma}^N + \sigma^N \bar{\sigma}^M \ri_{\alpha}^{\,\,\, \beta} = - 2 \eta^{MN} \delta_{\alpha}^{\,\,\, \beta}  \, , \qquad
\le \bar{\sigma}^M \sigma^N + \bar{\sigma}^N \sigma^M \ri_{\,\,\, \dot{\alpha}}^{ \dot{\beta}} = - 2 \eta^{MN} \delta_{\,\,\, \dot{\alpha}}^{ \dot{\beta}}    & 
\end{align}

\vskip 5pt
\subsection*{Spinorial derivatives}
In order to manipulate expressions with spinorial objects it is useful to adopt a notation where spinorial indices are manifest.
For the case of vectors and in particular for partial derivatives this is achieved by defining
\beq \label{conv_derivatives}
\p_{\alpha {\dot{\alpha}}} = (\sigma^M)_{\alpha {\dot{\alpha}}} \p_M \, , \qquad
\p^{\alpha {\dot{\alpha}}} = \epsilon^{\alpha \beta} \epsilon^{\dot{\alpha} \dot{\beta}} \p_{\beta \dot{\beta}} = (\bar{\sigma}^M)^{{\dot{\alpha}} \alpha} \p_M \, , \qquad
\p_M = - \frac12 (\bar{\sigma}_M)^{{\dot{\alpha}} \alpha} \p_{\alpha {\dot{\alpha}}}  
\eeq
which in particular imply
\beq
\square \equiv \p^M \p_M = - \frac12 \p^{\alpha \dot{\alpha}} \p_{\alpha \dot{\alpha}} \, , \qquad
\p^{\alpha \dot{\gamma}} \p_{\dot{\gamma} \beta} = - \delta^{\alpha}_{\,\,\, \beta} \, \square 
\eeq
We assign rules for the coordinates consistently with the requirement $ \p_M x^M = \p_{\alpha {\dot{\alpha}}} x^{\alpha {\dot{\alpha}}} =4 $,  that is
\beq
x^{\alpha {\dot{\alpha}}} = - \frac12 (\bar{\sigma}_M)^{{\dot{\alpha}} \alpha} x^M \, , \qquad
x^M = (\sigma^M)_{\alpha {\dot{\alpha}}} x^{\alpha {\dot{\alpha}}} 
\eeq
It follows that $x^2 \equiv x^M x_M = - 2 x^{\alpha \dot{\alpha}} x_{\alpha \dot{\alpha}}$.

Finally, we define partial spinorial derivatives acting on Grassmann variables as
\beq
\p_{\alpha} \theta^{\beta} = \delta_{\alpha}^{\,\,\, \beta} \, , \qquad
\p^{\beta} \theta_{\alpha} = - \delta_{\alpha}^{\,\,\, \beta} \, , \qquad
\bar{\p}_{\dot{\alpha}} \bar{\theta}^{\dot{\beta}} =  \delta_{\dot{\alpha}}^{\,\,\, {\dot{\beta}}} \, , \qquad
\bar{\p}^{\dot{\beta}} \bar{\theta}_{\dot{\alpha}} = - \delta^{\dot{\alpha}}_{\,\,\, {\dot{\beta}}} 
\eeq
Imposing the reality of $ \delta_M^{\,\, N} = [\p_M , x^N]  $ and $ \delta_{\alpha}^{\,\, \beta} = \lbrace \p_{\alpha} , \theta^{\beta} \rbrace $  we find that spacetime derivatives are anti-hermitian, $ (\p_{M})^{\dagger} = - \p_{M} , $ while the spinorial ones are hermitian, $ (\p_{\alpha})^{\dagger} = \bar{\p}_{\dot{\alpha}}$.  

\subsection*{Superspace}
 
The SUSY generators can be written as
\beq
P_{\alpha {\dot{\alpha}}} = - i \p_{\alpha {\dot{\alpha}}} \, , \qquad
{\cal Q}_{\alpha} = i \le \p_{\alpha} + \frac{i}{2} \bar{\theta}^{\dot{\alpha}}\p_{\alpha {\dot{\alpha}}} \ri \, , \qquad
\bar{\cal Q}_{\dot{\alpha}} =  - i \le \bar{\p}_{\dot{\alpha}} + \frac{i}{2} \theta^{\alpha} \p_{\alpha {\dot{\alpha}}} \ri  
\eeq
such that the algebra is
$\lbrace {\cal Q}_{\alpha}, \bar{\cal Q}_{\dot{\alpha}} \rbrace =
  i \p_{\alpha {\dot{\alpha}}} =  - P_{\alpha {\dot{\alpha}}}$.
The covariant differential operators which anticommute with the supercharges are
\beq
{\cal D}_{\alpha} =  \p_{\alpha} - \frac{i}{2} \bar{\theta}^{\dot{\alpha}}\p_{\alpha {\dot{\alpha}}} =
 -i {\cal Q}_{\alpha} - i \bar{\theta}^{\dot{\alpha}}\p_{\alpha {\dot{\alpha}}}  \, , \qquad
\bar{\cal D}_{\dot{\alpha}} =  \bar{\p}_{\dot{\alpha}} - \frac{i}{2} \theta^{\alpha} \p_{\alpha {\dot{\alpha}}} = 
i \bar{\cal Q}_{\dot{\alpha}} -  i \theta^{\alpha} \p_{\alpha {\dot{\alpha}}}   
\eeq
and they satisfy
\beq
\lbrace {\cal D}_{\alpha}, \bar{\cal D}_{\dot{\alpha}} \rbrace = -  i \p_{\alpha {\dot{\alpha}}} =   P_{\alpha {\dot{\alpha}}}  
\eeq
With these conventions, we obtain
\beq
\bar{\cal Q}_{\dot{\alpha}} = ({\cal Q}_{\alpha})^{\dagger} \, , \qquad
\bar{\cal D}_{\dot{\alpha}} = ({\cal D}_{\alpha})^{\dagger} 
\eeq

We define
\beq
{\cal Q}^2 \equiv \frac12 {\cal Q}^{\alpha} {\cal Q}_{\alpha} \, , \qquad
\bar{\cal Q}^2 \equiv \frac12 \bar{\cal Q}_{\dot{\alpha}} \bar{\cal Q}^{\dot{\alpha}} \, , \qquad
{\cal D}^2 \equiv \frac12 {\cal D}^{\alpha} {\cal D}_{\alpha} \, , \qquad
\bar{\cal D}^2 \equiv \frac12 \bar{\cal D}_{\dot{\alpha}} \bar{\cal D}^{\dot{\alpha}} 
\eeq

\vskip 5pt
\subsection*{Chiral superfields}
Chiral superfields satisfy
$\bar{\cal D}_{\dot{\alpha}} \Sigma (x^M, \theta^{\alpha}, \bar{\theta}^{\dot{\alpha}} ) = 0 $,
and can be written as
\beq
\Sigma (x_L, \theta, \bar{\theta}) = \phi (x_L) + \theta^{\alpha} \psi_{\alpha} (x_L) - \theta^2 F(x_L) \, ,
\qquad
x_L^{\alpha \dot{\alpha}} \equiv x^{\alpha \dot{\alpha}} - i \theta^{\alpha} \bar{\theta}^{\dot{\alpha}}  
\eeq
Similarly, anti-chiral superfields satisfy
${\cal D}_{\alpha} \bar{\Sigma} (x^M, \theta^{\alpha}, \bar{\theta}^{\dot{\alpha}} ) = 0 $,
whose solution is
\beq
\bar{\Sigma} (x_R, \theta, \bar{\theta}) = 
\bar{\phi}(x_R) +  \bar{\theta}_{\dot{\alpha}} \bar{\psi}^{\dot{\alpha}} (x_R) - \bar{\theta}^2 \bar{F}(x_R) \, ,
\qquad
x_R^{\alpha \dot{\alpha}} \equiv x^{\alpha \dot{\alpha}} + i \theta^{\alpha} \bar{\theta}^{\dot{\alpha}} 
\eeq
Using definitions $
\theta^2 \equiv \frac12 \theta^{\alpha} \theta_{\alpha}$, $\bar{\theta}^2 \equiv \frac12 \bar{\theta}_{\dot{\alpha}} \bar{\theta}^{\dot{\alpha}}$, 
we find the following compact expression for the components of a chiral superfield
\beq\label{components}
\phi = \Sigma| \, , \qquad
\psi_{\alpha} = {\cal D}_{\alpha} \Sigma| \, , \qquad
F= {\cal D}^2 \Sigma|  
\eeq
where $ | $ means that we evaluate the expression at $ \theta= \bar{\theta} = 0$. The anti-chiral components are simply given by the complex conjugated of 
these expressions.

\subsection*{Pauli matrices in light-cone coordinates}
Pauli matrices matrices in light-cone coordinates are
\bea
\sigma^{\pm} = \frac{1}{\sqrt{2}} (\sigma^{3} \pm \sigma^{0}) \, , &&\qquad
\bar{\sigma}^{\pm} = \frac{1}{\sqrt{2}} (\bar{\sigma}^{3} \pm \bar{\sigma}^{0}) 
\nl
\sigma^- = - \bar{\sigma}^+ = \sqrt{2}
\begin{pmatrix}
0 & 0 \\
0 & -1 
\end{pmatrix} \, , &&\qquad
\sigma^+ = - \bar{\sigma}^-= \sqrt{2}
\begin{pmatrix}
1 & 0 \\
0 & 0 
\end{pmatrix} 
\eea
Therefore, for instance we write (from \eqref{conv_derivatives})
\beq
\begin{aligned} \label{light-cone derivatives}
& \partial_{\alpha \dot{\alpha}} = (\sigma^+)_{\alpha \dot{\alpha}} \, \partial_+ +  (\sigma^-)_{\alpha \dot{\alpha}} \, \partial_- +  (\sigma^1)_{\alpha \dot{\alpha}} \, \partial_1 +   (\sigma^2)_{\alpha \dot{\alpha}} \, \partial_2 \\
& \partial^{\alpha \dot{\alpha}} = (\bar{\sigma}^+)^{\dot{\alpha}\alpha } \, \partial_+ +  (\bar{\sigma}^-)^{\dot{\alpha}\alpha } \, \partial_- +  (\bar{\sigma}^1)^{\dot{\alpha}\alpha }\, \partial_1 +   (\bar{\sigma}^2)^{\dot{\alpha}\alpha } \, \partial_2 
\end{aligned}
\eeq
with $\partial_{\pm} = \frac{1}{\sqrt{2}} (\partial_3 \pm \partial_0)$.
\\

\subsection*{Conventions in 2+1 dimensions}

Non-relativistic quantities in 2+1 dimensions are obtained from the null reduction of 3+1 dimensional Minkowski spacetime.
The prescription is to introduce light-cone coordinates $ x^M = (x^-, x^+, x^1, x^2) = (x^-, x^{\mu}) , $ compactify along a small circle in the $ x^- $ direction and perform the identifications
\beq
\p_- \rightarrow im \, , \qquad 
\p_+ \rightarrow \p_t \, , \qquad
 \phi (x^M) = e^{i m x^-} \varphi (x^{\mu}) 
\eeq
where $ m $ is the adimensional eigenvalue of the $ U(1) $ mass operator and $ \phi(x^M) $ is a local field.

Non-relativistic fermions in 2+1 dimensions are parametrized by two complex Grassmann scalars $\xi(x^{\mu})$ and $\chi(x^{\mu})$.
Under null reduction the identification with the 3+1 dimensional left-handed Weyl spinor $ \psi(x^M) $ works as follows
\beq
\psi_{\alpha} (x^M) = e^{i m x^-} \tilde{\psi}_{\alpha} (x^{\mu}) =e^{i m x^-} 
\begin{pmatrix}
\xi (x^{\mu}) \\
\chi (x^{\mu})
\end{pmatrix}
\eeq
Under complex conjugation we choose the prescription
\beq
\bar{\psi}_{\dot{\alpha}} =
 (\psi_{\alpha})^{\dagger} = 
 e^{- i m x^-} (\tilde{\psi}_{\alpha})^{\dagger} \equiv e^{- i m x^-}  \begin{pmatrix}
\bar{\xi} (x^{\mu}) \\
 \bar{\chi} (x^{\mu})
\end{pmatrix}
\eeq
Using identities \eqref{raise&lower} it is easy to infer the identification with the components of $\psi^\alpha$ and $\bar{\psi}^{\dot{\alpha}}$.

Taking the mass as a dimensionless parameter enforces the energy dimensions of the non-relativistic fermion to be
\beq
[\xi] = E^2 \, , \qquad
[\chi] = E 
\eeq
These assignments immediately follow when performing the null reduction of the Weyl Lagrangian
\beq
\mathcal{L} = i \psi^{\dagger} \bar{\sigma}^M \p_M \psi \rightarrow
\sqrt{2} m \bar{\xi} \xi + \sqrt{2} i \bar{\chi} \p_t \chi - i \bar{\xi} (\p_1 - i \p_2) \chi - i \bar{\chi} (\p_1 + i \p_2) \xi 
\eeq
We observe that the only dynamical component is $ \chi $, while $\xi$ turns out to be an auxiliary field that can be integrated out from the action.

Since null reduction affects only space-time coordinates without modifying the spinorial ones, we obtain ${\cal N}=2$ supersymmetry in three dimensions. According to the ordinary pattern for which the three dimensional ${\cal N}=2$ superspace is ``equal'' to the four-dimensional ${\cal N}=1$ superspace, all the properties related to manipulations of covariant derivatives and supercharges, \emph{e.g.} the D-algebra procedure, are directly inherited from the (3+1) relativistic theory under a suitable re-interpretation of the spinorial objects. 

In particular, the algebra of covariant derivatives reads
\beq
\{ D_\alpha , \bar{D}_\beta \} = - i \p_{\alpha \beta}\, ,  \qquad \qquad \{ D^\alpha , \bar{D}^\beta \} = - i \p^{\beta \alpha}
\eeq
where, as follows from \eqref{light-cone derivatives}, the three-dimensional derivatives are given by 
\beq\label{3d_derivatives}
\partial_{\alpha \beta} = \begin{pmatrix}
\sqrt{2} \p_t  &   \p_1 - i \p_2 \\
 \p_1 + i \p_2 & - i \sqrt{2} M
\end{pmatrix} 
\qquad \quad 
\partial^{\alpha \beta} = \begin{pmatrix}
- i\sqrt{2} M  &   -(\p_1 - i \p_2) \\
 -(\p_1 + i \p_2) & \sqrt{2} \p_t 
\end{pmatrix} 
\eeq
They satisfy the following identities
\beq 
\p^{\alpha \beta} = \epsilon^{\alpha \delta} \epsilon^{\beta \gamma} \p_{\gamma \delta} 
\qquad \qquad
\p_{\beta \alpha} = \epsilon_{\alpha \gamma} \epsilon_{\beta \delta} \, \p^{\gamma \delta}
\eeq
Therefore, we have for instance $\bar{\xi}_{\alpha} \p^{\alpha \beta} \chi_\beta = \bar{\xi}^{\alpha} \p_{\beta \alpha} \chi^\beta$.
Identities which turn out to be useful for the reduction of the action to components are
\beq \label{useful_identities}
\begin{aligned}
& [ D^\alpha , \bar{D}^2 ] = i \p^{\beta\alpha } \bar{D}_\beta \, , \qquad \quad [ \bar{D}^\alpha , D^2 ] = -i \p^{\alpha \beta} D_\beta \\
& D^2 \bar{D}^2 + \bar{D}^2 D^2 = (2iM\p_t + \p_i^2) + D^\alpha \bar{D}^2 D_\alpha = (2iM\p_t + \p_i^2)  + \bar{D}_\alpha D^2 \bar{D}^\alpha
\end{aligned}
\eeq

\subsection*{Spin connection}
\label{spinc}

The explicit expression for the spin connection is:
\bea
\omega_{MAB} &=& \frac{1}{2} \left[ e^{N}_{\,\,\,A} \left( \p_M  e_{NB} - \p_N   e_{MB} \right) 
-  e^{N}_{\,\,\,B}  \left( \p_M  e_{NA} - \p_N   e_{MA} \right)  \right. \nl
& & \left. 
- e^{N}_{\,\,\,A} e^{P}_{\,\,\,B}  \left( \p_N  e_{PC} - \p_P   e_{NC} \right)   e^{C}_{\,\,\,M} \right] \, .
\eea
We thus obtain the components:
\bea
\omega_{\underset{(M)}{-} AB} &=& - \frac{1}{2} e^{\mu}_{\,\,\,A} e^{\nu}_{\,\,\, B} \tilde{F}_{\mu\nu} \, , \quad
\omega_{\mu \underset{(A)}{-} A}   =  -\frac{1}{2}  e^{\nu}_{\,\,\,A}  \tilde{F}_{\mu\nu}  \, , 
\nl
\omega_{\mu \underset{(A)}{+} a} &=& \frac{1}{2} v^{\nu} \left(  \p_{\mu} e^{a}_{\,\,\, \nu} -  \p_{\nu} e^{a}_{\,\,\, \mu}  \right)  - \frac{1}{2}   e^{\nu}_{\,\,\, a} F_{\mu\nu} - \frac{1}{2} v^{\nu} e^{\rho}_{\,\,\, a}  \left[  A_{\mu}  \tilde{F}_{\nu\rho} + n_{\mu} F_{\nu\rho} +e^b_{\,\,\, \mu} \left( \p_{\nu} e^{b}_{\,\,\, \rho}-  \p_{\rho} e^{b}_{\,\,\, \nu}  \right)  \right] \, , 
\nl
\omega_{\mu  a b}  &=& \frac{1}{2} e^{\nu}_{\,\,\,a} \left(  \p_{\mu} e^{b}_{\,\,\, \nu} -  \p_{\nu} e^{b}_{\,\,\, \mu}  \right)  - \frac{1}{2}   e^{\nu}_{\,\,\, b} \left(  \p_{\mu} e^{a}_{\,\,\, \nu} -  \p_{\nu} e^{a}_{\,\,\, \mu}   \right) + 
\nl
& & - \frac{1}{2}e^{\nu}_{\,\,\,a} e^{\rho}_{\,\,\, b}  \left[  A_{\mu}  \tilde{F}_{\nu\rho} + n_{\mu} F_{\nu\rho} +e^c_{\,\,\, \mu} \left( \p_{\nu} e^{c}_{\,\,\, \rho}-  \p_{\rho} e^{c}_{\,\,\, \nu}  \right)  \right]  \, . 
\eea
Note that $\omega_{\underset{(M)}{-} -B}=0$.


\chapter{Explicit calculation of the heat kernel perturbative expansion}
\label{app-Explicit calculation of the heat kernel perturbative expansion}

In this Appendix we show the explicit calculation of the perturbative expansion of the heat kernel applied to a non-relativistic differential operator as in eq. (\ref{splitting of the non-relativistic heat kernel operator}).
The terminology of the insertion terms refers to the decomposition in eq. (\ref{Operatore heat kernel in termini di P,S,Q_i}), which induces the splitting at first and second order as in eqs. (\ref{1st order splitting heat kernel}),(\ref{2nd order splitting heat kernel}).
For simplicity, we will omit the subscript \emph{E} referred to the Euclidean version of the Schr\"odinger operator.

We start with the case where the insertion operators are time-independent and then we switch to the time-dependent case.


\section{First order expansion of the heat kernel operator}
\label{sect-First order expansion of the heat kernel operator}

We start with the simplest case, when we consider the single insertion of a multiplicative function $P(x).$
By definition
\beq
K_{1P} (\tau) = \int_0^{\tau} d \tau' \, \langle x, t | e^{(\tau-\tau') \bigtriangleup} P(x) e^{\tau' \bigtriangleup}  | x', t' \rangle \, .
\eeq
We insert a completeness relation in order to use the expression of the flat space heat kernel operator inside the previous integral
\beq
\begin{aligned}
K_{1P} (\tau) & = \int_0^{\tau} d \tau' \int d^d \tilde{x} \int d \tilde{t}  \, \langle x, t | e^{(\tau-\tau')\bigtriangleup} |\tilde{x}, \tilde{t}  \rangle   P(\tilde{x}) \langle  \tilde{x}, \tilde{t}   |  e^{\tau' \bigtriangleup}  | x', t' \rangle = \\
& = \frac{1}{(2 \pi)^2} \int_0^{\tau} d \tau'  \, 
\frac{1}{(4 \pi (\tau - \tau'))^{d/2}}  \frac{1}{(4 \pi \tau')^{d/2}} \int d \tilde{t} \, \frac{m(\tau-\tau')}{m^2 (\tau- \tau')^2 + \frac{(t-\tilde{t})^2}{4}} \, \frac{m\tau'}{m^2 \tau'^2 + \frac{(\tilde{t}-t')^2}{4}} \times  \\
& \times \int d^d \tilde{x} \, P(\tilde{x}) \, \exp{ \left[ - \frac{(x-\tilde{x})^2}{4(\tau-\tau')} - \frac{(\tilde{x}-x')^2}{4 \tau'} \right]} \, .
\end{aligned}
\eeq
We Fourier-transform the $P(\tilde{x})$ function
\beq
P(\tilde{x}) = \frac{1}{(2\pi)^{d/2}} \int d^d k \, e^{i k \tilde{x}} P(k)
\eeq
and we perform explicitly the time integration to get
\beq
\begin{aligned}
K_{1P} (\tau) & = \frac{1}{(2 \pi)^2} \int_0^{\tau} d \tau'  \, 
\frac{1}{(4 \pi (\tau - \tau'))^{d/2}}  \frac{1}{(4 \pi \tau')^{d/2}}  \frac{8 \pi m \tau}{4 m^2 \tau^2 + (t-t')^2} \times  \\
& \times \int d^d \tilde{x} \, \int \frac{d^d k}{(2 \pi)^{d/2}} \, P(k) \, \exp{ \left[ - \frac{(x-\tilde{x})^2}{4(\tau-\tau')} - \frac{(\tilde{x}-x')^2}{4 \tau'} + i k \tilde{x} \right]} \, .
\end{aligned}
\eeq
The Gaussian integral in the spatial coordinates can be performed exactly to find
\beq
\begin{aligned}
K_{1P} (\tau) & =
\frac{1}{(2 \pi)^2} \int_0^{\tau} d \tau'  \, 
 \frac{1}{(4 \pi \tau)^{d/2}}  \frac{8 \pi m \tau}{4 m^2 \tau^2 + (t-t')^2} \times \\
& \times \int \frac{d^d k}{2 \pi^{d/2}} \,
\exp{ \left[ - \frac{(x-x')^2}{4\tau}
+ i k \cdot \le x \frac{\tau}{\tau'} + x' \frac{\tau - \tau'}{\tau} \ri
 - k^2 \frac{\tau'}{\tau} (\tau - \tau') \right]} \,
P(k) \, .
\label{intermediate step spatial integral heat kernel single insertion P}
\end{aligned}
\eeq
Setting $x=x', t=t'$ means that we compute the trace of this insertion
\beq
\begin{aligned}
\mathrm{Tr} \, K_{1P} (\tau) & =
\frac{1}{2 \pi} \int_0^{\tau} d \tau' \, \frac{1}{(4 \pi \tau)^{d/2}} \frac{1}{m \tau}  \int \frac{d^d k}{2 \pi^{d/2}} \,
\exp{\left[ i k \cdot x - k^2 \frac{\tau'}{\tau} (\tau - \tau') \right]}  \, P(k) = \\
& = \frac{1}{2 \pi}  \frac{1}{(4 \pi \tau)^{d/2}} \frac{1}{m \tau} \int_0^{\tau} d \tau' \, \exp{\left[  \frac{\tau'}{\tau} (\tau - \tau') \p_x^2 \right]}  \, P(x) \, ,
\end{aligned}
\eeq
and a Taylor expansion of the exponential around $\tau=0$ gives
\beq
\mathrm{Tr} \, K_{1P} (\tau) =
\frac{2}{m (4 \pi \tau)^{d/2+1}} \le \tau P(x) + \frac{1}{6} \tau^2 \p_x^2 P(x) + \mathcal{O} (\tau^3) \ri \, .
\eeq
The single insertion of the operators $ S(x), Q_i(x) $ can be reduced to derivatives acting on the previous expression.
In particular we find
\beq
\begin{aligned}
K_{1S} (\tau) & = \int_0^{\tau} d \tau' \int d^d \tilde{x} \int d \tilde{t}  \, \langle x, t | e^{(\tau-\tau')\bigtriangleup} |\tilde{x}, \tilde{t}  \rangle   S(\tilde{x}) \sqrt{-\p_{\tilde{t}}^2} \langle  \tilde{x}, \tilde{t}   |  e^{\tau' \bigtriangleup}  | x', t' \rangle = \\
& = \sqrt{-\p_{t'}^2} \le \int_0^{\tau} d \tau' \int d^d \tilde{x} \int d \tilde{t}  \, \langle x, t | e^{(\tau-\tau')\bigtriangleup} |\tilde{x}, \tilde{t}  \rangle   S(\tilde{x}) \langle  \tilde{x}, \tilde{t}   |  e^{\tau' \bigtriangleup}  | x', t' \rangle   \ri = \\
& = \sqrt{-\p_{t'}^2} \, K_{1P} (\tau)|_{P(x)=S(x)} \, .
\end{aligned}
\eeq
In the last step, we recognized the expression for the multiplicative insertion $P(x),$ when calling in a different way the function $S(x)$ in the integrand.
In order to perform the differentiation, it is helpful to use the Fourier transform ($t, \omega$ are the conjugate variables)
\beq
\mathcal{F} \le \frac{1}{1+ \frac{t^2}{A^2}} \ri = \sqrt{\frac{\pi}{2}} \, A \, e^{-A |\omega|} 
\eeq
to obtain
\beq
\sqrt{-\p_{t}^2} \le \frac{1}{1+ \frac{t^2}{A^2}} \ri =
\frac{A(A^2 -t^2)}{(A^2 + t^2)^2} \, .
\eeq
Using this method and computing the trace, we similarly obtain
\beq
\mathrm{Tr} \, K_{1S} (\tau) =
\frac{2}{m(4 \pi \tau)^{d/2+1}} \, \mathrm{tr} \, \left( \frac{S}{2m} + \frac{\tau}{12m} \p_i^2 S + \frac{\tau^2}{120m} \p_i^4 S + \mathcal{O} (\tau^3) \right) \, ,
\eeq
where the $\mathrm{tr}$ on the r.h.s. refers to internal indices of the operators, while on the l.h.s $\mathrm{Tr}$ refers also to the sum over the spacetime coordinates.

A similar trick can be applied to the operator $Q_i (x)$ with the spatial derivative, by observing that
\beq
\begin{aligned}
K_{1Q_i} (\tau) & = \int_0^{\tau} d \tau' \int d^d \tilde{x} \int d \tilde{t}  \, \langle x, t | e^{(\tau-\tau')\bigtriangleup} |\tilde{x}, \tilde{t}  \rangle   Q_i(\tilde{x}) \p_{\tilde{x}_i} \langle  \tilde{x}, \tilde{t}   |  e^{\tau' \bigtriangleup}  | x', t' \rangle = \\
&  = - \p_{{x'}_i} \le \int_0^{\tau} d \tau' \int d^d \tilde{x} \int d \tilde{t}  \, \langle x, t | e^{(\tau-\tau')\bigtriangleup} |\tilde{x}, \tilde{t}  \rangle   Q_i(\tilde{x}) \langle  \tilde{x}, \tilde{t}   |  e^{\tau' \bigtriangleup}  | x', t' \rangle   \ri = \\
& = - \p_{{x'}_i} \, K_{1P} (\tau)|_{P(x)=Q_i(x)} \, .
\end{aligned}
\eeq
Similar computations give (no sum on the index $i$)
\beq
\mathrm{Tr} \, K_{1Q_i} (\tau) =
\frac{2}{m(4 \pi \tau)^{d/2+1}} \, \mathrm{tr} \, \left( - \frac{\tau}{2} \p_i Q_i - \frac{\tau^2}{12} \p_i \p^2_k Q_i + \mathcal{O} (\tau^3) \right) \, .
\eeq


\section{Second order expansion of the heat kernel operator}
\label{sect-Second order expansion of the heat kernel operator}

In this section we consider double insertions of the operators appearing in the heat kernel expansion in the terms $K_{2 X_1 X_2} (\tau),$ where
\beq
X_1=\left\{ P(x_1), S(x_1), Q_i(x_1)  \right\} \, , 
\qquad
X_2=\left\{ P(x_2), S(x_2), Q_j(x_2)  \right\} \, ,
\eeq
whose explicit expression is
\beq
K_{2 X_1 X_2}(\tau) =
\int_0^{\tau} d \tau_2 \int_0^{\tau_2} d \tau_1
\langle x' ,t' | e^{-(\tau-\tau_2)\triangle}
| x_2, t_2 \rangle 
\hat{X}_2
\langle x_2, t_2 | e^{-(\tau_2-\tau_1) \triangle}  | x_1, t_1 \rangle
\hat{X}_1
\langle 
x_1, t_1 | e^{-\tau_1 \triangle }
| x, t \rangle \, ,
\eeq
where
\beq
\hat{X}_1=\left\{ P(x_1), S(x_1) \sqrt{- \p_{t_1}^2} ,Q_i(x_1)  \p_{i, x_1} \right\} \, , 
\qquad
\hat{X}_2=\left\{ P(x_2), S(x_2) \sqrt{- \p_{t_2}^2} ,Q_j(x_2) \p_{j, x_2} \right\} \, . 
\eeq
Notice that the order in which the operators appear as subscripts is opposite to the order in which they enter the integral expression.

The strategy to follow is technically more difficult, but theoretically analog to the first order case:
\begin{itemize}
\item We insert some completeness identities in order to make the coordinate-basis representation of the flat heat kernel appearing explicitly
\item We Fourier-transform the inserted operators
\item We perform the integration along the inserted spacetime coordinates
\item We finally take the trace and we Taylor-expand the result in $\tau$ 
\end{itemize}
Following this method, we can always put the insertions in the form
\beq
K_{2 X_1 X_2}(\tau)=\int_0^{\tau} d \tau_2 \int_0^{\tau_2} d \tau_1
\frac{1}{(4 \pi (\tau-\tau_2))^{d/2} } \, 
\frac{1}{(4 \pi (\tau_2-\tau_1))^{d/2} }  \, 
\frac{1}{(4 \pi \tau_1)^{d/2} } \Xi^{X_1 X_2} \,  \Psi^{X_1 X_2} \, ,
\eeq
where $\Xi^{X_1 X_2}$ and $\Psi^{X_1 X_2}$ correspond to the space and time part of the integrals, respectively.
Following the previous prescription, it is useful to Fourier-transform
\beq
\Xi^{X_1 X_2}=
\int \frac{d^d k_1}{(2 \pi)^{d/2}}
\frac{d^d k_2}{(2 \pi)^{d/2}}
\tilde{\Xi}^{X_1 X_2} \, ,
\eeq
and to introduce the quantity
\beq
\Upsilon=
\exp \le  i k_1 x_1+i k_2 x_2 
 -\frac{(x'-x_2)^2}{4 (\tau-\tau_2)}
 -\frac{(x_2-x_1)^2}{4 (\tau_2-\tau_1)} 
-\frac{(x_1-x)^2}{4 \tau_1}\ri \, .
\label{Upsilon quantity heat kernel expansion}
\eeq
Since the spatial and temporal integrals factorize, we can treat them separately.
In particular we obtain the following Fourier transforms of the spatial part
\bea
 \tilde{\Xi}^{PP} &=&
\int d x_1 \int d x_2 \Upsilon P(k_1) P(k_2)  \, ,
\nl 
\tilde{\Xi}^{Q_i P} &=&
-\p_{x,i}\left[  \int d x_1 \int d x_2 
\Upsilon Q_i(k_1) P(k_2) \right] \, ,
\nl
\tilde{\Xi}^{P a_j } &=&
 \int d x_1 \int d x_2 
 \left[ -\frac{(x_2-x_1)_j}{2 (\tau_2-\tau_1)} \right]
\Upsilon P(k_1) Q_j(k_2)  \, ,
\nl
\tilde{\Xi}^{Q_i Q_j }&=&
-\p_{x,i} \left[
 \int d x_1 \int d x_2 
 \left[ -\frac{(x_2-x_1)_j}{2 (\tau_2-\tau_1)} \right]
\Upsilon Q_i(k_1) Q_j(k_2)  \right] \, ,
\eea
where $P(k)$ and $Q_i(k)$ are the Fourier transform of
$P(x)$ and $Q_i(x)$.
The temporal part, on the other hand, is given by
 \begin{eqnarray}
\label{time integral heat kernel expansion appendix}
\Psi^{PP} &=&
\frac{1}{(2  \pi)^3}
 \int \! d t_1  \int \!  d t_2
 \frac{m (\tau-\tau_2)}{m^2 (\tau-\tau_2)^2 +\frac{(t_2-t')^2}{4}} \,
 \frac{m (\tau_2-\tau_1)}{m^2 (\tau_2-\tau_1)^2 +\frac{(t_2-t_1)^2}{4}} \,
 \frac{m \tau_1}{m^2 \tau_1^2 +\frac{(t_1-t)^2}{4}} \, ,
\\
\Psi^{SP} &=&
\frac{1}{4  \pi^3}
 \int \! d t_1  \int \!  d t_2
 \frac{m (\tau-\tau_2)}{m^2 (\tau-\tau_2)^2 +\frac{(t_2-t')^2}{4}} \, 
 \frac{m (\tau_2-\tau_1)}{m^2 (\tau_2-\tau_1)^2 +\frac{(t_2-t_1)^2}{4}} \, 
 \frac{ 4m^2 \tau_1^2 -(t_1-t)^2}{(4 m^2 \tau_1^2 +(t_1-t)^2)^2} \, ,
\nl
\Psi^{PS} &=&
\frac{1}{4  \pi^3}
 \int \! d t_1  \int \!  d t_2
 \frac{m (\tau-\tau_2)}{m^2 (\tau-\tau_2)^2 +\frac{(t_2-t')^2}{4}} \, 
 \frac{ 4m^2 (\tau_2-\tau_1)^2 -(t_2-t_1)^2}{(4 m^2 (\tau_2-\tau_1)^2 +(t_2-t_1)^2)^2} \,
 \frac{ms_1}{m^2 \tau_1^2 +\frac{(t_1-t)^2}{4}} \, ,
 \nl
\Psi^{SS} &=&
\frac{1}{2  \pi^3}
  \int \! d t_1  \int \!  d t_2
 \frac{m (\tau-\tau_2)}{m^2 (\tau-\tau_2)^2 +\frac{(t_2-t')^2}{4}}  \, 
 \frac{ 4m^2 (\tau_2-\tau_1)^2 -(t_2-t_1)^2}{(4 m^2 (\tau_2-\tau_1)^2 +(t_2-t_1)^2)^2}  \,
 \frac{ 4m^2 \tau_1^2 -(t_1-t)^2}{(4 m^2 \tau_1^2 +(t_1-t)^2)^2} \, .
 \nonumber
\end{eqnarray}
While it seems that this list is not exaustive, one should observe that the insertion of the operator $S(x)$ must be considered at the same level of a $P(x)$ insertion when considering the spatial integration, while the insertion of $Q_i(x)$ must be treated as a $P(x)$ insertion for the time part.
These rules follow from the fact that $S(x)$ operators carry time derivatives which do not affect the spatial part, while $Q_i(x)$ carry spatial derivatives which do not influence the time part.

The basic building blocks for the previous computations are found to give 
\begin{eqnarray}
 \tilde{\Xi}^{PP} &=&
 (4 \pi)^{d} \le \frac{\tau_1 (\tau-\tau_2) (\tau_2-\tau_1)}{s} \ri^{d/2}
 \nl
 & &
\exp \le 
\frac{i k_1 \tau_1 x'}{\tau}+\frac{i k_2 \tau_2 x'}{\tau}-\frac{i k_1 \tau_1 x}{\tau}
-\frac{i k_2 \tau_2 x}{\tau}+\frac{k_1^2 \tau_1^2}{\tau}
+\frac{k_2^2 \tau_2^2}{\tau}
-k_1^2   \tau_1-2 k_1 k_2 \tau_1-k_2^2 \tau_2
\right.
\nl & &
\left.
+\frac{2 k_1 k_2 \tau_1 \tau_2}{\tau}+i k_1 x+i k_2 x
-\frac{x^2}{4 \tau}+\frac{x x'}{2 \tau}-\frac{\left(x'\right)^2}{4 \tau}
\ri P(k_1) P(k_2)  \, ,
\label{result spatial parts double insertions heat kernel}
\end{eqnarray}
\begin{eqnarray}
 \tilde{\Xi}^{P Q_j} &=&
 \exp \le
\frac{i k_1 \tau_1 x'}{\tau}+\frac{i k_2 \tau_2 x'}{\tau}-\frac{i k_1 \tau_1 x}{\tau}
-\frac{i k_2 \tau_2 x}{\tau}+\frac{k_1^2 \tau_1^2}{\tau}+\frac{k_2^2
   \tau_2^2}{\tau}-k_1^2 \tau_1-2 k_1 k_2 \tau_1-k_2^2 \tau_2
  \right.
\nl & &
  \left.
   + \frac{2 k_1 k_2 \tau_1 \tau_2}{\tau}+i k_1 x+i k_2 x+\frac{x x'}{2
   \tau}-\frac{\left(x'\right)^2}{4 \tau}-\frac{x^2}{4 \tau} \ri 
    (4 \pi)^{d}  \le \frac{ \tau_1 \left(\tau_1-\tau_2\right) \left(\tau_2- \tau \right)}{\tau} \ri^{d/2}
  \nl  & &
     \frac{ \left( i k_1 \tau_1+  i k_2 \tau_2 - i k_2 \tau
   +\frac{x-x'}{2} \right)_i}{\tau}   P(k_1)  Q_j(k_2) \, ,
\end{eqnarray}
while the expressions for 
$\tilde{\Xi}^{Q_i P} $ and $\tilde{\Xi}^{Q_i Q_j }$
can be obtained differentiating
  $\tilde{\Xi}^{PP}$ and $ \tilde{\Xi}^{P Q_j}$ with respect to $x_i$.

On the other hand, we obtain for the time integration
\bea
\Psi^{PP}&=&\frac{1}{\pi} \frac{2 m \tau}{4 m^2 \tau^2+\left(t-t'\right)^2} \, ,
\qquad
\Psi^{SS}=\frac{1}{\pi}
\frac{4  m \tau \left(4 m^2 \tau^2-3 \left(t-t'\right)^2\right)}{\left(4 m^2 \tau^2+\left(t-t'\right)^2\right)^3} \, ,
\nl
\Psi^{PS}&=&\Psi^{SP}=
\frac{1}{ \pi}
\frac{\left(4 m^2 \tau^2-\left(t-t'\right)^2\right)}{\left(4 m^2 \tau^2+\left(t-t'\right)^2\right)^2} \, .
 \eea
Combining all the expressions, putting $x=x', t=t'$ and Taylor-expanding around $\tau=0$ we obtain ($\mathrm{tr}$ is the trace over the internal indices, there is no sum over the index $i$ of the operator $Q_i$)
\beq
\mathrm{Tr} \, K_{2PP} = \frac{2}{m(4 \pi \tau)^{d/2+1}} \mathrm{tr} \left( \frac{\tau^2}{2} P(x)^2 + \mathcal{O}(\tau^3) \right)  \, ,
\eeq
\beq
\begin{aligned}
\mathrm{Tr} \, K_{2SS}  = & \frac{2}{m(4 \pi \tau)^{d/2+1}} \mathrm{tr} \left( \frac{S^2}{4m^2} + \frac{\tau}{12m^2} S \p^2 S +
\frac{\tau}{24m^2} \p_k S \p_k S + \frac{\tau^2}{120m^2} S \p^4 S + \right. \\
& \left. +\frac{\tau^2}{144m^2} \p^2 S \p^2 S +
\frac{\tau^2}{60m^2} \p_i\p^2 S \p_i S + \frac{\tau^2}{180m^2} \p_{ij} S \p_{ij} S  + \mathcal{O}(\tau^3) \right)  \, ,
\end{aligned}
\eeq
\beq
\mathrm{Tr} \, K_{2PS} = \tilde{K}_{2SP}   =  \frac{1}{m(4 \pi \tau)^{d/2+1}} \mathrm{tr} \left( \frac{\tau}{2m} SP + \frac{\tau^2}{12m} S \p^2 P + \frac{\tau^2}{12m}\p^2 S  P + \frac{\tau^2}{12m} \p_i S \p_i P + \mathcal{O}(\tau^3) \right)  \, ,
\eeq
\beq
\begin{aligned}
 \mathrm{Tr} \, K_{2 Q_j Q_i } & = \frac{2 }{m (4 \pi \tau)^{d/2+1}}  
\textrm{tr}  \left[ -  \frac{\tau}{4}  Q_i Q_i    
- \frac{\tau^2}{24} (\p_{j} Q_i) (\p_i a_j) \right. \\
& \left. +  \frac{\tau^2}{8} (\p_{i} Q_i) (\p_{j} a_j)   -  \frac{\tau^2}{12} Q_i (\p^2 Q_i) 
 -  \frac{\tau^2}{24} (\p_i a_j)^2 
+  \mathcal{O}(\tau^3) \right]  \, ,
\end{aligned}
\eeq
\beq
\mathrm{Tr} \, K_{2 Q_i P}  =  \frac{2}{m(4 \pi \tau)^{d/2+1}} \mathrm{tr} \left( - \frac{\tau^2}{3} P (\p_i Q_i)  - \frac{\tau^2}{6} (\p_i P) Q_i  + \mathcal{O}(\tau^3) \right)  \, ,
\eeq
\beq
\mathrm{Tr} \, K_{2  P Q_i}  =  \frac{2}{m(4 \pi \tau)^{d/2+1}} \mathrm{tr} \left(   \frac{\tau^2}{6} Q_i (\p_i P) - \frac{\tau^2}{6} (\p_i Q_i) P + \mathcal{O}(\tau^3) \right)  \, ,
\eeq
\beq
\begin{aligned}
& \mathrm{Tr} \, K_{2 Q_i S}  =  \frac{2}{m(4 \pi \tau)^{d/2+1}} \mathrm{tr} \left[ - \frac{\tau}{24 m^2} \le   S \p_k^2 S + \frac{1}{2} (\p_k S)^2 \ri + \right. \\ 
& \left.  -\frac{\tau^2}{80 m^2}  \le \frac{1}{2} S \p_k^2 \p_j^2  S + \frac{7}{12} (\p_k^2 S)^2 + \frac{13}{12} \p_k S (\p_k \p^2_j S) + \frac{1}{3} (\p_k \p_j S)^2 \ri + \mathcal{O}(\tau^3) \right]  \, ,
\end{aligned}
\eeq
\beq
\begin{aligned}
& \mathrm{Tr} \, K_{2 S Q_i}  =  \frac{2}{m(4 \pi \tau)^{d/2+1}} \mathrm{tr} \left[  \frac{\tau}{48 m^2} \le  (\p_k S)^2  - S \p_k^2 S \ri + \right. \\ 
& \left.  + \frac{\tau^2}{80m^2 }  \le - \frac{1}{3} S \p_k^2 \p_j^2  S - \frac{1}{4} (\p_k^2 S)^2 + \frac{1}{4} \p_k S (\p_k \p^2_j S) + \frac{1}{3} (\p_k \p_j S)^2 \ri + \mathcal{O}(\tau^3) \right]  \, .
\end{aligned}
\eeq

\section{Time-dependent insertion contributions to the heat kernel (first order)}
\label{app-Time-dependent insertion contributions to the heat kernel (first order)}

We generalize to the time-dependent case the insertion operators appearing in the heat kernel expansion needed to compute the trace anomaly for the NC background (\ref{NC background heat kernel with particle number}).

We start with the single insertion of a multiplicative operator $P(x,t).$ Since we have an additional time dependence, the Fourier transform is
\beq
P(x,t) = \int \frac{d^d k}{(2 \pi)^{d/2}} \int \frac{d \omega}{\sqrt{2 \pi}} \, P(k, \omega) e^{i(kx-\omega t)}  
\eeq
and it is required to compute
\beq
\begin{aligned}
K_{1 P} (\tau) = & \int_0^{\tau} d \tau' \int d^d \tilde{x} \int d \tilde{t} \, \langle x t | e^{(\tau- \tau')\bigtriangleup}| \tilde{x} \tilde{t} \rangle P (\tilde{x}, \tilde{t}) 
\langle \tilde{x} \tilde{t} | e^{\tau' \bigtriangleup} | x' t' \rangle = \\
=& \int_0^{\tau} d \tau'  \frac{1}{(2 \pi)^2}  \frac{1}{(4 \pi (\tau-\tau'))^{d/2}}   \frac{1}{(4 \pi \tau')^{d/2}}  \int \frac{d \omega}{\sqrt{2 \pi}} \int d \tilde{t} e^{-i \omega \tilde{t}} \frac{m(\tau-\tau')}{m^2 (\tau-\tau')^2 + \frac{(t-\tilde{t})^2}{4}} \\
 & \frac{m \tau'}{m^2 \tau'^2 + \frac{(\tilde{t}-t')^2}{4}}  \int \frac{d^d k}{(2 \pi)^{d/2}} e^{ik \tilde{x}} 
\exp \le - \frac{(x-\tilde{x})^2}{4(\tau- \tau')} - \frac{(\tilde{x}-x')^2}{4 \tau'} \ri P(k, \omega)   \, .
\end{aligned}
\label{general single insertion time dependent}
\eeq
We observe that, despite the additional time dependence, the time and spatial parts of the integral still decouple and factorize. We can then use the same intermediate step in eq. (\ref{intermediate step spatial integral heat kernel single insertion P})
\beq
\begin{aligned}
& \int d^d \tilde{x} \int \frac{d^d k}{(2 \pi)^{d/2}} e^{ik \tilde{x}} 
\exp \le - \frac{(x-\tilde{x})^2}{4(\tau-\tau')} - \frac{(\tilde{x}-x')^2}{4s'}  \ri  = \\
& = \int \frac{d^d k}{2 \pi^{d/2}}  \exp \le - \frac{(x-x')^2}{4\tau} + i k \cdot \le x \frac{\tau'}{\tau} + x' \frac{\tau-\tau'}{\tau}  \ri - k^2 \frac{\tau'}{\tau} (\tau-\tau')  \ri   \, ,
\end{aligned}
\eeq
which in the case $ x=x' $ gives
\beq
\int \frac{d^d k}{2 \pi^{d/2}}  \exp \le  i k \cdot x - k^2 \frac{\tau'}{\tau} (\tau-\tau')  \ri   \, .
\label{result of the spatial integration, time dep insertions}
\eeq
The time integral $I(\omega)$ is better evauated after studying the analytic structure in the complex plane of the integrand
\beq
\begin{aligned}
I(\omega)= &  \int d \tilde{t} e^{-i \omega \tilde{t}} \frac{m(\tau-\tau')}{m^2 (\tau-\tau')^2 + \frac{(t-\tilde{t})^2}{4}} \frac{m \tau'}{m^2 \tau'^2 + \frac{(\tilde{t}-t')^2}{4}}  =  \\
= & 4 \alpha \beta  e^{-i \omega t'}  \int d \tilde{t} \frac{ e^{-i \omega \tilde{t}}}{(\tilde{t}+ i \beta)(\tilde{t}-i \beta)(\tilde{t} - \Delta t+ i \alpha)(\tilde{t} - \Delta t - i \alpha)} \, ,
\end{aligned}
\eeq
where we sent $ \tilde{t} \rightarrow \tilde{t} + t' $  and we defined
\beq
\alpha = 2m (\tau-\tau') \, , \qquad   \beta = 2m \tau'  \, , \qquad  \Delta t = t - t' \, .
\eeq
The quantities $ \alpha, \beta $ are positive by construction. We use the residue theorem to find
\beq
\begin{aligned}
I(\omega) & =   4 \alpha \beta  e^{-i \omega t'} \theta(\omega) \left[ \frac{\pi e^{- \beta \omega}}{\beta ((\Delta t + i \beta)^2 + \alpha^2)} +
 \frac{\pi e^{- \alpha \omega-i \Delta t \omega }}{\alpha ((\Delta t - i \alpha)^2 + \beta^2)}    \right] +  \\
& + 4 \alpha \beta  e^{-i \omega t'}  \theta (- \omega) \left[ \frac{\pi e^{\beta \omega}}{\beta ((\Delta t - i \beta)^2 + \alpha^2)} +
 \frac{\pi e^{ \alpha \omega-i \Delta t \omega }}{\alpha ((\Delta t + i \alpha)^2 + \beta^2)}    \right]  \, .
\end{aligned}
\eeq
It can be found that the expression for $ \omega=0 $ gives the time-independent results found in Appendix \ref{sect-First order expansion of the heat kernel operator} when using the prescription $ \theta(0)= 1/2 $ for the Heaviside distribution:
\beq
I(0) =\frac{8\pi m s}{4 m^2 \tau^2 + (t-t')^2}=  \int d \tilde{t} \frac{m(\tau- \tau')}{m^2 (\tau-\tau')^2 + \frac{(t-\tilde{t})^2}{4}} 
  \frac{m \tau'}{m^2 \tau'^2 + \frac{(\tilde{t}-t')^2}{4}} \, .
\eeq
The trace of the insertion is found putting $ t =t' $ to obtain
\beq
I(\omega, t=t') =  \frac{2 \pi}{m \tau} \frac{1}{\tau-2 \tau'} \left[ e^{-2m \tau' |\omega|} (\tau-\tau') - \tau' e^{-2m(\tau-\tau') |\omega|}  \right] =
\frac{2 \pi}{m \tau} + \mathcal{O} (\tau) \, .
\label{equal time insertion time dependent integral}
\eeq
Using eqs. (\ref{result of the spatial integration, time dep insertions}) and (\ref{equal time insertion time dependent integral}) inside 
eq. (\ref{general single insertion time dependent}) and expanding in the auxiliary time $\tau$ we finally obtain the result
\beq
\mathrm{Tr} \,  K_{1 P} (\tau)= \frac{2}{m (4 \pi \tau)^{d/2+1}} \mathrm{tr} \, \le \tau P(x,t) + \frac{1}{6} \tau^2 \p_i^2 P(x,t) + \mathcal{O} (\tau^3)   \ri \, .
\eeq
This is the same result of the case without time-dependence because the first order of the expansion of exponential terms vanishes.

Next we consider the single insertion of an operator with a spatial derivative acting on the fields. 
It turns out that the same trick of the time-independent case works, \emph{i.e.}
\beq
K_{1 Q_i } (s)= -  \frac{\p}{\p x'_i} \left[ \int_0^{s'} ds' \int d^d \tilde{x} \int d \tilde{t} \, \langle x t | e^{(s-s')\bigtriangleup}| \tilde{x} \tilde{t} \rangle Q_i (\tilde{x}, \tilde{t}) 
\langle \tilde{x} \tilde{t} | e^{s' \bigtriangleup} | x' t' \rangle   \right] \, .
\eeq
Since the expression in parenthesis does not change if we add a time dependence to the operators of the heat kernel expansion, and since spatial and temporal parts of the integral factorize, we obtain an equivalent formula also for 
\beq
\mathrm{Tr} \, K_{1 Q_i} (\tau) = \frac{2}{m (4 \pi \tau)^{d/2+1} } \mathrm{tr} \, \le - \frac{\tau}{2} \p_i Q_i (x,t) - \frac{\tau^2}{12} \p_i \p^2 Q_i (x,t) + \mathcal{O} (\tau^3) \ri \, .
\eeq
Single insertions of operators with a time derivative applied to the dynamical fields $ S(x,t) $ can be modified by time dependence, but since they vanish on the background (\ref{NC background heat kernel with particle number}) we will not consider them.


\section{Time-dependent insertion contributions to the heat kernel (second order)}
\label{app-Time-dependent insertion contributions to the heat kernel (second order)}

We start with the double insertion of multiplicative operators of kind $ P(x,t)  $
\beq
\begin{aligned}
  K_{2 P} (\tau)  = &  \int_0^\tau d \tau_2 \int_0^{\tau_2} d \tau_{1} \int d^d x_1 \int d^d x_2 \int dt_1  \int dt_2  \, \langle x t |  e^{(\tau- \tau_2)\bigtriangleup} | x_2 t_2 \rangle  & \\
 & \qquad \qquad P(x_2, t_2) \langle x_2 t_2 |  e^{(\tau_2-\tau_1)\bigtriangleup} | x_1 t_1 \rangle  P(x_1, t_1) \langle x_1 t_1 | e^{\tau_1 \bigtriangleup} | x' t' \rangle =  \\
  = &  \int_0^\tau d \tau_2 \int_0^{\tau_2} d \tau_{1} \frac{1}{(2\pi)^3} \frac{1}{(4 \pi (\tau-\tau_2))^{d/2}} \frac{1}{(4 \pi (\tau_2-\tau_1))^{d/2}} \frac{1}{(4 \pi \tau_1)^{d/2}} \int d^d x_1 \int d^d x_2  & \\
&   \int \frac{d^d k_1}{(2 \pi)^{d/2}}  \int \frac{d^d k_2}{(2 \pi)^{d/2}} 
 \int \frac{d \omega_1}{\sqrt{2 \pi}}\int \frac{d \omega_2}{\sqrt{2 \pi}} \, \Upsilon \,  \Psi^{P P}   e^{-i \omega_1 t_1 - i \omega_2 t_2} P(k_2, \omega_2) P(k_1, \omega_1)  \, , &
\end{aligned}
\label{espansione time-dependent inserzione doppia di P}
\eeq 
where $\Upsilon$ was given in eq. (\ref{Upsilon quantity heat kernel expansion}), while the $\Psi^{PP}$ term in eq. (\ref{time integral heat kernel expansion appendix}).
The time and spatial parts factorize again; the latter was evaluated in eq. (\ref{result spatial parts double insertions heat kernel}) and at coincident points it is given by
\beq
\begin{aligned}
& \int d^d x_1 \int d^d x_2 \, \Upsilon \, 
= (4 \pi)^{d} \le \frac{\tau_1 (\tau-\tau_2) (\tau_2-\tau_1)}{\tau} \ri^{d/2} \times & \\
& \times \exp \le  ik_1 x_1 + i k_2 x_2 +
k_1^2 \le \frac{\tau_1^2}{\tau} - \tau_1 \ri + k_2^2 \le \frac{\tau_2^2}{\tau} - \tau_2 \ri +2 k_1 k_2 \le \frac{\tau_1 \tau_2}{\tau} - \tau_1 \ri  \ri & \, .
\label{parte spaziale inserzione doppia}
\end{aligned}
\eeq
The temporal part $\Psi^{PP}$ can be integrated along the $t_1$ coordinate using the same technique of the single insertion case.
Using the definitions
\beq
\alpha = 2m (\tau_2 - \tau_1) \, , \qquad   \beta = 2m \tau_1 \, , \qquad  \Delta t = t_2 - t' \, ,
\eeq
we obtain
\beq
\begin{aligned}
  I(\omega_1) & = \int dt_1   e^{- i \omega_1 t_1}\frac{m (\tau_2-\tau_1)}{m^2 (\tau_2-\tau_1)^2 + \frac{(t-\tilde{t})^2}{4}}  \frac{m \tau_1}{m^2 \tau_1^2 + \frac{(t-\tilde{t})^2}{4}} = \\
 & =   4 \alpha \beta  e^{-i \omega t'} \theta(\omega_1) \left[ \frac{\pi e^{- \beta \omega_1}}{\beta ((\Delta t + i \beta)^2 + \alpha^2)} +
 \frac{\pi e^{- \alpha \omega-i \Delta t \omega }}{\alpha ((\Delta t - i \alpha)^2 + \beta^2)}    \right] +  \\
& + 4 \alpha \beta  e^{-i \omega t'}  \theta (- \omega_1) \left[ \frac{\pi e^{\beta \omega}}{\beta ((\Delta t - i \beta)^2 + \alpha^2)} + \frac{\pi e^{ \alpha \omega-i \Delta t \omega }}{\alpha ((\Delta t + i \alpha)^2 + \beta^2)}    \right]  \, .
\end{aligned}
\eeq
The last step in the time integration consists in evaluating
\beq
\Psi (t,t', \omega_1, \omega_2) = \int dt_2 e^{-i \omega_2 t_2} \frac{m(\tau-\tau_2)}{m^2 (\tau-\tau_2)^2 + \frac{(t-t_2)^2}{4}} I (\omega_1) \, .
\eeq
The formal result is very cumbersome, but  it can be checked that, using the prescription $\theta(0)=1/2$, it  gives exactly the time-independent result in the limit of vanishing frequencies:
\beq
\Psi (t=t', \omega_1=\omega_2=0) = \frac{16 \pi^2 \theta^2(0)}{m s} = \frac{4 \pi^2}{m s}  \, .
\eeq
Moreover, in order to compute the insertions of time-dependent operators we only need the lowest orders of the expansion around $\tau=0$ of the solution at coincident points, which is
\beq
\Psi (t=t', \omega_1, \omega_2) =  \frac{4 \pi^2}{m s} e^{-i (\omega_1 + \omega_2)t} + \mathcal{O}  (s) \, .
\label{parte temporale inserzione doppia}
\eeq
The zeroth order in the variable $\tau$ vanishes.

Combining eqs. (\ref{parte spaziale inserzione doppia}) and (\ref{parte temporale inserzione doppia}) into (\ref{espansione time-dependent inserzione doppia di P}) we find the same result of the time-independent case
\beq
\mathrm{Tr} \, K_{2PP} = \frac{2}{m(4 \pi \tau)^{d/2+1}} \mathrm{tr} \, \left( \frac{\tau^2}{2} P(x,t)^2 + \mathcal{O} (\tau^3) \right)  \, .
\eeq
Additional new terms can contribute only to higher orders in $\tau,$ but they do not modify the $ a_4 $ coefficient.

Since time and space integrals factorize and there are no contributions to lower-order terms in the heat kernel expansion, we can similarly find that $ \mathrm{Tr} \, K_{2 X} $ have the same expressions of the time-independent case, if we choose among the set
\beq
X= \lbrace P(x,t) , Q_i (x,t)  \rbrace \, .
\eeq
Additional terms could appear in insertions concerning the operator $ S(x,t)  $. They  will not be considered here because $ S(x,t) $ vanishes in the background (\ref{NC background heat kernel with particle number}).

\chapter{Non-relativistic Wess-Zumino model in components} 
\label{app_Non-relativistic Wess-Zumino model in components}

In this Appendix we show an alternative way to study quantum corrections to the Galilean Wess-Zumino model using component field formalism, which is more used in the literature concerning non-relativistic physics.

In Section \ref{sect-The non-relativistic Wess-Zumino model} we applied null reduction to derive the action in terms of superfields, and after decomposing them into the components of the supermultiplet, we found the action in terms of fundamental fields.
The same expression can be found with a slightly different procedure:
\begin{itemize}
\item Take the relativistic WZ model (\ref{classical rel WZ model}) in superfield formalism
\item Express the action in terms of the relativistic component fields
\item Perform the null reduction on the component fields.
\end{itemize}
The fact that this procedure gives the same result of Section \ref{sect-The non-relativistic Wess-Zumino model} confirms that the prescription we gave to apply null reduction at the level of superfields works well.
Moreover, it is possible to apply the DLCQ prescription after integrating out the auxiliary fields appearing after the reduction of the WZ action in components, and we find again the same result. 

We start the analysis of quantum corrections considering the action in components \eqref{Lagrangiana finale WZ interagente potenziale cubico}. Scalars and fermions share the same kinetic operator and then the tree-level propagators are
\bea
&& \langle \varphi_{1}(\omega, \vec{p}) \bar{\varphi}_{1}(-\omega, -\vec{p}) \rangle = \langle \chi_{1}(\omega, \vec{p}) \bar{\chi}_{1}(-\omega, -\vec{p}) \rangle = 
\frac{i}{2m \omega - \vec{p}^{\, 2} + i \varepsilon}  \nonumber \\
&& \langle \varphi_{2}(\omega, \vec{p}) \bar{\varphi}_{2}(-\omega, -\vec{p}) \rangle = \langle \chi_{2}(\omega, \vec{p}) \bar{\chi}_{2}(-\omega, -\vec{p}) \rangle = 
\frac{i}{4m \omega - \vec{p}^{\, 2} + i \varepsilon}  
\eea

\begin{figure}[h]
\centering
\begin{subfigure}[b]{0.22\linewidth}
\includegraphics[scale=1]{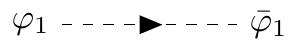}
\end{subfigure}
\medskip
\begin{subfigure}[b]{0.22\linewidth}
\includegraphics[scale=1]{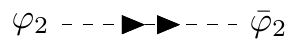}
\end{subfigure}
\begin{subfigure}[b]{0.22\linewidth}
\includegraphics[scale=1]{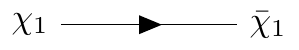}
\end{subfigure}
\begin{subfigure}[b]{0.22\linewidth}
\includegraphics[scale=1]{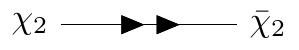}
\end{subfigure}
\caption{Propagators for the dynamical non-relativistic fields. The bosons are denoted by dashed lines, while the fermions with continuous lines. The number of arrows denote the particle number.} \label{fig17_propagatori_dinamici_componenti}
\end{figure}

\begin{figure}[h]
\begin{subfigure}[b]{0.5\linewidth}
\centering
\includegraphics[scale=1]{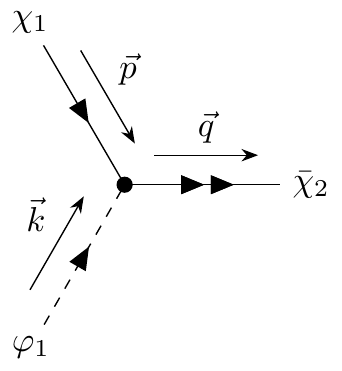}
\subcaption{$ -\sqrt{2} i g \left[ (q_1 - i q_2) -2 (p_1 - i p_2) \right] $}
\end{subfigure}
\begin{subfigure}[b]{0.5\linewidth}
\centering
\includegraphics[scale=1]{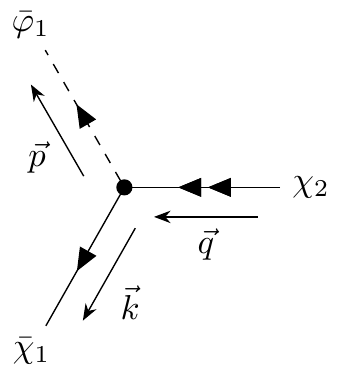}
\subcaption{$ \sqrt{2} i g^{*} \left[ (q_1 + i q_2) -2 (k_1 + i k_2) \right] $}
\end{subfigure}
\begin{subfigure}[b]{0.5\linewidth}
\centering
\includegraphics[scale=1]{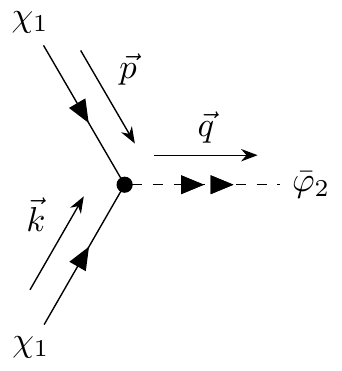}
\subcaption{$ -2 i g \left[ (k_1 - i k_2) - (p_1 - i p_2) \right] $}
\end{subfigure}
\begin{subfigure}[b]{0.5\linewidth}
\centering
\includegraphics[scale=1]{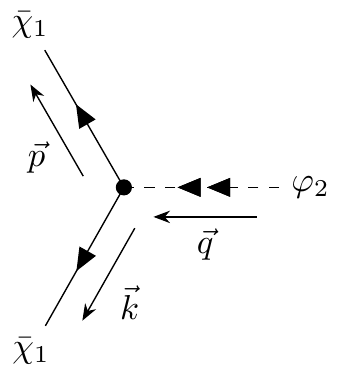}
\subcaption{$ 2 i g^{*} \left[ (k_1 + i k_2) - (p_1 + i p_2) \right] $}
\end{subfigure}
\caption{Feynman rules for three-point vertices. Scalars are denoted by dashed lines, while fermions by continuous lines. }
\label{fig18_Vertici_3_Lagrangiana_campi_dinamici_componenti}
\end{figure}

\begin{figure}[h]
\centering
\begin{subfigure}[b]{0.3\linewidth}
\includegraphics[scale=1]{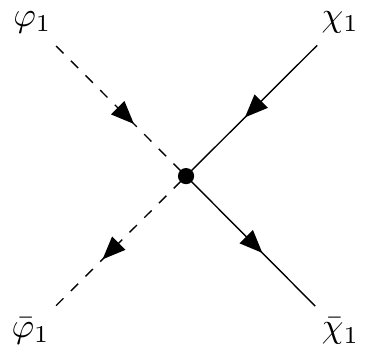}
\subcaption{$ -2 i |g|^2 $}
\end{subfigure}
\begin{subfigure}[b]{0.3\linewidth}
\includegraphics[scale=1]{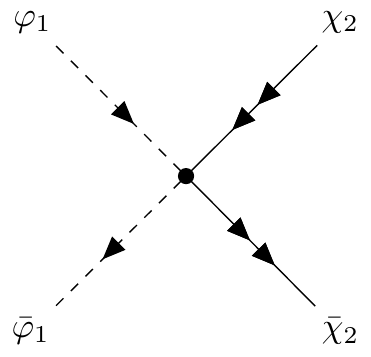}
\subcaption{$ -8 i |g|^2 $}
\end{subfigure}
\begin{subfigure}[b]{0.3\linewidth}
\includegraphics[scale=1]{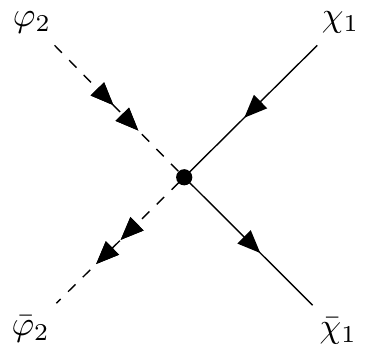}
\subcaption{$ 4 i |g|^2 $}
\end{subfigure}
\begin{subfigure}[b]{0.3\linewidth}
\includegraphics[scale=1]{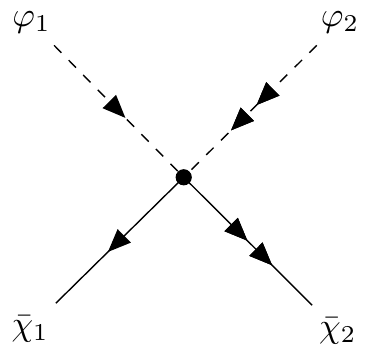}
\subcaption{$ 4 \sqrt{2} i |g|^2 $}
\end{subfigure}
\begin{subfigure}[b]{0.3\linewidth}
\includegraphics[scale=1]{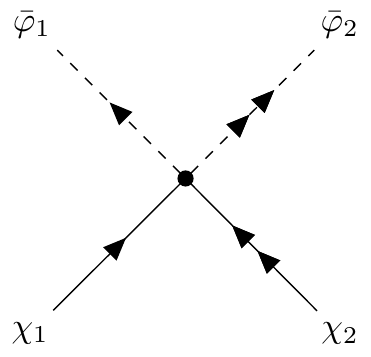}
\subcaption{$ 4 \sqrt{2} i |g|^2 $}
\end{subfigure}
\begin{subfigure}[b]{0.3\linewidth}
\includegraphics[scale=1]{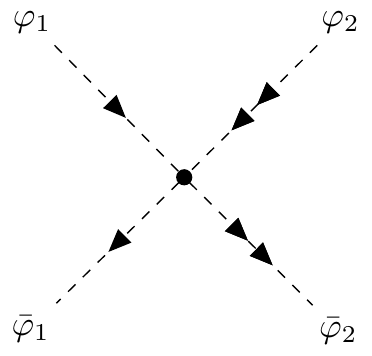}
\subcaption{$ -4  i |g|^2 $}
\end{subfigure}
\begin{subfigure}[b]{0.3\linewidth}
\includegraphics[scale=1]{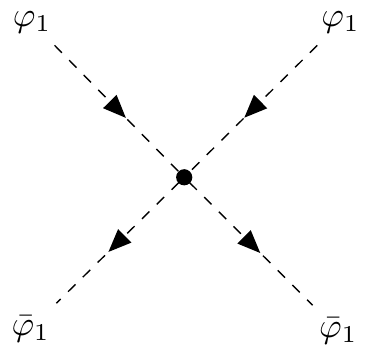}
\subcaption{$ -4 i |g|^2 $}
\end{subfigure}
\caption{Feynman rules for four-point vertices. Scalars are denoted by dashed lines, while fermions by continuous lines.}
\label{fig19_Vertici_4_Lagrangiana_campi_dinamici_componenti}
\end{figure}

Interaction vertices can be read directly from the lagrangian and are shown in fig. \ref{fig18_Vertici_3_Lagrangiana_campi_dinamici_componenti} and \ref{fig19_Vertici_4_Lagrangiana_campi_dinamici_componenti}, where we use dashed and continous lines to denote scalars and fermions, respectively.
The cubic vertices contain derivative interactions and then depend explicitly from the spatial momentum along the lines, while quartic vertices do not depend from the momentum.
 
In order to classify the admitted diagrams, we can take into account that the reduction in components does not affect the propagators as functions of  $\omega$ and $\vec{p}$. Therefore, the arguments that led to formulate the fundamental selection rule \ref{srule2} are still true. Moreover, the conservation of particle number at each vertex still provides the driving rule to select the admissible topologies and arrows configurations.
 
In order to properly define physical quantities and Green functions, we introduce renormalized fields and couplings defined as
\beq
\begin{cases}\label{ren_components}
\varphi_a = Z_a^{-1/2} \,  \varphi_a^{(B)} = \le 1- \frac12 \delta_{\varphi_a} \ri \varphi_a^{(B)} \qquad a=1,2\\
\chi_a = Z_a^{-1/2} \,  \chi_a^{(B)} = \le 1- \frac12 \delta_{\chi_a} \ri \chi_a^{(B)}  \\
m = Z^{-1}_m m^{(B)} = (1 - \delta_m) m^{(B)} \\
g = \mu^{-\varepsilon} Z^{-1}_g g^{(B)} = \mu^{-\varepsilon} (1 - \delta_g) g^{(B)} 
\end{cases}
\eeq
Spatial integrals are computed in dimension $ d = 2 - \varepsilon $ and we have introduced the mass scale $ \mu $ to keep the coupling constant dimensionless.


\subsubsection{One-loop corrections to the self-energies}

\begin{figure}[h]
\centering
\begin{subfigure}[b]{0.45\linewidth}
\centering
\includegraphics[scale=1.35]{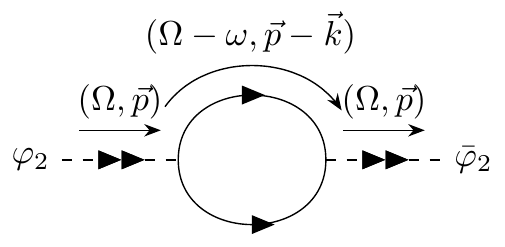}
\subcaption{$ i \mathcal{M}^{(2)}_{\mathrm{b}} (\varphi_2, \bar{\varphi}_2) $}
\end{subfigure}
\begin{subfigure}[b]{0.45\linewidth}
\centering
\includegraphics[scale=1.35]{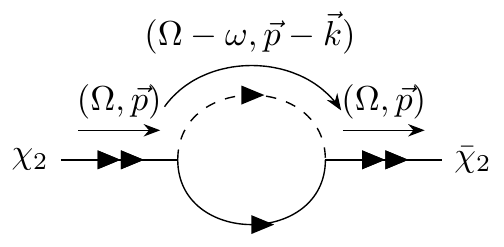}
\subcaption{$ i \mathcal{M}^{(2)}_{\mathrm{f}} (\chi_2, \bar{\chi}_2) $}
\end{subfigure}
\caption{1-loop correction to the scalar (a) and fermionic (b) self-energies in sector 2.}
\label{fig20_Correzione 1-loop settore 2 dinamico}
\end{figure}

By applying selection rule \ref{srule2} and particle number conservation we find that there are no admissible one-loop self-energy diagrams for particles in sector 1, while there is a non-vanishing contribution both for the scalar and the fermion in sector 2 corresponding to the diagrams in fig. \ref{fig20_Correzione 1-loop settore 2 dinamico}.
Direct inspection leads to the integral
\beq
i \mathcal{M}_{\mathrm{b}}^{(2)}=
i \mathcal{M}_{\mathrm{f}}^{(2)} = \frac{2 |g|^2}{(2 \pi)^3} \int d\omega \, d^2 k \,
\frac{(\vec{p} -2 \vec{k})^2}{\left[2 m \omega - \vec{k}^2 + i \varepsilon \right] \left[2m (\Omega - \omega) - (\vec{p} - \vec{k})^2 + i \varepsilon \right]}  
\eeq
Even if SUSY is not manifest in the component field formalism, we see that it shows via the equality of the quantum corrections of the fermionic and bosonic fields.
 
We solve the integration along the energy using the residue theorem 
\beq
\mathcal{M}^{(2)} =- \frac{ |g|^2}{m} \int \frac{d^2 k}{(2 \pi)^2} \, 
\frac{(\vec{p} - 2 \vec{k})^2}{2m \Omega - \vec{k}^2 - (\vec{p}-\vec{k})^2 + i \varepsilon} \, .
\eeq
The remaining integral is UV divergent and can be computed with standard techniques of dimensional regularization. 
In generic dimensions \emph{d} there exists a region in complex plane where the integral is convergent and we can translate the integration variable as
\beq
\vec{l} = \vec{k} - \frac{\vec{p}}{2} \, , \qquad
d \vec{l} = d \vec{k} \, ,
\eeq
giving
\beq
\mathcal{M}^{(2)} = \frac{2 |g|^2 \mu^{2(2-d)}}{(2 \pi)^d} \int d^d l \, \frac{4 l^2}{2 m \Omega - 2 l^2 - \frac{\vec{p}^2}{2}} = - \frac{4 |g|^2 \mu^{2(2-d)}}{(2 \pi)^d} \frac{2 \pi^{d/2}}{\Gamma(d/2)} \int_0^{\infty} dl \, \frac{l^{d+1}}{l^2 - m \Omega + \frac{\vec{p}^2}{4}} \, .
\eeq
After evaluating the remaining integral along the radial direction we find
\beq
\mathcal{M}^{(2)} =  \frac{|g|^2}{m} \, d \, \frac{ \mu^{2(2-d)}}{(4 \pi)^{d/2}} \, \Gamma \le 
- \frac{d}{2} \ri \, \le \frac{\vec{p}^2}{4} - m \Omega \ri^{\frac{d}{2}} =  \frac{|g|^2}{2 \pi m} \le 2 m \Omega - \frac{\vec{p}^2}{2}  \ri \frac{1}{\varepsilon} + \mathrm{finite}
\label{bubbola}
\eeq
In the minimal subtraction scheme the $1/\epsilon$ pole is cancelled by setting in  \eqref{ren_components}
\beq
\delta_{\varphi_2}^{\rm (1loop)}= \delta_{\chi_2}^{\rm (1loop)} = - \frac{|g|^2}{4 \pi m} \frac{1}{\varepsilon} \, , \qquad \; 
\delta_m^{\rm (1loop)} = 0  
\label{controtermini phi2}
\eeq
whereas $\delta_{\varphi_1}^{\rm (1loop)} = \delta_{\chi_1}^{\rm (1loop)} = 0$. This result is consistent with the one-loop renormalization \eqref{superfield_ren} that we have found in superspace.

\subsubsection{One-loop corrections to three-point vertices}

The action in components contains two kinds of three-point vertices (plus their complex conjugates). 
The vertex $ \mathbf{V}_3 ( \chi_1 , \chi_1 , \bar{\varphi}_2 )  $ and its complex conjugate are not corrected at one loop because we cannot build any diagram consistent with particle number conservation. It then follows immediately that
\beq
\le \delta_g + \delta_{\chi_1} + \frac12 \delta_{\varphi_2}^* \ri \Big|_{\rm (1loop)}   = 0 
\eeq
Combining this relation with eq. (\ref{controtermini phi2}), we find
\beq
\delta_g^{\rm (1loop)}  =   \frac{|g|^2}{8 \pi m} \frac{1}{\varepsilon}  
\label{controtermini primo vertice a 3}
\eeq

\begin{figure}[h]
\centering
\includegraphics[scale=1.5]{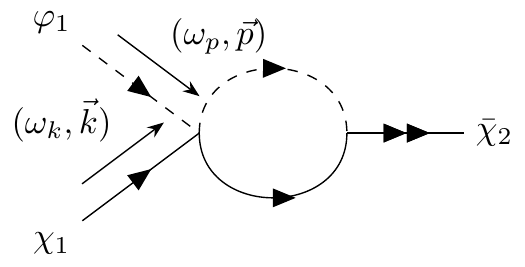}
\caption{1-loop correction to the 3-point vertex.}
\label{fig21_Correzione 1-loop vert a 3}
\end{figure}

On the other hand, the vertex  $ \mathbf{V}_3 ( \varphi_1 , \chi_1 , \bar{\chi}_2 ) $ has in principle a one-loop contribution shown in fig. \ref{fig21_Correzione 1-loop vert a 3}.

After the integration by residues of the $\omega$ variable, this diagram gives
\beq
\mathcal{M}^{(3)} (\varphi_1, \chi_1, \bar{\chi}_2) = - \frac{|g|^2}{m}  \frac{ \sqrt{2} g}{(2 \pi)^2}
\int d^2 l \, \frac{(p_1 + k_1) - i (p_2 + k_2) - 2 (l_1 - i l_2)}
{2 m (\omega_p + \omega_k) - \vec{l}^2 - (\vec{p} + \vec{k} - \vec{l})^2 + i \varepsilon}  
\eeq
We perform dimensional regularization along the spatial directions. 
Since the integrand contains in the numerator an expression which explicitly depends from the spatial momenta $ \vec{l} = (l_1, l_2), $ we should give a prescription to define in a covariant way the numerator.
For example we can assume that it comes from the contraction $ \vec{v} \cdot \vec{l} $ and that the first vector in \emph{d} dimensions is $ \vec{v}=(1,-i) , $ andis promoted in $d$ dimensions to a vector whose only the first two entries are non-vanishing.
In this way we obtain
\beq
\mathcal{M}^{(3)} (\varphi_1, \chi_1, \bar{\chi}_2) =- \frac{|g|^2}{m}  \frac{ \sqrt{2} g \mu^{3(2-d)}}{(2 \pi)^d}
\int d^d l \, \frac{ \vec{v} \cdot (\vec{p} + \vec{k} - 2  \vec{l} \, ) }
{2 m (\omega_p + \omega_k) - \vec{l}^2 - (\vec{p} + \vec{k} - \vec{l})^2 + i \varepsilon}  
\eeq
With the change of variables $\vec{q} = \vec{l} - \frac{\vec{p}+ \vec{k}}{2}$ we find
\beq
\mathcal{M}^{(3)} (\varphi_1, \chi_1, \bar{\chi}_2) = - \frac{|g|^2}{m} \frac{ \sqrt{2} g  \mu^{3(2-d)}}{(2 \pi)^d}
\int d^d q \, \frac{\vec{v} \cdot \vec{q}}{ q^2 - m (\omega_p + \omega_k) + \frac{(\vec{p} + \vec{k})^2 }{4} + i \varepsilon}  
\eeq
This integral vanishes because the range is even and the integrand is odd. 
This implies that the following relation holds:
\beq
\le \delta_g + \frac12 \delta_{\varphi_1} + \frac12 \delta_{\chi_1} + \frac12 \delta_{\chi_2} \ri \Big|_{\rm (1loop)}= 0 
\eeq
This condition is automatically satisfied by results in eqs. (\ref{controtermini phi2}) and (\ref{controtermini primo vertice a 3}).

We note that the one-loop result $\delta_g = - \frac12 \delta_{\chi_2}$ is the component version of the superspace non-renormalization theorem.
As for the self-energy, we find that quantum corrections do not break SUSY.

\subsubsection{One-loop corrections to four-point vertices}

In principle, the one-loop evaluation of self-energies and three-point vertices allows to solve for all the unknowns in lagrangian 
(\ref{Lagrangiana finale WZ interagente potenziale cubico}). Moreover, we have verified that the corrections are all consistent between themselves and with the superspace results.
However, we will provide further evidence of SUSY invariance at the level of component field formulation by considering the 1PI diagrams giving quantum corrections to the four-point vertices.
This will also show how the non-renormalization theorem works at the level of fundamental fields.

\begin{figure}[h]
\centering
\begin{subfigure}[b]{0.3\linewidth}
\includegraphics[scale=1]{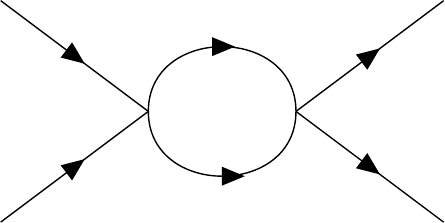}
\end{subfigure}
\begin{subfigure}[b]{0.3\linewidth}
\includegraphics[scale=1]{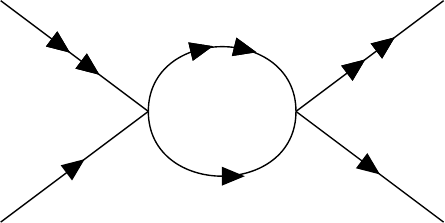}
\end{subfigure}
\begin{subfigure}[b]{0.3\linewidth}
\includegraphics[scale=1]{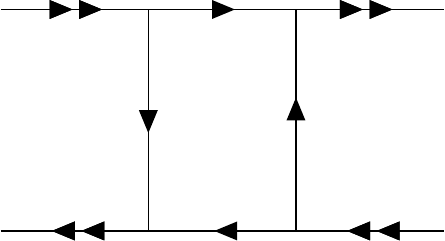}
\end{subfigure}
\begin{subfigure}[b]{0.3\linewidth}
\includegraphics[scale=1]{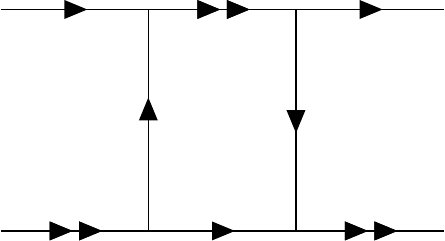}
\end{subfigure}
\begin{subfigure}[b]{0.3\linewidth}
\includegraphics[scale=1]{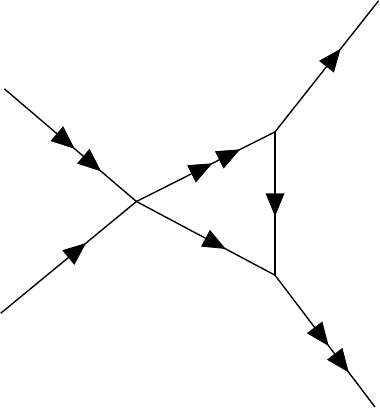}
\end{subfigure}
\begin{subfigure}[b]{0.3\linewidth}
\includegraphics[scale=1]{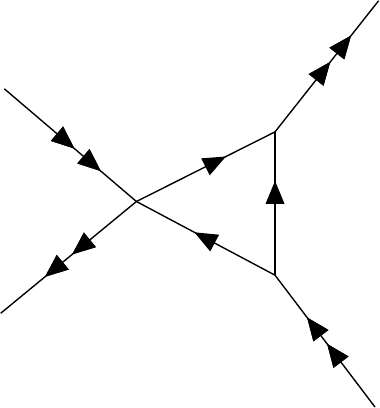}
\end{subfigure}
\begin{subfigure}[b]{0.3\linewidth}
\includegraphics[scale=1]{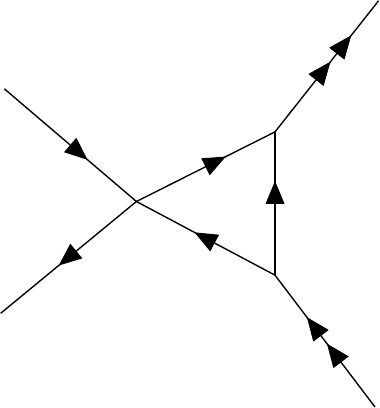}
\end{subfigure}
\caption{Possible topologies of one-loop corrections to four-point vertices for the dynamical fields.
In the picture we do not distinguish between bosonic and fermionic lines.}
\label{fig22_Classificazione_topologie_diagrammi_4}
\end{figure}

Compared to the previous cases, we have far more possibilities to build four-point diagrams with the vertices at our disposal (see figs. \ref{fig18_Vertici_3_Lagrangiana_campi_dinamici_componenti}, \ref{fig19_Vertici_4_Lagrangiana_campi_dinamici_componenti}). All the topologies of diagrams consistent with particle number conservation at each vertex are reported in fig. \ref{fig22_Classificazione_topologie_diagrammi_4}.

\begin{figure}[h]
\centering
\begin{subfigure}[b]{0.5\linewidth}
\includegraphics[scale=1.3]{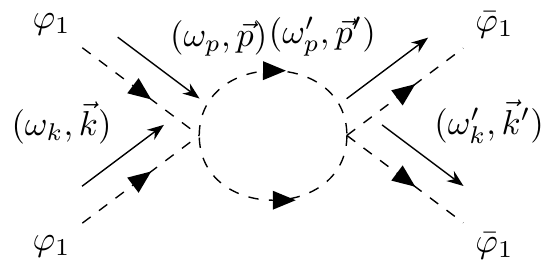}\caption{}
\end{subfigure}
\begin{subfigure}[b]{0.4\linewidth}
\includegraphics[scale=1.1]{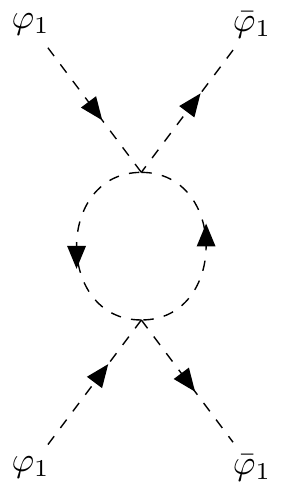}
\caption{}
\end{subfigure}
\begin{subfigure}[b]{0.4\linewidth}
\includegraphics[scale=1.1]{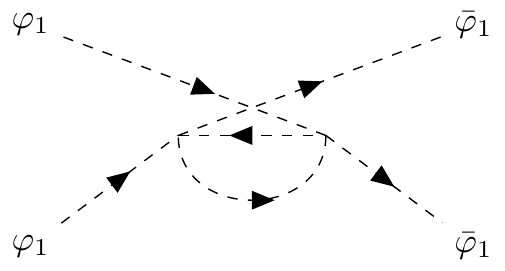}
\caption{}
\end{subfigure}
\caption{1-loop 1PI corrections to the 4-point vertex with external scalars from sector 1, coming from channels \emph{s}, \emph{t} and \emph{u} respectively.}
\label{fig23_Correzioni canali t e u 1-loop vert a 4}
\end{figure}

For example, we consider the first of such diagrams, \emph{i.e.} the one-loop correction to the vertex $ \mathbf{V}_4 (\varphi_1 , \varphi_1 , \bar{\varphi}_1 , \bar{\varphi}_1)$. This is the only graph among the many containing as external lines only fields from sector 1. 
We report the precise assignments of momenta and energy in fig. \ref{fig23_Correzioni canali t e u 1-loop vert a 4}.

The \emph{t} and \emph{u}-channel diagrams vanish because we have circulating arrows in the internal loop.
After integration in $\omega$ with the residue theorem, the integral corresponding to the \emph{s}-channel diagram is
\beq
\mathcal{M}^{(4)} (\varphi_1, \varphi_1, \bar{\varphi}_1, \bar{\varphi}_1) = - \frac{4 |g|^4}{m} \int \frac{d^2 l}{(2 \pi)^2} \,
\frac{1}{2m (\omega_p + \omega_k) - \vec{l}^2 - (\vec{p} + \vec{k} - \vec{l})^2 + i \varepsilon} 
\eeq
Performing the change of variables $\vec{q} = \vec{l} - \frac{\vec{p} + \vec{k}}{2}$, in dimensional regularization we can write
\beq
\mathcal{M}^{(4)} (\varphi_1, \varphi_1, \bar{\varphi}_1, \bar{\varphi}_1) =  \frac{4|g|^4}{m} \frac{ \mu^{4(2-d)}}{(4 \pi)^{d/2}} \, \frac{1}{\Gamma(d/2)} \int_0^{\infty} dq \,
\frac{q^{d-1}}{q^2 -m (\omega_p + \omega_k) + \frac{(\vec{p}+ \vec{k})^2}{4} + i \varepsilon} 
\eeq
After performing the last integration and expanding in $ \varepsilon=2-d $ we obtain
\beq
\mathcal{M}^{(4)} (\varphi_1, \varphi_1, \bar{\varphi}_1, \bar{\varphi}_1) =  \frac{|g|^4}{\pi m} \frac{1}{\varepsilon} + \mathrm{finite} 
\label{correzione 1 loop vertice a 4 scalari}
\eeq
The renormalization condition in minimal subtraction scheme requires  
\beq
\mathcal{M}^{(4)} (\varphi_1, \varphi_1, \bar{\varphi}_1 \bar{\varphi}_1) - 4  |g|^2 (2\delta_{g} + 2 \delta_{\varphi_1}) = 0 \, ,
\eeq
which means
\beq
\delta_g^{\rm (1loop)} =  \frac{|g|^2}{8 \pi m} \frac{1}{\varepsilon}  \, .
\eeq
This is consistent with \eqref{controtermini primo vertice a 3} and confirms that SUSY is preserved by quantum corrections.
Since the quantum corrections of the coupling constant  $g$ are completely determined by the wave-function renormalization, this is also a manifest way to see that the non-renormalization theorem works.

\subsubsection{Two-loop corrections to the self-energy}

We observe that in component field formalism the number of Feynman diagrams to study at every loop order is much greater than using the superfield approach.
This makes the check of SUSY invariance and the study of quantum corrections more involved when the number of loops increase.
However, the selection rules \ref{srule2} and \ref{srule3} help in decreasing the number of diagrams to consider in the non-relativistic case. In particular the 2-loop order for the self-energy is easily treatable and then we will consider it as an example of higher loop corrections in component field formalism.

\begin{figure}[h]
\centering
\includegraphics[scale=1.5]{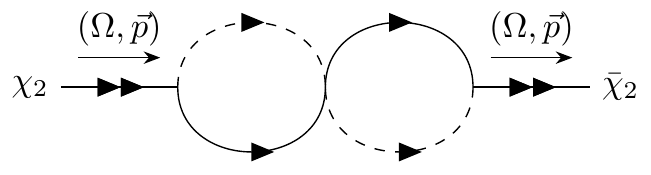}
\caption{Two-loop correction to the self-energy for the dynamical fermion in sector 2.}
\label{fig24_correzione 2 loop fermione dinamico}
\end{figure}

It is in fact possible to find only one \emph{a priori} non-vanishing diagram modifying the self-energy of the fermions in sector 2. This is depicted in Fig. \ref{fig24_correzione 2 loop fermione dinamico} and by consistency we expect to find that this contribution vanishes, because we do not find a diagram contributing to the bosonic superpartner.
We now prove that this is indeed the case.

Writing down the corresponding integral and first performing the $ \omega_k, \omega_l $ integrations by using the residue technique we find
\beq
\mathcal{M}^{(4)}_{\mathrm{f}} (\chi_2, \bar{\chi}_2) =
- \frac{|g|^4}{m^2} \int \frac{d^2 k \, d^2 l}{(2 \pi)^4} \, 
\frac{\vec{p}^{\, 2} + 4 \vec{l} \cdot \vec{k} + 2 (\vec{l}+ \vec{k}) \cdot \vec{p}}
{\left[2 m \Omega - \vec{k}^2 - (\vec{p}- \vec{k})^2 + i \varepsilon \right]\left[2 m \Omega - \vec{l}^2 - (\vec{p}- \vec{l})^2 + i \varepsilon \right]}  
\eeq
Performing the change of variables 
\beq
\vec{k}= \vec{K}+ \frac{\vec{p}}{2} \, , \qquad
\vec{l}= \vec{L}+ \frac{\vec{p}}{2}  
\eeq
and continuing the integral to $d=2-\epsilon$ dimensions we find
\beq
\mathcal{M}^{(4)}_{\mathrm{f}} (\chi_2, \bar{\chi}_2) =
- \frac{|g|^4}{m^2} \frac{\mu^{4(2-d)}}{(2 \pi)^{2d}} \int d^d K \, d^d L \, \frac{4 \vec{K} \cdot \vec{L}}
{\left[2 m \Omega - 2 K^2 - \frac{\vec{p}^2}{4} + i \varepsilon \right]\left[2 m \Omega - 2 L^2 - \frac{\vec{p}^2}{2} + i \varepsilon \right]} 
\eeq
The two integrals vanish for symmetry reasons.


\chapter{Additional details on the complexity computations}
\label{app-Additional details on the complexity computations}

In this appendix we collect various additional technical details useful to the computations of the complexity conjectures.

\section{An explicit model for WAdS black holes}
\label{app-An explicit model for WAdS black holes}

We start with the explicit model whose entropy satisfies the area law and admitting the metric eq.~(\ref{BHole}) as a solution \cite{Banados:2005da}, which we introduced in Section \ref{sect-An explicit realization in Einstein gravity}.
We will give a compendium of the dictionary required to match the notation used in the main text with the conventions of \cite{Banados:2005da}, and in this way we will find an alternative procedure to determine the conserved charges of the black hole.

In fact, in ref. \cite{Banados:2005da} the conserved charges associated to the asympthotic isometries of the black hole have been computed starting from the following form of the metric in the coordinates $ (\tilde{t}, \tilde{r}, \tilde{\theta}) $: 
\beq
ds^2 = p d\tilde{t}^2 + \frac{d\tilde{r}^2}{h^2-pq} + 2 h d\tilde{t} d\tilde{\theta} + q d \tilde{\theta}^2 \, ,
\label{forma metrica Godel}
\eeq
with functions given by
\beq
p(\tilde{r})= 8 G \mu  \, , \qquad 
q (\tilde{r}) = -\frac{4G \mathcal{J} }{\a} +2 \tilde{r} - 2 \frac{\gamma^2}{L^2} \tilde{r}^2 \, , \qquad
h (\tilde{r}) = -2 {\a} \tilde{r}  \, ,
\eeq
and $U(1)$ gauge field 
\beq
A=A_{\tilde{t}} d \tilde{t}+ A_{\tilde{\theta}} d \tilde{\theta} \, , \qquad
A_{\tilde{t}} (\tilde{r}) = \frac{{\a} ^2 L^2 -1}{\gamma {\a} L} + \zeta  \, , \qquad
A_{\tilde{\theta}} (\tilde{r})= - \frac{4G}{ {\a} } Q + \frac{2\gamma}{L} \tilde{r} \, ,
\eeq
where
\beq
\gamma = \sqrt{\frac{1-{\alpha}^2 L^2}{8 G \mu}} \, ,
\eeq
and $\zeta$ is a gauge constant.

We can put the metric (\ref{BHole}) in the form (\ref{forma metrica Godel}) by means of the coordinate change
\beq
\tilde{t}  = \sqrt{\frac{l^3}{\omega}} t  \, , \qquad
\tilde{r}=r - \frac{\sqrt{r_+ r_- (\nu^2 +3)}}{2 \nu} \, , \qquad
\tilde{\theta} = \frac{\sqrt{\omega l^3}}{2} \theta \, ,
\eeq
where
\beq
\omega = \frac{\nu^2 +3}{2 \nu l} \le \nu (r_+ + r_-) - \sqrt{r_+ r_- (\nu^2 +3)} \ri \, .
\eeq
The previous set of transformations is such that the gauge field in the coordinates $ (t,r,\theta) $ can be written as $  A= a dt + (b+cr) d \theta , $ motivating the ansatz (\ref{ansatz per campo di gauge}). 

The quantities $ \mu, \mathcal{J}, Q $ appearing in the previous solution are respectively 
identified with the mass, angular momentum and charge of the black hole.
The equations of motion and the change of coordinates do not uniquely fix the charge \emph{Q}, while we identify
\beq
\mu= \frac{\nu^2 +3}{16 G l^2} \le r_+ + r_- - \frac{\sqrt{r_+ r_- (\nu^2 +3)}}{\nu} \ri \, ,
\eeq
\beq
\mathcal{J} =  \frac{ 2 \nu  (r_+ + r_-) \sqrt{r_+ r_- (\nu^2 +3)} - (5\nu^2 +3)r_+ r_-}  
{8 G l  \le \nu  (r_+ + r_-)  - \sqrt{r_+ r_- (\nu^2 +3)}  \ri} \, .
\eeq
As it is pointed out in \cite{Banados:2005da}, the set $ \lbrace \mu, \mathcal{J} , Q \rbrace $ satisfies the first law of thermodynamics in the form
\beq
d\mu = T dS + \Omega \, d\mathcal{J} + \Phi_{\mathrm{tot}} dQ \, ,
\eeq
where the total electric potential is shown to be $ \Phi_{\mathrm{tot}} = 0, $ thus eliminating the contribution from the charge of the black hole.

This special form of the first law of thermodynamics is a consequence of the choice of the Killing vectors associated to mass and angular momentum in \cite{Banados:2005da}, since all the contributions coming from the charge are eliminated.

A direct match with the mass \emph{M} and angular momentum \emph{J} coming from the thermodynamic analysis in $ (t,r,\theta) $ coordinates gives:
\beq
\mu=\frac{M}{l^2} \, , \qquad \mathcal{J}=-\frac{4 J}{ \omega l^2} \, .
\eeq
In order to get the conserved charges associated to isometries in $ (t, r, \theta) $ coordinates, we need to adjust the normalization conditions:
\begin{itemize}
\item
The angular range $0 \leq \theta \leq 2 \pi $
corresponds to $0 \leq \tilde{\theta} \leq 2 \pi \frac{\sqrt{\omega l^3}}{2} $,
so extensive quantities, such as  mass, entropy and angular momentum
 in $ (t, r, \theta) $ coordinates get an extra  $\frac{\sqrt{\omega l^3}}{2}$ factor if we want to preserve the length of the integration along $ [0, 2 \pi] . $
\item Killing vectors are transformed as:
\beq
\frac{\p}{\p t} =\sqrt{\frac{l^3}{\omega}} \frac{\p}{\p \tilde{t} }  \, , \qquad
\frac{\p}{\p \theta} =\frac{\sqrt{\omega l^3}}{2 } \frac{\p}{\p \tilde{\theta} }  \, .
\eeq
\item In \cite{Banados:2005da} it is defined $ \Omega = - h(r_{+})/q(r_+) , $ 
while in eq. (\ref{M guess}),(\ref{J guess}) we followed the conventions of \cite{Anninos:2008fx}, where an additional factor of \emph{l} is put in the denominator both for the angular velocity and the Hawking temperature.
Choosing the last normalization amounts to modify $ \mu \rightarrow \mu/l , $ with the other conserved charges of the black hole unchanged.
\end{itemize}
Taking into account all these corrections, we get that the mass in $ (t,r, \theta) $ coordinates with the  Killing $\frac{\p}{\p t}$  is $M/2$ and the angular momentum  associated to the Killing $ -\frac{\p}{\p \theta}$ is   $J$. 
The $1/2$ factor in the normalization of the mass is reminiscent of Komar's anomalous factor and it is also pointed out for similar computations in \cite{Bouchareb:2007yx}.


\section{Another way to compute the asymptotic growth of action for WAdS black holes}
\label{another-way}

The asymptotic growth of the action of the WDW patch computed in section \ref{sect-Complexity=Action} can be derived in a different way following the procedure introduced in \cite{Brown:2015lvg}.
This is also a cross-check of our calculation.

 \begin{figure}[h]
\begin{center}
\includegraphics[scale=0.7]{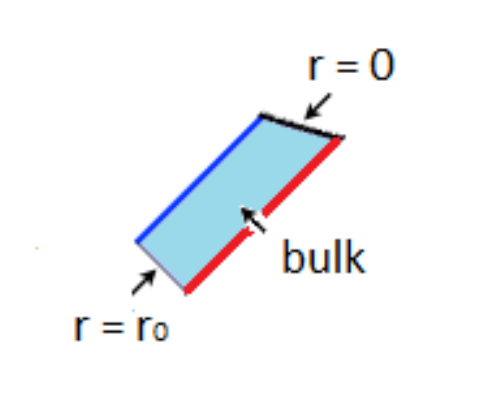}
\end{center}
\caption{Asymptotic contributions for the non-rotating case. In this picture we called the horizon radius $r_0 .$}
\label{SUS1}
\end{figure}

{\bf Non-rotating case:}
Following the argument in \cite{Brown:2015lvg}, the only relevant region of the WDW patch at late times is included between the horizon and the future singularity, as shown in fig. \ref{SUS1}.
The time derivative of the gravitational action evaluated in this region contains three contributions:
\begin{itemize}
\item The time derivative of the bulk contribution is given by eq. (\ref{bulk-late-time-non-rot}).
\item The time derivative of the GHY term nearby the singularity 
is given by eq. (\ref{GHY-late-time}).
\item The contribution from the joint at $r=r_m$ is replaced 
 by the GHY term nearby the horizon:
 \beq
\Delta I_{\rm GHY}^{r_h}  = 
\frac{(\nu^2 +3) l}{16 G}  \Delta t_b    \left[ 2r - r_h  \right]_{r=r_h} \, ,
\eeq
 which in the asymptotic limit gives the same contribution as the null joint.
\end{itemize}
In this way, summing these terms, we find the same result of eq. (\ref{non-rotating late time action growth}).

 \begin{figure}[h]
\begin{center}
\includegraphics[scale=0.7]{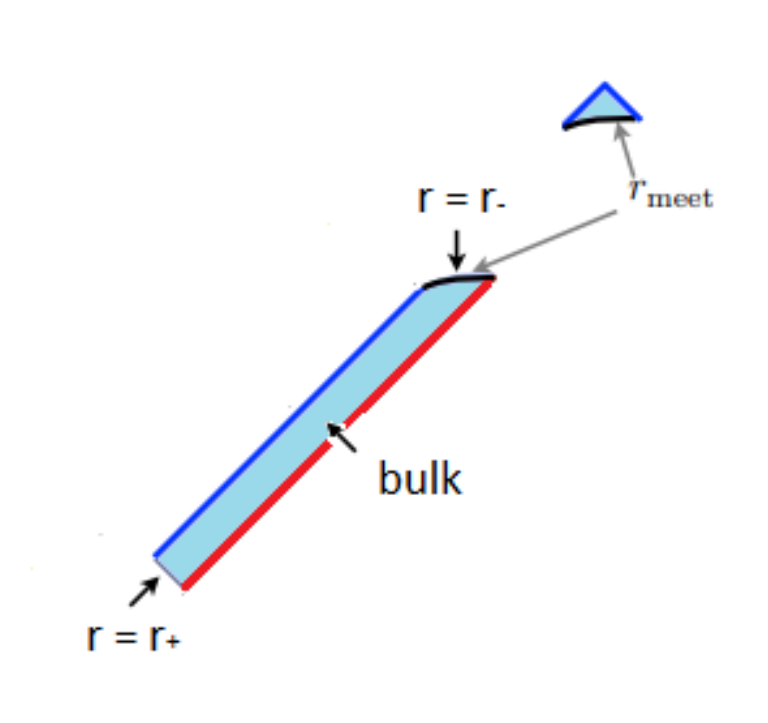}
\end{center}
\caption{Asymptotic contributions for the rotating case.}
\label{SUS2}
\end{figure}

{\bf Rotating case:} 
The region is depicted in figure \ref{SUS2}: in this case we need only the part of the WDW patch included between the inner and outer horizons. We consider the various contributions comparing with the computation in section \ref{sect-Complexity=Action}:
\begin{itemize}
\item The bulk contribution is still given by eq. (\ref{bulk-late-time-rot}).
\item The two null joints contributions are replaced by the GHY term evaluated
 on two constant-$r$ surfaces, one at $r \approx r_-$ and one at $r \approx r_+$.
 The induced metric on these constant-$r$ surfaces is:
 \beq
h_{ij} = l^2 \begin{pmatrix}
1 & \nu r -\frac12 \sqrt{(3+\nu^2) r_+ r_-} \\
\nu r -\frac12 \sqrt{(3+\nu^2) r_+ r_-} & \frac{r}{4} \Psi(r)
\end{pmatrix} \, , 
\eeq
\beq
\sqrt{h} = \frac{l^2}{2} \sqrt{(\nu^2 +3) (r_+-r) (r-r_-)} \, .
\eeq
 The normal vector to these slices is
\beq
n^{\mu} = \le 0 \, , - \frac{1}{l} \sqrt{(\nu^2 +3)(r_+  -  r)( r- r_- )} \, , 0 \ri \, ,
\qquad n^\a n_\a = -1 \, ,
\eeq
and the extrinsic curvature is
\beq
K = \frac{  \sqrt{\nu^2 +3} }{2 l} \,  \frac{2r - r_+ - r_-}{\sqrt{  (r_+  -  r)( r- r_- )  } } \, .
\eeq
 The GHY term nearby the inner horizon gives:
 \beq
 \frac{dI^{r_-}_{\rm GHY}}{d t_b} =-\frac{l}{4} \sqrt{\nu^2+3} \left[ 2 r -r_+ -r_- \right]_{r=r_-} \, ,
 \eeq
 while the term from the outer horizon
  \beq
 \frac{dI^{r_+}_{\rm GHY}}{d t_b} =\frac{l}{4} \sqrt{\nu^2+3} \left[ 2 r -r_+ -r_- \right]_{r=r_+} \, .
 \eeq
 These two contributions give the same result as the asymptotic contributions from the joints.
\end{itemize}
In this way, we find again a match with the late time result of eq. (\ref{late time action rate}).

\section{Divergence structure of the subregion complexity for WAdS black holes (non-rotating case)}
\label{appe-non-rot}

In this appendix we consider in detail the non-rotating case for the subregion complexity of asymptotically $\mathrm{WAdS}_3$ black holes considered in section \ref{sect-Subregion Complexity=Action} with $r_+=r_h$ and $r_-=0,$ and we check that the divergences
of complexity reproduce the appropriate limit from the rotating case.

\subsection{Total action}
\label{appe-total}
We recover the expression of the bulk action from eq. (\ref{bulk action CA conjecture}):
\beq
\begin{aligned}
I_{\mathcal{V}}^{\mathrm{tot}} = \frac{\mathcal{I}}{2G} \int_{0}^{\Lambda} dr \, (r^*_{\Lambda}- r^* (r)) = - \frac{l}{4G} (\nu^2 +3) \Lambda  r^*_{\Lambda} + \frac{l}{4G} (\nu^2 +3) \int_{_0}^{\Lambda} dr \, r^* (r)  \, .
\label{total bulk action before critical time}
\end{aligned}
\eeq
The GHY term can be recovered from eq. (\ref{GHY general})
\beq
I_{\mathcal{B}}= - \frac{(\nu^2 +3)l}{4G} (2 \varepsilon_0 - r_h) (r^*_{\Lambda} - r^*(\varepsilon_0)) = \frac{(\nu^2 +3)l}{4G} r_h \, 
\le r^*_{\Lambda} - r^* (0) \ri  \, , 
\eeq
where in the last step we performed the limit $ \varepsilon_0 \rightarrow 0  $ involving the IR cutoff.
The expression is divergent after sending $ \Lambda \rightarrow \infty $ due to the behaviour at infinity of the tortoise coordinate.

At $t_b=0$, the joints of the WDW patch are located at both the IR and UV cutoffs.
The former vanish as already observed, while the latter give a non-vanishing expression.
If we conventionally decide to take the flow of time in the bulk as increasing when going upwards, these joints take a negative sign $ \eta=-1 $ in eq. (\ref{jnn}) and we obtain
\beq
I_{\mathcal{J}} = 2 \times \frac{l}{4 G} \sqrt{\frac{\Lambda}{4} \Psi (\Lambda)} \log \left|\frac{l^2}{A^2} \frac{f(\Lambda)}{2 R (\Lambda)}  \right| =
\frac{l}{4 G} \sqrt{\Lambda \Psi (\Lambda)} \log \left|\frac{l^2}{A^2} \frac{(\nu^2 +3)(\Lambda - r_h)}{\Psi (\Lambda)}  \right|
 \, .
\eeq 
Finally, we have to add the counterterm which renders the action diffeomorphism-invariant:
\beq
I_{\rm ct}  = 4 \times
\frac{l}{4 G} \int_{\varepsilon_0}^{\Lambda} dr \, \frac{6(\nu^2 -1)r + (\nu^2 +3)r_h}{4 \sqrt{ r \Psi (r)}} \, \log \left| \frac{A \tilde{L}}{2 l^2} \frac{6(\nu^2 -1)r + (\nu^2 +3) r_h}{\sqrt{r \Psi (r)}}  \right| \, .
\eeq
The integration can be done analytically and we can also perform  the usual limit $ \varepsilon_0 \rightarrow 0 $ (where it was not evaluated yet), finding
\begin{align}
I_{\mathrm{ct}}=\frac{l}{4G}&\left[\frac{2(\nu^2+3)r_h}{\sqrt{3(\nu^2-1)}} \arctan \left(\frac{\sqrt{3(\nu^2-1)\Lambda}}{\sqrt{(\nu^2+3)r_h+3(\nu^2-1)\Lambda}}\right) \right. \notag\\&\left.-\sqrt{\Lambda \Psi (\Lambda)}\log\left|\frac{4l^4}{A^2\tilde{L}^2}\frac{\Lambda \Psi (\Lambda)}{((\nu^2+3)r_h+6(\nu^2-1)\Lambda)^2}\right| \right] \, .
\end{align}
Putting all these results together we obtain the expression for the total action in the WDW patch
\beq
\begin{aligned}
I^{\rm tot}  &=   \frac{l}{4G} (\nu^2 +3) \int_{0}^{\Lambda} dr \, r^* (r) 
- \frac{l}{4G} (\nu^2 +3) \Lambda  r^*_{\Lambda} + \frac{(\nu^2 +3)l}{4G} r_h \, \le r^*_{\Lambda} - r^* (0) \ri \\
& + \frac{l}{2G} \frac{(\nu^2 +3)r_h}{\sqrt{3(\nu^2 -1)}} \arctan \le \frac{\sqrt{3(\nu^2 -1)\Lambda}}{\sqrt{(\nu^2 +3)r_h + 3 (\nu^2 -1)\Lambda}} \ri \\
& + \frac{l}{4G} \sqrt{\Lambda \Psi (\Lambda)} \log \left| \frac{\tilde{L}^2}{4 l^2} \frac{(\nu^2 +3)(\Lambda-r_h)\left[(\nu^2 +3)r_h +6(\nu^2 -1)\Lambda \right]^2}{\Lambda \Psi^2(\Lambda)} \right| \, .
\label{total action before t_C}
\end{aligned}
\eeq

The divergent parts of the total complexity are:
\beq
I^{\rm tot} = 
\frac{l}{4G} 
\sqrt{3(\nu^2 -1)} \Lambda \left( \log \left| \frac{\tilde{L}^2}{l^2} (\nu^2 +3) \right| -1  \right) 
- \frac{l}{8G} \frac{\nu^2 +3}{\sqrt{3(\nu^2 -1)}} \, r_h \log \Lambda
 + \mathcal{O} (\Lambda^0) \, . 
 \label{UV behaviour total action}
\eeq
This reproduces eq. (\ref{azione-totale}) in the $r_- \rightarrow 0$ limit.

\subsection{External action}
\label{appe-exte}

The bulk and the counterterm action can be obtained in the same way as in the previous section \ref{appe-total}:
\beq
I^{\rm out}_{\mathcal{V}} = - \frac{l}{4G} \sqrt{3(\nu^2 -1)} \Lambda - \frac{l}{8G} \frac{7 \nu^2 -3}{\sqrt{3(\nu^2 -1)}} \, r_h \log \Lambda + \frac{l}{4G} (\nu^2 +3) r_h \, r^*_{\Lambda}
+ \mathcal{O} (\Lambda^0) \, , 
\eeq
\beq
I^{\rm out}_{\rm ct} = - \frac{l}{4G} \sqrt{\Lambda \psi (\Lambda)} \log \left| \frac{4l^4}{A^2 \tilde{L}^2}  \frac{\Lambda \Psi(\Lambda)}{\left[ 6(\nu^2 -1)\Lambda + (\nu^2 +3) r_h \right]^2} \right| + \mathcal{O} (\Lambda^0)
 \, .
\eeq
There is no spacelike or timelike boundary, then  there is no contribution from the GHY term.

As in the rotating case, we need to be careful
 with the regularization of the joints at the horizon; we use again 
 the same method as in \cite{Agon:2018zso}.
From (\ref{jjoints}) in this situation, we find
\begin{align}
I^{\mathrm{out}}_{\mathcal{J}} =  & - \frac{ l}{4 G} \sqrt{r_h \Psi(r_h)}  \left[-\log \left|\frac{l^2}{A^2} \frac{f(r_{\epsilon_U,\epsilon_V})}{2R(r_h)}  \right|+\log \left|\frac{l^2}{A^2} \frac{f(r_{U_0, \epsilon_V})}{2R(r_h)}  \right|
+\log \left|\frac{l^2}{A^2} \frac{f(r_{\epsilon_U, V_0})}{2R(r_h)}  \right| \right] \notag\\& 
+ \frac{l}{4 G} \sqrt{\Lambda
\Psi(\Lambda)}\log \left| \frac{l^2}{A^2} \frac{f(\Lambda))}{2R(\Lambda)} \right| \, .
\end{align}
In this case it is convenient to add and subtract the joint term $\frac{ l}{2 G} \nu r_h \log \left|\frac{l^2}{A^2} \frac{f(r_{\epsilon_U,\epsilon_V})}{2\nu r_h}  \right|$ and to use the relation (\ref{difference of logarithms at horizon}) to get
\beq
\begin{aligned}
I^{\mathrm{out}}_{\mathcal{J}} = &  -\frac{l}{2 G} \nu r_h \left[ \log(U_0 V_0)+\log \left|\frac{l^2}{A^2} \frac{f(r_{\epsilon_U,\epsilon_V})}{2\nu r_h}  \right|-\log(\epsilon_U\epsilon_V)  \right]  \\ & + \frac{l}{4 G} \sqrt{\Lambda
\Psi(\Lambda)}\log \left| \frac{l^2}{A^2} \frac{(\nu^2 +3)(\Lambda - r_h)}{\Psi(\Lambda)} \right|  \, .
\end{aligned}
\eeq
Finally, the expression simplifies by means of eqs. (\ref{log UV coordinates}) and (\ref{logarithm at the horizon}):
\begin{align}
I^{\mathrm{out}}_{\mathcal{J}} = &-\frac{ l}{2 G}\left[ \nu r_h \left(\frac{\nu^2 +3}{2 \nu} r^{*}_{\Lambda} +F(r_h)  \right) - \frac12 \sqrt{\Lambda
\Psi(\Lambda)}\log \left| \frac{l^2}{A^2} \frac{(\nu^2 +3)(\Lambda - r_h)}{\Psi(\Lambda)} \right| \right] \, .
\end{align}
The function $ F(r) $, which can be obtained from eq. (\ref{BIGF}), is finite
 and it is not needed to find the divergence structure.
Adding all the terms outside the horizon, we finally obtain
\beq
I^{\rm out}  =  \frac{l}{4G} \sqrt{3(\nu^2 -1)} \Lambda  \le \log \left| \frac{\tilde{L}^2}{ l^2} (\nu^2 +3)  \right| -1 \ri 
 - \frac{l}{8G} \frac{7 \nu^2 -3}{\sqrt{3(\nu^2 -1)}} \, r_h \log \Lambda + \mathcal{O} (\Lambda^0) \, .
 \label{final outside action}
\eeq
This results reproduces eq. (\ref{external-action})  in the $r_- \rightarrow 0$ limit.

\subsection{Volume} 
\label{appe-volume}

The non-rotating case of the subregion volume computation has to match with the limit $r_- \rightarrow 0$ of the computation in section \ref{sect-Subregion Complexity=Volume}. 
The volume of the extremal slice anchored at the boundary and bounded by the RT surface is given by the induced metric computed from the non-rotating metric
\beq
V (L)= \int_0^{2 \pi} d \theta \int_{r_h}^{\Lambda} dr \, \sqrt{h} =
2 \pi l^2 \int_{r_h}^{\Lambda} dr \,
\sqrt{\frac{3(\nu^2 -1)r + (\nu^2 +3)r_h}{4 (\nu^2 +3)(r-r_h)}} \, .
\eeq
We introduce the convenient coordinate parametrization $ R= r/r_h $ and we obtain
\beq
V (L)= 2 \pi l^2 r_h \int_{1}^{\Lambda/r_h} dR \, \sqrt{\frac{3(\nu^2 -1)R + (\nu^2 +3)}{4 (\nu^2 +3)(R-1)}} \, .
\eeq
This expression can be analytically solved, giving a primitive function
\beq
\begin{aligned}
V (L) & = 2 \pi l^2 r_h \left[ \sqrt{\frac{(\nu^2 +3)+ 3R(\nu^2 -1)}{\nu^2 +3}} \frac{\sqrt{R-1}}{2}  + \right. \\
& \left.  + \frac{2 \nu^2 \log\le \frac{\sqrt{3 (\nu^2 -1)(R-1)} + \sqrt{3+ \nu^2 + 3R(\nu^2 -1)}}{2 \nu}  \ri }{\sqrt{3 (\nu^2 -1)(\nu^2 +3)}}  \right]_{R=1}^{R=\Lambda/r_h} \, ,
\end{aligned}
\eeq
and consequently the result
\beq
\begin{aligned}
V (L) & = \pi l^2 \sqrt{\frac{3(\nu^2 -1)}{\nu^2 +3}} \Lambda +
\frac{2 \pi l^2 \nu^2 r_h}{\sqrt{3(\nu^2 -1)(\nu^2 +3)}} \, \log \le \frac{\Lambda}{r_h} \ri + \\
& + \pi l^2 r_h \frac{(3- \nu^2) + 2 \nu^2 \log \left[ \frac{3(\nu^2 -1)}{\nu^2} \right]}{2 \sqrt{3(\nu^2 -1)(\nu^2 +3)}} + \mathcal{O} (\Lambda^{-1}) \, . 
\end{aligned}
\eeq
The divergent parts of this expression reproduce eq. (\ref{divergenza-volume}) in the $r_- \rightarrow 0$ limit, as expected.


\section{Subsystem complexity and temperature}
\label{appe-temperature}

In this Appendix we give the details for the computation of the temperature dependence of subregion complexity for the $\mathrm{WAdS}$ black holes as given in section \ref{sect-Temperature behaviour}.
Let us compute the temperature dependence of $M$ at constant $J$, which is the specific heat at constant $J$:
\beq
C_J=\frac{\p M}{\p T} \Big|_J=\frac{\p M}{\p r_+} \frac{\p r_+}{\p T} + 
\frac{\p M}{\p r_-} \frac{\p r_-}{\p T} \, .
\eeq
The quantities $\frac{\p r_+}{\p T}$ and  $\frac{\p r_-}{\p T}$
can be computed from the inverse of the matrix 
\beq
\begin{pmatrix}
 \frac{\p T}{\p r_+}&  \frac{\p T}{\p r_-} \\
 \frac{\p J}{\p r_+}&  \frac{\p J}{\p r_-}
\end{pmatrix} \, ,
\eeq
 which can be directly calculated from eqs. (\ref{T,Omega}) and (\ref{J guess}).
 This gives (here we define $\epsilon=r_-/r_+ $ ):
\beq
 C_J=\frac{\pi l r_+}{4 G} \, 
 \frac{ \nu  (\epsilon -1) \left(\epsilon  \left(-3 \nu ^2+2 \nu 
   \sqrt{\left(\nu ^2+3\right) \epsilon }+3\right)-2 \nu  \sqrt{\left(\nu ^2+3\right)
   \epsilon }\right)}{ \epsilon  \left(\nu ^2 (4 \epsilon -1)-3\right)} \, .
\eeq
The quantity $C_J$ is negative for
 $ 0< \epsilon< \frac{\nu^2+3}{4 \nu^2}$ and positive for $   \frac{\nu^2+3}{4 \nu^2} < \epsilon <1$.
For $\epsilon=0$  and $\epsilon=\frac{\nu^2+3}{4 \nu^2}$, $C_J$ is diverging and there is a second order phase transition,
similar to the one which occurs for Kerr and Reissner-Nordstr\"om black holes in flat spacetime
\cite{Davies:1978mf}.

With a similar method, one can compute the temperature dependence of
 $K_+$ and $K_-$. The result is:
\beq
\frac{\p K_+}{\p T} \Big|_J=\frac{\hat{a}}{\hat{b}} \, , \qquad \frac{\p K_-}{\p T} \Big|_J=\frac{\hat{c}}{\hat{b}} \, ,
\eeq
where
\bea
\hat{a}&=&2 \pi  l r_+ \left(\sqrt{\left(\nu ^2+3\right) r_+^2 \epsilon }-2 \nu  r_+\right){}^2
   \left(\nu  \left(\nu ^2 ((\epsilon -18) \epsilon -7) 
      +3 \epsilon  (\epsilon
   +6)+3\right) 
   \right.
  \nl && \left. 
\sqrt{\left(\nu ^2+3\right) r_+^2 \epsilon }-r_+ \epsilon  \left(-31 \nu
   ^4+6 \nu ^2+\left(\nu ^2+3\right)^2 \epsilon +9\right)\right) \, ,
\eea
\bea
\hat{b}&=&3 \left(\nu ^2-1\right) \sqrt{\left(\nu ^2+3\right) r_+^2 \epsilon } \left(4 \nu 
   \sqrt{\left(\nu ^2+3\right) r_+^2 \epsilon }+\left(\nu ^2+3\right) r_+ (-\epsilon
   -1)\right) \nl
  &&  \left(2 \nu  (\epsilon +1) \sqrt{\left(\nu ^2+3\right) r_+^2 \epsilon
   }-\left(5 \nu ^2+3\right) r_+ \epsilon \right) \, ,
\eea
\bea
\hat{c}&=&2 \pi  l r_+ \left(\sqrt{\left(\nu ^2+3\right) r_+^2 \epsilon }-2 \nu  r_+\right){}^2
   \left(\nu  \left(\nu ^2 (\epsilon  (7 \epsilon +18)-1) 
     -3 (\epsilon  (\epsilon
   +6)+1)\right)
   \right. 
\nl
&&   \left. 
 \sqrt{\left(\nu ^2+3\right) r_+^2 \epsilon }
+r_+ \epsilon 
   \left(\left(\nu ^2+3\right)^2+\left(-31 \nu ^4+6 \nu ^2+9\right) \epsilon
   \right)\right) \, .
\eea


\section{Another regularization for the subregion action of one segment in the BTZ background}
\label{app:other-reg}

\begin{figure}[h]
\center
\begin{tabular}{cc}
\includegraphics[scale=0.5]{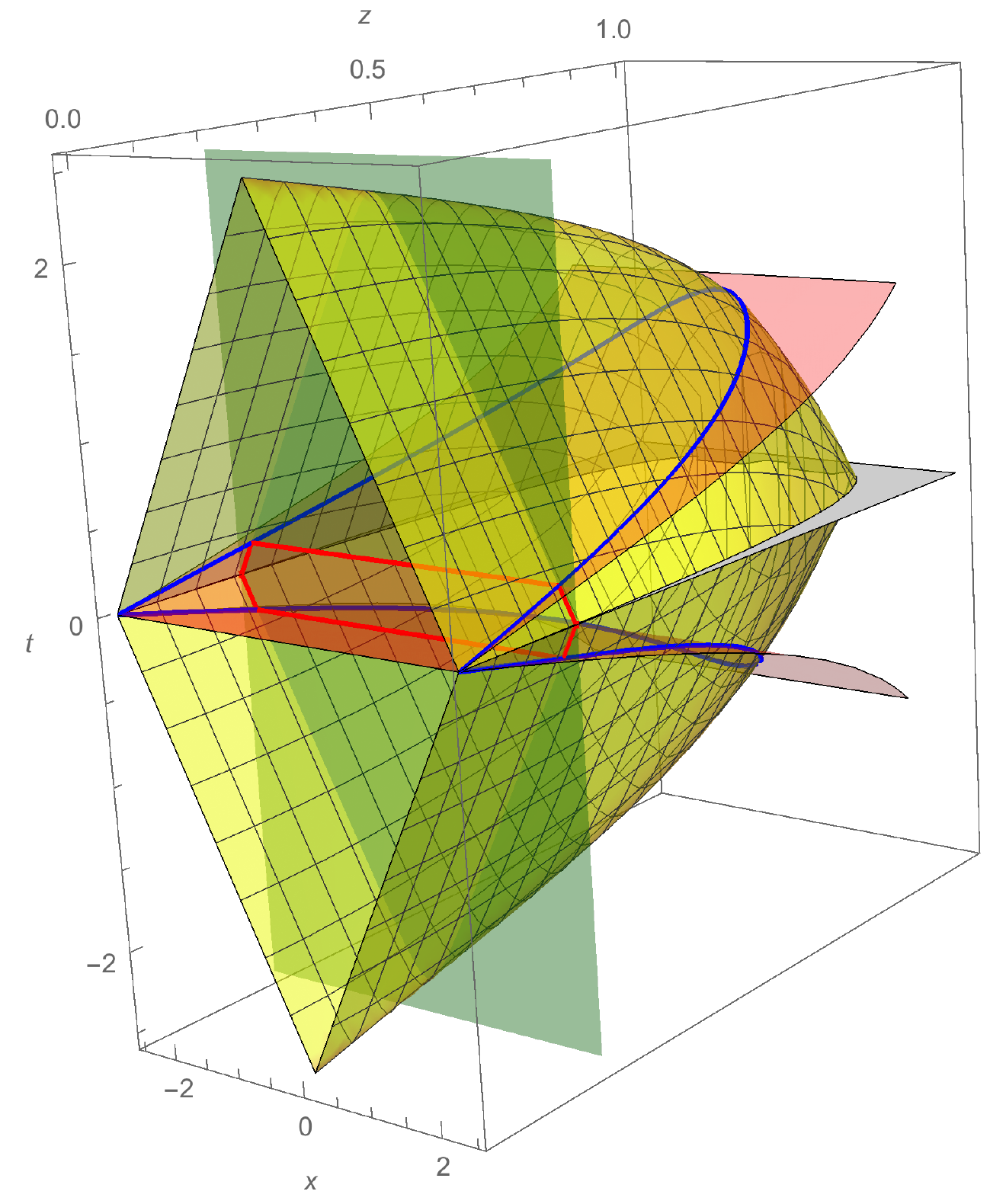} & \includegraphics[scale=0.55]{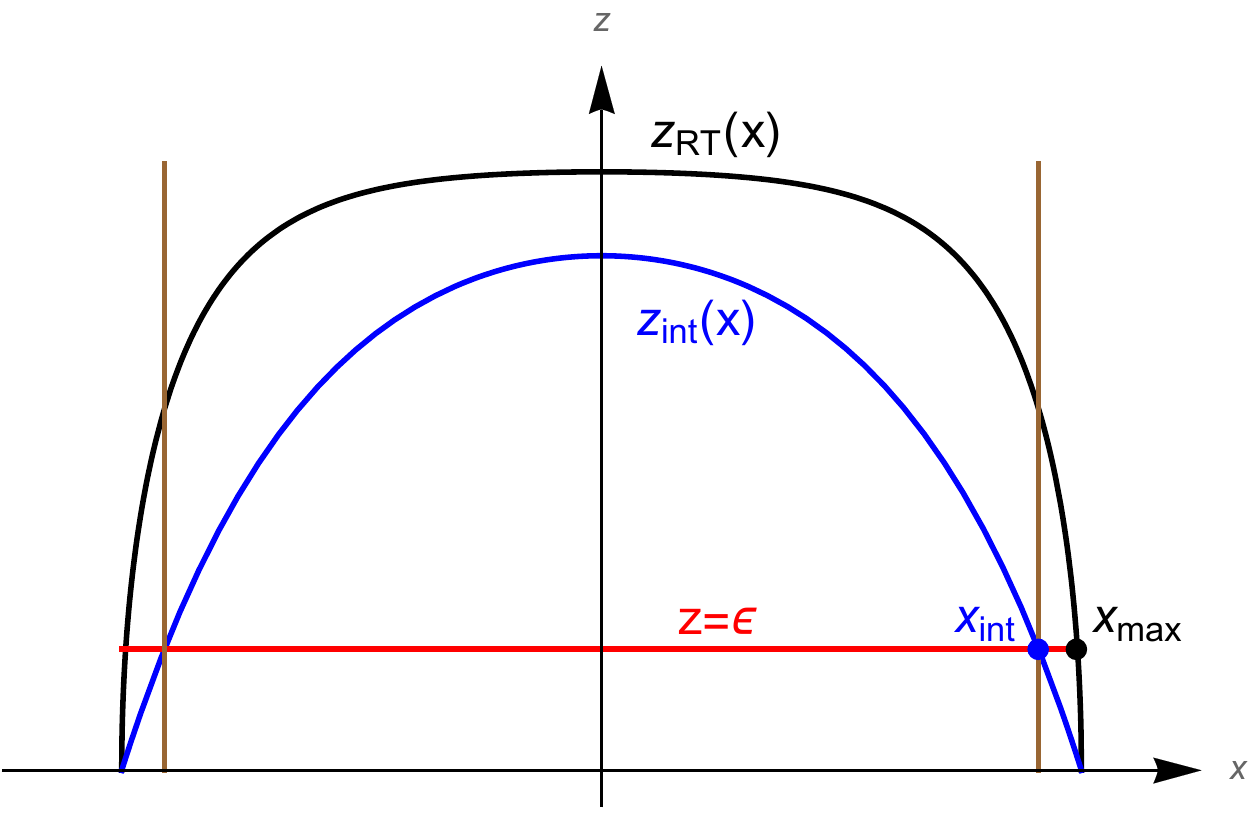}
\end{tabular}
\caption{ Another regularization for the BTZ case.}
\label{fig-BTZ-another}
\end{figure}

 In this Appendix we follow the prescription $A$ introduced in section \ref{sect:AdS} to regularize
  the action, where the null boundaries of the WDW patch are sent from
   the true boundary $z=0$ and we add a timelike cutoff surface
    at $z=\varepsilon$ cutting the bulk structure we integrate over. 
   The geometry of the region is shown in figure \ref{fig-BTZ-another}. 
 
The geometric data are slightly different than the ones introduced in Section \ref{sect:BTZ}.
The RT surface and the corresponding entanglement wedge are the same, see eqs. (\ref{RT surface}) and (\ref{twedge}), but
the WDW patch starts from the true boundary $z=0$ and then the null lines which delimit it are parametrized by
\beq
t_{\rm WDW} = \pm \frac{z_h}{4} \log \le \frac{z_h + z}{z_h - z} \ri^2 \, ,
\eeq
where $\pm$ refers to positive and negative times, respectively.
The intersection curve between the WDW patch and the entanglement wedge is given in this case by
\beq
z_{\rm int} = \coth \le \frac{l}{2 z_h} \ri - \cosh \le \frac{x}{z_h} \ri  \mathrm{csch}  \le \frac{l}{2 z_h} \ri \, .
\label{zint 1st regularization}
\eeq
The null normals to the boundaries of the WDW patch and the entanglement wedge are unchanged. 

Unlike the case of the other regularization, the intersection curve and the
 RT surface do not meet at $z=\varepsilon,$ but at the true boundary $z=0.$
For this region, there are no codimension-three joints. 
The intersection curve between the boundaries of the WDW patch and the entanglement wedge
meets the cutoff surface at:
\beq
x_{\rm int}= \mathrm{arccosh} \, \left[ \cosh \le \frac{l}{2 z_h} \ri - \frac{\varepsilon}{z_h} \sinh  \le \frac{l}{2 z_h} \ri \right] \, .
\eeq
This expression is found by inverting eq. (\ref{zint 1st regularization}) and imposing $z=\varepsilon.$
In the following sections we compute all the terms entering the gravitational action.

\subsection{Bulk contribution}
\label{Bulk contribution 1st regularization}

We split the contributions as follows
\begin{equation}
I_{\cal V} = 4 \left( I_{\cal V}^{1} + I_{\cal V}^{2} + I_{\cal V}^{3} \right) \, ,
\end{equation}
where
\bea
I_{\cal V}^{1} &=& - \frac{L}{4 \pi G} \int_{0}^{x_{\rm int} } dx \int_{\varepsilon}^{z_{\rm int}} dz \int_{0}^{t_{\rm WDW}} dt \, \frac{1}{z^{3}} \, ,
\nl
I_{\cal V}^{2} &=& -  \frac{L}{4 \pi G} \int_{0}^{x_{\rm int} } dx \int_{z_{\rm int}}^{z_{\rm RT}} dz \int_{0}^{t_{\rm EW}} dt \, \frac{1}{z^{3}} \, ,
\nl
I_{\cal V}^{3} &=& -  \frac{L}{4 \pi G} \int_{x_{\rm int}}^{{x_{\rm max}} } dx \int_{\varepsilon}^{z_{\rm RT}} dz \int_{0}^{t_{\rm EW}} dt \, \frac{1}{z^{3}} \, .
\eea
In this case the sum of bulk terms obtained by
 splitting the spacetime region with the intersection between the boundaries of the WDW 
 patch and the entanglement wedge does not give the entire bulk action.
We need to add  $I_{\cal V}^{3}$ which accounts 
for the region between the values $x_{\rm int}$ and $x_{\rm max}$ of the transverse coordinate.

A direct evaluation gives
\beq
\begin{aligned}
 I^1_{\cal V}  + I_{\cal V}^{2}  = \frac{L}{16 \pi G z_h} & \int_{0}^{x_{\rm int}(\varepsilon)} dx \, \left\lbrace \coth \le \frac{x}{z_h} \ri \log \left| \frac{\sinh \le \frac{l-2x}{2 z_h} \ri  \sinh^2 \left[ \frac{l+2x}{4 z_h} \right]}{\sinh \le \frac{l+2x}{2 z_h} \ri  \sinh^2 \left[ \frac{l-2x}{4 z_h} \right]}  \right| \right. \\
& \left.  +  \frac{2 \sinh \le \frac{l}{2 z_h} \ri}{\cosh \le \frac{l}{2 z_h} - \cosh \le \frac{x}{z_h} \ri \ri} - \frac{2 z_h}{\varepsilon}  + \le \frac{z_h^2}{\varepsilon^2} - 1 \ri \log \left| \frac{z_h - \varepsilon}{z_h + \varepsilon}  \right|
   \right\rbrace \, .
 \end{aligned}
\eeq
\beq
I^{3}_{\cal V}   = - \frac{L}{16 \pi G} \, .
\label{bulk 3 1st reg}
\eeq

\subsection{Gibbons-Hawking-York contribution}
The Gibbons-Hawking-York (GHY) surface term in the action for timelike and spacelike boundaries is
\begin{equation}
I_{GHY} = \frac{1}{8 \pi G} \int_{\partial \mathcal{B}} d^{2} x \sqrt{- \det  h_{\mu \nu}} \, K
\end{equation}
with $h_{\mu \nu}$ the induced metric on the boundary and $K$ the trace of the extrinsic curvature. 
The only contribution of this kind comes from the timelike regularizing surface at $z = \varepsilon$.

The GHY contribution is conveniently splitted into by two parts: the first one involves the WDW patch, while the second one involves the entanglement wedge
\begin{equation}
I_{\rm GHY}^{1} = \left[ \frac{L}{8 \pi G}  \int_{0}^{x_{\rm int} } dx \int_{0}^{t_{\rm WDW} } dt \, \le  \frac{2}{z^2} - \frac{1}{z_h^2} \ri \right]_{z = \varepsilon} = \frac{L}{8 \pi G} \, \frac{l}{\varepsilon} - \frac{L}{4 \pi G}  \, ,
\label{GHY sotto xint 1st reg}
\end{equation}
\begin{equation}
I_{\rm GHY}^{2} = \left[ \frac{L}{8 \pi G}  \int_{x_{\rm int} }^{x_{max} } dx \int_{0}^{t_{\rm EW} } dt \,  \le  \frac{2}{z^2} - \frac{1}{z_h^2} \ri \right]_{z = \varepsilon} = \frac{L}{8 \pi G} \, .
\label{GHY sopra xint 1st reg}
\end{equation}
The total GHY contribution is
\begin{equation}
\label{ghy1}
I_{\rm GHY} = 4 \left( I_{\rm GHY}^{1} + I_{\rm GHY}^{2} \right) = \frac{L}{2 \pi G} \left( \frac{l}{\varepsilon} - 1 \right) \, .
\end{equation}

\subsection{Null boundaries counterterms}

The details of calculation are very similar to the ones in section \ref{sect-BTZ-countertems}.
The contribution in eq. (\ref{null-bou}) and  the counterterm on the boundary of entanglement
wedge again vanish. The counterterm on the boundary of the WDW patch gives:
\beq
\begin{aligned}
I_{\rm ct}^{\rm WDW} & = - \frac{L}{2 \pi G} \int_0^{x_{\rm int} } dx \int_{\varepsilon}^{z_{\rm int} } dz \, \frac{1}{z^2} \log \left| \frac{\tilde{L}}{L^2} \, \alpha z \right| = \\
& =  \frac{L}{2 \pi G} \int_0^{x_{\rm max}} dx  \, \left\lbrace \frac{1 + \log \left| \frac{\tilde{L}}{L^2} \, \alpha \varepsilon \right|}{\varepsilon} + \frac{\sinh \le \frac{l}{2 z_h}  \ri}{z_h \left[ \cosh \le \frac{x}{z_h} \ri - \cosh \le \frac{l}{2 z_h} \ri   \right]}  \times \right. \\
& \left. \times \le 1 + \log \left| \frac{\tilde{L} z_h \alpha}{L^2} \frac{\cosh \le \frac{l}{2 z_h} \ri - \cosh \le \frac{x}{z_h} \ri}{\cosh \le \frac{l}{2 z_h}  \ri} \right|  \ri  \right\rbrace \, .
\end{aligned}
\eeq

\subsection{Joint terms}

The joint contribution to the gravitational action coming from a codimension-two 
surface given by the intersection of a codimension-one null surface and a codimension-one timelike (or spacelike) surface is 
\begin{equation}
\label{null-time joint}
I_{\mathcal{J}} = \frac{\eta}{8 \pi G} \int_{\mathcal{J}} dx \, \sqrt{\sigma} \log \left| \mathbf{k} \cdot \mathbf{n} \right| \, ,
\end{equation}
where $\sigma$ is the induced metric determinant on the codimension-two surface and $\mathbf{n}$ and $\mathbf{k}$ are the outward-directed normals to the timelike (or spacelike) surface and the null one respectively. Moreover, the sign in front of the expression can be determined with the rule
\begin{equation}
\label{sign null-time}
\eta = - \, \mathrm{sign} \left( \mathbf{k} \cdot \mathbf{n} \right) \, \mathrm{sign} \left( \mathbf{k} \cdot \hat{t} \right) 
\end{equation} 
in which $\hat{t}$ is the auxiliary unit vector in the tangent space of the boundary region, 
orthogonal to the joint and outward-directed from the region of interest \cite{Carmi:2016wjl}.

The unit normal vector $n^{\mu}$ to the $z = \varepsilon $ surface is
\begin{equation}
\label{nvector}
n^{\mu} = \left( 0, \, - \frac{z}{L} \sqrt{f(z)}, \, 0 \right) 
\end{equation}
where the sign must be chosen so that the vector is outward-directed from the region of interest. 

The joints give the following contributions:
\begin{itemize}
\item
The joint involving the WDW patch boundary and the cutoff surface:
\begin{equation}
I_{\mathcal{J}}^{\rm cutoff1} = - \, \frac{L}{2 \pi G} \int_{0}^{x_{\rm int} } \frac{dx}{\varepsilon} \log \left( \frac{\alpha \, \varepsilon}{L \sqrt{f(\varepsilon)}} \right) =
- \frac{L}{4 \pi G} \frac{l}{\varepsilon} \log \le \frac{L}{\alpha \varepsilon} \ri - \frac{L}{2 \pi G} \log \le \frac{L}{\alpha \varepsilon} \ri \, .
\label{joint cutoff 1 sotto xint 1st reg}
\end{equation}
\item
The joint involving the cutoff surface and the entanglement wedge boundary:
\begin{equation}
I_{\mathcal{J}}^{\rm cutoff2} = \mathcal{O} \left( \varepsilon \log \varepsilon \right) \, .
\label{joint cutoff 1 sopra xint 1st reg}
\end{equation}
\item The null-null joint contribution coming from the RT surface 
is  the same as in the previous regularization, see eq. (\ref{joint-BTZ-RT}).
\item
The joints coming from the intersection between the null boundaries of the 
WDW patch and the ones of the entanglement wedge 
give a similar contribution  as in eq. (\ref{joint-BTZ-EW}),
The main difference is that the integral is in  the range $ [0, x_{\rm int} (\varepsilon)] $ 
and the intersection is slightly different, because the WDW patch starts from $z=0$ in the present regularization:
\begin{equation}
I_{\mathcal{J}}^{\rm int} =  \frac{L}{2 \pi G z_h } \int_{0}^{x_{\rm int }} dx \, 
\frac{\sinh \le  \frac{l}{2 z_h} \ri}{\cosh \le \frac{l}{2 z_h} \ri - \cosh \le \frac{x}{z_h} \ri } \,
\log \left| \frac{\alpha \beta z_h^2}{2 L^2} \frac{\left(\cosh \le \frac{l}{2 z_h} \ri - \cosh \le \frac{x}{z_h} \ri \right)^2}{\cosh \le \frac{x}{z_h} \ri \cosh \le \frac{l}{2 z_h} \ri -1}  \right| \, .
\end{equation}
\end{itemize}

\subsection{Complexity}
Adding all the contributions and performing the integrals we finally get
\beq
 \mathcal{C}_{A}^{\rm BTZ} = \frac{l}{\varepsilon} \frac{c}{6 \pi^2} \le 1 +\log \left(\frac{\tilde{L}}{L} \right) \ri
- \log  \left(\frac{2\tilde{L}}{L} \right) \frac{S^{\rm BTZ}}{\pi^2} - \frac{c}{3 \pi^2} \le \frac12 + \log \le \frac{\tilde{L}}{L} \ri \ri + \frac{1}{24} c \, .
\label{ACTION-C 1st reg}
\eeq
The difference with expression (\ref{ACTION-C}) consists 
only in the coefficient of the divergence $1/\varepsilon$ 
and in a finite piece proportional to the counterterm scale $\tilde{L}$ via a logarithm.

Recently other counterterms were proposed to give a universal behaviour
 of all the divergences of the action \cite{Akhavan:2019zax}.
In particular, with this regularization we need to insert a codimension-one 
boundary term at the cutoff surface:
\beq
I_{\rm ct}^{\rm cutoff} = - \frac{1}{16 \pi G} \int d^{d-1} x \, dt \, \sqrt{-h} \, 
\le \frac{2(d-1)}{L} + \frac{L}{d-2} \tilde{R} \ri  \, ,
\label{akhavan}
\eeq
being $\tilde{R}$ the Ricci scalar on the codimension-one surface.
Adding the extra counterterm in eq. (\ref{akhavan}), we find
\beq
  \mathcal{C}_{A}^{\rm BTZ} = \frac{l}{\varepsilon} \frac{c}{6 \pi^2} \log \left(\frac{\tilde{L}}{L} \right) 
- \log  \left(\frac{2\tilde{L}}{L} \right) \frac{S^{\rm BTZ}}{\pi^2} - \frac{c}{3 \pi^2} \log \le \frac{\tilde{L}}{L} \ri  + \frac{1}{24} c \, .
\label{ACTION-C 1st reg plu new counterterm}
\eeq
The numerical coefficient of all the divergences is the same as in eq. (\ref{ACTION-C}).
The two regularizations differ only by a finite piece dependent
 from the counterterm length scale $\tilde{L}.$

\end{appendices}


\bibliography{at}
\bibliographystyle{at}

\end{spacing}


\printthesisindex 

\end{document}

%% file: thesis-info.tex
\title{Developments in non-relativistic field theory and complexity}

\author{Stefano Baiguera}

\dept{Department of Physics}

\university{University of Milan-Bicocca}
\crest{\includegraphics[width=0.2\textwidth]{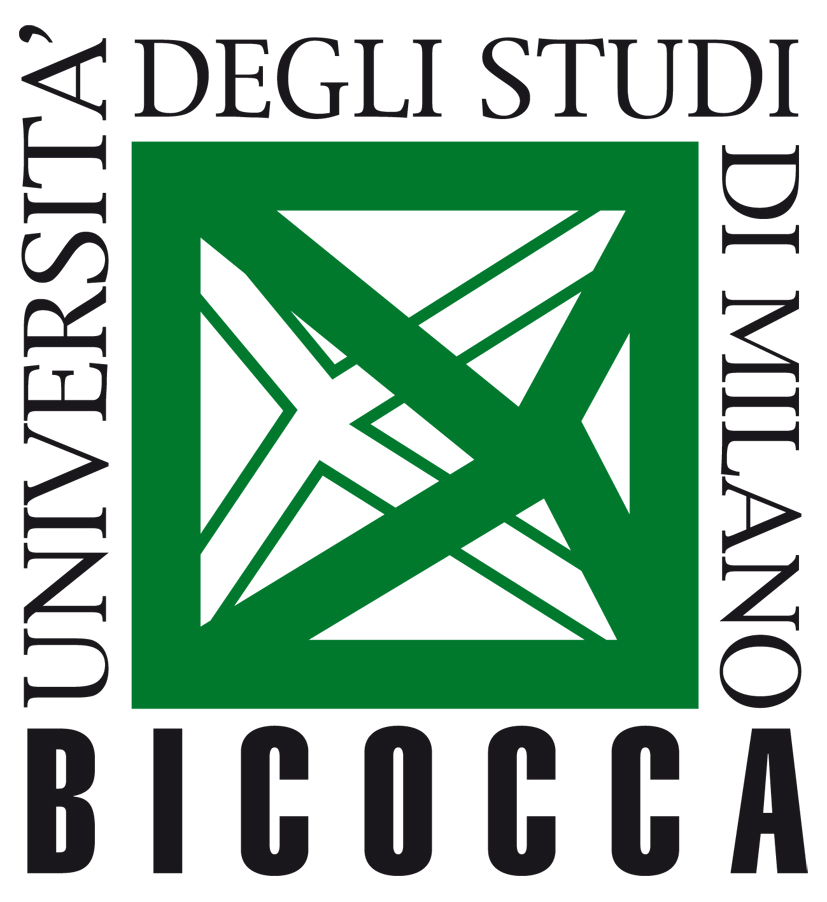}}

 \supervisor{Silvia Penati} \supervisorrole{Advisor: \,}
\advisor{Roberto Auzzi} \advisorrole{Co-advisor: \,}

\degreetitle{Doctor of Philosophy}

\college{PhD coordinator: Marta Calvi \par
Curriculum in Theoretical Physics
\par
Cycle XXXII \par
Matricola: 814546 \par
Academic year: 2018/2019
}

\degreedate{October 2019} 

\subject{} \keywords{{} {PhD Thesis} {Physics} {University of Milano-Bicocca}}